%% file: pap.tex
\documentclass[prd, twocolumn,superscriptaddress, nofootinbib, notitlepage, floatfix]{revtex4-1}
\usepackage{xcolor}
\usepackage{graphicx}
\usepackage{wrapfig}
\usepackage{amsmath}
\usepackage{amsfonts}
\usepackage{amssymb}
\usepackage{multirow}
\usepackage{slashed}
\usepackage{physics}
\usepackage[colorlinks=true, linkcolor=blue, citecolor=blue, urlcolor=blue]{hyperref}
\newcommand{\msbar}{{\overline{\mathrm{MS}}}}
\newcommand\befs{\begin{figure*}}
\newcommand\eefs[1]{\label{fig:#1}\end{figure*}}
\newcommand\bef{\begin{figure}}
\newcommand\eef[1]{\label{fig:#1}\end{figure}}
\newcommand\beq{\begin{equation}}
\newcommand\eeq[1]{\label{#1}\end{equation}}
\newcommand\beqa{\begin{eqnarray}}
\newcommand\eeqa[1]{\label{#1}\end{eqnarray}}
\newcommand\bet{\begin{table}}
\newcommand\eet[1]{\label{tb:#1}\end{table}}
\newcommand\bets{\begin{table*}}
\newcommand\eets[1]{\label{tb:#1}\end{table*}}
\newcommand\fgn[1]{Fig.\ \ref{fig:#1}}
\newcommand\eqn[1]{Eq.\ (\ref{#1})}
\newcommand\scn[1]{Section \ref{sec:#1}}
\newcommand\apx[1]{Appendix \ref{sec:#1}}
\newcommand\tbn[1]{Table \ref{tb:#1}}

\begin{document}
\widetext

\title{
Valence parton distribution of pion from lattice QCD: Approaching continuum
}
\author{Xiang Gao}
\email{xgao@bnl.gov}
\affiliation{Physics Department, Brookhaven National Laboratory, Upton, NY 11973, USA}
\affiliation{Physics Department, Tsinghua University, Beijing 100084, China}
\author{Luchang Jin}
\affiliation{Physics Department, University of Connecticut, Storrs, Connecticut 06269-3046, USA}
\affiliation{RIKEN-BNL Research Center, Brookhaven National Lab, Upton, NY, 11973, USA}
\author{Christos Kallidonis}
\affiliation{Department of Physics and Astronomy, Stony Brook University, Stony Brook, NY 11794, USA}
\affiliation{Department of Physics, College of William \& Mary, 300 Ukrop Way, Williamsburg, VA 23185, USA}
\author{Nikhil Karthik}
\email{nkarthik@bnl.gov}
\affiliation{Physics Department, Brookhaven National Laboratory, Upton, NY 11973, USA}
\author{Swagato Mukherjee}
\affiliation{Physics Department, Brookhaven National Laboratory, Upton, NY 11973, USA}
\author{Peter Petreczky}
\affiliation{Physics Department, Brookhaven National Laboratory, Upton, NY 11973, USA}
\author{Charles Shugert}
\affiliation{Physics Department, Brookhaven National Laboratory, Upton, NY 11973, USA}
\affiliation{Department of Physics and Astronomy, Stony Brook University, Stony Brook, NY 11794, USA}
\author{Sergey Syritsyn}
\affiliation{Department of Physics and Astronomy, Stony Brook University, Stony Brook, NY 11794, USA}
\affiliation{RIKEN-BNL Research Center, Brookhaven National Lab, Upton, NY, 11973, USA}
\author{Yong Zhao}
\affiliation{Physics Department, Brookhaven National Laboratory, Upton, NY 11973, USA}
\begin{abstract}
  We present a high-statistics lattice QCD determination of the
  valence parton distribution function (PDF) of the pion,  with a mass
  of \(300\)~MeV, using two very fine lattice spacings of \(a=0.06\)~fm
  and \(0.04\)~fm. We reconstruct the \(x\)-dependent PDF, as well
  as infer the first few even moments of the PDF using leading-twist 1-loop perturbative 
  matching framework.  Our
  analyses use both RI-MOM and ratio-based schemes to renormalize
  the equal-time bi-local quark-bilinear matrix elements of pions
  boosted up to \(2.4\)~GeV momenta. We use various model-independent and
  model-dependent analyses to infer the large-\(x\) behavior of the
  valence PDF. We also present technical studies on lattice spacing
  and higher-twist corrections present in the boosted pion matrix elements.
\end{abstract}
\date{\today}
\maketitle

\section{Introduction}\label{sec:intro}

QCD factorization implies that the cross-sections of hard inclusive
hadronic processes can be written in terms of convolution of  partonic
cross-section and parton distribution functions (PDF)~\cite{Collins:1989gx}.
Field theoretically~\cite{Soper:1976jc,Collins:1989gx}, 
the quark PDF $f(x,\mu)$ of a hadron $H$ is defined in 
terms of quark fields $\psi$ as
\beqa
&& f(x,\mu)=\int \frac{d\nu}{2\pi} e^{-i \nu x} {\cal M}(\nu,\mu);\quad\text{\ where}\quad\nu=P^+z^-\quad\text{and},\cr
&& 2P^+{\cal M}(P^+z^-,\mu)=\langle H(P)|\overline{\psi}(z^-)\gamma^+ W_+(z^-,0)\psi(0)|H(P)\rangle.\cr &&\quad
\eeqa{pdfdef}
The above definition involves quark and anti-quark displaced by
$z^-$ along the light-cone (and made gauge-invariant by the Wilson-line
$W_+(z^-,0)={\cal P}\exp\left(i\int_0^{z^-}dz'^- A^+\right)$ that runs along the
light-cone. The dimensionless light-cone distance $\nu$ is referred
to as the Ioffe-time and the matrix element ${\cal M}(\nu,\mu)$,
renormalized in the $\msbar$ scheme by convention, is referred to
as the Ioffe-time distribution (ITD).  Notwithstanding such a
straight-forward definition of PDF, the unequal Minkowski time separation in
$z^-$ posed a challenge to the  Euclidean lattice computation until
recently.

Previously, lattice computations have been able to access the moments of 
PDFs using local twist-two hadron matrix elements (c.f.,~\cite{Martinelli:1987zd} for an early work).
A recently proposed 
method to obtain the $x$-dependent PDF is the
quasi-PDF (qPDF), which is defined from matrix elements of equal-time
bilocal quark bilinear operators and can be related to the PDF for
large hadron momenta~\cite{Ji:2013dva}. This method was then developed into LaMET
which provides the framework to calculate all parton physics~\cite{Ji:2014gla}.
Later, 
there was suggestion to use the so-called pseudo-PDF approach~\cite{Radyushkin:2017cyf,Orginos:2017kos},
which relates the same matrix elements to the light-cone
correlations for PDFs at small distances. The hadron matrix element
that is central to both LaMET and the pseudo-PDF approaches is
\beqa
\langle E,P_z|\overline{\psi}(z)W_z(z,0)\gamma_t \psi(0)|E,P_z\rangle&\equiv& 2E(P_z) {\cal M}(z P_z, z^2,\mu_R).\cr&&\quad
\eeqa{pitd}
It is very similar to \eqn{pdfdef}, except that quark and anti-quark
are at equal-time and separated by spatial distance $z$ and evaluated
in an on-shell hadron state at large spatial momentum $P_z$. Such
a matrix element can be easily computed on the
lattice~\cite{Ji:2013dva,Ji:2014gla}. In the literature, the matrix
element ${\cal M}$ is also referred to as the Ioffe-time pseudo
distribution (pITD)~\cite{Radyushkin:2017cyf}, wherein by considering
the arguments of ${\cal M}$ as the Lorentz invariants $p.z$ and
$z^2$, the difference between ITD and pITD become a choice of the
4-vectors $p$ and $z$. 
A similar idea was also considered in an
earlier work~\cite{Braun:2007wv} related to distribution amplitudes.
In the literature, the Lorentz invariant 
$p.z$ is sometimes termed as the Ioffe time regardless of the frame used for the 
sake of simplicity~\cite{Braun:1994jq}. In this work, we will refer to this invariant simply as 
$z P_z$, thereby, bring attention to actual values of $z$ and $P_z$ used to reach the value of the
invariant.
The bilocal bilinear matrix element was also considered earlier in~\cite{Musch:2010ka}, albeit
in a different context of studying transverse momentum distributions.
As a crucial step in the UV regulated field theory, the multiplicative
renormalizability of the bilocal operator was recently demonstrated
to all orders of perturbation
theory~\cite{Ji:2017oey,Ishikawa:2017faj,Green:2017xeu}.  The
renormalized matrix element (and its Fourier transforms with respect
to $z$ or $zP_z$) can be systematically related to the PDF within
the perturbative twist-2 framework (i.e., Large-Momentum Effective Theory
(LaMET)~\cite{Ji:2014gla,Ji:2020ect} or short distance factorization (SDF)~\cite{Radyushkin:2017cyf,Orginos:2017kos}
depending on the limits being taken).  The matching factors
from various intermediate renormalization schemes for the equal-time
bilocal bilinear matrix element at some renormalization scale $\mu_R$
to the $\msbar$ PDF at a factorization scale $\mu$ are known to 1-loop
accuracy~\cite{Constantinou:2017sej,Alexandrou:2017huk,Stewart:2017tvs,Izubuchi:2018srq,Liu:2018uuj,Radyushkin:2017lvu,Zhang:2018ggy},
and recently, papers related to 2-loop matching have also
appeared~\cite{Chen:2020arf,Chen:2020iqi,Li:2020xml}.  A related
good lattice cross-sections~\cite{Ma:2014jla,Ma:2017pxb} approach
has also been recently proposed to calculate PDF on the lattice.
In practice, the lattice calculations and the perturbative factors
are at fixed order, the different methods may have different
advantages and drawbacks. The status of these calculations is
summarized in recent review
papers~\cite{Zhao:2018fyu,Cichy:2018mum,Monahan:2018euv,Ji:2020ect}.  We
also note that other methods to extract PDFs and their moments have
also been proposed~\cite{Liu:1993cv,Detmold:2005gg}.

In this paper, we study the valence pion PDF. The study of pion PDF
is interesting for several reasons, both technical as well as with
interesting physics issues.  The most interesting reason being that
the pions are the pseudo-Nambu-Goldstone bosons of QCD and it is
important to study its structure in order to understand the relation
between hadron mass and hadron structure. Closely related to this,
is the question of how fast the PDF vanishes as $x$ approaches 1.
This issue of whether the vanishing behavior is $(1-x)^2$ or slower
is being vigorously debated with various non-perturbative
approaches~\cite{Nguyen:2011jy,Chen:2016sno,Bednar:2018mtf,Aicher:2010cb,RuizArriola:2002wr,Broniowski:2017wbr,deTeramond:2018ecg,Ding:2019qlr},
now including lattice
QCD~\cite{Sufian:2019bol,Sufian:2020vzb,Izubuchi:2019lyk}.
There have been LO and NLO analyses of the
experimental
data~\cite{Badier:1983mj,Betev:1985pf,Conway:1989fs,Owens:1984zj,Sutton:1991ay,Gluck:1991ey, Gluck:1999xe, Wijesooriya:2005ir,Aicher:2010cb}, but
the results are less constrained than the nucleon PDF due to
availability of experimental data and therefore, the lattice
calculations can have large impact here.  The other interesting
reasons for studying pion in particular are technical. First, the
smallness of the pion mass means that it is easier to have highly
boosted hadronic states required in the qPDF approach. Second, the
excited state contamination for pions is less problematic due to
larger gaps at typical momenta of 1-2 GeV.  There has been lattice
calculations of pion PDF using the quasi/pseudo-PDF
frameworks~\cite{Chen:2018fwa,Izubuchi:2019lyk,Joo:2019bzr,Lin:2020ssv},
and also using the good lattice cross-section
approach~\cite{Sufian:2019bol,Sufian:2020vzb}.  

In our previous work~\cite{Izubuchi:2019lyk}, we studied the valence
pion PDF in 2+1 flavor QCD using the mixed action with lattice
spacing $a=0.06$ fm and LaMET approach. In the sea, we used Highly
Improved Staggered Quark (HISQ) action, while in the valence quark
sector we used clover improved action with hypercubic (HYP) smearing
\cite{Izubuchi:2019lyk}.  We extend this study in three ways in
this paper. First, we perform calculations at another smaller lattice
spacing, namely $a=0.04$ fm. Second, we increase the statistics in
the $a=0.06$ fm ensemble by more than two-fold. Third, we combine
the analysis of the bilocal bilinear matrix element renormalized
in RI-MOM scheme~\cite{Chen:2017mzz} with the ratio scheme (also
referred to as reduced ITD~\cite{Orginos:2017kos}), and also
propose and use generalizations of the ratio scheme with the promise
of lesser higher-twist contamination.  At a practical level, it has
been conventional in the lattice calculation that used quasi-PDF
formalism to use an intermediate RI-MOM scheme, while those using
pseudo-PDF formalism to use an intermediate ratio scheme.  We do
not make such distinctions, and simply refer to matrix elements of
operator in \eqn{pitd} that is made gauge-invariant with a straight
Wilson-line as bilocal bilinear matrix elements, or simply as the
matrix elements for the sake of brevity, in various renormalization schemes;
RI-MOM matrix element or ratio matrix element, for example. Also, in this work,
we simply label the methodology to be that of twist-2 perturbative matching,
so as to encompass LaMET and SDF approaches. This is because in the absence 
of any actual large momentum or short-distance limits being taken, 
the combined 
analysis of a sample of data that spans a range of distances and momenta
in either real or Fourier space are equivalent~\cite{Izubuchi:2018srq},
up to choices of approaching the inverse problem to relate 
the PDF to the matrix elements. Therefore, the readers of this paper 
can approach the contents presented in one way or another 
equivalently, depending on their preference.

The plan of the paper is as follows. In \scn{setup}, we discuss the
details of the lattice ensembles, statistics and other computational
specifics.  In \scn{2pt}, we elaborately describe the extraction
of ground- and excited-states of pion from the boosted two-point
functions. In \scn{3pt}, we describe the extraction of the boosted pion matrix element from
three-point function via excited-state extrapolations.  In \scn{ren},
we discuss the various renormalization schemes used.
Readers not interested in the details of the lattice calculation
can skip Sections~\ref{sec:setup}-\ref{sec:ren}.  In \scn{matching},
we describe the twist-2 perturbative matching 
formulation which forms the basis of
the results presented in the following sections. We also present a
study of higher-twist contamination in this section. The \scn{mom}
contains the direct extraction of the valence moments of pion from
the $P_z z$ and $z^2$ dependences of bilocal bilinear matrix element. 
In \scn{pdf}, we reconstruct the $x$-dependent valence PDF
at $\mu=3.2$ GeV based on fits to the pion matrix elements using phenomenology
motivated ansatz for the PDFs.  In \scn{largex}, we address the
issue of large-$x$ exponent of the valence pion PDF based on model
dependent fits as well as from a novel model-independent method we
introduce here.  In \scn{cont}, we speculate the continuum results
based on our observation at two fine lattice spacings. 
The conclusion and comparisons with other analyses are given in
\scn{conclusion}. More technical details are present in the 
appendices.

\section{Lattice setup}\label{sec:setup}

In this work, we use two different $L_t\times L^3$ lattice gauge
ensembles both of them with relatively small lattice spacings ---
(1) ensemble with lattice spacing $a=0.06$ fm with lattice extents
$48\times 64^3$, and (2) a finer ensemble with $a=0.04$ fm with
extent $64\times 64^3$. These gauge ensembles were generated by the
HotQCD collaboration~\cite{Bazavov:2014pvz} using 2+1 flavor Highly
Improved Staggered Quark (HISQ) action~\cite{Follana:2006rc} in the
sea. In both these ensembles, the sea quark mass was tuned such
that the pion mass was 160 MeV. On these gauge field ensembles, we
used 1-HYP tadpole improved Wilson-Clover valence quarks.  That is,
we used the Wilson-Clover quark propagator in the Wick contractions
required in the computations of the three-point and two-point
functions, and the gauge links that went into the construction of
the propagator were smoothened using 1 step of HYP
smearing~\cite{Hasenfratz:2001hp}. We set the clover coefficient $c_{\rm sw}=u_0^{-3/4}$, 
where $u_0$ is the average plaquette with 1-HYP smearing; we used 
$c_{\rm sw}=1.02868$ and 1.0336 for $a=0.06$ fm and 0.04 fm respectively. 
We tuned the Wilson-Clover quark
mass $m_q a$  in both the ensembles so that the valence pion mass,
$m_\pi$, is 300 MeV. Through an initial set of tuning runs we
determined $m_q a=-0.0388$ for $a=0.06$ fm and $m_q a=-0.033$ for
$a=0.04$ fm lattices. For this pion mass, the values of $m_\pi L_t$
on the $a=0.06$ fm and 0.04 fm lattices are 5.85 and 3.89 respectively.
Thus it would be more important to take care of wrap around effects
in the finer lattice and we do so in the analysis.  With the usage
of 1-HYP smeared gauge links in the Wilson-Clover operator, we did
not find any exceptional configurations at both the lattice spacings,
as noted by absence of any anomaly in the convergence of the Dirac operator
inversions.
We used the $a=0.06$ fm ensemble in our previous analysis of the
valence PDF of pion~\cite{Izubuchi:2019lyk}. With this work, we
have increased the statistics used in this ensemble by more than
two times.

\bet
\centering
\begin{tabular}{|c|c|c|c|c|c|c|}
\hline
\hline
 ensemble & $m_q a$ & $m_\pi L_t$  & $n_z$ & $z$ range  &  \#cfgs & (\#ex,\#sl) \cr
$a,L_t\times L^3$ &  & & & & & \cr
\hline
$a=0.06$ fm,        &-0.0388  & 5.85 & 0,1 & [0,15] & 100 & $(1, 32)$\cr
    \cline{4-7}
$64\times 48^3$     &         &      & 2,3,4,5  & [0,8]  & 525 & $(1,32)$\cr  
          &         &         &      & [9,15]  & 416 & $(1,32)$\cr  
          &         &         &      & [16,24]  & 364 & $(1,32)$\cr  
\hline
$a=0.04$ fm, & -0.033 & 3.90 & 0,1 & [0,32] & 314 & $(3, 96)$\cr
    \cline{4-7}
$64\times 64^3$          &         &      & 2,3  & [0,32]  & 314 & $(4,128)$\cr  
    \cline{4-7}
                         &         &      & 4,5  & [0,32]  & 564 & $(4,128)$\cr
\hline
\hline
\end{tabular}
\caption{Details of the measurements on two lattice ensembles used
in this paper. For each ensemble, we have specified the bare Wilson
fermion quark mass $m_q a$ corresponding to a $300$ MeV pion mass
$m_\pi$, the temporal extent $L_t$ of the lattice in $m_\pi$ units.
We specify the number of gauge configurations used (\#cfgs) and the
number of exact and sloppy inversions per configurations (\#ex,\#sl)
for different Wilson-line lengths $z$ used in three-point functions
and the pion momentum $P_z=2\pi n_z/(L a)$. 
}
\eet{setup}

The most basic element of this computation is the Wilson-Dirac quark
propagator inverted over boost smeared sources and
sinks~\cite{Bali:2016lva} as we discuss more in the next section
on two-point functions.  We used the multigrid
algorithm~\cite{Brannick:2007ue} for the Wilson-Dirac
operator inversions to get the quark propagators.  
These calculations were performed on GPU using the QUDA suite~\cite{Clark:2009wm,Babich:2011np, Clark:2016rdz}.

We used boosted quark source~\cite{Bali:2016lva} and sink with
Gaussian profile, as we discussed in detail in~\cite{Izubuchi:2019lyk}.
Instead of using the gauge-covariant Wuppertal
smearing~\cite{Gusken:1989ad} to implement the Gaussian profiled
quark sources, we gauge-fixed the configurations in the Coulomb
gauge to construct the sources as we found it to be computationally
less expensive. We fixed the radius of the Gaussian profile on
$a=0.06$ fm and $a=0.04$ fm ensembles to be 0.312 fm and 0.208 fm
respectively.  We discussed the details of tuning the Gaussian
smearing parameters in the Appendix of~\cite{Izubuchi:2019lyk}.
Using these quark propagators, we are able to compute hadron two-point
and three-point functions in hadrons boosted to momentum $P_z=2\pi
n_z/(La)$.

We tabulate the details of the statistics used in the two ensembles
in \tbn{setup}. We increased the statistics in two ways (a) using
statistically uncorrelated gauge field configurations, which are
labeled as \#cfg in \tbn{setup}, and (b) by using All Mode Averaging
(AMA)~\cite{Shintani:2014vja} on each gauge configuration.  In order
to mitigate the reduction in the signal-to-noise ratio in both the
three-point and two-point functions as one increases $P_z\propto
n_z$, we used more gauge field configurations for larger $n_z$ than
at smaller ones. In $a=0.06$ fm ensemble, we effectively increased
the statistics 32 times by using 1 exact Dirac operator inversion
and 32 sloppy inversions in the AMA per configuration.  In the
$a=0.04$ fm ensemble, we increased the number of exact and sloppy
solves for $n_z=2,3$ and more for $n_z=4,5$.  We used a stopping
criterion of $10^{-10}$ and $10^{-4}$ for the exact and sloppy
inversions respectively.

\section{Analysis of Excited states in the two-point function of boosted pion}\label{sec:2pt}

\bet
\centering
\begin{tabular}{|c|c|c|c|}
\hline
\hline
    $n_z$ &  \multicolumn{2}{c|}{$P_z$ (GeV)} & $\zeta$ \cr
    \cline{2-3}
    & $a=0.06$ fm & $a=0.04$ fm & \cr
\hline
    0  &   0  & 0 & 0\cr
    1  &  0.43 & 0.48 & 0\cr
    2  &  0.86 & 0.97 & 1 \cr
    3  &  1.29 & 1.45 & 2/3 \cr
    4  &  1.72 & 1.93 & 3/4 \cr
    5  &  2.15 & 2.42 & 3/5 \cr
\hline
\hline
\end{tabular}
\caption{Table of momenta $P_z$ in GeV at the two lattice spacings.
The values of the $\zeta$ used in the boosted Gaussian sources used
for each $P_z$ is also shown.}
\eet{tabmom}

\bef
\includegraphics[scale=0.6]{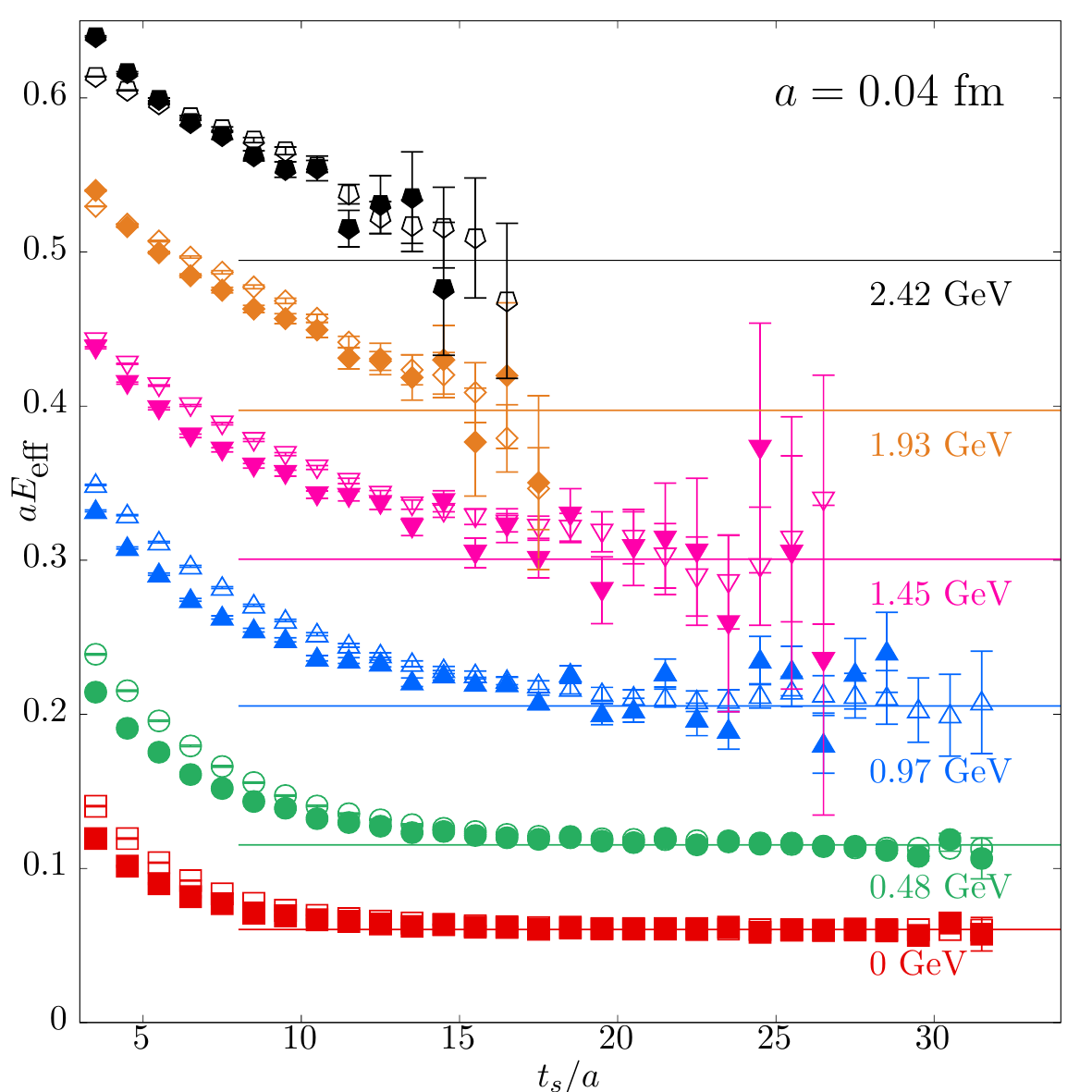}
\caption{The effective mass $E_{\rm eff}$ is shown as a function
of source-sink separation $t_s$ for the $a=0.04$ fm lattice. The
filled and open symbols are obtained from SS and SP correlators
respectively.
}
\eef{64c64meff}

In this section, we discuss the computation of boosted pion correlators 
and the extraction of the excited state contributions. Using 
a smeared (s) pion source $\pi_s(\mathbf{P},t)$ 
\beq
\pi_s(\mathbf{P},t)=\sum_{\mathbf{x}}\overline{d}_s(\mathbf{x},t)\gamma_5 u_s(\mathbf{x},t) e^{-i \mathbf{P}.\mathbf{x}},
\eeq{pionop}
for pion $\pi^+$ that is moving with spatial momentum 
$\mathbf{P}=(0,0,P_z)$ along the $z$-direction,
we computed 
the two-point function of pions 
\beq
C^{ss'}_{\rm 2pt}(t_s;P_z)=\left\langle \pi_{s'}(\mathbf{P},t_s) \pi^\dagger_{s}(\mathbf{P},0) \right\rangle.
\eeq{2ptdef}
In this computation, we used momenta on a periodic lattice
\beq
P_z=\frac{2\pi n_z}{La},
\eeq{pdef}
for $n_z=0,1,2,3,4$ and 5 at both lattice spacings.  These values
of $n_z$ correspond to $P_z$ up to 2.15 GeV and 2.42 GeV on the
$a=0.06$ fm and 0.04 fm lattices respectively.  For ease of reference,
we have tabulated the physical values of $P_z$ for the two lattices
in \tbn{tabmom}. Such large momenta are central to the applicability
of the leading-twist perturbative matching framework. It is important
that we are able to suppress the excited state contributions to the
two-point function within smaller source-sink separations $t_s$ to
deal with the signal-to-noise ratio at larger $t_s$. This is the
reason for the smeared pion source and sink, $\pi_s$, that are
constructed out of smeared quark fields, $u_s$ and $d_s$. We
constructed two-kinds of two-point functions: smeared-source ($s=S$)
point-sink ($s'=P$) correlators referred to as SP, and smeared-source
($s=S$) smeared-sink ($s'=S$) correlators referred to as SS henceforth.
For smeared sources, we used boost smeared Gaussian profiled sources,
as is now standard in the lattice PDF computations.  We have tabulated
the values of the tunable parameter $\zeta$ for the boost
smearing~\cite{Bali:2016lva} at different $P_z$ in \tbn{tabmom}.

The two-point functions enter the PDF determination in two ways;
for determining the excited state spectrum of the boosted pion on
the two lattices, which in turn will enable us to extract the boosted pion
matrix elements. Below, we will discuss the excited state analysis
of the two-point function.  In our previous
publication~\cite{Izubuchi:2019lyk}, we discussed the extraction
of the pion spectrum in detail for the $a=0.06$ fm lattice. Since
the only difference in this paper is the increased statistics for
this ensemble, we focus on the pion spectrum in the finer $a=0.04$
fm lattice in this section.  In \fgn{64c64meff}, we show the effective
mass $E_{\rm eff}(t_s)$ of pion at different $P_z$ as a function
of source-sink separation $t_s$ for the SP (open symbols) and SS
correlators (filled symbols) respectively. For comparison, the
values of $E(P_z)$ for the ground state pion based on its dispersion
relation are shown by the horizontal lines.  One can notice that
the signal-to-noise ratio gets poorer at shorter $t_s$ as $P_z$ is
increased. Therefore, we are forced to work with $t_s/a=9,12,15$
and 18 corresponding to physical distances of $0.36$ fm to 0.72 fm for the 
case of three-point functions.
The largest operator insertion times $\tau$, which we will discuss
in the next section, are $t_s/2$. In this range of $t_s$, the
effective mass is not plateaued and careful consideration of excited
states becomes important.  Up to $n_z=3$, it is clear that the
effective masses plateau at the the dispersion values for the pion.
One can also note that SS correlator approaches the plateau faster
than SP as expected.  The difference between SS and SP correlators
is due to the differences in the amplitudes of the states in the
two, and we will use this advantageously in the extraction of first
and second excited states of the pion.

\bef
\centering
\includegraphics[scale=0.33]{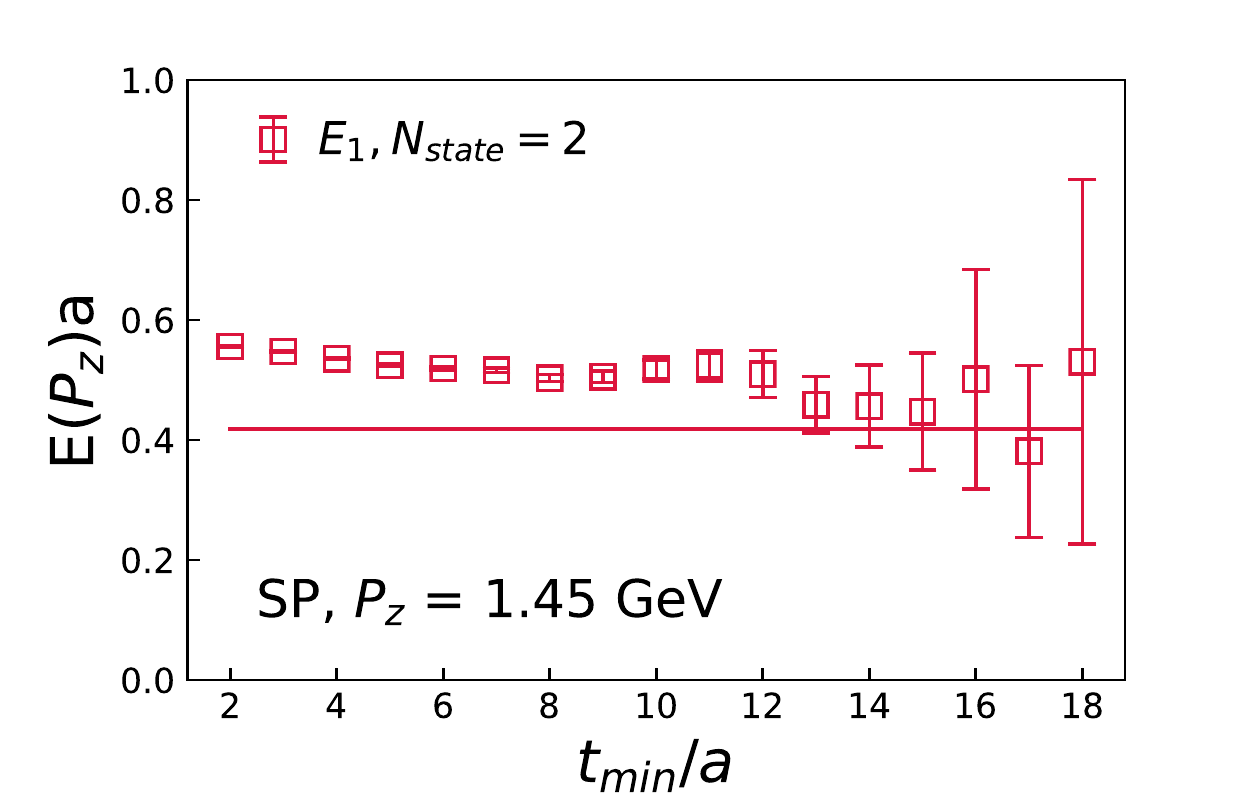}
\includegraphics[scale=0.33]{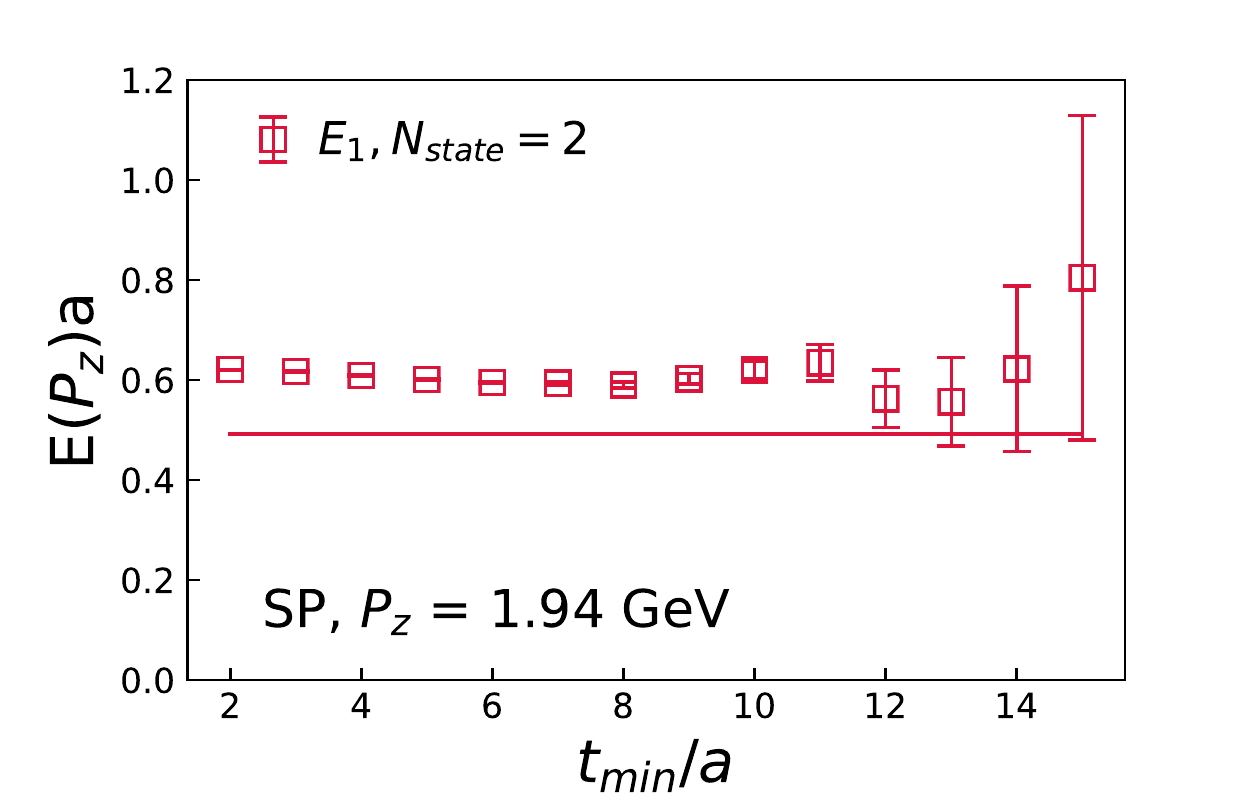}

\includegraphics[scale=0.33]{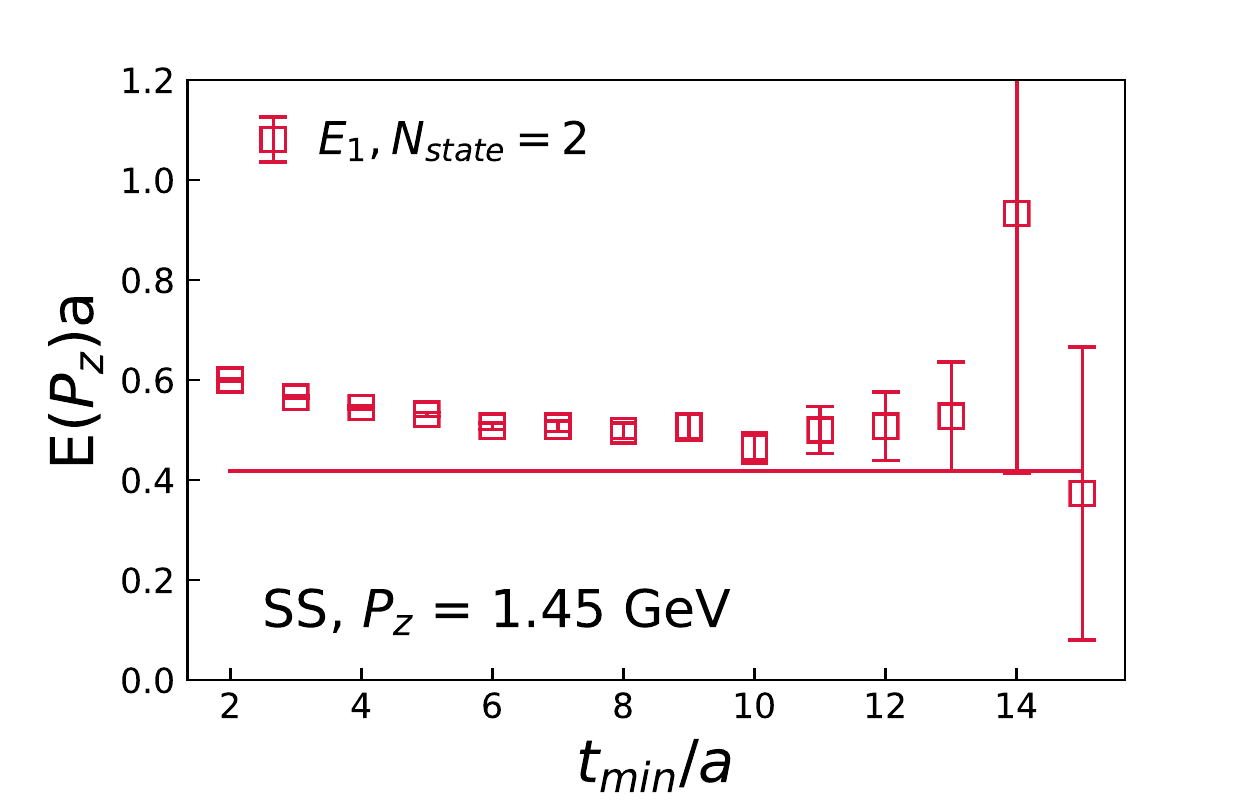}
\includegraphics[scale=0.33]{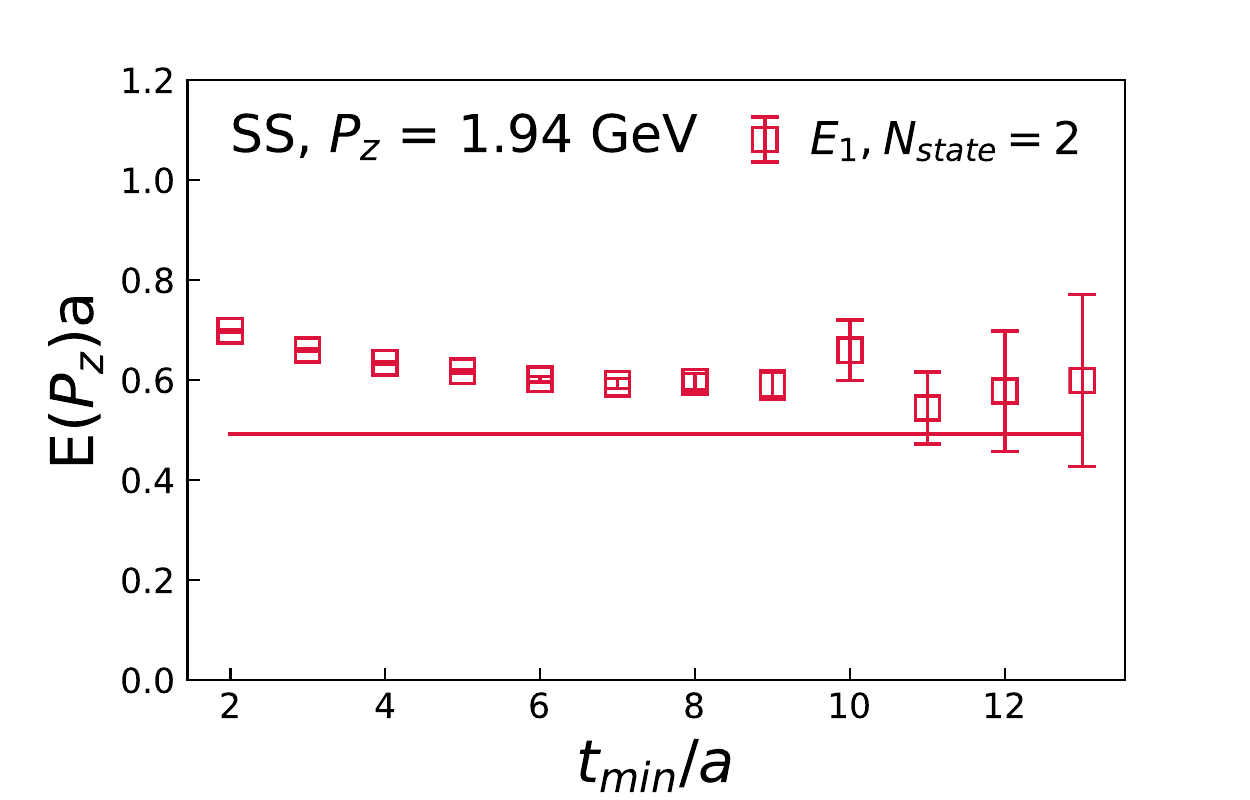}

\includegraphics[scale=0.33]{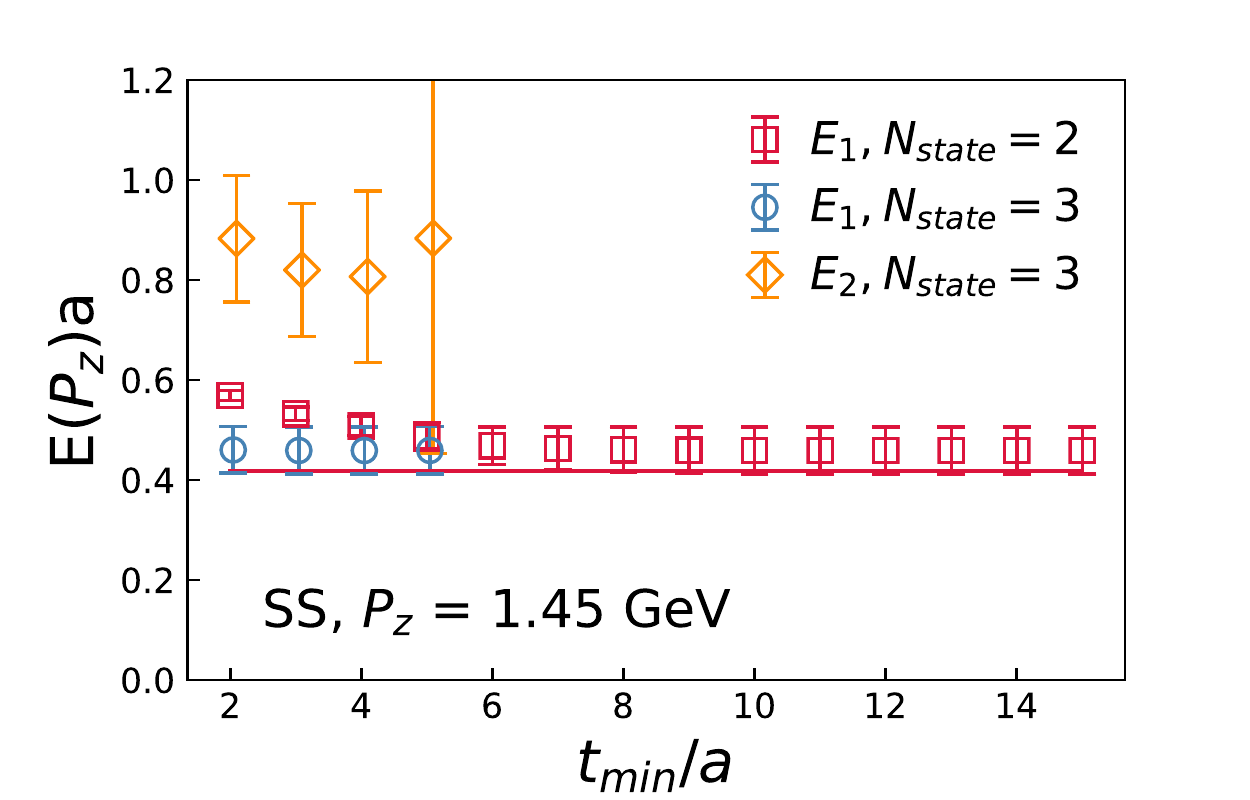}
\includegraphics[scale=0.33]{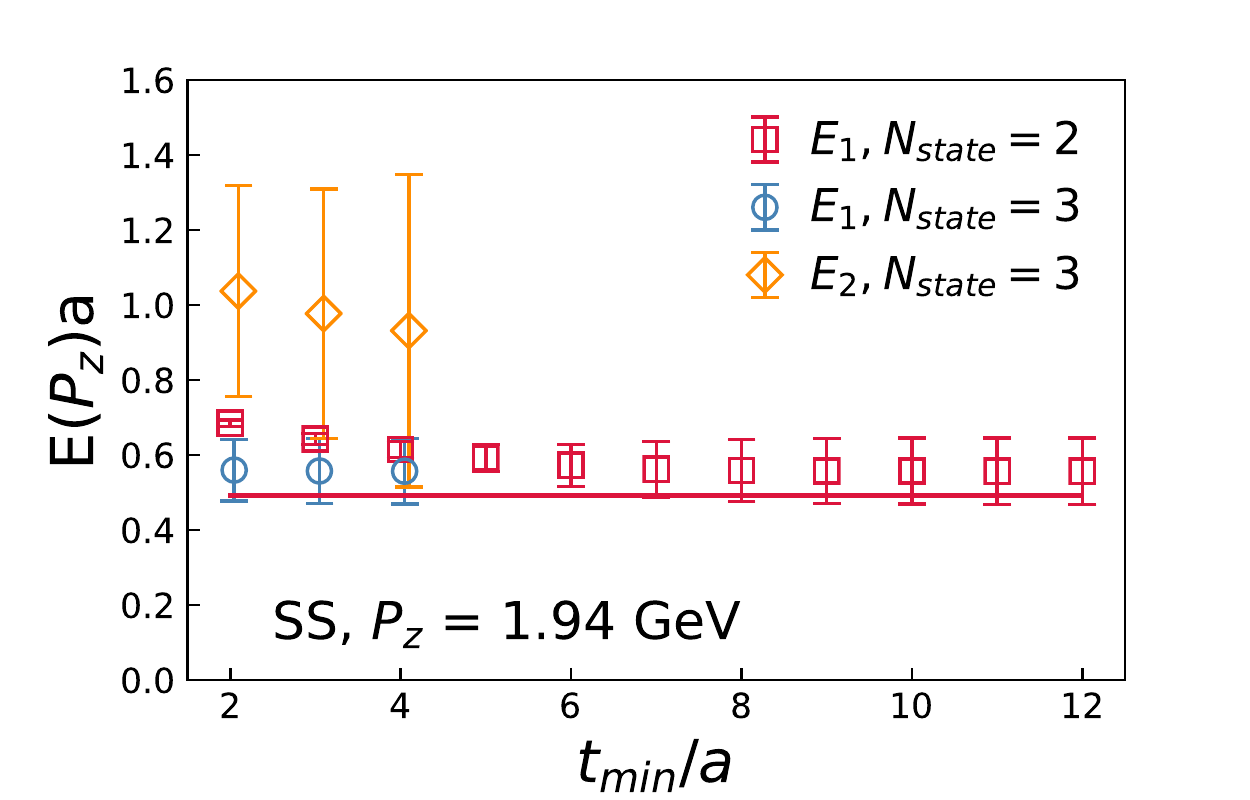}
\caption{The dependence of the fitted values of energy levels on
the fit range $[t_{\rm min},32a]$ is shown. The top-left and top-right
panels show this dependence for the first excited level $E_1$ as
obtained from two-state fits to SP correlator at $P_z=1.45$ GeV and
$P_z=1.94$ GeV respectively. Similar results using SS correlator
are shown in the two middle panels. The results for $E_1$ and $E_2$
obtained using three-state fits to the SS correlator 
(with prior on $E_0$ and $E_1$) are
shown in the two bottom panels.  }
\eef{64c64E1}

\bef
\centering
\includegraphics[scale=0.8]{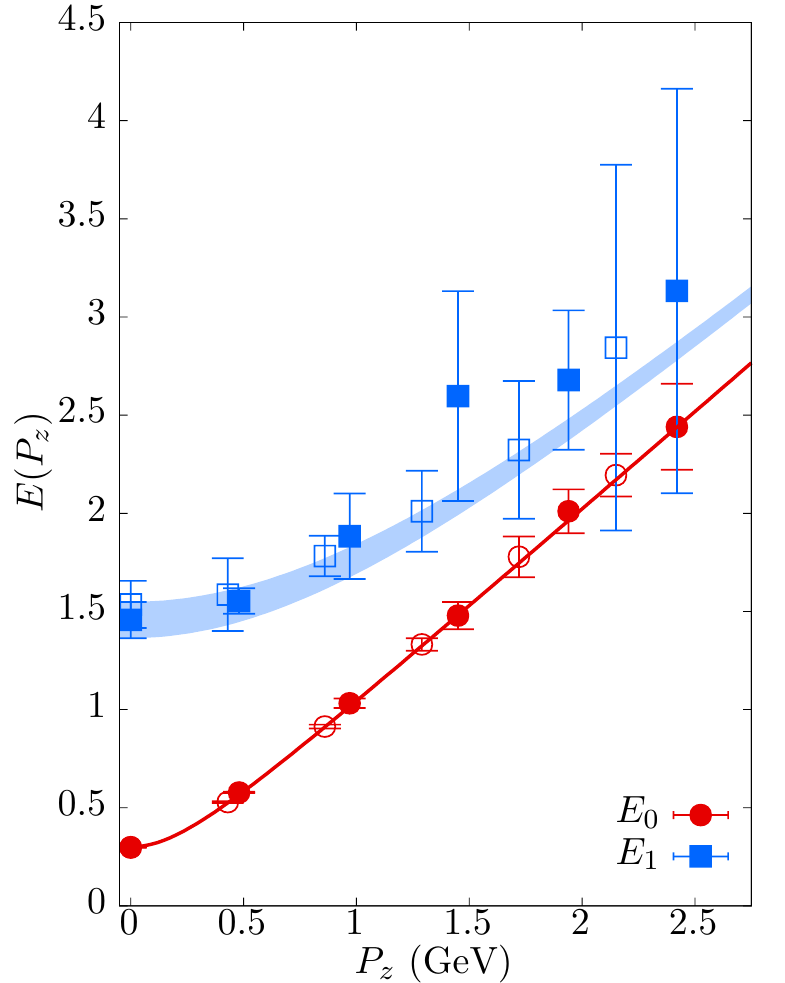}
\caption{ The energy of the ground state ($E_0$) and the first
excited state ($E_1$) are shown as a function of $P_z$.  The results
from $a=0.04$ fm are shown as filled symbols and those from $a=0.06$
fm as the open symbols.  The results shown in the plot for $E_0$
were obtained from an unconstrained two-state fit, while $E_1$ were
obtained by fixing $E_0$ to its dispersion values.  }
\eef{disp}

In order to determine the energy levels $E_0,E_1,\ldots$, 
we fit the spectral decomposition of $C_{\rm 2pt}(t_s)$,
\begin{equation}
    C_{\rm 2pt}(t_s)=\sum_{n=0}^{N_{\rm state}-1} A_n (e^{-E_n t_s}+e^{-E_n (aL_t-t_s)}),
\end{equation}
with $E_{n+1}>E_n$. The above expression is 
truncated at $N_{\rm state}$ to both the SS and SP two-point function
data over a range of values of $t_s$ between $[t_{\rm min},a L_t/2]$.
We performed this fitting with one-state ($N_{\rm state}=1$),
two-state ($N_{\rm state}=2$), and  three-state  ($N_{\rm state}=3$)
ansatz.  As evident from the behavior of effective mass in
\fgn{64c64meff}, in order for the 1-state fits to work, we had to
use $t_{\rm min}>0.56$ fm and the results were consistent with the
one from dispersion relation $E_0(P_z)=\sqrt{m_\pi^2+P_z^2}$ with
$m_\pi=300$ MeV. When we performed an unconstrained 4-parameter
2-state fit to both the SS and SP correlators, we found the
approach to the expected $E_0(P_z)$ to be at even shorter $t_{\rm min}
\sim 0.2$ fm.  Since we were able to obtain the ground state energy
$E_0(P_z)$ reliably from one and two exponential fits to both the
SS and SP correlators and they agree with the expectation from the
dispersion relation well, we then fixed the value of $E_0$ to its
dispersion value to perform a more stable two and three exponential
constrained fits with one less free parameter.  

The results for the
first excited state $E_1(P_z)$ using different $t_{\rm min}$ in
such a constrained two-state fits for $n_z=3$ and $n_z=4$ are shown
in the top and middle panels of \fgn{64c64E1};  the top panel is for
SP and the middle one for SS. One
can notice that for $t_{\rm min}/a>10$, it is possible to reliably
estimate the first excited state in both SP and SS correlators, and
the two estimates are also consistent with each other giving more
confidence in the results.  The horizontal lines in the figures
correspond to the expected result for $E_1(P_z)$ based on a single
particle type dispersion relation $E_1(P_z)=\sqrt{P_z^2+E^2_1(P_z=0)}$.
As $t_{\rm min}$ is increased, the fitted values of $E_1(P_z)$ are
actually the dispersion values. We observed this behavior at different
$P_z$ as well. 
We will address this more in the end of this section.
Having understood the actual spectral decomposition of the pion
correlator, it has been found to be better practically
to use the effective value of $E_1$ and the corresponding amplitude $A_1$ in the range of $\tau\approx t_s/2$
used in the two-state fits to three-point function~\cite{Fan:2020nzz}.  By doing this, we
effectively take care of excited states higher than $E_1$ that could be present at $\tau \le t_s/2$ in the
two-state fits to the three-point function.  We follow this procedure 
here and take the value of $E_1$ and $A_1$
in the pseudo-plateau region for $E_1$ seen in middle panels of 
\fgn{64c64E1} for $t_{\rm min} \in [5a,10a]$.

We also performed constrained 3-state fits on the SS two-point
function.  Besides fixing $E_0$, we also imposed a prior on $E_1$
using its best estimate from the SP correlators with the corresponding
errors~\cite{Lepage:2001ym}.  The results for $E_1$ and $E_2$ from
this analysis are shown in the bottom panel of \fgn{64c64E1} for
$n_z=3$ and 4.  As a consistency check, the 3-state prior fit is
able to reproduce the input prior for $E_1$ starting from $t_{\rm
min}/a=2$. It also results in an estimate for $E_2$ which is large
and noisy, and it is likely that it is an effective third state
capturing several higher excited states. For our excited state
extrapolations, such an effective estimate is sufficient.  We
repeated the above set of analysis for the $a=0.06$ fm lattice and
we were able to obtain the ground and first excited state reliably.

In \fgn{disp}, we show the first two energy levels for both $a=0.04$
fm and 0.06 fm lattices, as a function of $P_z$. It is not very
surprising that the ground state, which is the pion, follows the
particle dispersion well even up to $P_z=2.4$ GeV on the fine lattices
we use. But, it is remarkable that the first excited state $E_1$
also follows a single particle dispersion relation. We noted this
also in our discussion of \fgn{64c64E1}. To solidify the claim, we
observed the same behavior in both SS and SP channel. Also, the
difference between $E_1$ on the two physical volumes $24.9\text{\
fm}^3$ and $16.78\text{\ fm}^3$ for the $a=0.06$ fm and $a=0.04$
fm lattices is not seen.  Thus, it is likely not a multi-particle
state with a gapped finite volume spectrum that mimics a single
particle state.  In order to account for the 300 MeV pion mass, we
added 0.16 GeV to the PDG value~\cite{Tanabashi:2018oca} of the
first pion radial excitation, $\pi_1(1300)$ to estimate a value of
1.46 GeV. This value agrees well with our estimates of $E_1(P_z=0)$
at both the lattice spacings. Therefore, we find it reasonable to
conclude that the ground state is the pion and the first excited
state is the radial excitation of pion, $\pi_1$, with $E_1(P_z=0)$
being identified with its mass.

\section{Extraction of bare matrix elements from excited state extrapolations}\label{sec:3pt}

\befs
\centering
\includegraphics[scale=0.58]{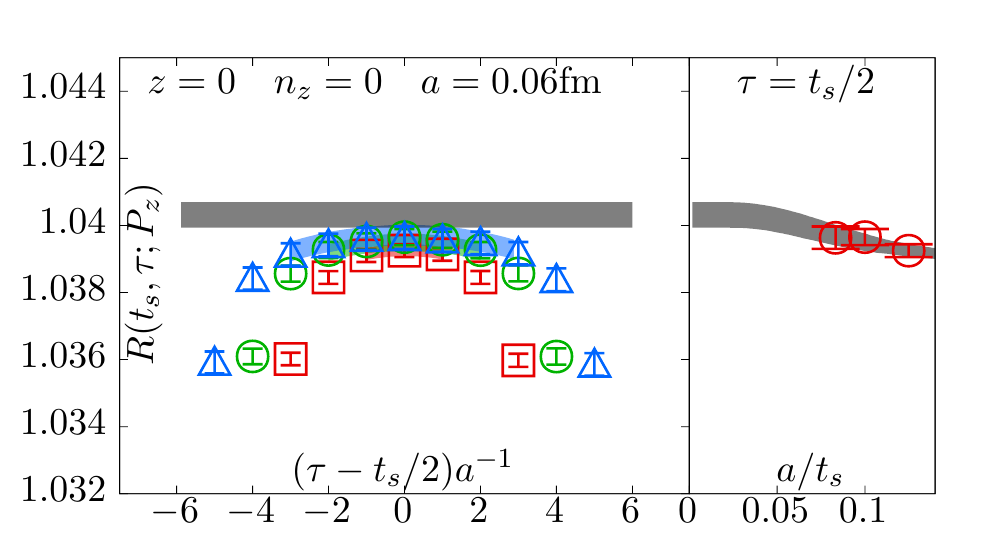}
\includegraphics[scale=0.58]{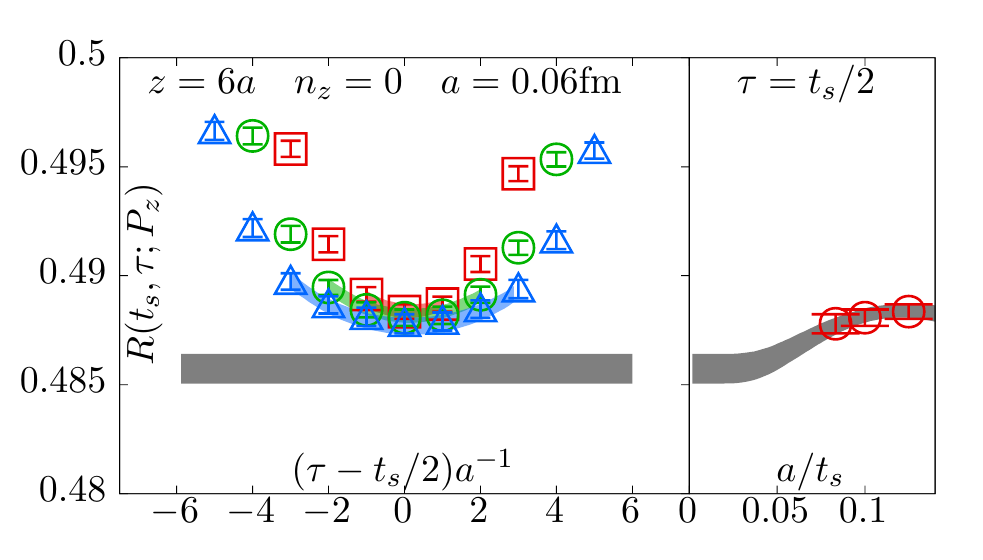}
\includegraphics[scale=0.58]{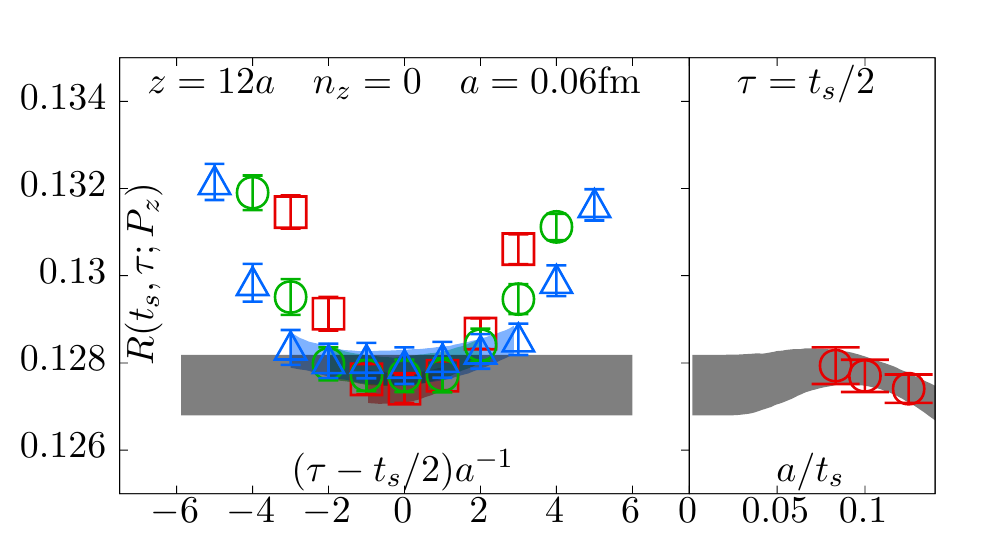}

\includegraphics[scale=0.58]{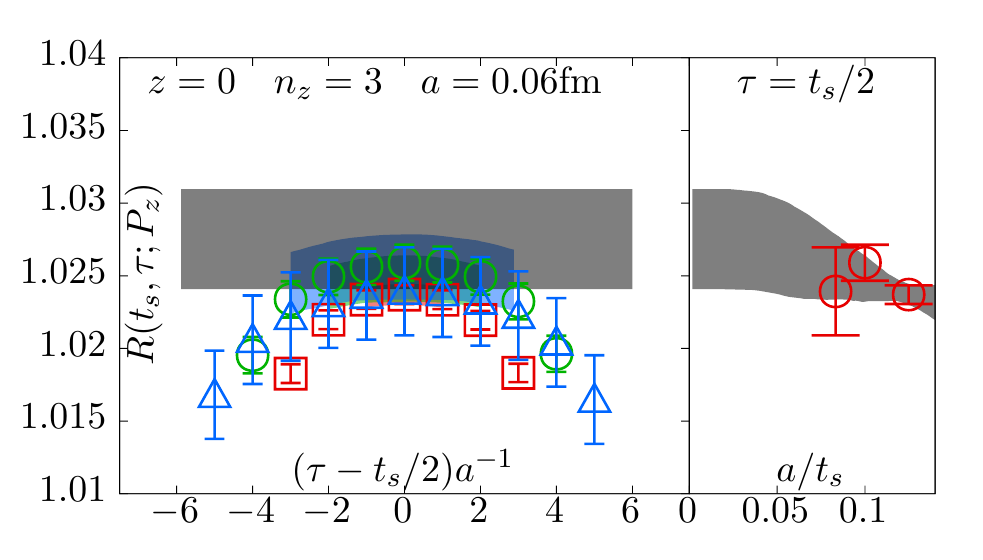}
\includegraphics[scale=0.58]{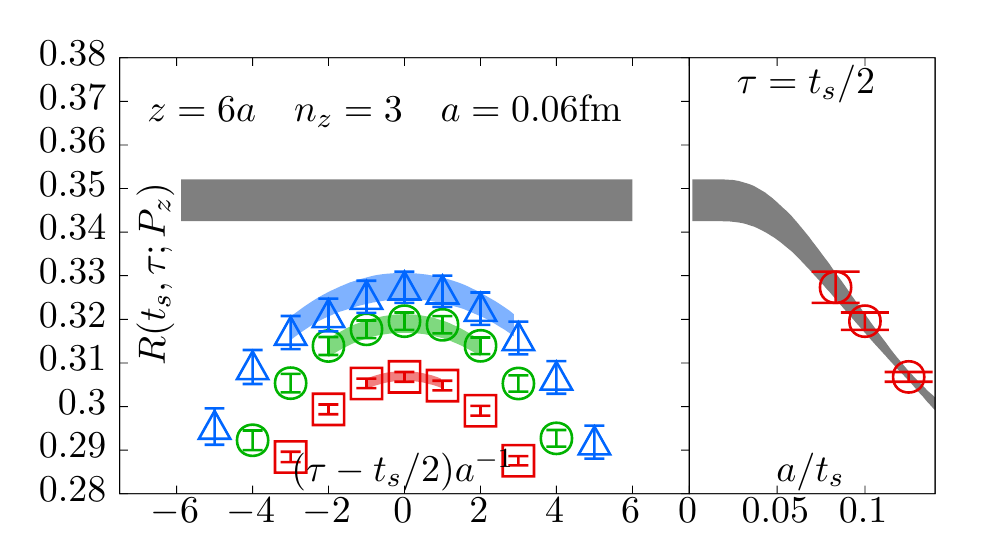}
\includegraphics[scale=0.58]{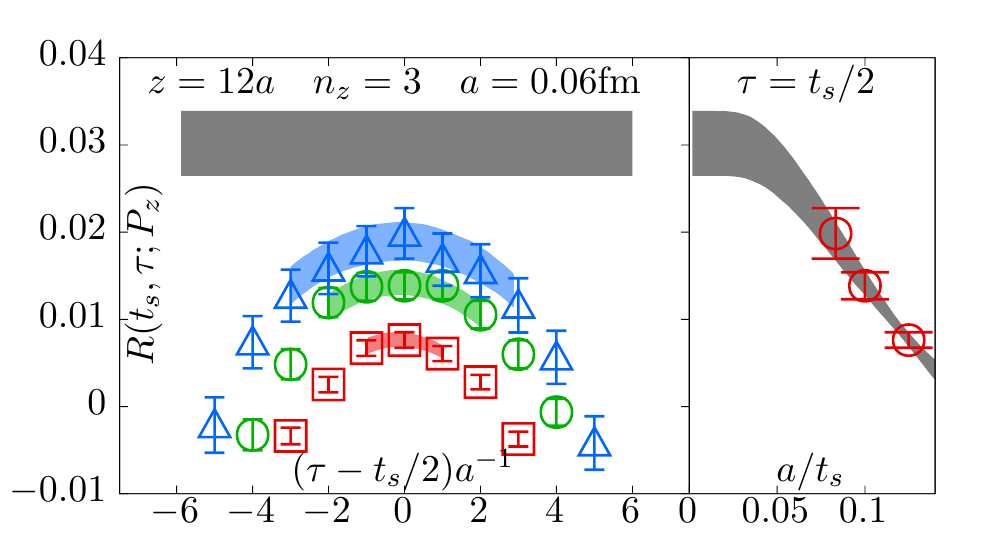}

\includegraphics[scale=0.58]{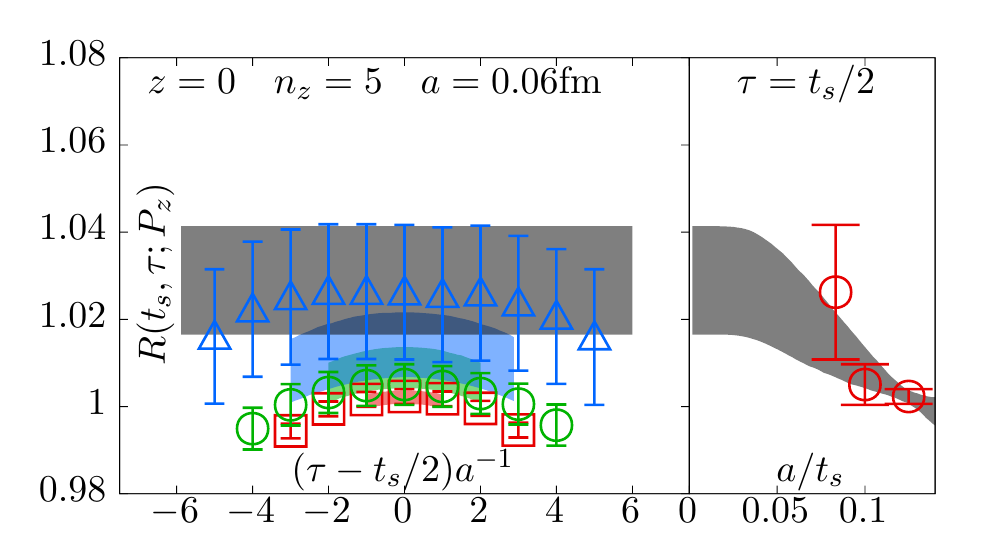}
\includegraphics[scale=0.58]{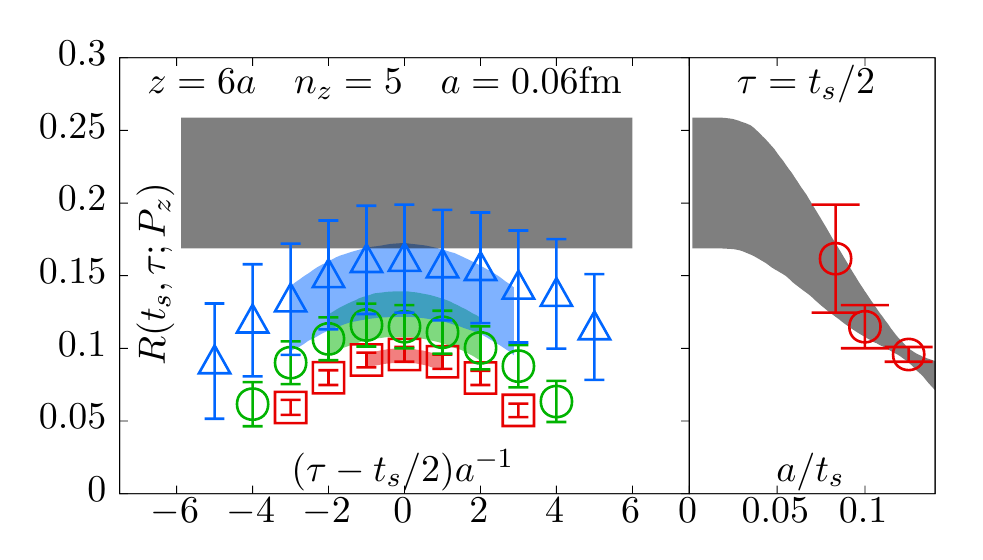}
\includegraphics[scale=0.58]{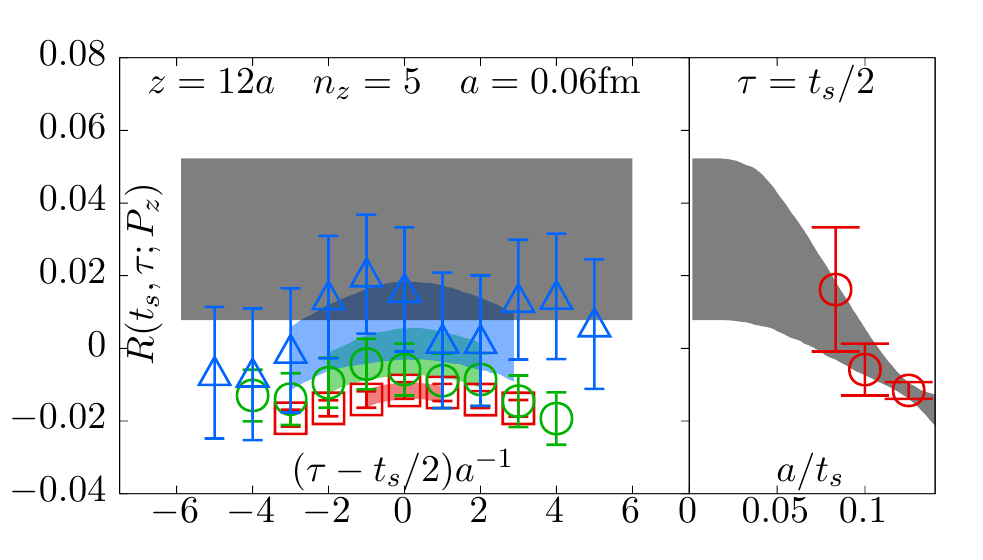}
\caption{
    The source-sink $t_s$ and operator insertion $\tau$ dependence
    of the ratio $R(t_s,\tau; z,P_z)$, at fixed $z$ and $P_z=2\pi
    n_z/L$ are shown for the lattice spacing $a=0.06$ fm.  The
    top-rows are for $n_z=0$, middles ones for $n_z=3$ and bottom
    ones for $n_z=5$. The left panels are for $z=0$, middle panels
    for $z=6a$ and right ones for $z=12a$ respectively. Each plot
    has left and right sub-panels.  In the left sub-panels, the
    $t_s-\tau/2$ dependence is shown at $t_s=8a$ (red squares),
    $10a$ (green circles) and $12a$ (blue triangles).
    The corresponding colored bands are the 1-$\sigma$
    errorbands from Fit$(2,3)$ (see text).
    On the right sub-panels, the extrapolation (grey band) to
    $t_s\to\infty$ is shown as a function of $a/t_s$ at fixed
    $\tau=t_s/2$.
}
\eefs{3ptfitsa6}
The next ingredient in the extraction of the pion matrix element
is the three-point function
\beq
C_{\rm 3pt}(z,\tau,t_s)=\left\langle \pi_S(\mathbf{P},t_s) O_\Gamma(z,\tau) \pi_S^\dagger(\mathbf{P},0)\right\rangle,
\eeq{3pt}
involving the insertions of smeared pion source
$\pi_S^\dagger(\mathbf{P},0)$ and smeared sink $\pi_S(\mathbf{P},t_s)$
separated by an Euclidean time $t_s$ and projected to spatial
momentum $\mathbf{P}=(0,0,P_z)$. The operator $O_\Gamma(z;\tau)$
is the isospin-triplet operator that involves a quark and anti-quark
that are spatially separated by distance $z$
\beqa
O_\Gamma(z,\tau)&=&\sum_{\mathbf{x}}\bigg{[}\overline{u}(x+{\cal L})\Gamma W_z(x+{\cal L},x) u(x)-\cr &&\overline{d}(x+{\cal L})\Gamma W_z(x+{\cal L},x) d(x) \bigg{]},
\eeqa{qpdfop}
where $x=(\mathbf{x},\tau)$ with $\tau$ being the time-slice where
the operator is inserted, and the quark-antiquark being displaced
along the $z$-direction by ${\cal L}=(0,0,0,z)$.  The operator is
made gauge-invariant through the presence of the straight Wilson-line
of length $z$, $W_z(x+{\cal L},x)$, that connects the lattice sites
at $x+{\cal L}$ to $x$.  The Wilson-line is constructed out of
1-level HYP smeared gauge links to get better signal to noise ratio.
The matrix $\Gamma$ is either the Dirac $\gamma$-matrix $\gamma_z$
or $\gamma_t$ for the unpolarized PDFs that we will study in this
paper. For the case of lattice Dirac operators that break the chiral
symmetry explicitly at finite lattice spacings, it was shown
perturbatively in that $O_{\gamma_z}$ mixes with the scalar operator
$O_1$ due to renormalization~\cite{Chen:2017mzz,Constantinou:2017sej}.
Such mixing is absent in the case of $O_{\gamma_t}$. In addition
to this mixing, we also found in our previous work~\cite{Izubuchi:2019lyk}
for the case of pion that $O_{\gamma_z}$ is comparatively noisier
compared to $O_{\gamma_t}$ with same statistics, and also suffered from larger 
excited state contamination. Another pertinent advantage of $O_{\gamma_t}$ over 
$O_{\gamma_z}$ is the absence of additional higher-twist effects proportional 
to separation vector $z_\mu$. Therefore, we
resort to only the usage of $\Gamma=\gamma_t$ in this paper. The
pure multiplicative renormalization of $O_{\gamma_t}$ also allows
us to explore the renormalization group invariant ratios  in addition
to RI-MOM scheme as an advantage, and we will explain this in detail
in the next section.
The above $u-d$ three-point function 
is purely real in the case of pion, and the real part is symmetric about 
$z=0$. Therefore, we symmetrized the data by averaging over $\pm z$. 
Further, the matrix element depends only on the Lorentz invariant 
$\nu=P_z z$. Therefore, one can average over the matrix elements determined with
$\pm P_z$; to reduce computational cost, we only used positive $P_z$. In the plots that follow, we 
will display the three-point function in the positive $z$ direction only. In addition, 
only the quark-line connected piece contributes to the isotriplet three-point function.
We refer the reader to the Appendix of~\cite{Izubuchi:2019lyk}
for detailed proofs of the above characteristics.

\befs
\centering
\includegraphics[scale=0.58]{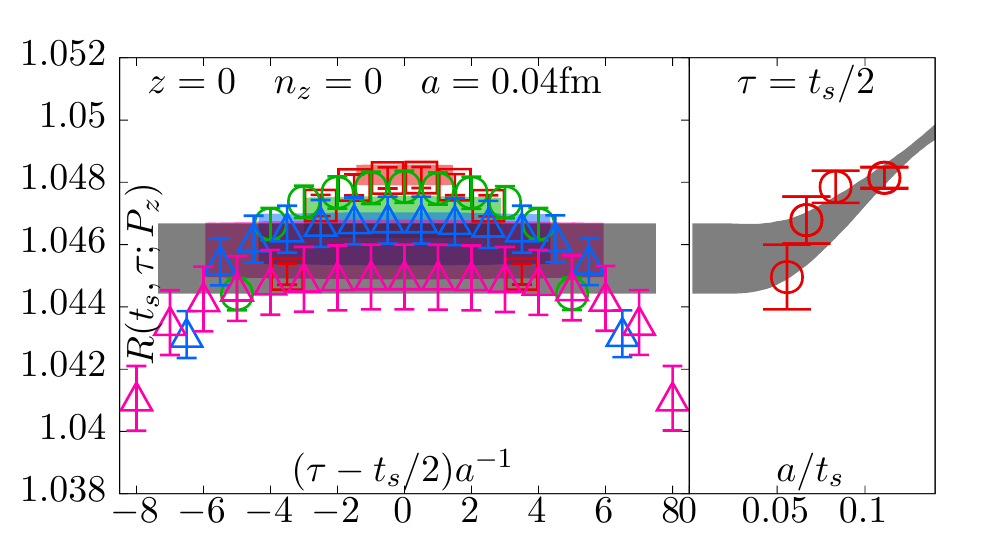}
\includegraphics[scale=0.58]{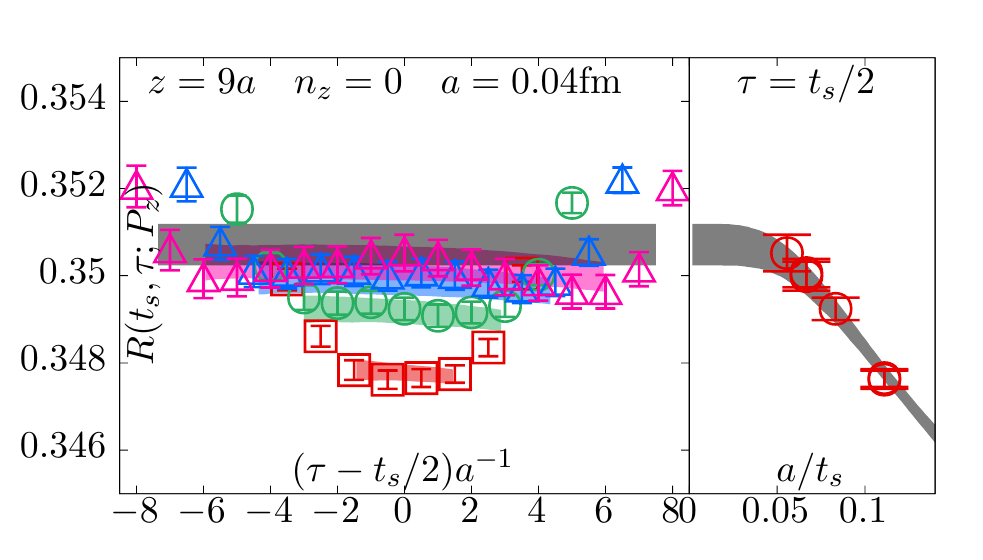}
\includegraphics[scale=0.58]{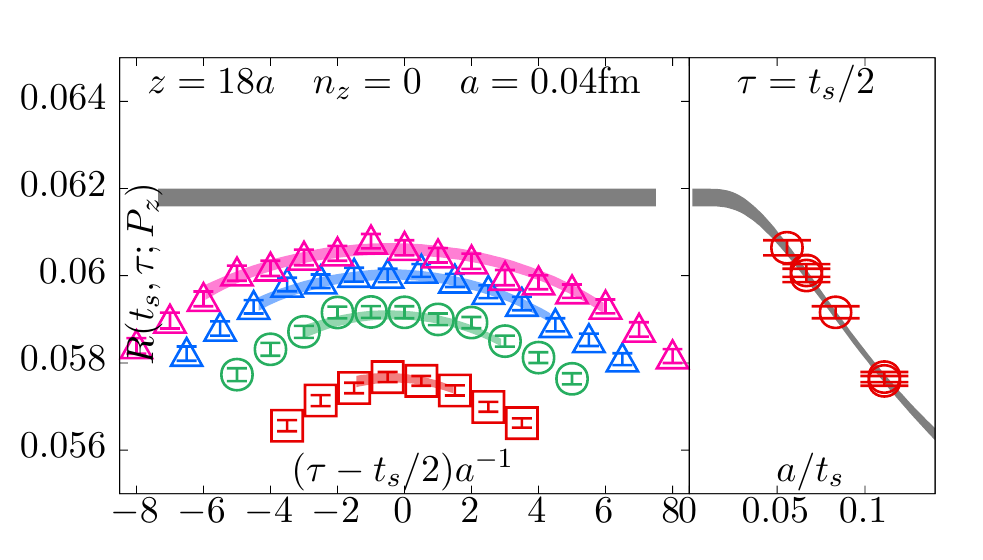}

\includegraphics[scale=0.58]{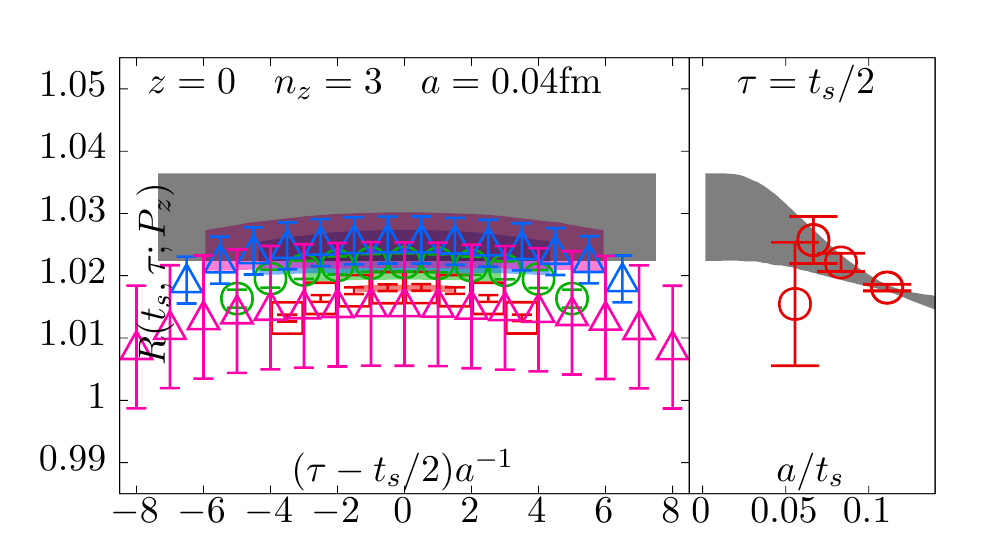}
\includegraphics[scale=0.58]{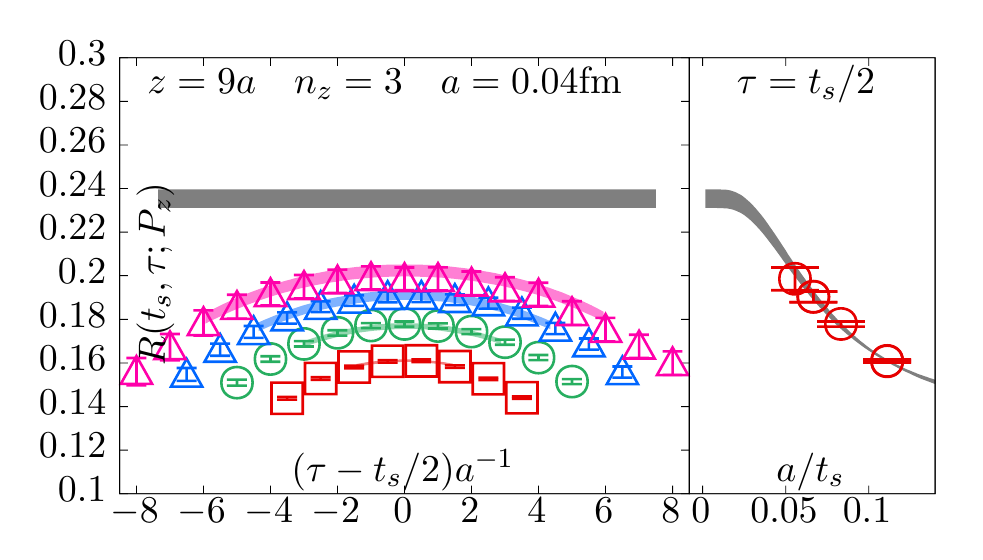}
\includegraphics[scale=0.58]{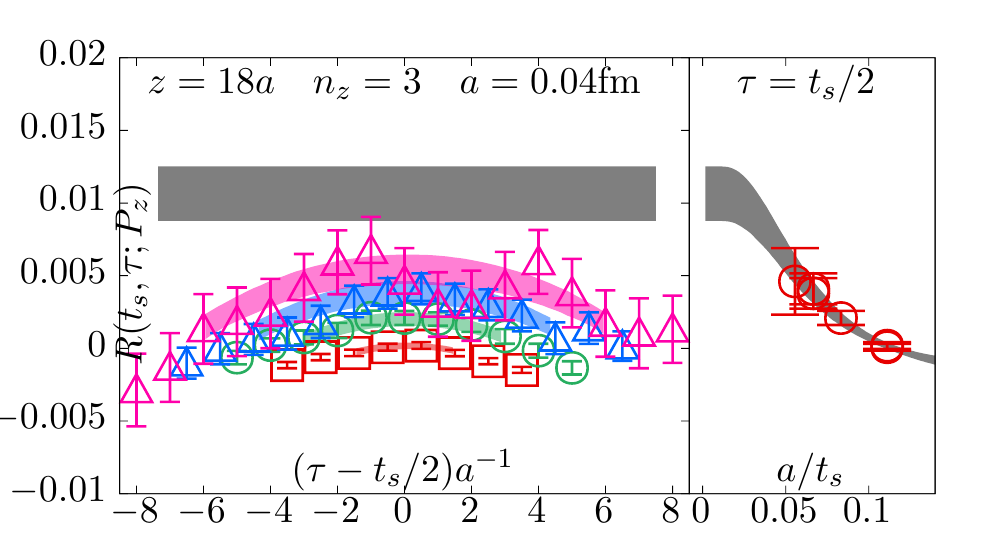}

\includegraphics[scale=0.58]{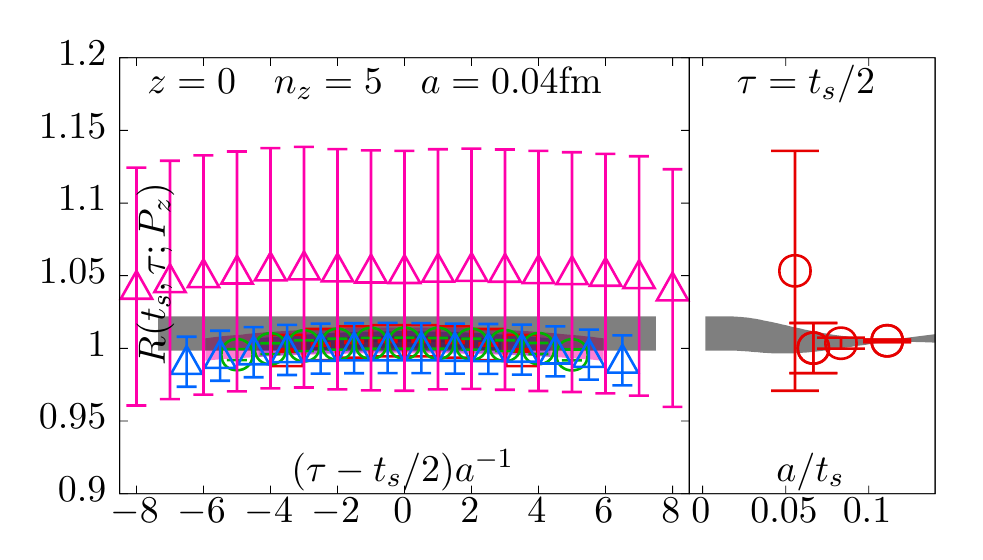}
\includegraphics[scale=0.58]{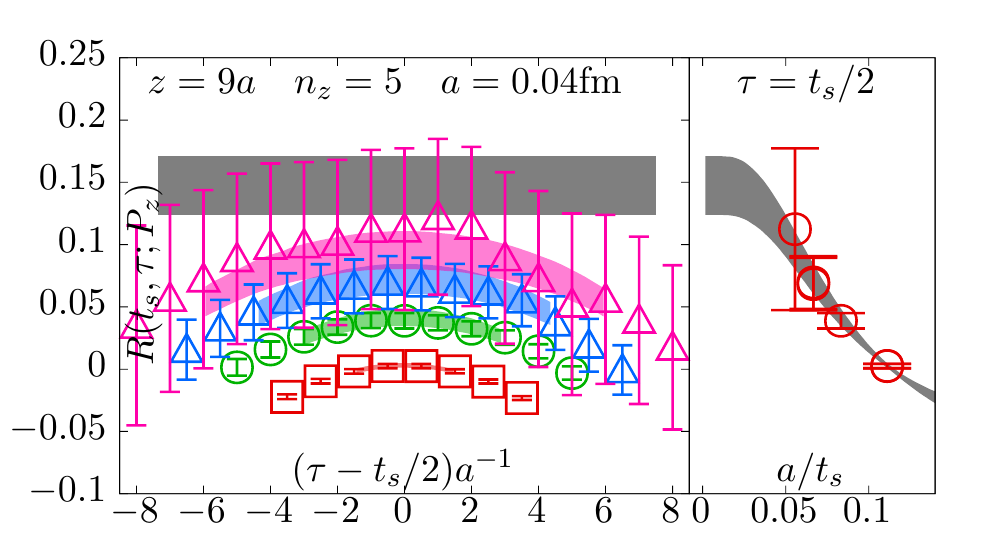}
\includegraphics[scale=0.58]{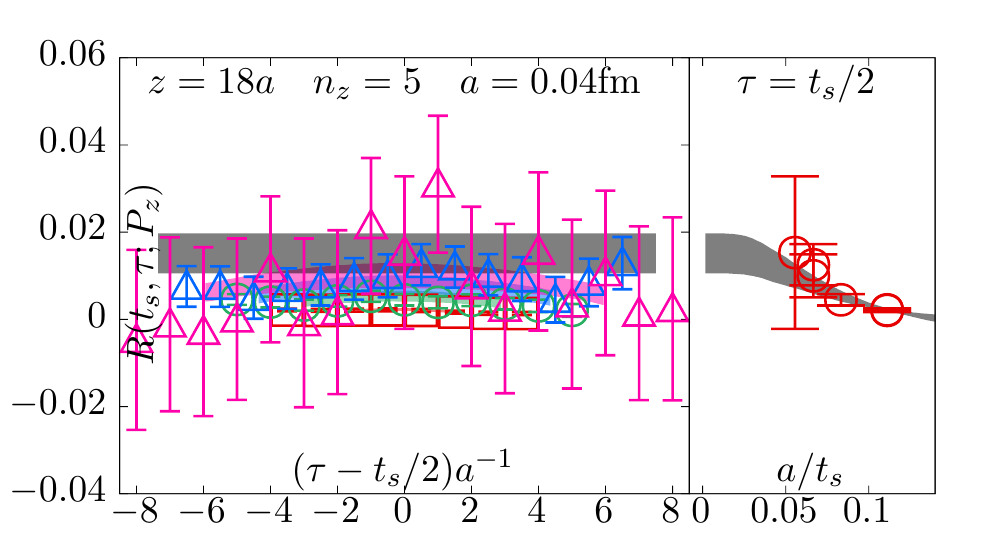}

\caption{
    The source-sink $t_s$ and operator insertion $\tau$ dependence
    of the ratio $R(t_s,\tau; z,P_z)$, at fixed $z$ and $P_z=2\pi
    n_z/L$ are shown for the lattice spacing $a=0.04$ fm.  The
    top-rows are for $n_z=0$, middles ones for $n_z=3$ and bottom
    ones for $n_z=5$. 
    The left panels are for $z=0$, middle panels
    for $z=9a$ and right ones for $z=18a$ respectively. Each plot
    has left and right sub-panels.  In the left sub-panels, the
    $t_s-\tau/2$ dependence is shown at $t_s=9a$ (red squares),
    $12a$ (green circles), $15a$ (blue triangles) and $18a$ (pink
    inverted-triangles).
    The corresponding colored bands are the 1-$\sigma$
    errorbands from Fit$(2,3)$ (see text).
    On the right sub-panels, the extrapolation (grey band) to
    $t_s\to\infty$ is shown as a function of $a/t_s$ at fixed
    $\tau=t_s/2$.
}
\eefs{3ptfitsa4}

From the three-point function and the two-point function, the central
quantity from which the bare matrix element can be obtained from,
is the ratio
\beq
R(t_s,\tau;z,P_z)\equiv \frac{C_{\rm 3pt}(t_s,\tau;z,P_z)}{C_{\rm 2pt}(t_s;P_z)}.
\eeq{rdef}
In order to take care of the wrap-around effect due to the finite
temporal lattice extent $L_t$, we replace $C_{\rm 2pt}(t_s;P_z)$
with $C_{\rm 2pt}(t_s;P_z)-A_0\exp\left(-E_0 (a L_t-t_s)\right)$
where $A_0$ and $E_0$ are the amplitude and energy of the ground
state obtained via fits to the two-point function in the last
section. This is especially important to take care of at $P_z=0$
on our lattices. In the above equation, the variables are $t_s$ and
$\tau$ at fixed $z$ and $P_z$, and hence we will keep $z$ and $P_z$
implicit in the discussion of $R$ below.  Through the spectral
decomposition of $R$, it is easy to see that~\footnote{Wrap-around 
effects in three-point function are ignored in the expression. 
We discuss this in \apx{wrap}.}
\beqa
&&R(t_s,\tau)=\cr
&&\frac{\sum^N_{n,n'}A_nA_{n'}^* \mel**{E_n,P}{O_{\gamma_t}(z)}{E_{n'},P)} e^{-(E_{n'}-E_n)\tau-E_n t_s}}{\sum^N_m |A_{m}|^2 e^{-E_m t_s}}.\cr &&
\eeqa{spect}
with $E_{n+1} \ge E_n$, $E_0=E_\pi$ and $A_n=\mel**{\Omega}{\pi}{\pi}$.
In the infinite $t_s$ limit, $R(t_s, \tau; z,P_z)$ is equal to the
bare matrix element
$h^B(z,P_z)=\mel**{\pi}{O_{\gamma_t}(z)}{\pi}$. In practice,
we obtain $h^B(z,P_z)$ by fitting the right-hand side of \eqn{spect}
to the $t_s$ and $\tau$ dependence of the lattice data for the ratio
$R$.  The fit parameters are the matrix elements
$\mel**{E_n,P}{O_{\gamma_t}(z)}{E_{n'},P)}$.  We take
fixed values of $E_n$ and $A_n$ from our analysis of $C_{\rm 2pt}$ that we discussed in 
the last section; 
namely, in two state fits, values of $E_1,A_1$ were taken from the 
pseudo-plateau seen in \fgn{64c64E1} that covers the typical 
range of $\tau$ used here, while in the three state fits,
the values of $E_1,A_1$ were fixed to the actual dispersion
values of $\pi_1$ and $E_2,A_2$ effectively captured the tower of higher excited states. 
We truncated the number of states $N$ entering the fit ansatz in
\eqn{spect} at $N=2$ and 3.  To reduce the excited state contamination,
we excluded cases where operator insertion is too close to either
the source or sink by using only values of $\tau\in [n_{\rm
sk}a,t_s-n_{\rm sk}a]$.  We used $n_{\rm sk}=1,2$ for $N=3$ and
$n_{\rm sk}=2,3$ for $N=2$.  We denote such $N$-state fits as
\texttt{Fit}$(N,n_{\rm sk})$.

In \fgn{3ptfitsa6} and \fgn{3ptfitsa4}, we show some sample results
of the extrapolations using \texttt{Fit}$(2,3)$ for the $a=0.06$
fm and $a=0.04$ fm lattices respectively.  Each panel in the plot
has two sub-panels. Let us first focus on the larger left sub-panels
which show the dependence of $R(\tau,t_s)$ on $\tau-t_s/2$. The
lattice data for $R$ are shown as the symbols with the colors
distinguishing the different $t_s$.  For the $a=0.06$ fm lattice,
we used $t_s/a=8,10$ and 12 (i.e., $t_s=0.48$ fm, 0.6 fm and 0.72
fm) in the fits. Similarly, we used $t_s/a=9,12,15$ and 18 for
$a=0.04$ fm ensemble, which corresponds to similar physical values
of $t_s=0.36$ fm, 0.48 fm, 0.6 fm and 0.72 fm respectively. Along
with the data for $R(t_s,\tau)$, we have also shown the results
from \texttt{Fit}$(2,3)$ as the similarly colored bands. The result
for the matrix element $h^B$, i.e., $t_s\to\infty$ limit of the
fit, is shown by the grey horizontal band in the figures. The degree
to which extrapolation differs from the actual data in the range
of $t_s<1$ fm can be seen from the smaller right sub-panels, where
we have shown the $1/t_s$ dependence of the data (points) as well
as the fit (grey band) with $\tau=t_s/2$, the maximal distance of
operator from source and sink.  In general, one can see that the
extrapolations get steeper as the value of $z$ increases. However,
given the small errors at smaller $z$, the extrapolation again plays
a significant role at smaller $z$. From the agreement of the two-state
fits with the actual data, one can gain confidence in the extrapolations.

\bef
\centering
\includegraphics[scale=0.71]{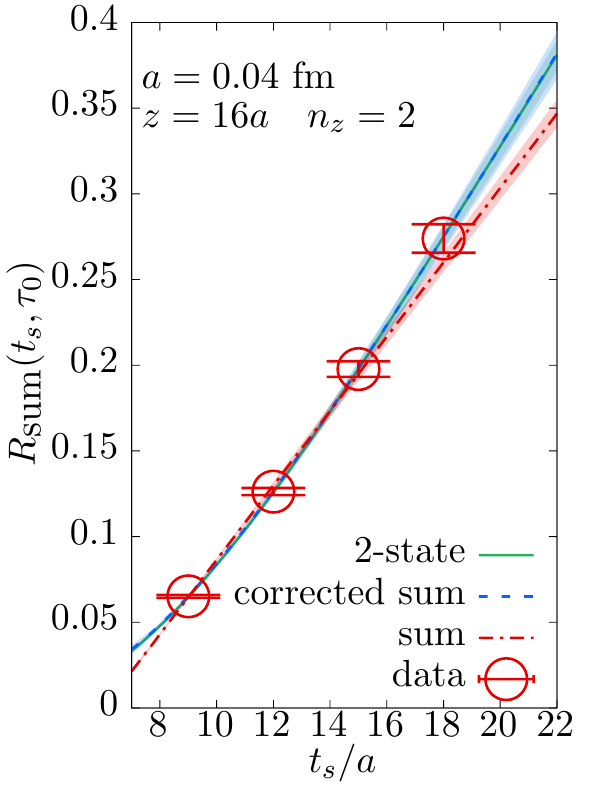}
\includegraphics[scale=0.71]{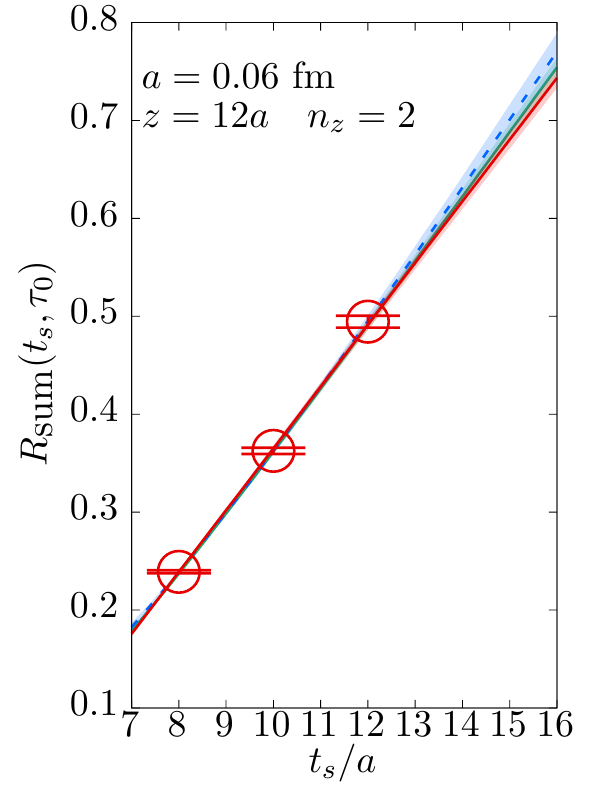}
\caption{
    An example for summation method, Sum$(\tau_0)$, that uses only
    insertion points $\tau>\tau_0=2a$ are shown as a function of
    $t_s/a$ at fixed $z$ and $P_z$. The left panel is at $a=0.04$
    fm with $z=16a$ and $n_z=2$. The right panel is at $a=0.06$ fm
    with $z=12a$ and $n_z=2$.  In each panel, the red circles are
    the lattice data. In addition, there are three different curves
    --- the red one is the straight line summation fit to the data,
    the blue one is the {\sl corrected} summation fit \texttt{SumExp} that includes
    $\exp\left[-(E_1-E_0)t_s\right]$ correction, and the green one
    is the expected summation curve from the two-state fit Fit$(2,3)$.
}

\eef{sumex}

In addition to the $N$-state fits, which are sensitive to the values
of $E_n,A_n$, we also used the summation technique~\cite{Maiani:1987by}
which does not require inputs of the spectral details of the two-point
function. For this, we use the standard definition
\beq
  R_{\rm{sum}}(t_s) =
  \sum_{\tau=n_{\rm sk}a}^{t_s - n_{\rm sk}a}R(t_s, \tau).
\eeq{sumdef}
For large $t_s$, we would find a linear behavior in $t_s$ of $R_{\rm sum}$ as
\beq
R_{\text{sum}}(t_s) =
   (t_s - 2 n_{\rm sk}a) h^B(z,P_z)+B_0+{\cal O}(e^{-(E_1-E_0) t_s}).
\eeq{sumfit}
We refer to this method where we ignore ${\cal O}(e^{-(E_1-E_0) t_s})$ corrections
and fit only $h^B(z,P_z)$ and $B_0$ as \texttt{Sum}($n_{\rm sk}$). 
Since our source-sink separations are less than $1$ fm, we also included 
the additional $e^{-(E_1-E_0) t_s}$ correction in the fitting ansatz as
\beq
R_{\text{sum}}(t_s) =
   (t_s - 2 n_{\rm sk}a) h^B(z,P_z)+B_0+B_1 e^{-(E_1-E_0) t_s}.
\eeq{sumfit2}
We refer to this method as \texttt{SumExp}($n_{\rm sk}$). In
\fgn{sumex}, we show a sample result for the summation fits. In the
left and right panels of the figure correspond to $a=0.04$ fm and
0.06 fm lattice ensembles.  We have used momenta $P_z=2\pi n_z/(La)$
with $n_z=2$ in both the cases at an intermediate separation $z=0.72$
fm in both ensembles. The lattice data for $R_{\rm sum}$ are shown
as the red circles.  The result from a linear fit to the data is
shown as the red band. The slope of the fit is the estimator of the
matrix element $h^B$. One can see in both the cases that the straight
line fit is able to describe the data. However, one can certainly
see deviations from the straight line fit at $t_s=18a$ for the
$a=0.04$ fm case. For comparison, the expectation for $R_{\rm
sum}(t_s)$ from the 2-state fit described above is shown as the
green band. Here, the curve is able to describe the data at all
$t_s$ well and can be seen be seen to approach a straight line with
larger slope only for $t_s > 0.72$ fm. In order to account for these
discrepancies, we also show the result from \texttt{SumExp} as the
blue dashed line. This result does deviate from the simple \texttt{Sum}
and agrees better with the expected result from \texttt{Fit}. This
shows that there are residual ${\cal O}(e^{-(E_1-E_\pi) t_s})$
effects which cannot be ignored in the summation fits in the ranges
of $t_s$ we are working with. While we have picked an example case
where we observe this discrepancy to be larger, similar discrepancy
could be seen in other values of $P_z$ and $z$ as well in the case
of $a=0.04$ fm data. The  \texttt{Sum} data agreed better with
expectation from \texttt{SumExp} and \texttt{Fit} for the $a=0.06$
fm data.  Therefore, we use the results from \texttt{Fit}, and only
use \texttt{Sum} and \texttt{SumExp} to serve as cross-checks on
the results.

\befs
\centering
\includegraphics[scale=0.72]{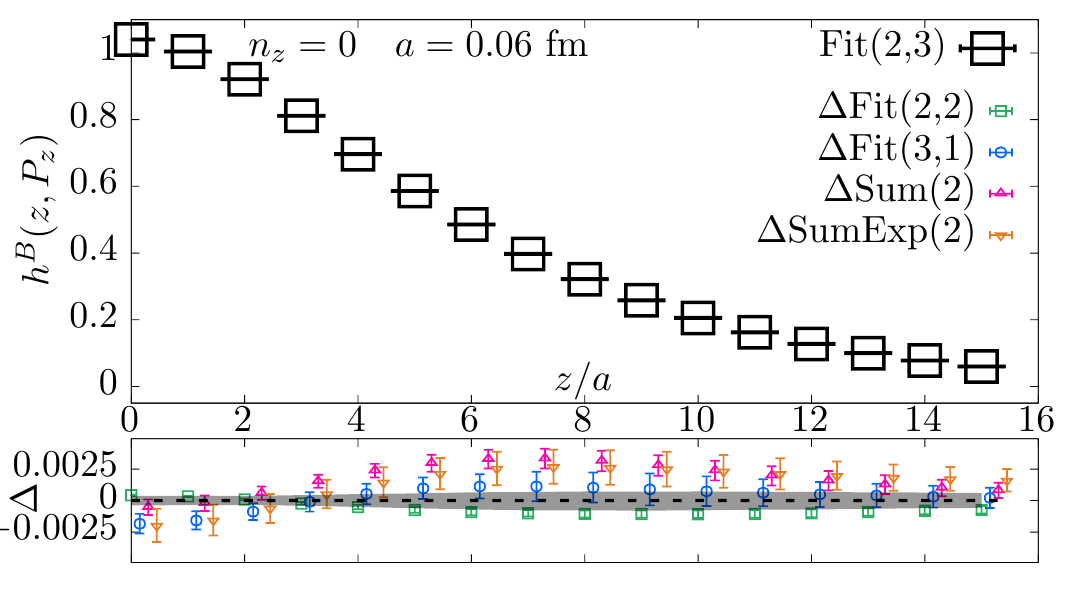}
\includegraphics[scale=0.72]{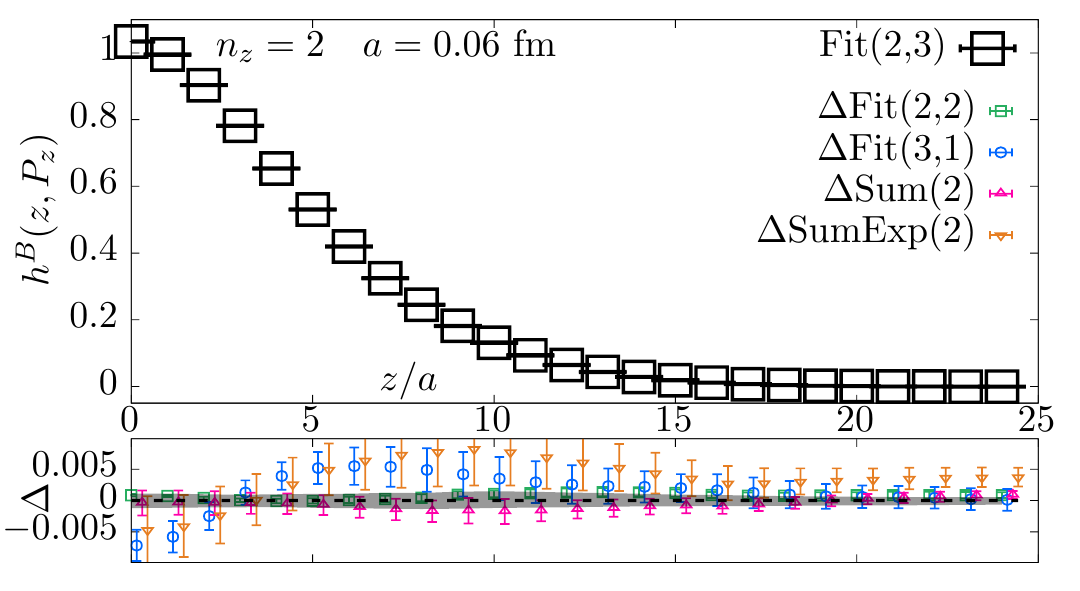}
\vskip 0.5cm
\includegraphics[scale=0.72]{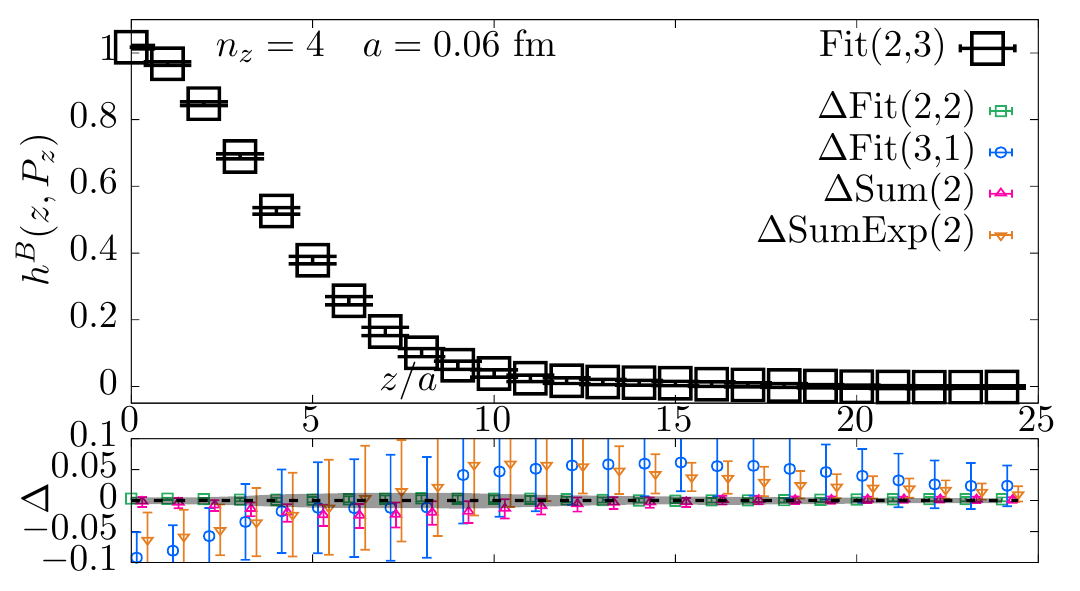}
\includegraphics[scale=0.72]{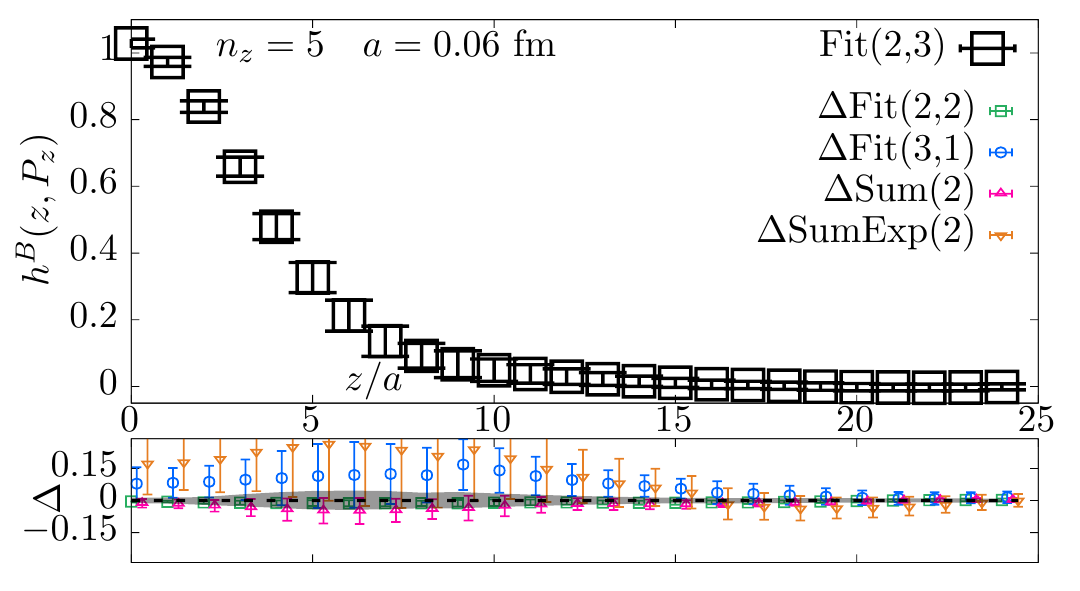}

\caption{
    The bare matrix elements $h^B(z,P_z)$ from excited state
    extrapolations are shown as a function of $z/a$ for the $a=0.06$
    fm ensemble. The results from $n_z=0, 2, 4$ and 5 are shown in
    the top-left, top-right, bottom-left and bottom-right are shown.
    The top part of each panel shows the $z/a$ dependence using
    two-state extrapolation Fit(2,3) using $t_s/a=8, 10$ and 12.
    The bottom part of each panel shows the deviation, $\Delta$,
    of the different extrapolation methods Fit(2,2), Fit(3,1),
    Sum(2), SumExp(3) from the method Fit(2,3). The scatter of these
    differences from 0 (shown by dashed line) characterizes the
    robustness of the extrapolation.
}
\eefs{hba06}

\befs
\centering
\includegraphics[scale=0.74]{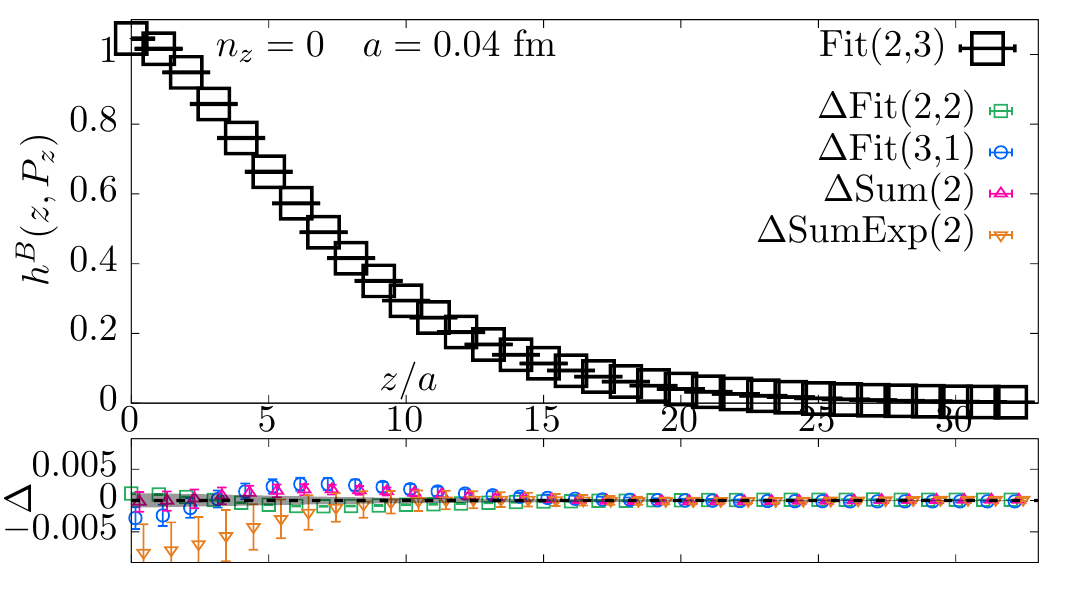}
\includegraphics[scale=0.74]{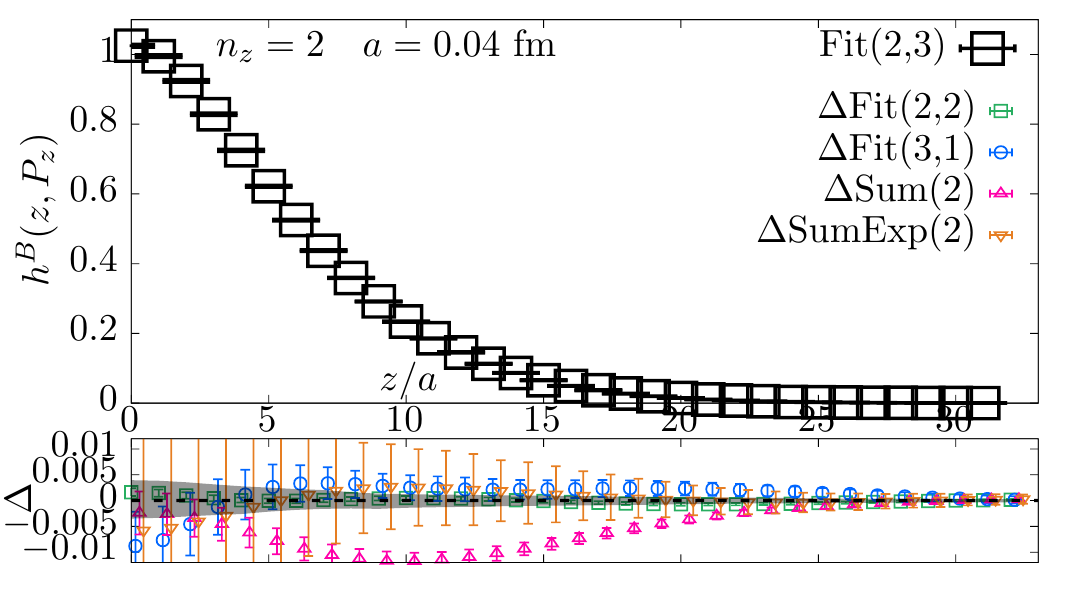}
\vskip 0.5cm
\includegraphics[scale=0.74]{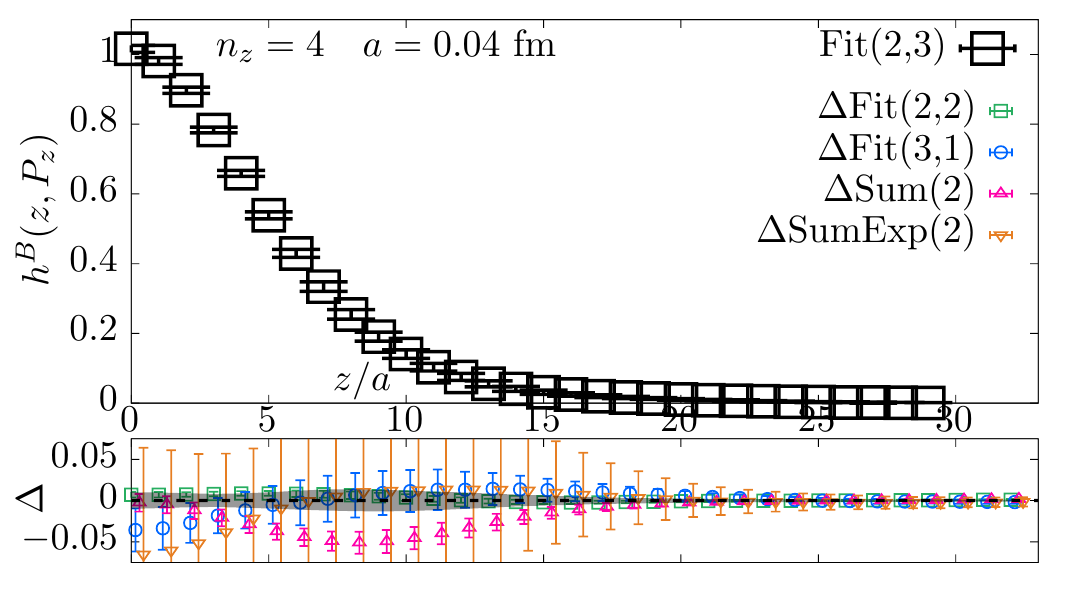}
\includegraphics[scale=0.74]{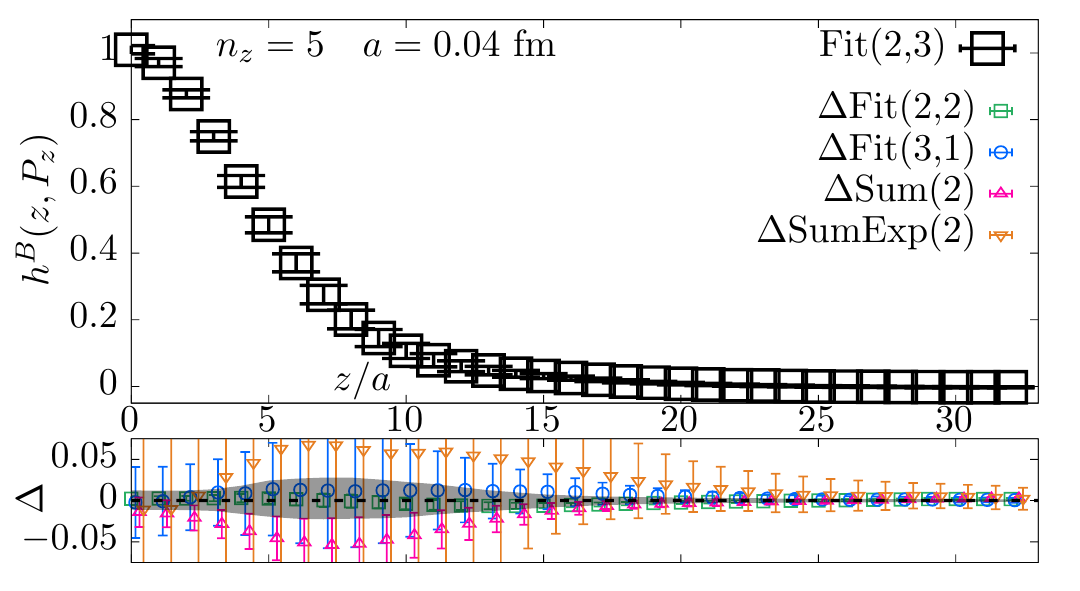}
\caption{
    The bare matrix elements $h^B(z,P_z)$ from excited state
    extrapolations are shown as a function of $z/a$ for the $a=0.04$
    fm ensemble. The results from $n_z=0, 2, 4$ and 5 are shown in
    the top-left, top-right, bottom-left and bottom-right are shown.
    The top part of each panel shows the $z/a$ dependence using
    two-state extrapolation Fit(2,3) using $t_s/a=9, 12, 15$ and 18.
    The bottom part of each panel shows the deviation, $\Delta$,
    of the different extrapolation methods as explained in \fgn{hba06}.
    Deviations of results from Sum(2) from 2- and 3-state fit results
    are seen. But we find that {\sl corrected} sum SumExp(2) is
    consistent with the fit results. This is a result of the
    observation in \fgn{sumex}.
}
\eefs{hba04}

As we demonstrated above, the $t_s\to\infty$ extrapolations lead
to values of $h^B$ which are not simply obtained from plateau
values of $R(t_s,\tau)$ even for the largest $t_s=0.72$ fm we use.
Therefore, a way to reasonably justify the correctness of our
extrapolations is by adapting the multiple fitting schemes, namely
\texttt{Fit}, \texttt{Sum} and \texttt{SumExp}, and show consistency
among them.  This is what we show in \fgn{hba06} and \fgn{hba04},
for the $a=0.06$ fm and 0.04 fm lattices respectively. The different
panels show the results for four different values of $P_z=2\pi
n_z/(La)$. In the top part of the different panels, we have shown
the bare matrix element $h^B(z,P_z)$, obtained by \texttt{Fit}(2,3)
as the black open squares, as a function of the length of Wilson-line
$z$. Since we are working with iso-triplet matrix element for the
pion, only the real part of $h^B$ is non-zero. One should remember
that the bare matrix element at any finite lattice spacing has the
Wilson-line self-energy divergence, $\exp(-c z/a)$, which causes
the rapid decay of $h^B(z,P_z)$ as a function of $z$ in the figures.
With the increased statistics used in our computation, one can note
that we are able to obtain matrix elements with good signal to noise
ratio even up to momenta corresponding to $n_z=5$ in both the lattice
spacings. Below the top part of each panel in \fgn{hba06} and
\fgn{hba04}, we show the deviations, $\Delta(z)$, of different
extrapolation methods from values obtained with \texttt{Fit}(2,3)
as a function of $z$. That is,
\beq
\Delta(z)\equiv h^B_{\rm method}(z,P_z)-h^B_{\texttt{Fit}(2,3)}(z,P_z),
\eeq{deltadef}
where $h^B_{\rm method}$ is the bare matrix element obtained using
an extrapolation technique {\sl method}, which could be \texttt{Fit}(2,2),
\texttt{Fit}(3,2), \texttt{Sum}(2), or \texttt{SumExp}(2), in
\fgn{hba06} and \fgn{hba04}. If the extrapolations are perfect,
then we would find $\Delta(z)$ to be consistent with zero at all
$z$ and $P_z$. For comparison, we also show the statistical error
in $h^B_{\texttt{Fit}(2,3)}(z,P_z)$ as the grey error band along
with the values of $\Delta(z)$. For $a=0.06$ fm case shown in
\fgn{hba06}, we find $\Delta(z)$ is consistent with zero within
error for larger $P_z$ while there is little tension at smaller
$P_z$ in the top two panels. The small but visible deviations of
\texttt{Fit}(3,2) is less than $2\sigma$. The deviation of
\texttt{Sum}(2) is comparatively larger, but when we supplement
\texttt{Sum}(2) with the exponential corrections, i.e., \texttt{SumExp}(2),
the $\Delta(z)$ moves towards zero and becomes consistent with zero.
This again points to the importance of excited state effects that
cannot be neglected in summation fits on our lattices. This effect
is more apparent in the case of $a=0.04$ fm lattice shown in
\fgn{hba04}. Thus we understand the deviation of \texttt{Sum} from
the rest as an excited state effect, and we find that the
\texttt{Fit}(2,3), \texttt{Fit}(2,2), \texttt{Fit}(3,2) and
\texttt{SumExp}(2) are all consistent among themselves. Thus, we
are able to demonstrate the goodness of our extrapolations. Henceforth,
we will use \texttt{Fit}(2,3) for both $a=0.04$ fm and 0.06 fm
ensembles in discussing our further analysis.

\bef
\centering
\includegraphics[scale=0.9]{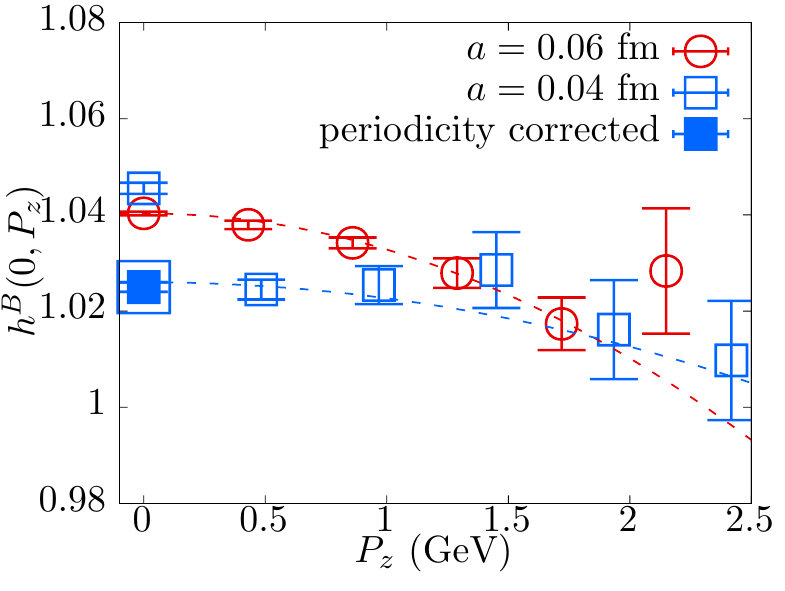}
\caption{
    The result for the local bare matrix element $h^B(z=0,P_z)$ is shown
    as a function of $P_z$. The red and blue open symbols are the 
    estimates of the bare matrix elements on $a=0.06$ fm and $a=0.04$ fm
    lattices.  The estimated value of $h^B(z=0,P_z=0)$ for $a=0.04$ fm
    after correcting for wrap-around effect (see text) is shown as the 
    filled blue square. The red and the blue dashed curves are the 
    modeled lattice spacing effects using an ansatz, 
    $h^B(z=0,P_z)=h^B(z=0,P_z=0)+b (P_z a)^2$ for $a=0.06$ fm
    and $0.04$ fm respectively, with fixed $b=-0.0813$ in both cases.
}
\eef{z0dep}

A well determined matrix element that can be used to cross-check
our results is the value of bare matrix element at $z=0$, which in
the continuum limit will be the total isospin of pion, which is 1.
At any finite $a$, the bare matrix element suffers from ${\cal
O}(\alpha_s(\mu=a^{-1}))$ correction to 1, which under finite
renormalization will be canceled by $Z_V$.  If the excited-state
extrapolations were perfect and the finite volume effects were
negligible, the estimates of $h^B(z=0,P_z)$ cannot change with $P_z$
up to possible finite $a$ corrections at non-zero $P_z$.  In order
to check for this,  we show the behavior of $h^B(z=0,P_z)$ as a
function of $P_z$ in \fgn{z0dep}. For $a=0.06$ fm lattice, the value
of $h^B(0,0)$ is 1.0404(4) and the values of $h^B(0,P_z)$ get smaller
than this value gradually at larger $P_z$, albeit only by less than
2\% by $P_z=2.15$ GeV. This $P_z$ dependence is likely to arise due
to increasing lattice spacing effect at higher momenta, and
empirically, it was possible to fit the $P_z$ dependence to an
ansatz $h^B(z=0,P_z)=h^B(z=0,P_z=0)+b (P_z a)^2$.

For $a=0.04$ fm lattice, the value of $h^B(0,0)$ is 1.045(1) which
is higher than value of $h^B(0,0)$ at $a=0.06$ fm.
However, one expects $h^B(0,0)$ to decrease and approach 1 as $a\to 0$~\cite{Bhattacharya:2015wna}.
One observes a sharp decrease in the value of matrix elements at
non-zero $P_z$ to values around 1.025 and changes little with
$P_z>0$.  We were able to understand this anomalous behavior at
$P_z=0$ to arise from larger periodicity effects ($\sim
e^{-M_\pi(L_t-t_s)}$) in the $P_z=0$ three-point function for the
finer $a=0.04$ fm lattice (which is in addition to such wrap-around
effects in two-point function that we corrected for in \eqn{rdef}).
We discuss this further in \apx{wrap}, and we estimate the value
of $h^B(0,0)$ after correcting for the wrap-around effect to be
1.024(1). For the $a=0.06$ fm case, this effect is negligible.  The
(approximate) corrected estimate for $h^B(0,0)$ is shown as the
filled blue square in \fgn{z0dep}, which shows surprisingly good
agreement with the estimates at other non-zero $P_z$. We used the
same fitted $(P_z a)^2$ ansatz that we discussed above, with only
the value of $a$ changed from 0.06 fm to 0.04 fm, and the result
is shown as the blue dashed curve in \fgn{z0dep}. This nice agreement
gives credence to our explanation of lattice spacing effect being
the cause of the mild $P_z$ dependence in $a=0.06$ fm $h^B(0,P_z)$
estimates and the even milder $P_z$ dependence in $a=0.04$ fm
estimates.  We discuss the estimation of $Z_V$ within the RI-MOM
framework in \apx{zv} which give results consistent with the values
from the bare pion matrix element in \fgn{z0dep}.

\section{Renormalization}\label{sec:ren}

The bare matrix element $h^B(z,P_z)$ obtained in the last section
needs non-perturbative renormalization in order for it to have a
well defined continuum limit.  The non-perturbative renormalization
removes the UV self-energy divergence of the Wilson-line which is
inherently non-perturbative and can only be captured by methods
such as the ab-initio lattice QCD heavy-quark potential computations
(c.f.~Ref~\cite{Bazavov:2018wmo} for the ensembles used here).  With
the removal of this non-perturbative piece, one would expect the
remaining renormalized matrix element to be describable within the
perturbative large momentum effective theory framework. Therefore, a judicious
choice of the nonperturbative renormalization scheme for the bilocal quark bilinear 
operator that is implementable on an Euclidean lattice and at the same time
reduces the higher-twist corrections to the matrix element in any given small
values of $z$ is important.

RI-MOM is one such renormalization scheme that uses renormalization
conditions at off-shell space-like external quark four-momentum
$P^R$. A more careful description of the calculation of RI-MOM
factor as applied to our work can be found in~\cite{Izubuchi:2019lyk}.
The RI-MOM renormalized matrix element is defined as
\beq
h^R_0(z,P_z,P^R)=Z_q Z_{\gamma_t\gamma_t}(z,P^R) h^B(z,P_z),
\eeq{rime0}
where $Z_q$ is the quark wavefunction renormalization factor
(c.f.~Ref~\cite{Alexandrou:2010me}) and $Z_{\gamma_t\gamma_t}$ is
the renormalization factor for ${\cal O}_{\gamma_t}(z)$ defined via
the condition imposed using the amputated matrix element evaluated with quark external
states at momentum $p$, $\Lambda(p)$, as
\beq
Z_{\gamma_t\gamma_t}(z,P^R){\rm Tr}\left(\slashed{p}\Lambda(p)\right)_{p=P^R}\equiv 12 P^R_t e^{i P^R_z z}.
\eeq{zricond}
The above condition is referred to as the $\slashed{p}$-projection
scheme within the RI-MOM scheme~\cite{Stewart:2017tvs,Chen:2017mzz}.
The operator $O_{\gamma_t}$ does not mix with any other
operator, unlike $O_{\gamma_z}$~\cite{Constantinou:2017sej,Chen:2017mie}.  We used
the Landau gauge fixed configurations to determine
$Z_{\gamma_t\gamma_t}(z)$ non-perturbatively in both $a=0.06$ fm and
$a=0.04$ fm ensembles. We will refer to the component of $P^R$ along
the direction of Wilson-line as $P^R_z$ and the norm of the component
perpendicular to $z$-direction as $P^R_\perp$. Since the value of
$h^R_0(z=0,P_z,P^R)=1$ for the pion, we impose this condition through
a redefinition
\beq
h^R(z,P_z,P^R)\equiv \frac{h^R_0(z,P_z,P^R)}{h^R_0(0,P_z,P^R)}.
\eeq{rime}
This implicitly takes care of the effect of $Z_q$ and at the same
time reduces the statistical errors in $h^R$ at the other non-zero
values of $z$ through their correlation with $h^R(z=0)$.

\bef
\centering
\includegraphics[scale=0.7]{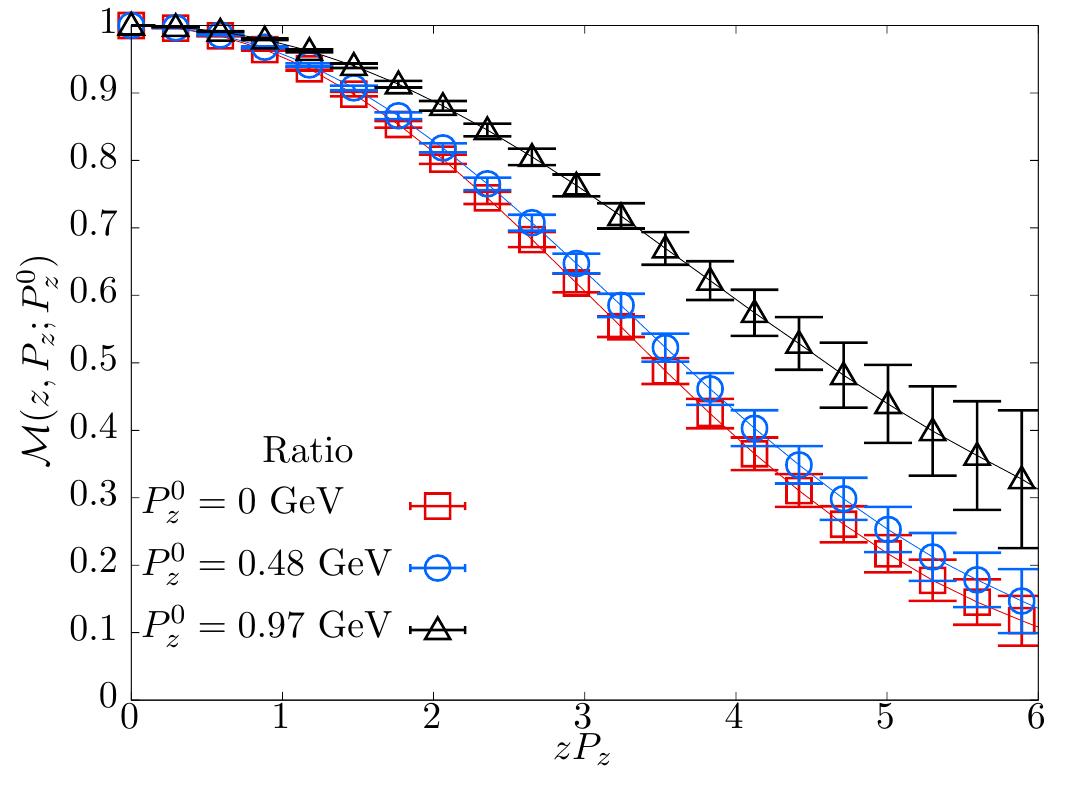}
\caption{
Comparison of renormalized matrix elements at fixed $P_z=1.45$ GeV
in the ratio scheme with generalized non-zero values of the reference
momentum $P_z^0$. The different colored symbols correspond to
different $P_z^0=0,0.48$ and 0.97 GeV.  The effect of changing
$P_z^0$ is significant, but it does not cause a big difference in
signal-to-noise ratio at smaller $z$ that we are interested in. It
is the expectation that matrix elements with $P_z,P_z^0 >\{ \Lambda_{\rm QCD},m_\pi\}$
suffer from lesser higher-twist contamination.
}
\eef{ratgen}

Instead of using quark external states, it is possible to cancel
the UV divergence in $h^B(z,P_z)$ using the pion matrix element at
a different fixed reference momentum $P_z^0$, that is, $h^B(z,P_z^0)$.
Such a procedure to remove the UV divergences via renormalization
group invariant ratios is referred to as the {\sl ratio}
scheme~\cite{Izubuchi:2018srq,Ji:2020ect}. With this, we can define
a renormalized matrix element,
\beq
{\cal M}_0(z,P_z,P_z^0)=\frac{h^B(z,P_z)}{h^B(z,P^0_z)}.
\eeq{ratme0}
The choice $P_z^0=0$ has been used in literature and the resulting
matrix element ${\cal M}(z,P_z,0)$ is also referred to as the {\sl
reduced} ITD~\cite{Orginos:2017kos,Izubuchi:2018srq}.  Non-zero
$P_z^0$ was applied to proton in~\cite{Fan:2020nzz}.  Similar to
the RI-MOM matrix element, we can reduce the statistical errors by
redefining the matrix element as
\beq
{\cal M}_0(z,P_z,P_z^0)\rightarrow {\cal M}(z,P_z,P_z^0)=\frac{{\cal
M}_0(z,P_z,P_z^0)}{{\cal M}_0(0,P_z,P_z^0)},
\eeq{ratme}
so that the condition ${\cal M}(z,P_z,P_z^0)=1$ is automatically
fulfilled.
We use values of $P_z>P_z^0$ in this work. The preference for using
$P_z,P_z^0>{\Lambda_{\rm QCD}}$ will become clearer with the
discussion on perturbative matching in the next section. In
\fgn{ratgen}, we compare the result of ${\cal M}(z,P_z,P_z^0)$ for
three different $P_z^0=2\pi n_z^0/(La)$ for $n_z^0=0, 1$ and 2 on
$a=0.06$ fm lattice.  These values of $n_z^0$ correspond to 0, 0.48
and 0.97 GeV respectively, and thus using even the lowest $n_z^0$
available makes sure $P_z^0 > \Lambda_{\rm QCD}$.  The effect of
using $P_z^0$ as a new scale leads to significant changes
to the $P_z z$ and $z^2$ dependence, which will be taken care of
the corresponding twist-2 expressions. But one should note that we
do not significantly compromise on the quality of signal by choosing
non-zero values of $P_z^0<1$ GeV, and hence, they are as good choices
of the the reference momentum scale in the ratio scheme as $P^0_z=0$ GeV.

\bef
\centering
\includegraphics[scale=0.7]{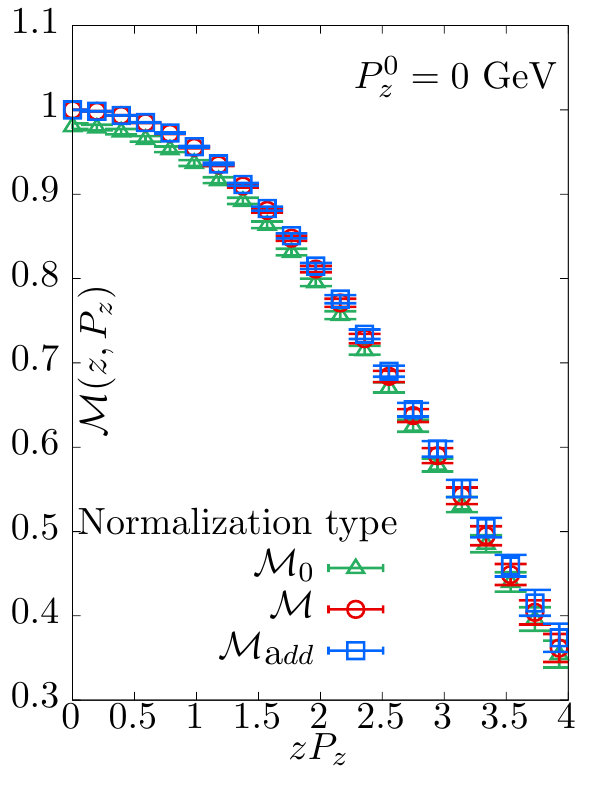}
\includegraphics[scale=0.7]{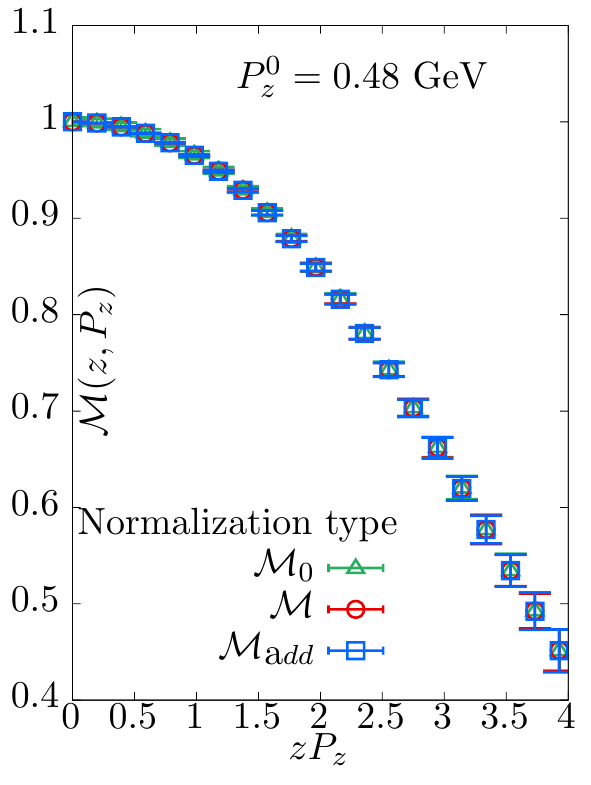}
\caption{
Quantifying the residual systematic effects after significant
statistical error reduction by the process of normalizing $z=0$
matrix element to 1 (see \eqn{rime} and \eqn{ratme}).  The plot
compares the different normalization types, ${\cal M}$ and ${\cal
M}_{\rm add}$, and the matrix element without any imposed normalization ${\cal
M}_0$.  In the above plot, we show results for the $a=0.04$ fm
ensemble.
}
\eef{ratnorm}

Our choice of the normalization conditions in \eqn{rime} and
\eqn{ratme} such that the value of pion matrix element at $z=0$ is 1, assumes
implicitly that our estimates of the matrix elements at $z=0$ do
not suffer from any systematic corrections.  In the discussion
around \fgn{z0dep}, we found about $1\%$ systematic errors at $z=0$
due to deviations of the matrix element as a function of $P_z$.
Below, we justify that the imposition of the normalization conditions
\eqn{rime} and \eqn{ratme} also reduces some of these systematic errors. Instead
of imposing the normalization multiplicatively as in 
\eqn{ratme}, an equally good choice is additively through
\beq
{\cal M}_{\rm add}(z,P_z,P_z^0) \equiv {\cal M}_0(z,P_z,P_z^0)-{\cal M}_0(0,P_z,P_z^0)+1.
\eeq{add}
The multiplicative and additive normalization are equivalent, only
provided ${\cal M}_0(0,P_z,P_z^0)$ is itself
exactly 1. In \fgn{ratnorm}, we compare the result of
${\cal M}_{\rm add}(z,P_z,P_z^0)$ and ${\cal M}(z,P_z,P_z^0)$ at
$P_z=1.29$ GeV on $a=0.04$ fm lattice. The left and right panels
are for $P_z^0=0$ and 0.48 GeV respectively.  For comparison, we
have also shown the matrix element ${\cal M}_0$ before imposing the
normalization.  First, one can note the error reduction due to the
normalization at smaller values of $z$.  As we discussed in the
last section, the $z=0$ matrix element at $P_z=0$ for $a=0.04$ fm
suffers from larger systematic effects than the rest. From the left
panel which shows the result for $P_z^0=0$, we surprisingly find
that the difference between ${\cal M}_{\rm add}$ and ${\cal M}$ is
absent within the errors at all $z$. On the right panel, which uses
$P_z^0=0.48$ GeV, the agreement is perfect between all the estimates
of ${\cal M}$. Through this, we demonstrated that the systematic
effects in our matrix element determination are further reduced due
to the ratios using the prior knowledge that the local matrix element at $z=0$ is 1
for pion.

\bef
\centering
\includegraphics[scale=0.6]{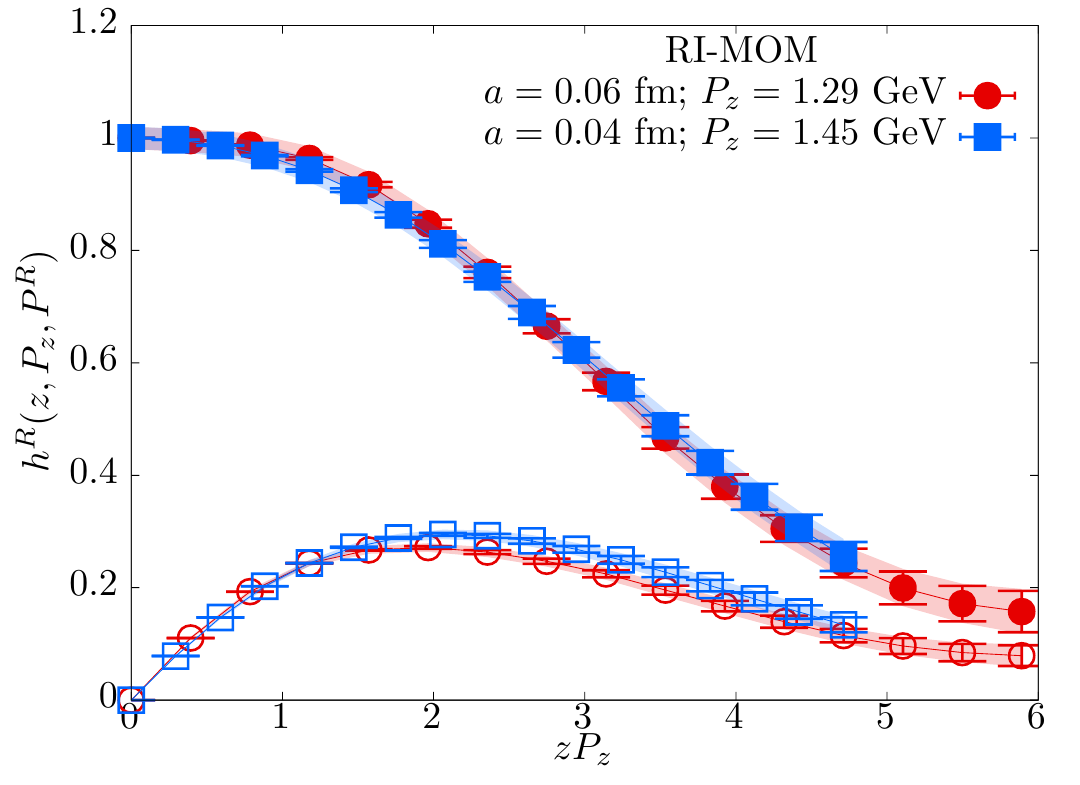}

\includegraphics[scale=0.6]{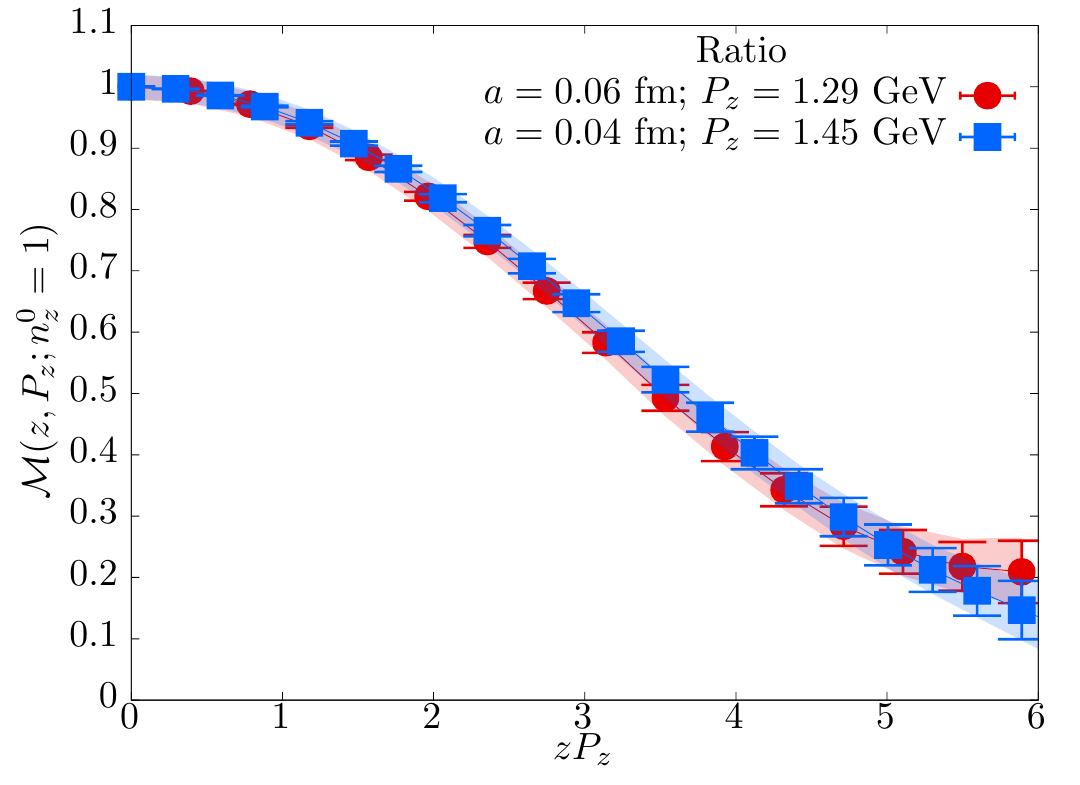}
\caption{
Comparison of renormalized matrix elements at two lattice spacings
$a=0.06$ fm (red circles) and 0.04 fm (blue squares) plotted as a
function of $\nu=z P_z$. The top panel uses RI-MOM
renormalization scheme.  The real and imaginary parts of the
renormalized RI-MOM matrix element are shown as closed and open
symbols.  The bottom panel uses ratio scheme with $n_z^0=1$. 
The variation of data by $\pm 2\%$ is shown as the red and blue 
bands.
}
\eef{ritdcompare}

Finally, we address the lattice corrections to the renormalized
matrix elements.  In \fgn{ritdcompare}, we have shown the comparison of
renormalized matrix elements at two lattice spacings $a=0.06$ fm (red
circles) and 0.04 fm (blue squares) plotted as a function of
$\nu=z P_z$. The top and bottom panels show the comparison
using RI-MOM and $n_z^0=1$ ratio scheme respectively.  Due to lattice
periodicity constraints, we could only choose pion momentum $P_z$
that are approximately the same at the two lattice spacings; namely,
$P_z=1.29$ GeV for $a=0.06$ fm and $P_z=1.45$ GeV for $a=0.04$
lattices.  By looking at the pion matrix elements at the two lattice spacing as a
function of $P_z z$, such small mismatch between $P_z$ should affect
results only logarithmically in this discussion.  For the RI-MOM
scheme, we have chosen a comparable set of renormalization momenta
$(P_z^R,P_\perp^R)=(1.93,2.23)$ GeV for $a=0.06$ fm lattice and
$(1.93,2.51)$ GeV for $a=0.04$ fm lattice. In the bottom panel, we
have used matrix element in ratio scheme with $n_z^0=1$. We find only
a little difference between the matrix elements at the two fine
lattice spacings. To aid the eye, we have also shown bands that
cover $\pm 2\%$ variation on the $a=0.06$ fm and $a=0.04$ fm data.
Within this band, the real parts of the data are consistent with
perhaps little more correction to the imaginary part of RI-MOM at
intermediate $\nu$. Thus, we can bound the lattice corrections in
our data to be at the level of 1 to 2\%.  In the RI-MOM data, perhaps there
are residual lattice spacing effects of about 1\% at different $z$.
Even though the this lattice spacing effect is only about a percent,
we will see that $a^2 P^2_z$ corrections become important in the
analysis at smaller $z$ due to to their very small errors ensured
by the normalization process.

\section{Perturbative matching from the renormalized boosted hadron matrix element to $\msbar$ PDF}\label{sec:matching}

\subsection{Leading twist expressions to match equal time hadron matrix elements to PDF}

The computation of renormalized pion matrix element is the final step as far as
the non-perturbative lattice input is concerned. The perturbative
matching lets us make the connection between the renormalized boosted hadron matrix element
with the light-cone $\msbar$ PDF, $f(x,\mu)$.  Since the renormalization
factors for the RIMOM $h^R(z,P_z,P^R)$ and the ratio  ${\cal M}(z,P_z,P^0_z=0)$
do not depend on the PDF of the hadron itself, they lead to simpler
factorized expressions and hence let us consider them first. Using
such expressions, we will consider ${\cal M}(z,P_z,P^0_z)$ for
non-zero $P^0_z$.  Taking Ji's proposal~\cite{Ji:2013dva,Ji:2014gla}
of quasi-PDF in the RI-MOM scheme, $\tilde{q}(x,P_z,P^R)$, which
is the Fourier transform of the $z$-dependent matrix element
\beq
\tilde{q}(x,P_z,P^R)=\int_{-\infty}^\infty dz e^{-i x P_z z} h^R(z,P_z,P^R),
\eeq{qpdfdef}
the perturbative matching is expressed
as a convolution
\beqa
&&\tilde{q}(x,P_z,P^R)=\int_{-\infty}^\infty \frac{dy}{|y|}C_{\rm RI}\left(\frac{x}{y},yP_z,\mu,P^R\right) f(y,\mu)\cr
&&\quad +{\cal O}\left(\Lambda^2_{\rm QCD}/(1-x)x^2 P^2_z, m^2_\pi/P_z^2\right).
\eeqa{qpdfmatch}
The kernel of the convolution $C_{\rm RI}(x,\ldots)$ is of the
form (with dependence other than $x$ being implicit),
\beq
C_{\rm RI}(x)=\delta(x-1)+\left[C^{(1)}_{\rm RI}(x)\right]_{+}+{\cal O}(\alpha_s^2),
\eeq{ker}
where $C^{(1)}(x)$ is the 1-loop
contribution~\cite{Stewart:2017tvs,Izubuchi:2018srq,Constantinou:2017sej,Alexandrou:2017huk},
and the notation $[\ldots]_{+}$ represents the standard
plus-function~\footnote{$\int^\infty_{-\infty}[f(x)]_+ g(x)dx\equiv
\int^\infty_{-\infty}f(x)(g(x)-g(1))dx$.}.  Though we have used
RI-MOM scheme in the above equations, one can use the
matrix element in ratio scheme, ${\cal M}(z,P_z,P^0_z=0)$, as well with a corresponding
$C_{\rm ratio}(x,P_z,\mu)$.

An equivalent approach, that is suitable for our analysis in the
real space $z$, instead of performing a Fourier transform in
\eqn{qpdfdef} to the conjugate $x$, is through the formulation of
operator product expansion of the renormalized boosted hadron 
matrix element~\cite{Izubuchi:2018srq} using
only the twist-2 operators. That is, for the case of ${\cal M}(z,P_z,P^0_z=0)$
computed at large $P_z$ and with $z$ in the perturbative regime,
its OPE that is dominated by twist-2 terms is
\beqa
{\cal M}(z,P_z,P^0_z=0)&=&\sum_n c_n(z^2\mu^2) \langle x^n \rangle(\mu) \frac{(-i P_z z)^n}{n!},\cr&&\quad
\eeqa{t2ope}
up to ${\cal O}\left(z^2\Lambda_{\rm QCD}^2\right)$ corrections.
Here, $\langle x^n \rangle(\mu)$ are the $n$-th moments of the PDF~\footnote{Our notation is trivially 
different from a convention of naming $\langle x^{n-1}\rangle$ as $n$-th moments.} at a factorization
scale $\mu$,
\beq
\langle x^n \rangle(\mu) = \int^1_{-1} x^n f(x,\mu) dx,\quad \text{with}\quad \langle x^0 \rangle=1.
\eeq{xmomdef}
The coefficients $c_n(z^2\mu^2)$ are the perturbatively computable
Wilson-coefficients defined as the ratio of $\msbar$ Wilson-coefficients,
$c_n(z^2\mu^2)=c^\msbar_n(z^2\mu^2)/c^\msbar_0(z^2\mu^2)$.  The
1-loop expressions for $c_n(z^2\mu^2)$ can be found in
Refs.~\cite{Izubuchi:2018srq}.  These
Wilson-coefficients are related to the matching kernel $C_{\rm
ratio}$ through the relation~\cite{Izubuchi:2018srq},
\beq
\sum_n c_n(\mu^2 z^2) \frac{(-i z P_z)^n}{n!}=\int_{-\infty}^\infty dx C_{\rm ratio}(x) e^{-i x z P_z}.
\eeq{cntokernel}
The corrections denoted as ${\cal O}\left(z^2\Lambda_{\rm QCD}^2\right)$
arise from the operators in the OPE that are of twist higher than two.

For the RI-MOM scheme, a similar OPE that is valid up to ${\cal
O}\left(z^2\Lambda_{\rm QCD}^2\right)$ corrections is
\beqa
h^R(z,P_z,P^R)&=&\sum_n c^{\rm RI}_n(z^2,\mu^2,P^R) \langle x^n \rangle(\mu) \frac{(-i P_z z)^n}{n!},\cr&&\quad
\eeqa{t2operi}
where the RI-MOM Wilson-coefficients are $c^{\rm RI}_n$. Using the
multiplicative renormalizability of the bilocal operator $O_{\gamma_t}$, we can deduce that
\beq
c^{\rm RI}_n(z^2,\mu^2,P^R) = Z_{\rm{ratio}\to\rm{RI}}(z,P^R,\mu) c_n(z^2\mu^2),
\eeq{zrirat}
where $Z_{\rm{ratio}\to\rm{RI}}$ is the perturbatively computable
$P_z$-independent conversion factor from RI-MOM to ratio
scheme~\cite{Constantinou:2017sej,Zhao:2018fyu}.  By taking the
ratio of \eqn{cntokernel} for the ratio scheme and a corresponding
similar expression for the RI-MOM scheme involving $c^{\rm RI}_n$
and $C_{\rm RI}$, we can work out the conversion factor
$Z_{\rm{ratio}\to\rm{RI}}$ to be
\beqa
&&Z_{\rm{ratio}\to\rm{RI}}(z,P^R,\mu)= 1+\cr&&\ \int_{-\infty}^\infty dx \left[C^{(1)}_{\rm ratio}(x,P_z,\mu)-C^{(1)}_{\rm RI}(x,P_z,P^R)\right]\times\cr&&\qquad \left(e^{-i (x-1)z P_z}-1\right),
\eeqa{zconv}
up to 1-loop order. Though there is an explicit $P_z$ present in the above expression, its dependence gets 
canceled in the final expression, as expected.

We can now consider the ratio scheme for general values of $P_z^0$.
Noting that ${\cal M}(z,P_z,P^0_z)={\cal M}(z,P_z,0)/{\cal
M}(z,P^0_z,0)$, we can write the twist-2 expression as
\beq
{\cal M}(z,P_z,P^0_z)=\frac{\sum_n c_n(z^2\mu^2) \langle x^n \rangle(\mu) \frac{(-i P_z z)^n}{n!}}{\sum_n c_n(z^2\mu^2) \langle x^n \rangle(\mu) \frac{(-i P^0_z z)^n}{n!}},
\eeq{ratgenexp}
up to ${\cal O}(z^2 \Lambda_{\rm QCD}^2)$ corrections. Such an expression
cannot be written in a factorized form involving a convolution of
a perturbative kernel and PDF. As we noted in the beginning of this
section, we anticipated this since the ``renormalization factor"
is $({\cal M}(z,P^0_z,0))^{-1}$ for the ratio scheme at non-zero
$P_z^0$ and hence by itself dependent on the hadron PDF, unlike the
RI-MOM or the $P_z^0=0$ ratio schemes. However, as far as the
practical implementation of the analysis is concerned, the
non-factorizability of \eqn{ratgenexp} is not a hindrance, and the
analysis proceeds in exactly the same way for all the schemes
considered, i.e., by extracting the moments $\langle x^n\rangle$
from the boosted hadron matrix elements either in a model independent way or by modeling the
PDFs to phenomenology inspired ansatz. The reader can refer to Ref~\cite{Fan:2020nzz} for this
method implemented for the nucleon.

Finally, the above discussion ignored any presence of lattice spacing
corrections present at smaller $z$ at the order of few lattice
spacings that could spoil the applicability of the twist-2 expression
as it is. As discussed in \scn{3pt}, we found indications of $(P_z
a)^2$ corrections to matrix element at $z=0$. Such lattice corrections
were removed at $z=0$ by taking the ratio and making $z=0$ renormalized
matrix elements to be one by construction. However, such a procedure
will not ensure cancellation of $(P_z a)^2$ corrections at any
non-zero $z$.  We will take care of such correction by including
fit terms, $r a^2 P_z^2$, by hand in the twist-2 expressions above,
with $r$ being an extra free parameter.  As a concrete example, we
will modify \eqn{ratgenexp} to
\beq
{\cal M}(z,P_z,P^0_z)=\frac{\sum_n c_n(z^2\mu^2) \langle x^n \rangle(\mu) \frac{(-i P_z z)^n}{n!}+r (a P_z)^2}{\sum_n c_n(z^2\mu^2) \langle x^n \rangle(\mu) \frac{(-i P^0_z z)^n}{n!}+r (a P_z^0)^2},
\eeq{ratgenexp2}
to accommodate for any short-distance lattice artifacts.  It is
easy to see that the effect of such a $(P_z a)^2$ correction is to
shift the second moment in $(z/a)^{-2}$ manner,
\beq
\langle x^2\rangle \to \langle x^2\rangle-\frac{2 r}{c_2(\mu^2 z^2)} \frac{1}{(z/a)^2},
\eeq{secmom}
in all the twist-2 expressions above.  Indeed, we will present
evidence for the presence of such $(P_z a)^2$ corrections, and we
defer that discussion to \scn{mom}. One should note that the above
ansatz for correcting $(aP_z)^2$ effects is strictly true only for
$z>0$, since the ratio has to be exactly 1 at $z=0$. The actual
form of lattice correction would automatically ensure this, but we
found this simpler form to be practically enough to describe the
$z>0$ data, starting from $z=a$.

Before ending this subsection, we should remark that the LaMET
approach tries to suppress the higher-twist by taking the $P_z\to\infty$
limit, whereas the short-distance factorization approach aims to
remove the higher-twist effects by taking $z^2\to 0$ limit.  However,
in a practical implementation where one analyses the data at set of finite
momenta and finite $z^2$ as presented in this paper, 
one can think of the analyses being presented in either way, simply related by
\eqn{cntokernel}. For example, without loss of generality, one can
think of the analysis to be presented in the next part of the paper in 
the following way --- one starts from a model PDF
ansatz, to which one applies the LaMET kernel in \eqn{ker} to obtain
a model quasi-PDF, which is Fourier transformed to real
space to be fitted to the real-space matrix element determined
on the lattice.  Keeping this in mind, we will simply use the OPE
expressions, such as \eqn{ratgenexp} for our twist-2 perturbative
matching analysis.

\subsection{Numerical investigation of higher-twist effect in pion matrix element at low momenta}

\bef
\centering
\includegraphics[scale=0.6]{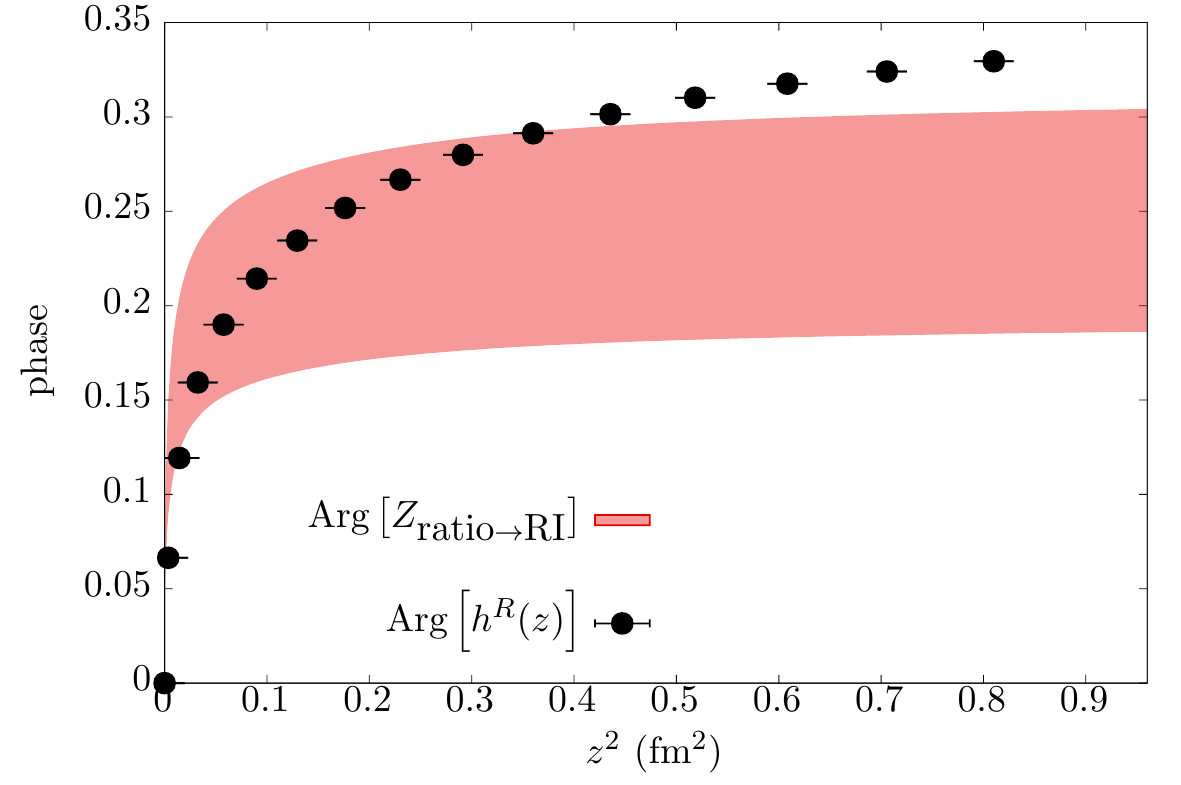}
\includegraphics[scale=0.6]{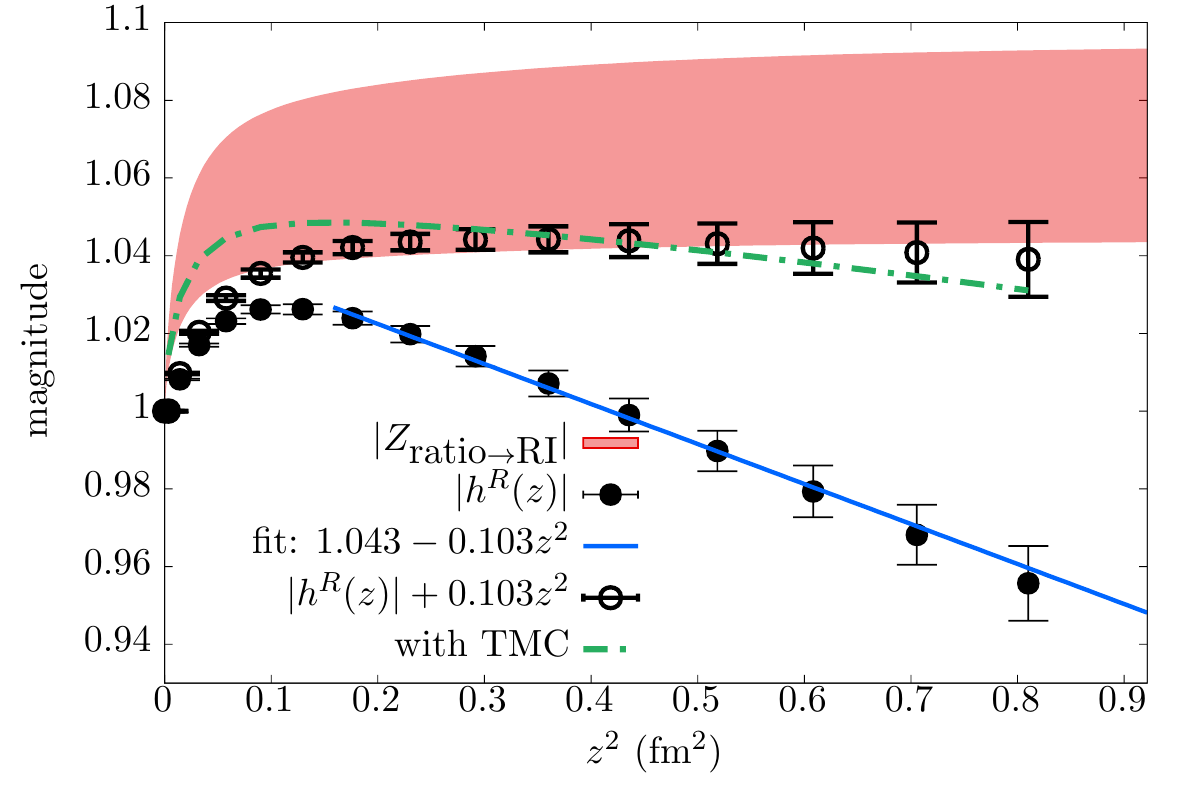}
\caption{
The phase (top) and the magnitude (bottom) of the $P_z=0$ matrix
element in RI-MOM scheme with $(P_z^R,P_\perp^R)=(1.29,2.98)$ GeV
are shown.  The expectations from the 1-loop leading twist results
are shown as the red bands.  In the bottom panel, the actual lattice
data is shown using filled black circles.  The absolute part clearly
suffers from a leading $z^2$ correction shown as the blue straight
line. The twist-2 target mass correction is shown as the green curve
for comparison. The lattice data after subtracting the $z^2$
correction term is shown using open circles.
}
\eef{hightwist}

In the remaining part of this section on perturbative matching, we
discuss a way to use the hadron matrix elements at smaller momenta to understand the
importance of higher twist effects at {\sl intermediate} values of
$z\sim$ 0.3 fm to 1 fm, and thereby, understand the rationale for
the ratio scheme which hitherto had been discussed using a conjectured
separation of higher twist effects and leading-twist terms into two separate factors~\cite{Orginos:2017kos}.  
The
\eqn{t2ope}, \eqn{t2operi} and \eqn{ratgenexp} are valid only up
to ${\cal O}(z^2 \Lambda_{\rm QCD}^2)$ higher twist effects. 
At any large value of $P_z$, the twist-2 terms become larger compared to the ${\cal O}(z^2 \Lambda_{\rm
QCD}^2)$ higher twist terms.
This is the basis of the large momentum effective theory.  As a
corollary, the matrix element where the higher twist effects show
up significantly is the $P_z=0$ matrix element. This is not a useful
observation when applied to the ratio ${\cal M}(z,P_z,P_z^0=0)$,
for which $c_0$ has the value 1 at all $z$ when $P_z=0$.
This agrees with the twist-2 expectation by construction, but the
corrections could show up at other non-zero $P_z$.  Therefore, we
use $h^R(z,P_z=0,P^R)$ in the RI-MOM scheme for this study where
we compare the lattice result with the non-trivial $z$ dependence
from twist-2 term. Also, since the wrap-around effects in $P_z=0$
matrix elements are negligible only for the $a=0.06$ fm lattice,
we use this case for this study.

For $P_z=0$, the only non-zero twist-2 contribution is from the
local current operator because all other terms have explicit factors of $P_z$ and they become zero. 
Its $z$ dependence comes from the
Wilson-coefficient $c_0^{\rm RI}(z)$, which is the conversion factor
$Z_{{\rm ratio}\to{\rm RI}}(z,P^R,\mu)$. We use 1-loop expressions
to calculate $Z_{{\rm ratio}\to{\rm RI}}$.  We vary the scale of
$\alpha_S(\mu)$ that enters $Z_{{\rm ratio}\to{\rm RI}}$ from $\mu/2$
to $2\mu$ with $\mu=3.2$ GeV, and gives an estimate of the expected error on perturbative
result. It is convenient to separate $Z_{{\rm ratio}\to{\rm RI}}$
and the lattice result $h^R(z,P_z=0,P^R)$ into their magnitudes and
phases. The phase ${\rm Arg}\left[h^R(z,P_z=0,P^R)\right]$ is the
same as ${\rm Arg}\left[Z_{\gamma_t\gamma_t}(z,P^R)\right]$, which
is a property of the RI-MOM scheme itself. On the other hand, the
magnitude $|h^R|$ depends on the pion matrix element.

In the top panel of \fgn{hightwist}, we compare the $z^2$ dependence
of the phase ${\rm Arg}\left[h^R(z,P_z=0,P^R)\right]$ with the
perturbative twist-2 phase ${\rm Arg}\left[Z_{{\rm ratio}\to{\rm
RI}}(z,P^R)\right]$.  We have chosen a renormalization scale
$(P^R_z,P^R_\perp)=(1.29,2.98)$ GeV on the $a=0.06$ fm ensemble as
a sample case, but the observations hold for other cases as well.
We find a good agreement within the perturbative uncertainties up
to 0.7 fm, and the lattice data slightly overshoots the 1-loop
result for larger $z$. Nevertheless, the overall qualitative agreement
validates the 1-loop perturbation theory as applied to quark external
states used in RI-MOM $Z$-factor.  This should serve as a companion
observation to the studies on RI-MOM $Z$-factor presented in our
previous work~\cite{Izubuchi:2019lyk}.

In the bottom panel of \fgn{hightwist}, we compare the $z^2$
dependence of the magnitude $\left|h^R(z,P_z=0,P^R)\right|$ with
$\left|Z_{{\rm ratio}\to{\rm RI}}(z,P^R)\right|$. The actual lattice
data is shown as the filled circles. It is clear that the non-perturbative
result disagrees with the near constant behavior of the twist-2
term at larger $z$, and that this disagreement comes from a striking
$z^2$ dependence at larger $z$. The coefficient, $k$, of the $z^2$
dependence is $-(63 \text{\ MeV})^2$ (with little variations around
this value with $P^R$), and thus it is reasonable to identify such
a term to arise from a higher twist operator or an effective
contribution of a number of higher twist operators. There could also be
corrections to the leading twist result coming from the twist-2
target mass correction (TMC)~\cite{Chen:2016utp,Radyushkin:2017ffo}
(and discussed in \apx{tmc}). The 1-loop result with TMC is shown
as the green dashed line in the bottom panel of \fgn{hightwist}, which
is visibly small compared to the observed discrepancy. Numerically,
the coefficient of $z^2$ from twist-2 target mass correction term
is $-m^2_\pi \langle x^2\rangle_v/8=-(35.2\text{\ MeV})^2$, which
about one-third of the observed value (assuming $\langle
x^2\rangle_v\approx 0.11$ as we will see later).  In addition, when
we correct for the $z^2$ effect by subtracting it from the lattice
data, shown by the open black circles in the figure, we find a nice
agreement with the 1-loop, twist-2 expectation.  It is quite
remarkable that such a simple ${\cal O}(\Lambda_{\rm QCD}^2 z^2)$
effect is enough to describe the non-perturbative data even up to
1 fm.

\bef
\centering
\includegraphics[scale=0.5]{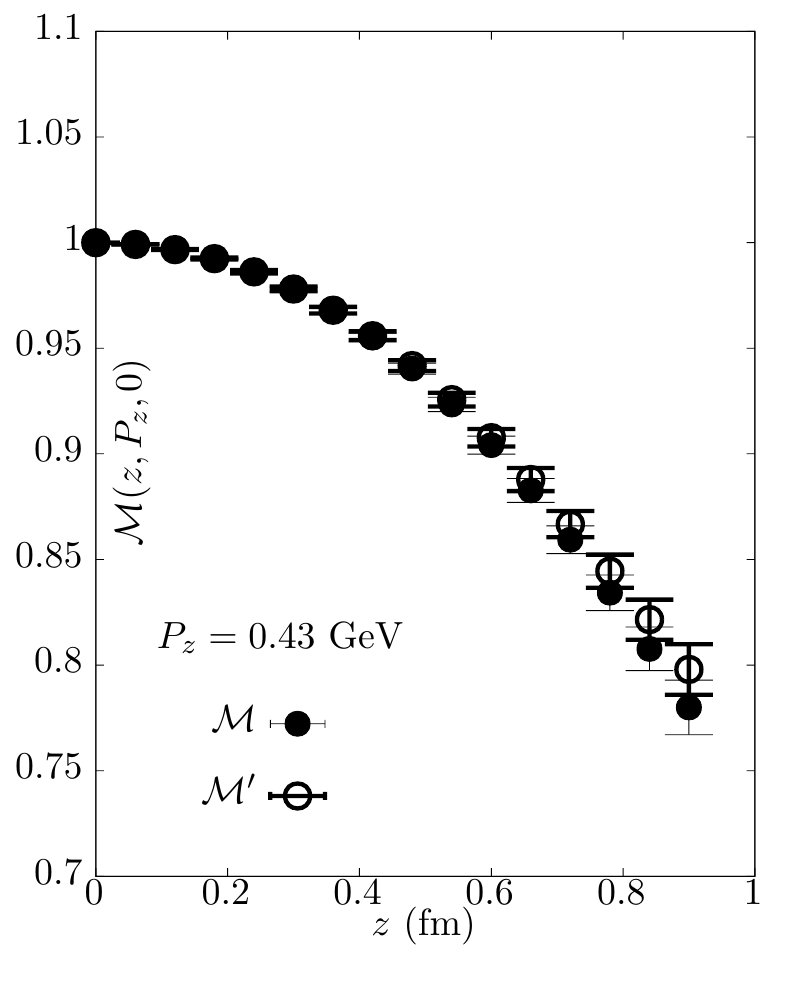}
\includegraphics[scale=0.5]{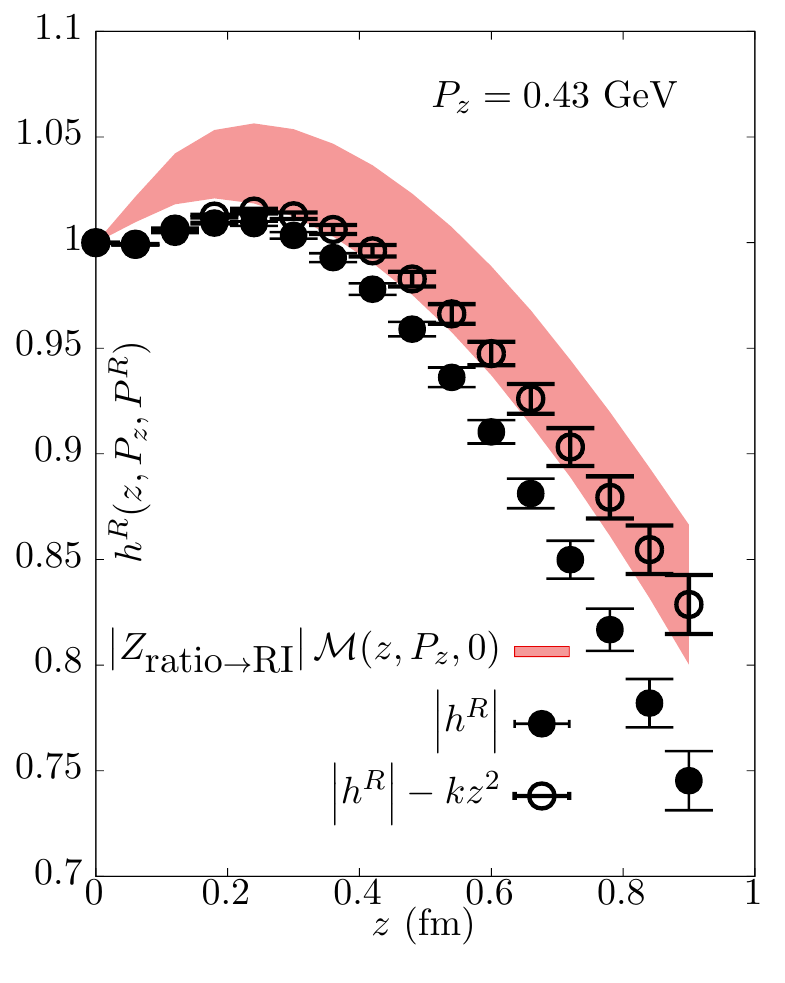}
\caption{
The effect of higher-twist term $k z^2$ on the pion matrix element at fixed small
$P_z=0.43$ GeV is discussed.  The left panel compares the usual
ratio, ${\cal M}(z,P_z,P_z^0=0)$, and the ratio ${\cal
M'}$ derived from RI-MOM matrix element after the subtraction of $k z^2$ (see
\eqn{imprat}). The right panel compares the actual RI-MOM matrix element and
the RI-MOM matrix element with $k z^2$ subtraction with the result expected
by applying the conversion factor $Z_{{\rm ratio}\to{\rm RI}}$ to
the ratio ${\cal M}$ (see \eqn{convertri}).
}
\eef{ratiotwist}

Now, we take the hypothesis that the observed $z^2$ effect in
$h^R(z,P_z=0,P^R)$ is the dominant higher twist effect, and try to
understand its effect on the matrix element in the ratio scheme. Perturbatively, the
ratio scheme is defined via the subtraction of the UV divergence
by a division with $n=0$ $\msbar$ Wilson coefficient, $c_0^\msbar$.
On the lattice, we identify this procedure as the division by $P_z=0$
matrix element, and hence the equality in \eqn{t2ope}. The underlying
assumption is that the higher twist effect in $P_z=0$ matrix element
is negligible or somehow cancels with the higher twist effect present
also in the non-zero $P_z$ matrix elements. In order to understand
this, we can redefine the ratio scheme that better agrees with the
assumptions that go into twist-2 matching framework; 
namely form the ratio after subtracting
off the higher-twist effects
\beq
{\cal M}'(z,P_z,P^0_z=0)\equiv\frac{|h^R(z,P_z,P^R)|-k z^2}{|h^R(z,P_z=0,P^R)|-k z^2},
\eeq{imprat}
where $k$ is the coefficient we determined using the analysis of
$h^R(z,P_z=0,P^R)$ and assume the same $kz^2$ correction is present
at non-zero $P_z$ as well. For $k=0$, ${\cal M}'={\cal M}$.  The
result of this {\sl improved} ratio ${\cal M}'(z,P_z,0)$ is compared with the usual ratio
${\cal M}(z,P_z,0)$ in the left panel of \fgn{ratiotwist} for the
first non-zero momentum $P_z=0.43$ GeV on $a=0.06$ fm lattice. The
difference between the two ways of defining the ratio are consistent
within errors, with perhaps very little difference at larger $z$.
This provides a better understanding of how the ${\cal O}(\Lambda_{\rm
QCD}^2 z^2)$ corrections in the numerator and denominator of
\eqn{imprat} almost cancel each other without resorting to any
factorwise separation of higher-twist corrections, and instead, results
simply from the smallness of $k$. Having demonstrated the inconsequential
role of higher twist effects in ${\cal M}$ for $z<1$ fm given the
errors in the data, we now look closely at the RI-MOM $h^R(z,P_z,P^R)$
at the same small momentum $P_z=0.43$ GeV. Within the twist-2 framework,
we can obtain $h^R$ from ${\cal M}$ via
\beq
h^{R'}(z,P_z,P^R)\equiv Z_{{\rm ratio}\to{\rm RI}}(z,P^R){\cal M}(z,P_z,0).
\eeq{convertri}
In the right panel of \fgn{ratiotwist}, we compare $h^{R'}$, shown
as the red band, with $h^R$ which are the black filled symbols.  We
find a deviation from the twist-2 expectation $h^{R'}$ for $z>0.3$
fm.  When we correct for the $z^2$ effect using $|h^R(z,P_z,P^R)|-k
z^2$, shown as the open symbols, we find a very good agreement with
$h^{R'}$.  Putting together the above results, we self-consistently
justified that the observed $kz^2$ effect in $P_z=0.43$ GeV is
almost the same as in $P_z=0$ as we assumed, and that ${\cal M}$
is least affected by such corrections. At higher momenta $P_z$,
such higher-twist effect will play even lesser role for $z<1$ fm.

\section{A model-independent computation of the even moments of valence pion PDF}\label{sec:mom}

\bef
\centering
\includegraphics[scale=0.9]{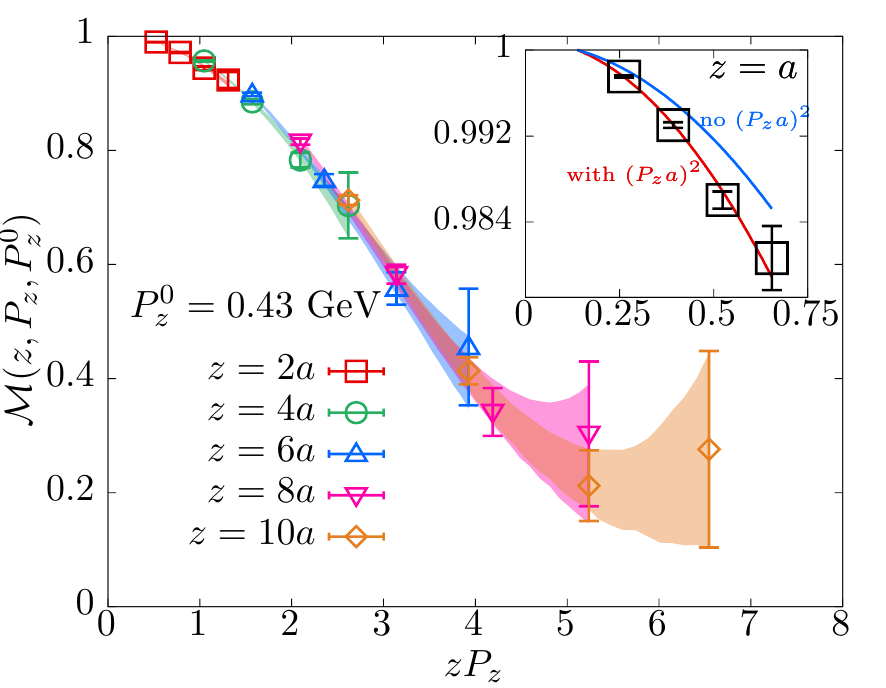}
\caption{Plot of the ratio, ${\cal M}(z,P_z,P_z^0)$, shown as a function
of $zP_z$ using $a=0.06$ fm data with $P_z^0=0.43$ GeV.  The
points of same colored symbols have the same value of $z$, and hence
the $zP_z$ dependence comes from the variation in $P_z$.  The
corresponding colored bands are the fits of \eqn{ratgenexp} to data
at each fixed $z$.  Inset: The data points are at fixed $z=a$. The
blue curve is the expectation for the $z P_z$ dependence at $z=a$
based on the moments obtained at $z=5a$ without correcting for $(P_z
a)^2$ lattice terms. The red curve is obtained after accounting for
such lattice corrections.
}
\eef{momfit}

\bef
\centering
\includegraphics[scale=0.7]{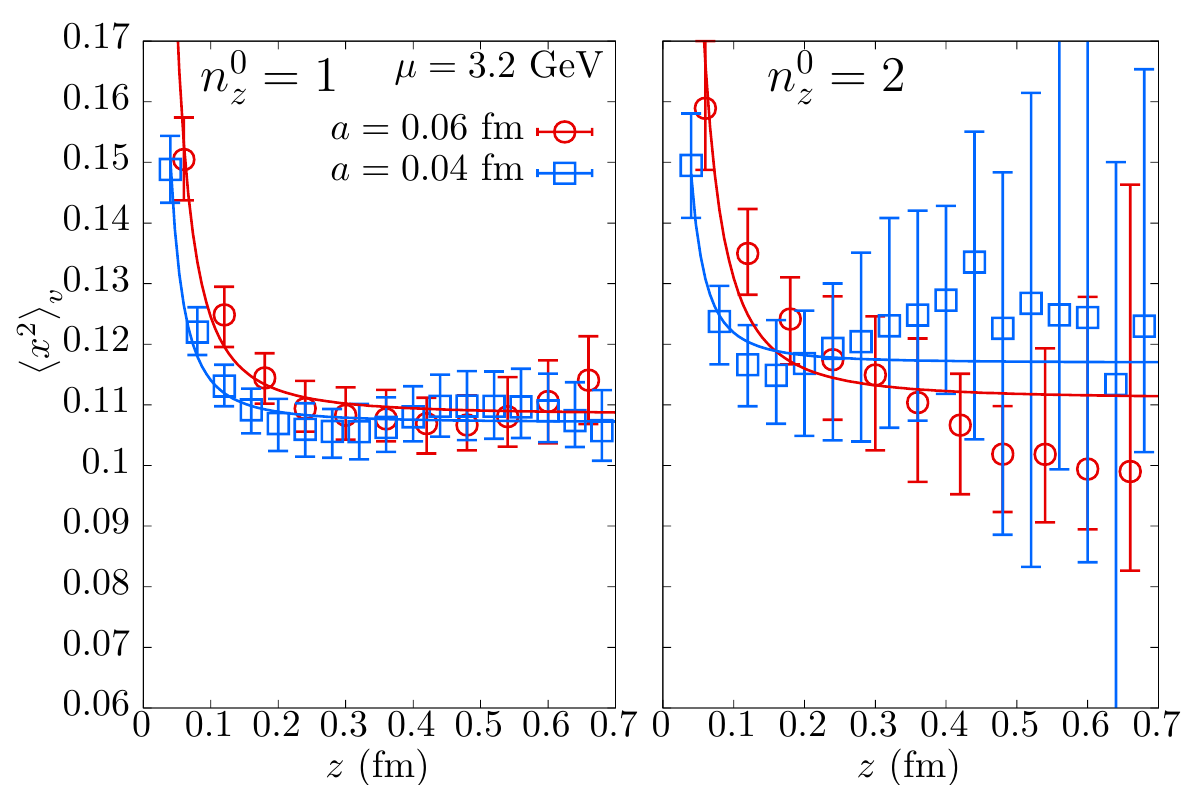}
\caption{The $z$ dependence of $\langle x^2\rangle$ obtained by
fitting rational polynomial functions in $z P_z$ to matrix elements
${\cal M}(z,P_z,P_z^0)$ at different fixed values of
$z$, with $P_z^0=2\pi n_z^0/L$.  The left panel shows results as
obtained with $n_z^0=1$ and the right panel for $n_z^0=2$.  In each case,
the red and blue points correspond to $a=0.06$ fm and 0.04 fm lattice
spacings. The curves are fits to a functional form $\langle
x^2\rangle+2r(z/a)^{-2}$ to capture possible $(aP_z)^2$ corrections (refer text).
}
\eef{x2}

In this section, we apply the twist-2 perturbative matching 
formalism, that we discussed
in \scn{matching}, to our lattice data for the isovector $u-d$ PDF
of pion. For this, we will use the boosted pion matrix element in
the ratio scheme with non-zero reference momentum $P_z^0=2\pi
n^0_z/(La)$ with $n_z^0=1$ and 2.  This way, we expect to suffer
from smaller non-perturbative corrections and also avoid the larger
periodicity effect in zero momentum matrix elements (we also discuss
results using $n_z^0=0$ for $a=0.06$ fm lattice, where wrap-around
effect was small,  in \apx{kx0}).  Through \eqn{ratgenexp}, we can
find the values of the moments $\langle x^n \rangle$ by fitting
them as free parameters such as to best describe the $zP_z$ and
$z^2$ dependence of the ${\cal M}(z,P_z,P_z^0)$ data.  Such a method,
usually referred to as {\sl OPE without OPE}~\cite{Martinelli:1998hz}, has been previously
applied to the case of reduced ITD ($n_z^0=0$) for pion~\cite{Joo:2019bzr}
and nucleon~\cite{Karpie:2018zaz,Joo:2019jct,Joo:2020spy}.

\subsection{Connection between isovector PDF and valence PDF of pion}
First, it is important to recall as to how the $u-d$ isovector pion matrix element
that we compute on the lattice relates to the valence PDF of pion.
Let $f_u(x)$ and $f_d(x)$ are the $u$ and $d$ quark PDF with support
in $x\in[-1,1]$ and a convention that includes the quark distribution
for $x>0$ and anti-quark distribution for $x<0$ via the relation
$f_u(-x)=-f_{\overline{u}}(x)$ and $f_d(-x)=-f_{\overline{d}}(x)$.
The $u-d$ isovector matrix element relates to the $f_{u-d}(x)$,
\beq
f_{u-d}(x)=f_u(x)-f_d(x),\quad x\in[-1,1].
\eeq{uminusd}
Due to isospin symmetry in $\pi^+$,  $f_{u-d}(x)=f_{u-d}(-x)$. The
moments that occur in the OPE expressions  \eqn{t2ope}, \eqn{t2operi}
and \eqn{ratgenexp} when applied to the isovector matrix element are the $u-d$
PDF moments, $\langle x^n\rangle=\langle x^n \rangle_{u-d}$.  Due
to the symmetry of $f_{u-d}(x)$ about $x=0$, only the
moments $\langle x^n\rangle_{u-d}$ for even $n$ are non-vanishing
for pion.  For the pion $\pi^+$ with the valence structure
$u\overline{d}$, the valence PDF is
\beq
f_v^\pi(x)\equiv f_u(x)-f_{\overline{u}}(x),\quad x\in[0,1].
\eeq{vpdf}
This could be understood from the fact that $\overline{u}$ parton
could only be produced radiatively in $\pi^+$, and hence
$f_{\overline{u}}$ only has a sea quark distribution, which thereby
cancels the sea quark distribution of $f_u$  in the above definition.
Due to the isospin symmetry present in our QCD computation that
does not include QED corrections, $f_{\overline{u}}(x)=f_d(x)$.
Thus, $f_v^\pi(x) = f_u(x)-f_d(x),\quad x\in[0,1]$.  Unlike the
$u-d$ PDF of pion, both even and odd valence moments $\langle
x^n\rangle_v$ of the pion are non-vanishing.  By comparing the above
equivalent definition of the valence PDF in terms of $u$ and $d$
quark PDFs in pion with the $u-d$ PDF in \eqn{uminusd}, one can
deduce that
\beq
f_{u-d}(x)=\begin{cases}f_v(x),\quad x\in [0,1]\cr f_v(|x|),\quad x\in [-1,0], \end{cases}
\eeq{vpdf3}
and that for the moments
\beq
\langle x^{n}\rangle_{u-d} = \begin{cases} \langle x^n\rangle_v,\quad n \text{\ is even},\cr 0 ,\quad\quad\ \ n \text{\ is odd}.\end{cases}
\eeq{vmom}
Thus, the OPE expression in \eqn{ratgenexp} for $u-d$ pion matrix element has
only even powers $n$, from which we could obtain the values of
$\langle x^{n}\rangle_{u-d}$ for even values of $n$, which as we
discussed is the valence moment $\langle x^{n}\rangle_v$.  Unfortunately,
the $u-d$ matrix element does not directly let us access the odd valence
moments, but we will later try to determine them based on models
of valence PDF $f_v^\pi$ itself.

\subsection{Method for model independent fits}

We performed model independent determinations of $\langle x^n\rangle_v$
by fitting the rational functional form in \eqn{ratgenexp}, which
we denote as ${\cal M}^{\rm OPE}(z,P_z,P_z^0)$ here, with the even
moments $\langle x^n\rangle_v$ as the fit parameters, over a range
of $z_1\le z \le z_2$ and $P_z^0\le P_z\le P_z^{\rm max}$.  The
possibility of larger lattice corrections at very short separation
, $z_1$, has to be accounted for. Therefore, we tried fits including
or excluding $(P_z a)^2$ correction term in ${\cal M}^{\rm
OPE}(z,P_z,P_z^0)$ as discussed in \scn{matching}.  For a larger
range $[z_1,z_2]$, there is a larger curvature in the data for
${\cal M}$, which makes the fits sensitive to the higher order terms
of $\nu$ in \eqn{ratgenexp}. On the other hand, by using a larger
$z_2$, there is  the undesired possibility of working in a
nonperturbative regime of QCD. We strike a balance between the two
by choosing the maximum, $z_2$, over range of values from 0.36 fm
to 0.72 fm.  We choose the factorization scale ${\mu}$ to be 3.2
GeV in the following determinations.  Since the Wilson coefficients
$c_n$ are known only to 1-loop order, the scale of strong coupling
constant $\alpha_s$ is still unspecified.  We take care of this
perturbative uncertainty by using the variation in \eqn{ratgenexp}
when the scale of $\alpha_s$ is changed from $\mu/2$ to $2\mu$ as
part of error, where $\mu$ is the factorization scale at which
$\langle x^n\rangle_v$ are determined. Concretely, we minimize the
following $\chi^2$ to determine the moments:
\beqa
&&\chi^2\equiv \sum_{z=z_1}^{z_2}\sum_{P_z=P_z^0}^{P_z^{\rm max}}\frac{\left({\cal M}(z,P_z,P_z^0)-{\cal M}^{\rm OPE}(z,P_z,P_z^0)\right)^2}{\sigma_{\rm stat}^2(z,P_z,P_z^0)+\sigma_{\rm sys}^2(z,P_z,P_z^0)},\cr
&&\sigma_{\rm sys}(z,P_z,P_z^0)=\frac{1}{2}\big{(}{\cal M}^{\rm OPE}_{\alpha_s(\mu/2)}-{\cal M}^{\rm OPE}|_{\alpha_s(2\mu)}\big{)}(z,P_z,P_z^0).\cr&&\quad
\eeqa{chisqmom}
While the above expression is a convenient way to include the
perturbative error in the analysis, it comes at the cost of missing
the covariance matrix. We take care of it by using the same set of
bootstrap samples for all $z$ and $P_z$.  We use the factorization
scale $\mu=3.2$ GeV to determine $\alpha_s$ used in the twist-2
expressions; for this, we used the values $\alpha_s=0.33, 0.24$ and
0.19 at scales $\mu/2,\mu$ and $2\mu$ respectively, by interpolating
the running coupling data compiled by the PDG~\cite{Tanabashi:2018oca}.
Since we take the variation of $\alpha_s$ with scale into account in the error budget of 
our analysis, a precise input of $\alpha_s$ is not necessary.
We can also improve the estimate of higher moments by imposing
priors on $N_{\rm prior}$ lower moments by using
\beq
\chi^2=\chi^2+\sum_{i=1}^{N_{\rm prior}}\frac{(\langle x^i\rangle_v-\langle x^i\rangle_{\rm prior})^2}{(\sigma_i^{\rm prior})^2},
\eeq{prior}
where $\langle x^i\rangle_{\rm prior}$ and $\sigma_i^{\rm prior}$
are the prior on $i$-th moment and error on the prior respectively.
We used this method only to determine $\langle x^6\rangle_v$ with
prior imposed on only $\langle x^2\rangle$, or both $\langle
x^2\rangle$ and $\langle x^4\rangle_v$.  For the prior, we used the
result of fits with $z_2=0.5$ fm and the error on that estimate as
$\sigma_{\rm prior}$. In the future, it would be interesting to use
estimates of lower moments from the other twist-2 local-operator
techniques on the same gauge ensemble as priors in the twist-2 matching 
methodologies in order to determine higher moments. 

We point out an improved way to implement the fit for valence pion
PDF.  Naively, one might expect that including more terms in
\eqn{ratgenexp} will lead to unstable results and larger errors due
to the increase in the number of fit parameters, $\langle x^n\rangle_v$.
For the case of valence PDF pion, we can use an additional fact to
constrain the moments --- that of the positivity of $f^\pi_v(x)$,
and hence of $f_{u-d}(x)$ for all $x\in[-1,1]$.  The positivity of
$f^\pi_v(x)$ is usually implicit in simple ansatz such as $f^\pi_v(x)\sim
x^{\alpha}(1-x)^\beta$.  This stems from the fact that the $u$-quark
is present at the order ${\cal O}(\alpha^0_s)$ due to its valence
nature while the $d$-quark is only in the sea and hence its
distribution can start only at ${\cal O}(\alpha_s)$.  Thus, it is
a well justified expectation that $f_u(x)>f_d(x)$. The positivity
of $f^\pi_v(x)$ leads to the conditions that the even derivatives
of $\langle x^n\rangle_v$ with respect to $n$ are positive (i.e.,
$\frac{d^m\langle x^t\rangle_{v}}{dt^m}|_{t=n}=\langle x^n
\log^m(x)\rangle_{v}>0$ for even $m$) and that the odd derivatives
(i.e., $m$ is odd) are negative.  The interesting consequences are
the inequalities
\beqa
&&\langle x^{n+2}\rangle_{u-d}< \langle x^{n}\rangle_{u-d} \quad\text{and}, \cr
&&\langle x^{n+2}\rangle_{u-d}+\langle x^{n-2}\rangle_{u-d}-2\langle x^n\rangle_{u-d} >0.
\eeqa{momconst}
These inequalities lead to strong constraints on the fitted moments
and lead to the stabilization of the estimates (and their errors)
of the lower moments as one increases the number of terms in
\eqn{ratgenexp} to larger values, thereby eliminating the order of
\eqn{ratgenexp} as a tunable parameter and prevents over-fitting the
data. The two inequalities in \eqn{momconst} can be easily implemented
through a change of variables
\beq
\langle x^n\rangle_v
\equiv \sum^N_{i=n}\sum^N_{j=i} e^{-\lambda_j},
\eeq{cov}
where the sum runs over even $i$ and $j$ for the pion.
The parameters $\lambda_j>0$, and $N$ being the largest even moment
used in the fit. In the discussions below, we used even moments
up to $\langle x^8\rangle_v$ in the fits over multiple $z$. In
cases where certain higher moments were irrelevant to the fits,
they promptly converged to values very close to zero without affecting
the relevant smaller moments. In this way, we do not have to choose
the order of the polynomial to be used in the fits.

\subsection{Determining an estimate, its statistical and systematic error}
Since the various estimates in this section and the rest depend on
the range $[z_1,z_2]$ and the value of $n_z^0$ used, we define the
central estimate of a quantity $A$ and its systematic error as
$\overline{{\rm Mean}(A)}$ and $\overline{{\rm SD}}(A)$ respectively;
here, ${\rm Mean}(A)$ is the mean over different estimates (variations
in fit range etc.,) in a given bootstrap sample, and ${\rm SD}(A)$
is the standard deviation of various estimates of $A$ within the
same bootstrap sample. The notation $\overline{{\rm Mean}(A)}$ and
$\overline{{\rm SD}}(A)$ stand for average of those mean and standard
deviation over the bootstrap sample.  In this way, we obtain the
statistical error on $\overline{{\rm Mean}(A)}$ also in the
standard bootstrap procedure. We will use this procedure in the
later sections too, and the extra dependences on model ansatz, 
and renormalization schemes (ratio, RI-MOM scheme and
their various scales) will also enter in evaluating the systematic
error.

\bef
\centering
\includegraphics[scale=0.7]{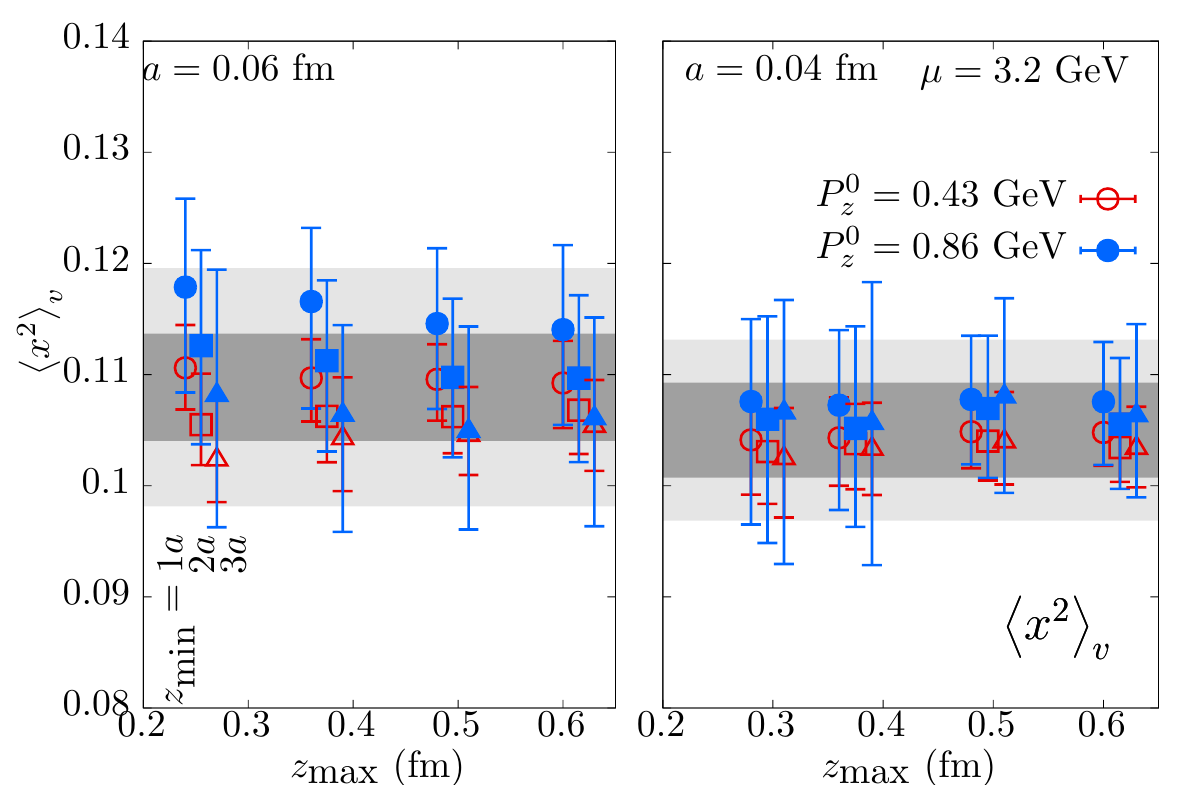}
\caption{$\langle x^2 \rangle$ from combined fits of the rational polynomial
function in $zP_z$ to ${\cal M}(z,z P_z,P_z^0)$  for the data
$z\in[z_{\rm min},z_{\rm max}]$. The dependence of $\langle x^2
\rangle$ on $z_{\rm max}$ is shown. For each $z_{\rm max}$, values
from three different $z_{\rm min}$ are shown.  The results for
$a=0.06$ fm and 0.04 fm are shown in the left and right panels. The
grey bands are the estimates --- the inner band includes only
statistical error, and outer one includes both statistical and
systematic error (see text). }
\eef{combinedx2}

\bef
\centering
\includegraphics[scale=0.7]{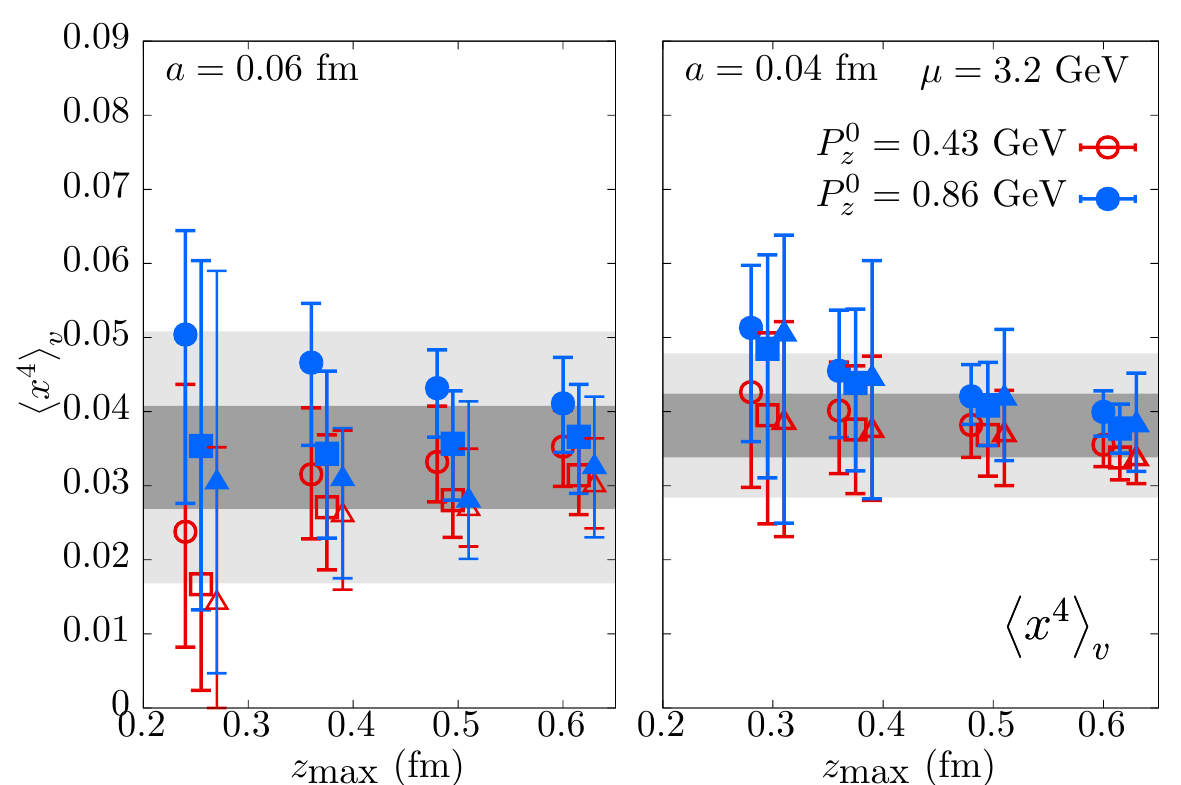}
\caption{$\langle x^4 \rangle$ from combined fits of the rational polynomial
function in $z P_z$ to ${\cal M}(z,z P_z,P_z^0)$  for the data
$z\in[z_{\rm min},z_{\rm max}]$. The description of the points are 
the same in \fgn{combinedx2}.}
\eef{combinedx4}

\bef
\centering
\includegraphics[scale=0.7]{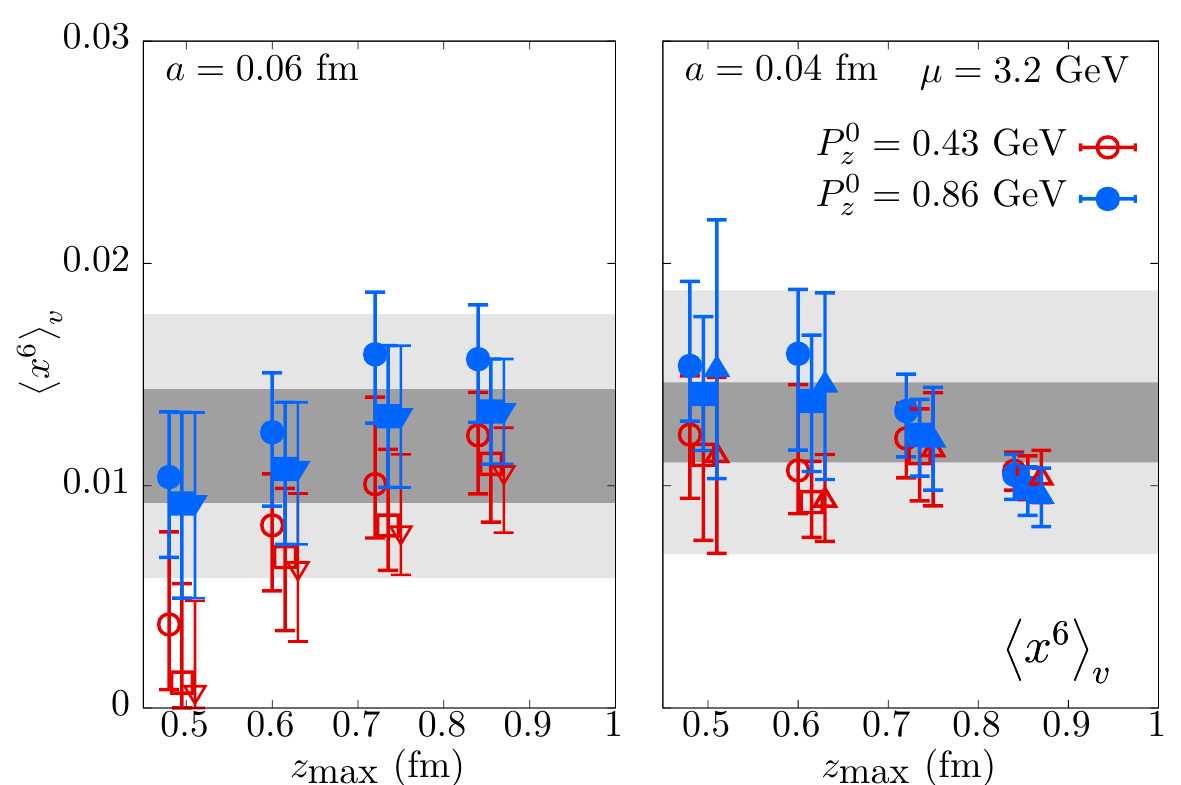}
\caption{$\langle x^6 \rangle$ from combined fits of the rational polynomial
function in $z P_z$ to ${\cal M}(z,z P_z,P_z^0)$  for the data
$z\in[z_{\rm min},z_{\rm max}]$ using priors $\langle x^2\rangle_v$
and $\langle x^4\rangle_v$.  The dependence of $\langle x^6\rangle$
on $z_{\rm max}$ is shown. At each each $z_{\rm max}$, values from
$z_{\rm min}=a,2a$ and 3$a$ are shown.}
\eef{combinedx6}

\subsection{Model independent analysis of moments at fixed $z^2$}
The $\nu=z P_z$ dependence can come from either the $z$ variation
at fixed $P_z$ or from $P_z$ variation at fixed $z$. We first look
at the latter case.  In \fgn{momfit}, we show the result of fitting
the rational polynomial in $\nu$ given by \eqn{ratgenexp} to the
$a=0.06$ fm data at different fixed $z$. For the case shown,
$P_z^0=0.43$ GeV. In this analysis at fixed $z$, we did not 
take any $(P_z a)^2$ correction into account.
The results of the fits at various fixed $z$ are shown as the bands
having the same color as the corresponding data points.  Since we
only have five different values of $P_z$, the smaller $z$ data cover
shorter ranges in $\nu$ compared to the larger fixed $z$ data. It
is clear from the data that in order to be sensitive to deviations
from simple $\nu^2$ term, we need to resort to data at larger
$z>8a=0.48$ fm as well.  We repeated this analysis with $n_z^0=2$
also.  In \fgn{x2}, we show the value of $\langle x^2\rangle_v$
that is extracted from the fits as a function of the fixed values
of $z$ used in the fits.  The results as obtained from both $n_z^0=1$
and 2 are shown in the left and right panels. In order to look for
lattice spacing effects, we have shown results from the two lattice
spacings (but keeping in mind that the two $n_z^0$ at two lattice
spacings lead to slightly different $P_z^0$ in physical units).
The inferred moments are more precise for $n_z^0=1$ than for $n_z^0=2$,
as one would expect from deteriorating signal as momentum is
increased.  One can see a plateau in $\langle x^2\rangle_v$ starting
from $3a$ even up to $z=0.7$ fm. This shows that the $z^2$
dependence in the pion matrix element is canceled to a good accuracy by the
perturbative Wilson coefficients $c_n(\mu^2 z^2)$.

There is a clear tendency for the fitted $\langle x^2\rangle_v$ to
increase at very short  lattice distance $z/a\sim 1$ which is most
likely a result of increased lattice corrections at smaller $z$.
One can see this by comparing the results from the two lattice
spacings and noting that at fixed short physical distance $z$, there
is tendency for the $a=0.04$ fm data to lie closer to the plateau
than the $a=0.06$ fm data. If the lattice spacing effect is coming from
$(P_z a)^2$ corrections, then we should find the $(z/a)^{-2}$
behavior of $\langle x^2\rangle$ as we outlined in \scn{matching}
(where we ignore the logarithmic dependence in $z$ present in $c_2$
for this first analysis in this subsection.) The fits to such
$\langle x^2\rangle+ 2r(z/a)^{-2}$ are shown as the corresponding
colored curves in \fgn{x2}. Indeed, we find a very nice description
of the observed data, thereby, show the importance of $(P_z a)^2$
corrections at first few lattice separations $z/a$.  Also, as a
consistency check, the values of $r$ from the fits on the two lattice
spacings were about the same, namely, 0.021 and 0.022 on the 0.04
fm and 0.06 fm lattice spacings respectively.  It could be
counter-intuitive to find a rather large lattice spacing effect
affecting the moments when we do not find anything unusual about
the small $z/a$ in \fgn{momfit}, or in \fgn{ritdcompare} where we
compared the data at two different lattice spacings.  In order to
understand this, we take the values of moments as obtained from
$z=5a$ (which lies in the plateau of $\langle x^2\rangle_v$) and
reconstruct the expected $\nu$ dependence at fixed $z=a$ using
\eqn{ratgenexp}, without including any $(P_z a)^2$ corrections in
the expression. In the inset of \fgn{momfit}, we compare this
expected curve (blue) with the actual $z=a$ data points. The clear
disagreement between the two is the cause of the anomalously large
$\langle x^2 \rangle_v\approx 0.15$ at $z=a$ in \fgn{x2}. One should
note the rather enlarged scale on the $y$-axis of the inset, and
the disagreement is actually sub-percent.  But, the data at small
$z/a$ is so precise that such small lattice spacing effects show
rather clearly in the extracted moments. This is the crux of the
problem. After accounting for the $r a^2 P^2_z$ correction, the
expected curve is shown in red, which agrees perfectly with the
data and gives $\langle x^2\rangle_v$ that is consistent with the
one extracted from larger $z/a$. In the analyses henceforth, we
will use the correction term $r (P_z a)^2$ term in the fits as
outlined in \scn{matching} with $r$ being an extra fit parameter,
and this way, we were able to use $z_1=a,2a,3a$ in the fits and
obtain no contradictory strongly $z_1$-dependent results.

\subsection{Model independent combined analysis of moments}
In order to estimate $\langle x^n\rangle_v$, it is better to to fit
both $z P_z$ and $z^2$ dependence using all the data within $z\in
[z_1,z_2]$ and $P_z >P_z^0$.  In \fgn{combinedx2}, we show the best
fit values of $\langle x^2\rangle_v$ as a function of the maximum
of the range of $z$, i.e., $z_2$.  The left and the right panels
are for the $a=0.06$ fm and $a=0.04$ fm data.  Along with $z_2$
dependence, we have also shown $\langle x^2\rangle_v$ from the three
different values of $z$-range minimum, $z_1=1a, 2a$ and $3a$ (as
we noted, we include a $r (P_z a)^2$ term in the fits in order to
be able to use $z_1=a$ and $2a$). The two different colored symbols
differentiate the reference momenta $n_z^0=1$ and 2. These combined fits with
moments being the fit parameters lead to typical $\chi^2/{\rm dof}\approx 0.7$ in all the 
cases. For both the
lattice spacings, we find the various estimates to be consistent
with each other.  The scatter of values at $a=0.04$ fm seems to be
centered around a slightly lower value than at $a=0.06$ fm, pointing
to a small lattice spacing dependence.  Using the convention for
summarizing the various estimates, we find
\beq
\langle x^2\rangle_v = \begin{cases}0.1088^{+(48)(58)}_{-(48)(58)},\quad a=0.06\text{\ fm}\cr 
                       0.1050^{+(43)(39)}_{-(43)(39)},\quad a=0.04\text{\ fm},\end{cases}
\eeq{estx2}
at $\mu=3.2$ GeV, with the first error being statistical and the
second one being systematic.  We input the fit results from
$z_1=1a,2a,3a$, $z_2\in[0.24,0.6]$ fm, and $n_z^0=1,2$ to obtain
the above single estimate. These estimates with statistical error
band, and with both statistical and systematic error band are shown
in \fgn{combinedx2}.  For comparison, the estimate of $\langle
x^2\rangle_v$ from JAM collaboration~\cite{Barry:2018ort} at $\mu=3.2$
GeV is at a slightly lower value, 0.095. The soft-gluon resummed ASV
result~\cite{Aicher:2010cb} is even lower at about 0.086 at the same scale $\mu$.

In \fgn{combinedx4}, we show a similar plot for $\langle x^4\rangle_v$
at $\mu=3.2$ GeV.  At each $z_2$, we show determinations with
$z_1=1a, 2a$ and $3a$.  We find  consistent determinations with various
fit ranges and renormalization procedures. We estimate
\beq
\langle x^4\rangle_v = \begin{cases}0.0346^{+(50)(73)}_{-(57)(73)},\quad a=0.06\text{\ fm}\cr 
                       0.0382^{+(43)(54)}_{-(44)(54)},\quad a=0.04\text{\ fm},\end{cases}
\eeq{estx4}
These estimates are the bands in \fgn{combinedx4}.  The JAM estimate,
$\langle x^4\rangle_v=0.032$, is slightly lower than for the 300
MeV pion studied here~\cite{Barry:2018ort}. Whereas the ASV result for the fourth moment can be inferred to be  
about 0.023.  For both $\langle x^2\rangle_v$ and $\langle x^4\rangle_v$, we did not use priors.
On the other hand, it was not possible to obtain a good estimate
of $\langle x^6\rangle_v$ without inputting the knowledge of the
lower moments using the procedure we outlined previously.  We
obtained results for $\langle x^6\rangle_v$ at the two lattice
spacings by inputting prior for only $\langle x^2\rangle_v$,  and
by using priors for both $\langle x^2\rangle_v$ and $\langle
x^4\rangle_v$.  We display the results for the latter case in
\fgn{combinedx6}.  In addition, we have to use $z_2>0.5$ fm in order
for $\langle x^6\rangle_v$ to be a relevant parameter in the fit.
As we noted in the previous section, the $\Lambda_{\rm QCD}^2 z^2$
corrections seem to be canceled effectively even in the ratio scheme
with $P_z^0=0$, and the error we commit by using values of $z_2$
up to 1 fm might not be large and also further reduced by non-zero
$P_z^0$ we use in the modified ratio scheme.  Perhaps this is the
reason, we find the estimates to be independent of $z_2$ and $P_z^0$
to a good degree. We estimate
\beq
\langle x^6\rangle_v = \begin{cases}0.0117^{+(26)(33)}_{-(26)(33)},\quad a=0.06\text{\ fm}\cr 
                       0.0126^{+(20)(41)}_{-(15)(41)},\quad a=0.04\text{\ fm}.\end{cases}
\eeq{estx6}
To compare, the JAM and ASV estimates as inferred from their fits are 0.015 and 0.009 respectively.
In the above fits, we obtained the coefficient $r$ of the $(P_z
a)^2$ correction to be $-0.026(7)(10)$ and $-0.018(8)(8)$ for $0.06$
fm and $0.04$ fm lattice spacings, which are quite consistent with
each other as expected, and with our rough estimate in the last
subsection. We should also point out that in the above discussion, we did not 
include any target mass correction (trace) terms in the OPE used in fits since 
we did not find any significant change by including such additional terms due to
the smallness of pion mass.

\section{Valence PDF of pion by fits to boosted pion matrix elements in real space}\label{sec:pdf}

\bef
\centering
\includegraphics[scale=0.6]{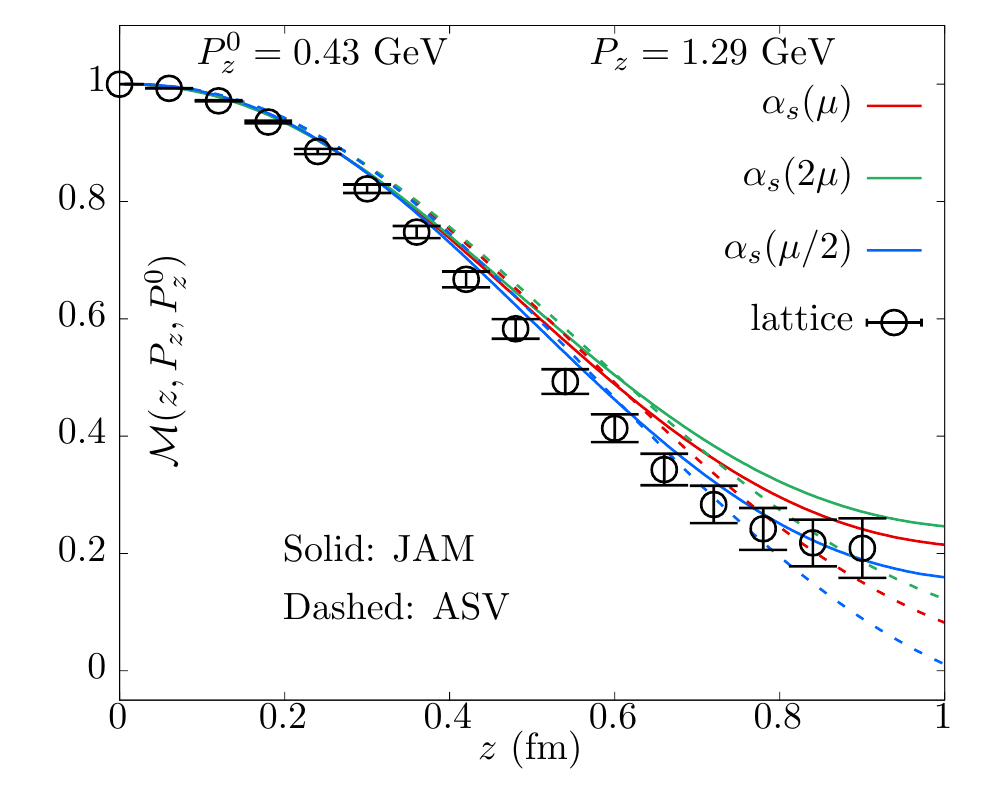}
\caption{
The plot shows the model boosted pion matrix element ${\cal M}(z,P_z,P_z^0)$ at $P_z=1.29$
GeV, $P_z^0=0.43$ GeV as a function of $z$, constructed based on
the JAM valence PDF~\cite{Barry:2018ort} (solid curves) and ASV
result~\cite{Aicher:2010cb} (dashed curves) at $\mu=3.2$ GeV.  The
red, green and blue curves are the matrix elements constructed using
\eqn{ratgenexp} using values of $\alpha_s(\mu)$, $\alpha_s(2\mu)$
and $\alpha_s(\mu/2)$ respectively.  The lattice data for ${\cal
M}(z,P_z,P_z^0)$ from $a=0.06$ fm lattice are also shown.
}
\eef{jamitd}

\befs
\centering
\includegraphics[scale=0.65]{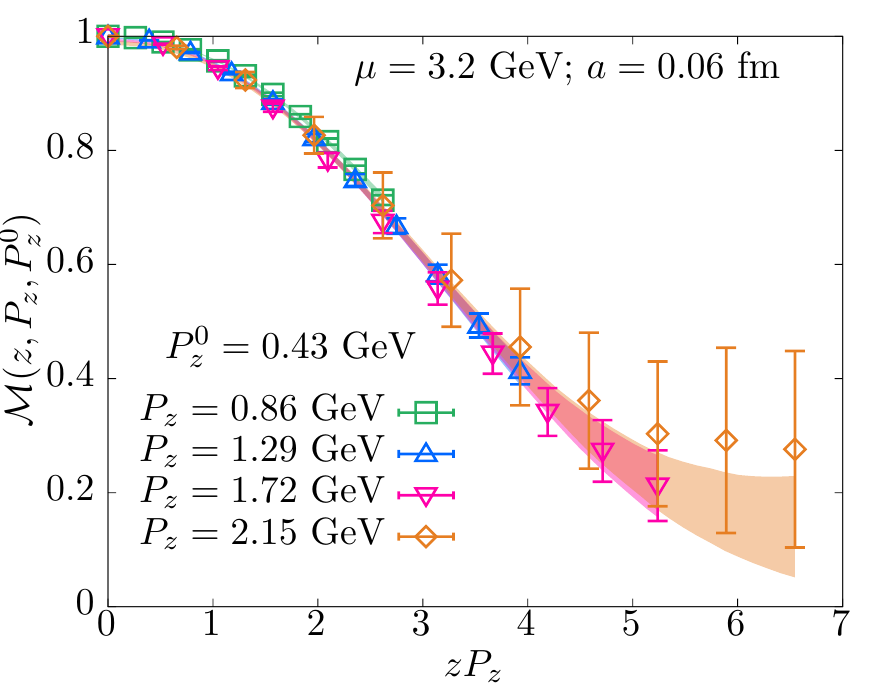}
\includegraphics[scale=0.65]{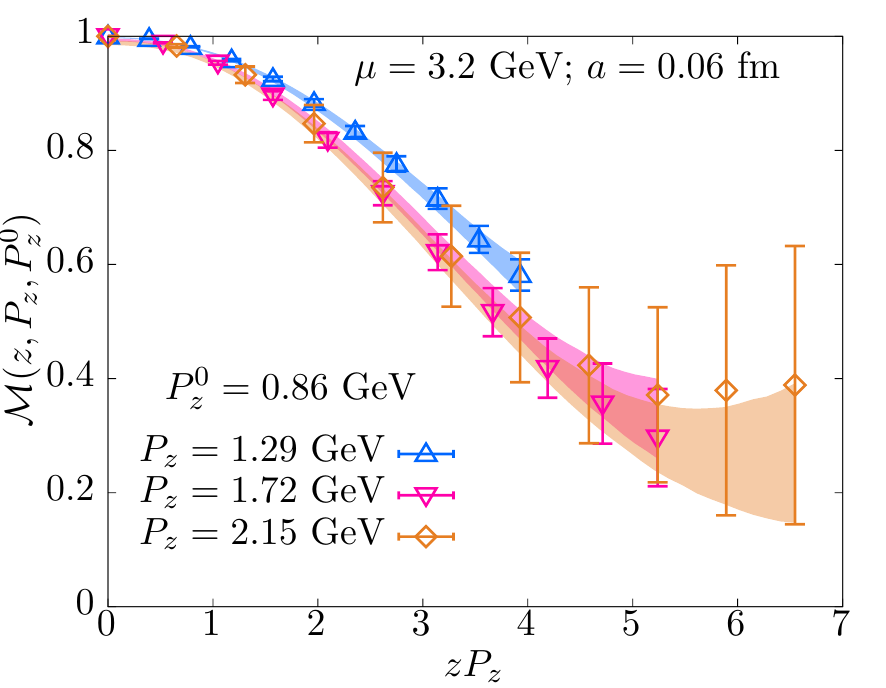}
\includegraphics[scale=0.65]{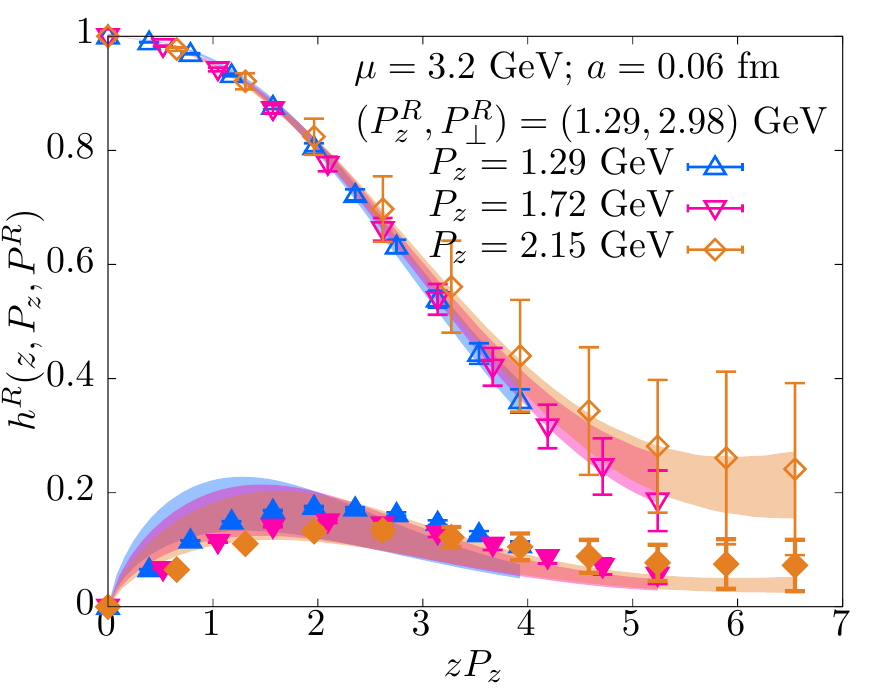}

\includegraphics[scale=0.65]{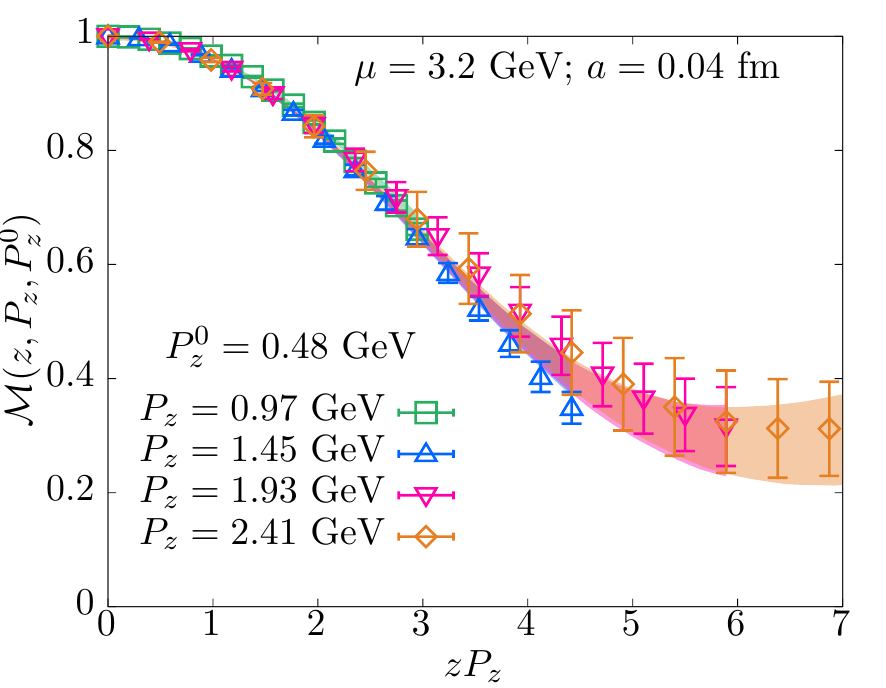}
\includegraphics[scale=0.65]{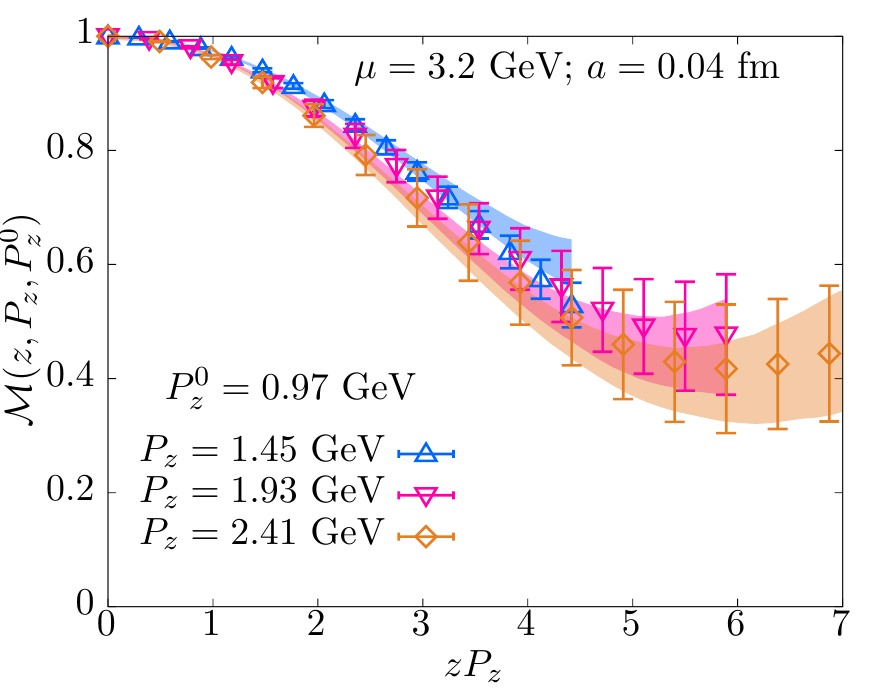}
\includegraphics[scale=0.65]{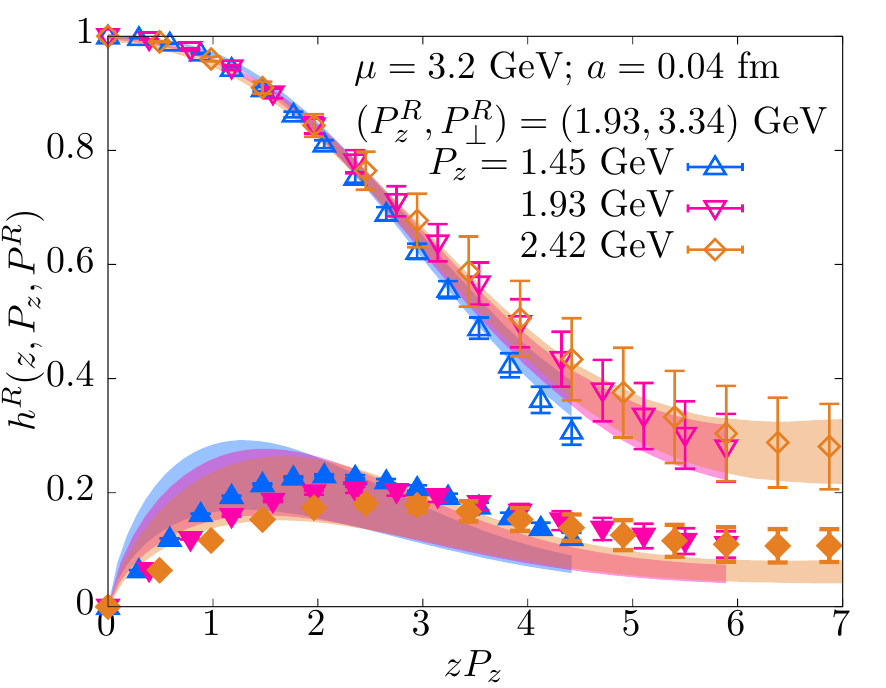}

\caption{Renormalized pion matrix elements in various schemes are shown as a function of $z P_z$ 
along with the best fits using PDFs $f_v^\pi(x)$ of the form
$x^\alpha(1-x)^\beta\left(1+s\sqrt{x}+tx\right)$.  The top and
bottom rows show the results for $a=0.06$ fm and $a=0.04$ fm
respectively. The leftmost panels are using the ratio renormalization
scheme with the reference momentum $P_z^0=2\pi/(La)$, which is
$P_z^0=0.43$ GeV for $a=0.06$ fm and $P_z^0=0.48$ GeV for $a=0.04$
fm. The middle panels use $P_z^0=4\pi/(La)$, which is $P_z^0=0.86$
GeV for $a=0.06$ fm and $P_z^0=0.97$ GeV for $a=0.04$ fm.  The
rightmost panels are in the RI-MOM scheme with the renormalization
momentum $(P_z^R,P_\perp^R)=(1.29,2.98)$ GeV for $a=0.06$ fm case
and $(1.93,3.34)$ GeV for $a=0.04$ fm case. Here, the real and
imaginary parts are shown.  In each panel, the different colored
symbols are the actual lattice data at different pion
momentum $P_z$. The corresponding similarly colored bands are the
results of the combined fit over the fixed range $z\in[2a,0.5]$fm
from different $P_z>P_z^0$ in the ratio scheme, and $n_z=3,4,5$ in
the case of RI-MOM scheme.}
\eefs{ritdfit}

\befs
\centering
\includegraphics[scale=0.85]{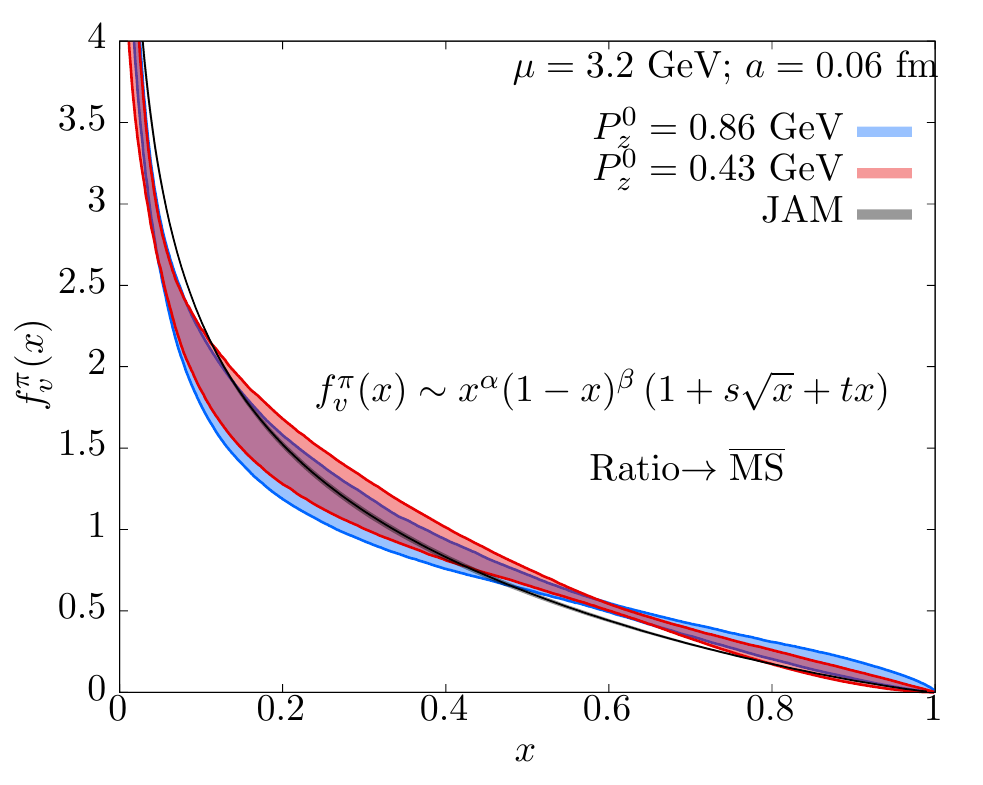}
\includegraphics[scale=0.85]{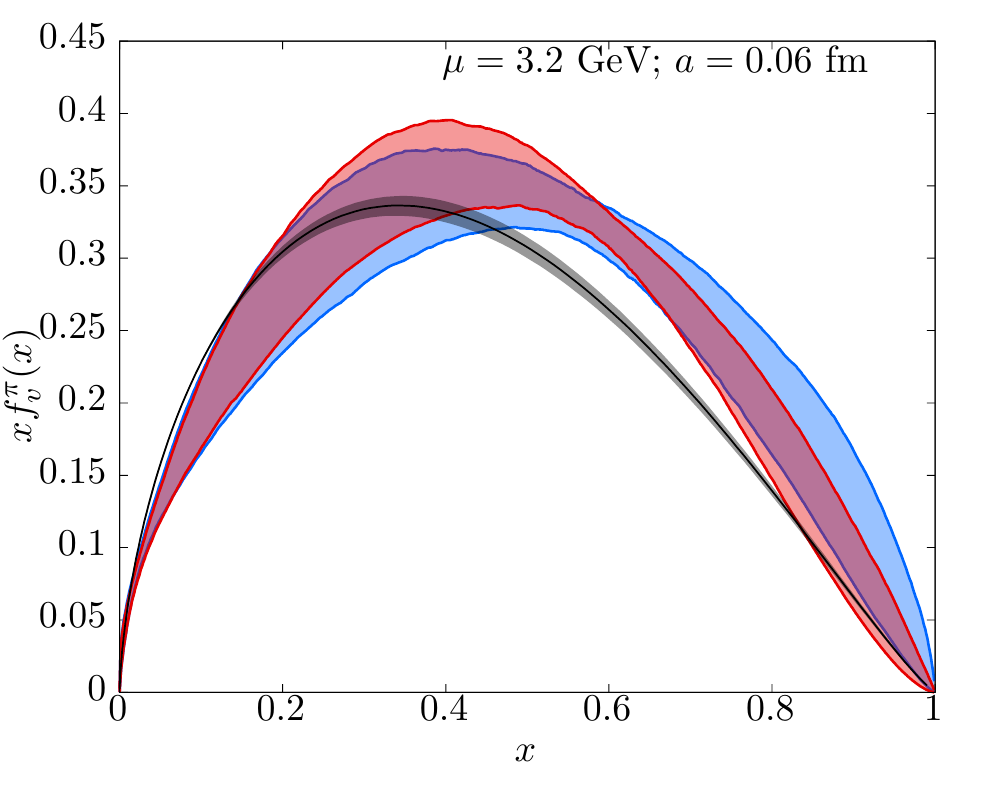}

\includegraphics[scale=0.85]{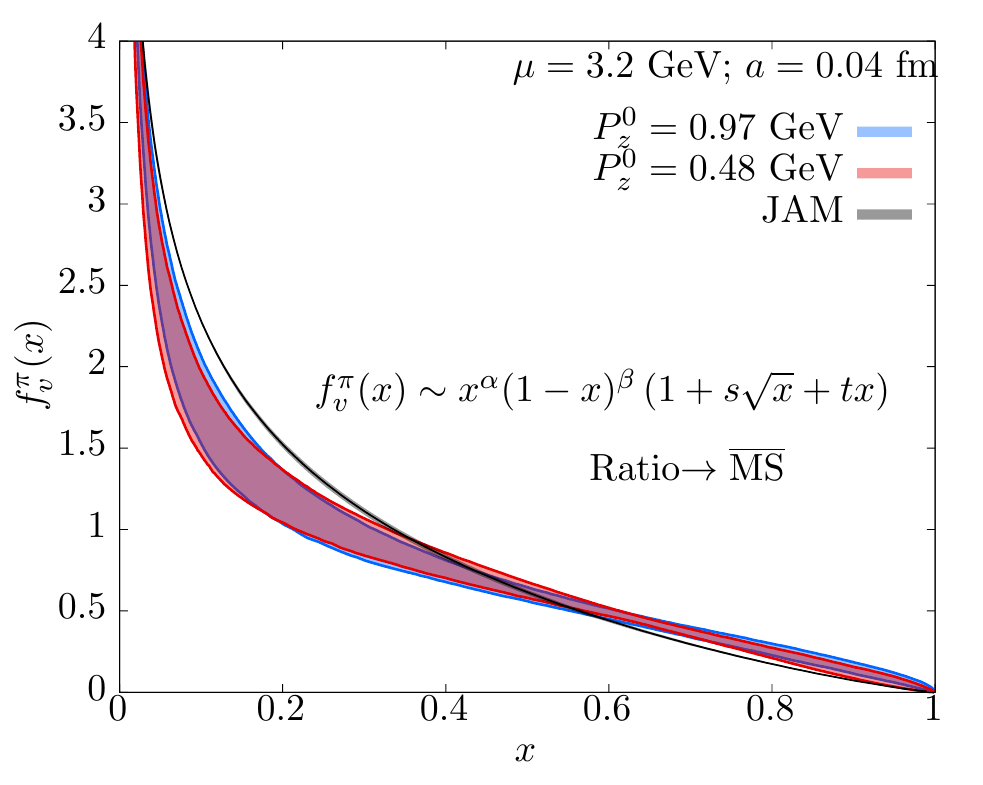}
\includegraphics[scale=0.85]{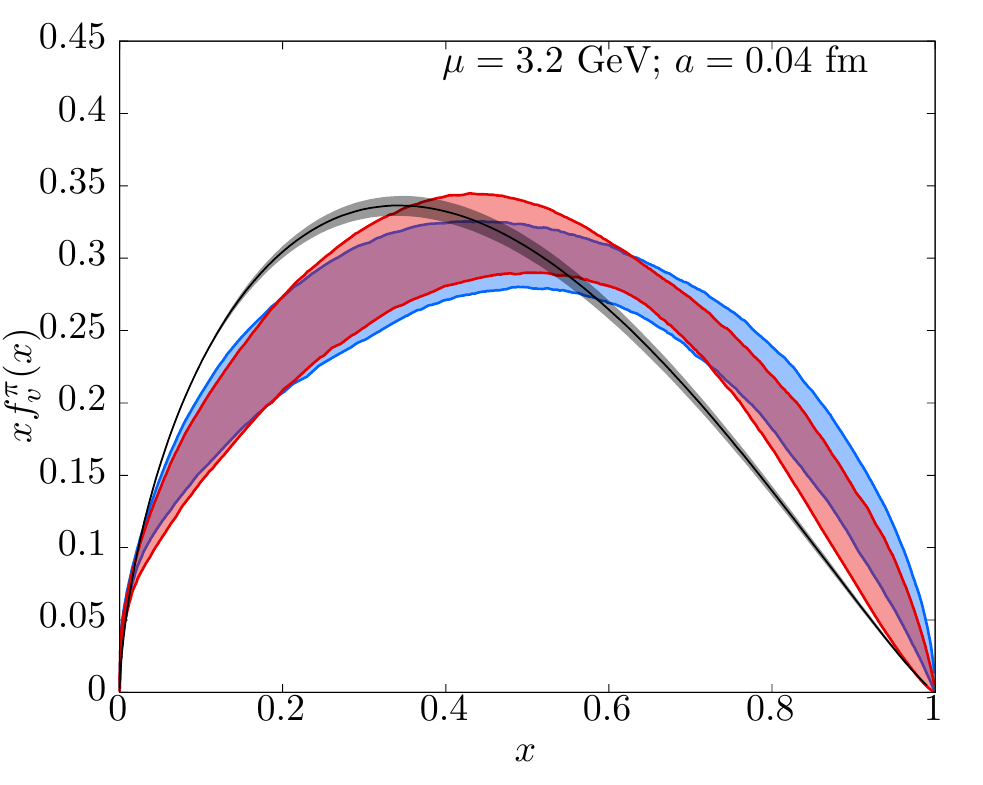}

\caption{The valence PDF of pion $f_v^\pi(x)$ at $\mu=3.2$ GeV. The
top and bottom panels are for $a=0.06$ fm and $a=0.04$ fm respectively.
The left panels show $f_v^\pi(x)$ and the right panel re-plot the
same data as $x f_v^\pi(x)$. The ansatz $f_v^\pi(x)={\cal N}x^\alpha
(1-x)^\beta(1+s\sqrt{x}+t x)$ was used for this reconstruction of
valence PDF via their ability to describe pion matrix elements in real space in different ratio
schemes involving ranges of quark-antiquark separation $z\in[2a,0.5]$
fm.  In each panel, such results for $f_v^\pi(x)$ based on ratio
schemes with reference momenta $P_z^0=2\pi n^0_z/L$ with $n^0_z=1,2$
are shown as different colored bands. For comparison, the JAM result
at the same $\mu$ is shown as the black band.  }
\eefs{pdffit}

In the last section, we estimated the even moments directly from
the equal-time boosted pion matrix elements. However, it is not possible to reconstruct an
$x$-dependent PDF using only the knowledge of the first few even
moments.  One way of PDF reconstruction from  the boosted pion matrix element is through
data interpolation over the range of $z$ where lattice data
is available and then extrapolate it to zero smoothly at larger
$z$~\cite{Alexandrou:2019lfo,Liu:2018uuj}.  
Instead, as in our
previous work, we adopt the method of using phenomenology motivated
ansatz for $f_v^\pi(x)$ and fit the ansatz to our lattice matrix element over
ranges of $z$ smaller than 1 fm. In this way, we avoid the usage
of data with $z\gtrapprox 1$ fm which could be deep in the
nonperturbative regime, and might not be consistent with the
perturbative framework that we rely on. There are also 
other methods of PDF reconstruction that have been investigated in the 
literature~\cite{Karpie:2019eiq,Bhat:2020ktg}.

\subsection{PDF ansatz and analysis method}
As is typical in the global analysis of valence PDFs, we use two 
different valence pion PDF ansatz 
\beqa
f_v^\pi(x;\alpha,\beta)&=&{\cal N}x^\alpha\left(1-x\right)^\beta,\cr
f_v^\pi(x;\alpha,\beta,s,t)&=&{\cal N}' x^\alpha\left(1-x\right)^\beta\left(1+s\sqrt{x}+tx\right),
\eeqa{pdfansatz}
with the first one being a special case of the second and hence
more restrictive.  The normalization factors ${\cal N}, {\cal N}'$
are chosen such that $\int_0^1 f_v^\pi(x) dx=1$.  The parameters
$\alpha,\beta,s,t$ are the tunable fit parameters. These model PDFs
enter the analysis via their corresponding moments, for example
$\langle x^n\rangle(\alpha,\beta)=\int_0^1 x^n f_v^\pi(x;\alpha,\beta)dx$,
which appear in the OPE expressions; \eqn{ratgenexp} for the ratio
scheme and \eqn{t2operi} for RI-MOM scheme. In both the schemes,
we corrected for $(P_z a)^2$ lattice artifacts that affect smaller
$z$ by using a term $r (P_z a)^2$ in the OPE expressions with $r$
being a fit parameter, as we did in our model independent fits.
Through this, we can construct the model matrix elements ${\cal M}_{\rm
model}(z,P_z,P_z^0;\alpha,\beta,\ldots)$ and $h^R_{\rm
model}(z,P_z,P^R;\alpha,\beta,\ldots)$.

Let us first consider the ratio scheme. In addition to the statistical
error $\sigma_{\rm stat}(z,P_z,P_z^0)$ for the lattice data point
${\cal M}(z,P_z,P_z^0)$, there is also the perturbative uncertainty
resulting from the 1-loop truncation of the twist-2 Wilson coefficients.
We quantify this error through the arbitrary nature of the scale
$\mu$ of the strong coupling $\alpha_s(\mu)$, as we did in \scn{mom}.
We use $\mu$ in the $\alpha_s$ to be the same as the factorization
scale of the PDF, and quantify the error we commit through the
systematic error $\sigma_{\rm sys}(z,P_z,P_z^0)$ which we define
as the change in ${\cal M}_{\rm model}(z,P_z,P_z^0;\alpha,\beta,\ldots)$
when $\alpha_s$ is changed from $\alpha_s(\mu/2)$ to $\alpha_s(2\mu)$.
That is,
\beqa
\sigma_{\rm sys}(z,\ldots) &=&\frac{1}{2}\bigg{(}{\cal M}_{\rm model}(z,\ldots)|_{\alpha_s(\mu/2)} -\cr &&\quad {\cal M}_{\rm model}(z,\ldots)|_{\alpha_s(2\mu)}\bigg{)}.
\eeqa{syserror}
Let us take the JAM data and the ASV analysis data at $\mu=3.2$ GeV as a specific case. The JAM
data can be described to a very good accuracy by the form \eqn{pdfansatz}
with $\alpha=-0.37$ and $\beta=1.20$. In \fgn{jamitd}, we show the result for ${\cal
M}_{\rm model}(z,P_z,P_z^0)$ at $P_z=1.29$ GeV, $P_z^0=0.43$ GeV using the JAM valence
PDF~\cite{Barry:2018ort} with solid curves, and using ASV result~\cite{Aicher:2010cb} using dashed curves. 
For each case, we plot three different curves for ${\cal
M}_{\rm model}$ as obtained using $\alpha_s(\mu/2)$, $\alpha_s(\mu)$
and $\alpha_s(2\mu)$.  For comparison, the actual lattice data and
the error for ${\cal M}_{\rm model}(z,P_z,P_z^0)$ is also shown.  We
can see that the spread in ${\cal M}_{\rm model}$ for both JAM and ASV get especially
important for $z>0.4$ fm, and become comparable to the statistical
error in the data. Therefore, given the significant perturbative
uncertainty that is unavoidable at present, it would be misleading
to favor or rule out models of PDF (such as JAM and ASV results in the example here) 
simply based on the statistical precision of the lattice data.  
Therefore, we select the model PDFs
that best describes the shape of the lattice matrix element that takes
$\sigma_{\rm sys}$ into account, by minimizing,
\beqa
&&\chi^2 \equiv\cr&& \sum_{P_z>P_z^0}^{P_z^{\rm max}}\sum_{z=z_1}^{z_2}\frac{\left({\cal M}(z,P_z,P_z^0)-{\cal M}_{\rm model}(z,P_z,P_z^0;\alpha,\ldots)\right)^2}{\sigma^2_{\rm stat}(z,P_z,P_z^0)+\sigma_{\rm sys}^2(z,P_z,P_z^0)}.\cr&&\quad
\eeqa{chisqsys}
The correlations between the lattice data at different $z$ and $P_z$
are partly taken into account by picking ${\cal M}(z,P_z,P_z^0)$
from the same bootstrap samples. Similarly, in the case of RI-MOM
matrix element, we fit only the real part, $\text{Re}\left[h^R(z,P_z,P^R)\right]$,
and the imaginary part is obtained as an outcome. In the case of
RI-MOM matrix element, we found taking care of $\sigma_{\rm sys}$
to be even more important as the $\alpha_s$ dependence starts from
$c_0^{\rm RI}$, unlike in ratio scheme.

\befs
\centering
\includegraphics[scale=0.8]{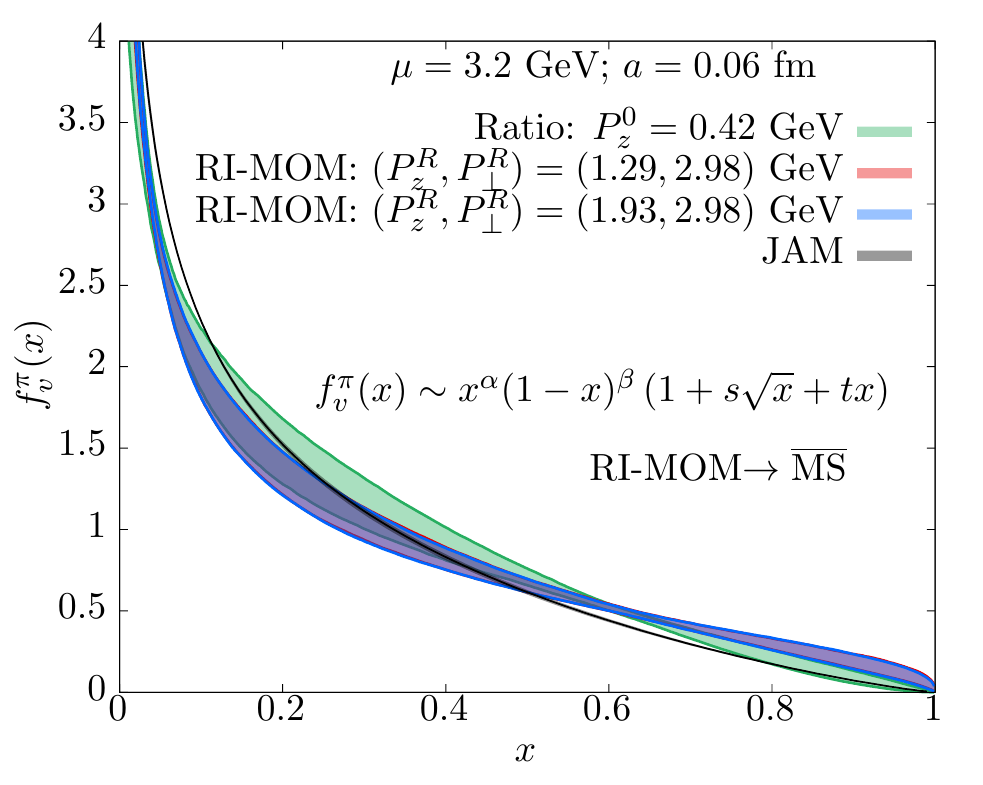}
\includegraphics[scale=0.8]{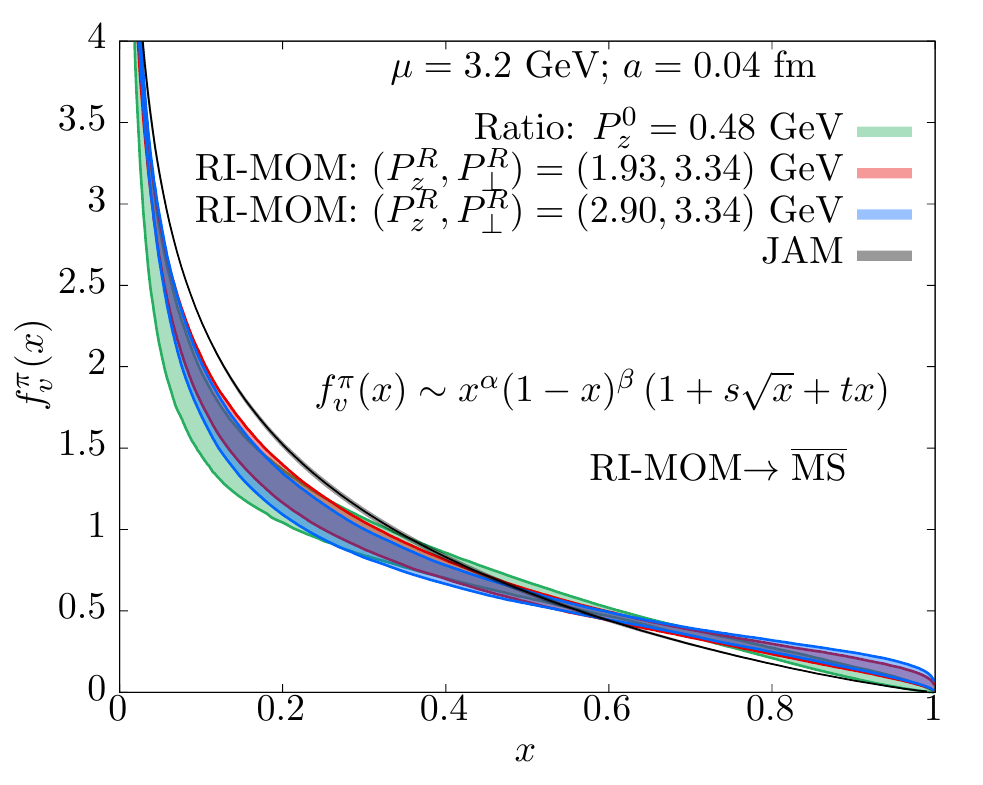}
\caption{Valence PDF of pion at $\mu=3.2$ GeV extracted from RI-MOM
renormalized matrix elements. The left and the right panels show data for
$a=0.06$ fm and $a=0.04$ fm. The red and blue bands are the PDFs
that best describes the RI-MOM data at two different RI-MOM scales
$(P^R_z,P_\perp^R)$.  For comparison, the PDF as extracted
using ratio scheme with $P^0_z=2\pi n^0_z/(La)$ with
$n_z^0=1$ is shown as the green band. The JAM estimate~\cite{Barry:2018ort}
of $f_v^\pi(x)$ is shown as the black band.}
\eefs{pdfri}

\subsection{Results for $f_v^\pi(x)$}
In \fgn{ritdfit}, we show the resulting best fit model matrix elements along
with the actual lattice data.  We have used the 4-parameter ansatz
$f_v^\pi(x;\alpha,\beta,s,t)$ for the fits shown.
For the fits, we used $\mu=3.2$ GeV in the perturbative Wilson
coefficients, and hence the model PDF corresponds to this factorization
scale.  In the results shown in \fgn{ritdfit}, we used only the
boosted matrix elements with the quark-antiquark separations $z\in[2a,0.5\text{\ fm}]$,
but we performed the analysis also with $z_1=a,2a,3a$ and $z_2\in[0.36,0.72]$
fm.  The matrix elements at different fixed $P_z$ are differentiated (by their
color and symbols). In the top and bottom panels we have shown the
results for $a=0.06$ and 0.04 fm respectively. We have shown the
results for the ratio scheme with $P_z^0=2\pi n_z^0/(La)$ for
$n_z^0=1$ and 2 in the left and middle panels of \fgn{ritdfit}.
For $n_z^0=1$ ratio scheme, we used the momenta with $n_z=2,3,4,5$,
and for $n_z^0=2$, we used $n_z=3,4,5$.  For the RI-MOM scheme, we
used only the larger set of momenta corresponding $n_z=3,4,5$. 
This is to avoid the larger ${\cal O}(\Lambda_{\rm QCD}^2 z^2)$ corrections 
in the RI-MOM scheme observed in~\scn{ren}.
The results
of the fit to the RI-MOM matrix elements at renormalization scales
$(P^R_z,P^R_\perp)=(1.29,2.98)$ GeV and (1.93,3.34) GeV for the
$a=0.06$ fm and $a=0.04$ fm lattices respectively are shown in the
rightmost panels. The fit is performed only on the real part of
$h^R$.  But, the non-zero imaginary part of $h^R$ also compares
well with the resultant imaginary part of the fit. The fits in 
all the cases gave good $\chi^2/{\rm dof}$ between 0.5 and 1, and 
we discuss this in \apx{chisq}. We refer the reader to~\apx{kx0} for
a similar discussion on fits to $P_z^0=0$ ratio matrix elements (i.e., reduced ITD).

Each of the best fit model matrix elements in \fgn{ritdfit} correspond to valence
PDFs, $f_v^\pi(x;\alpha,\beta,s,t)$ at $\mu=3.2$ GeV. In \fgn{pdffit},
we have shown the results of the valence PDFs,
$f_v^\pi(x;\alpha,\beta,s,t)$, that are reconstructed from ${\cal
M}(z,P_z,P_z^0)$. The left and the right panels of \fgn{pdffit}
show $f_v^\pi(x)$ and $x f_v^\pi(x)$ as functions of $x$.  The red
and blue bands are for the two values of $n_z^0=1,2$ respectively.
For comparison, the JAM valence PDF~\cite{Barry:2018ort} at the
same $\mu$ is shown as the black band. At a qualitative level, it
is reassuring that the PDFs we determined compares well with the
phenomenological result. At both lattice spacings, the results from
different $P_z^0$ differ only by a little, and such variations belong
in the systematic error budget.  However, when we look closely, one
can find that the best fit PDFs always have a tendency to be above
the JAM result for $x>0.6$.  This ties back to the PDF moment
determination in the last section where we found $\langle x^2\rangle_v$
and other higher moments also to be consistently higher than the
phenomenological result. In \fgn{pdfri}, we show similar results
for PDF as obtained using the RI-MOM $h^R$. The results using
two different renormalization scales $P^R$ are consistent with
each other as one would expect.  One can also note that the RI-MOM
results also agree overall with the one from ratio scheme. When we
focus on specific details of the PDF, as we would do next, the
difference across renormalization schemes and renormalization scales
will become easier to notice.

\bef
\centering
\includegraphics[scale=0.9]{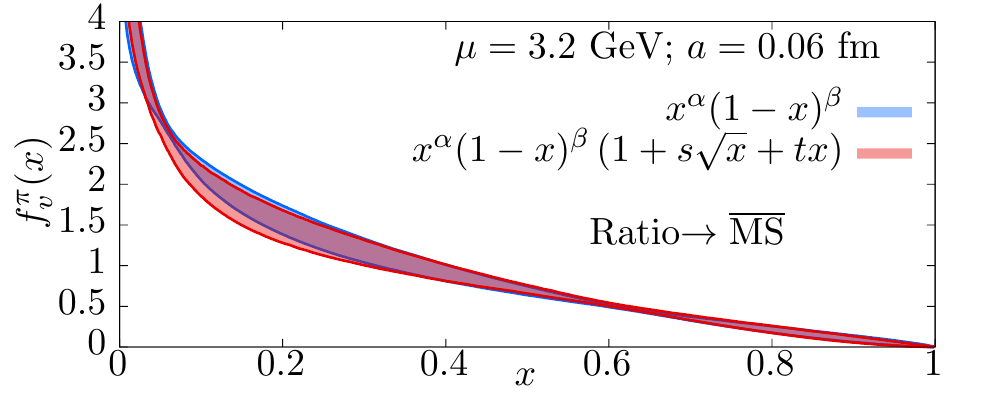}

\includegraphics[scale=0.9]{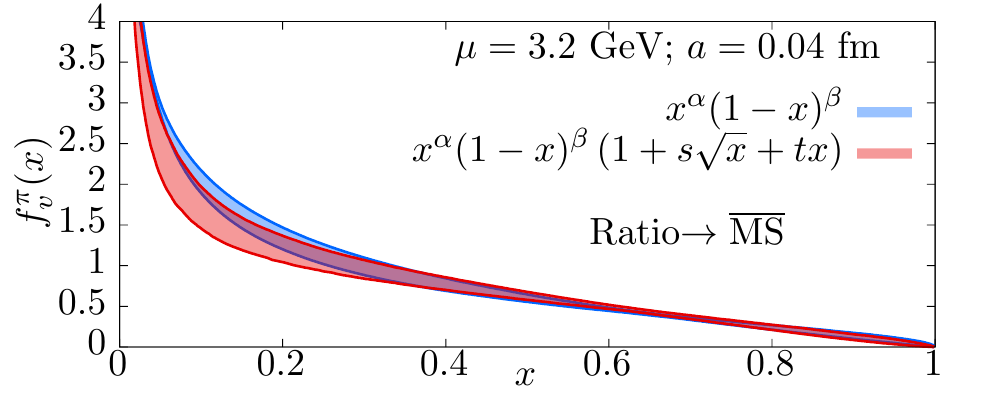}
\caption{Comparison of PDF as extracted using a simple two-parameter
$x^\alpha(1-x)^\beta$ ansatz and from a four-parameter
$x^\alpha(1-x)^\beta(1+s\sqrt{x}+t x)$ ansatz.  }
\eef{pdfansatz}

For the results we discussed above, we limited ourselves to a
specific fit range in $z$ from $2a$ up to $0.5$ fm using an ansatz
$f_v^\pi(x;\alpha,\beta)$. The obvious addendum to this discussion
is to also specify what happens when we change the various choices
we used in the fits. First, we check that the constructed PDF is not
sensitive to the PDF ansatz. We used both the ansatz in \eqn{pdfansatz}
in our analysis, and in fact, the simpler ansatz $f_v^\pi(x;\alpha,\beta)$
by itself is sufficient to describe our pion matrix elements in real space; in all the cases $\chi^2/{\rm dof}
$ varied between 0.5 to 0.9. The ansatz $f_v^\pi(x;\alpha,\beta,s,t)$ includes terms that
affect only the small-$x$ behavior and therefore more flexible. In
\fgn{pdfansatz}, we compare the best fit PDFs using the two ansatz
for a sample case that used ratio scheme with $n_z^0=1$.  It is
clear that the ansatz dependence is very little, and the effect of
including more free parameters in $f_v^\pi(x;\alpha,\beta,s,t)$ is
to increase the uncertainties in the fitted PDFs without changing
the overall shape.

It would be cumbersome to describe one dependence after another in
terms of the resulting PDFs. Therefore, we summarize the results
of fitted PDFs using various choices for $z$-range $[z_1,z_2]$ and
the renormalization schemes via their first four moments $\langle
x\rangle_v$,  $\langle x^2\rangle_v$, $\langle x^3\rangle_v$ and
$\langle x^4\rangle_v$ in \fgn{fitx1}.  It is noteworthy that even
though we cannot access the odd moments directly, we can obtain
them indirectly from model PDFs. Let us focus on one of the panels
in \fgn{fitx1} to unpack the details.  Each point is an estimate
of the moment labeled in the $x$-axis of the panel. The red and
blue points result from using $f_v^\pi(x;\alpha,\beta)$ and
$f_v^\pi(x;\alpha,\beta,s,t)$ respectively, and demonstrates the
variation due to fitted ansatz. For each of the ansatz (i.e., red
or blue), the variation due to the range of $z$ used, $[z_1,z_2]$,
is shown as one moves up along the $y$-axis. The results with the
same $[z_1,z_2]$ are enclosed within the dashed lines.  For each
$[z_1,z_2]$, the variation coming from the renormalization scheme
used for the equal time matrix elements is shown. We have shown four such renormalized
results with each set of $[z_1,z_2]$ --- ratio scheme with $n_z^0=1,2$
(denoted as ratio-1 and ratio-2 in the figure), and RI-MOM scheme
at two different $(P_z^R,P_\perp^R)$ (denoted as RI-1 and RI-2).
For $a=0.06$ fm, the two RI-MOM scales are (1.29,2.98) GeV and
(1.93,2.98) GeV, and for $a=0.04$ fm, the two scales are (1.93,3.34)
GeV and (2.9,3.34) GeV.  It is satisfactory that the results for
$\langle x^2\rangle_v$ and $\langle x^4 \rangle_v$ obtained here
indirectly agrees well with the direct determination in the last
section, and serves as a cross-check. One sees comparatively larger
renormalization dependence in the more stringent two-parameter
ansatz, but it seems to be reduced and accounted for by using a
more flexible four-parameter ansatz.  As we change $z_2$ from 0.42
fm to 0.72 fm, the results remain almost intact.  From these fits,
we estimate the moments and their statistical and systematic errors
(coming from $[z_1,z_2]$, RI-MOM and ratio renormalization schemes, and the two PDF
ansatz) as
\beqa
\langle x\rangle_v&=&\begin{cases} 0.2491^{+(77)(61)}_{-(81)(61)}\ ,\ \ a=0.06\text{\ fm}\cr
                     0.2296^{+(79)(57)}_{-(87)(57)}\ ,\ \ a=0.04\text{\ fm}.
       \end{cases}\cr
\langle x^2\rangle_v&=&\begin{cases} 0.1174^{+(50)(71)}_{-(44)(71)}\ ,\ \ a=0.06\text{\ fm}\cr
                     0.1122^{+(45)(57)}_{-(52)(57)}\ ,\ \ a=0.04\text{\ fm}.
       \end{cases}\cr
\langle x^3\rangle_v&=&\begin{cases} 0.0698^{+(52)(80)}_{-(48)(80)}\ ,\ \ a=0.06\text{\ fm}\cr
                     0.0690^{+(52)(60)}_{-(52)(60)}\ ,\ \ a=0.04\text{\ fm}.
       \end{cases}\cr
\langle x^4\rangle_v&=&\begin{cases} 0.0470^{+(52)(76)}_{-(47)(76)}\ ,\ \ a=0.06\text{\ fm}\cr
                     0.0478^{+(44)(58)}_{-(51)(58)}\ ,\ \ a=0.04\text{\ fm}.
       \end{cases}
\eeqa{momfitdat}
The indirect determination of the first moment $\langle x\rangle_v$
shows that each of the two valence quarks carry about a quarter of
the pion energy as has been seen before.  Especially here, one
certainly sees a lattice spacing effect that tends to make $\langle
x\rangle_v$ closer to the JAM value of 0.223 at $\mu=3.2$ GeV. Such
a lattice spacing effect is seen to a lesser extent in $\langle
x^2\rangle_v$, and difficult to see in the higher moments.

\befs
\centering
\includegraphics[scale=0.65]{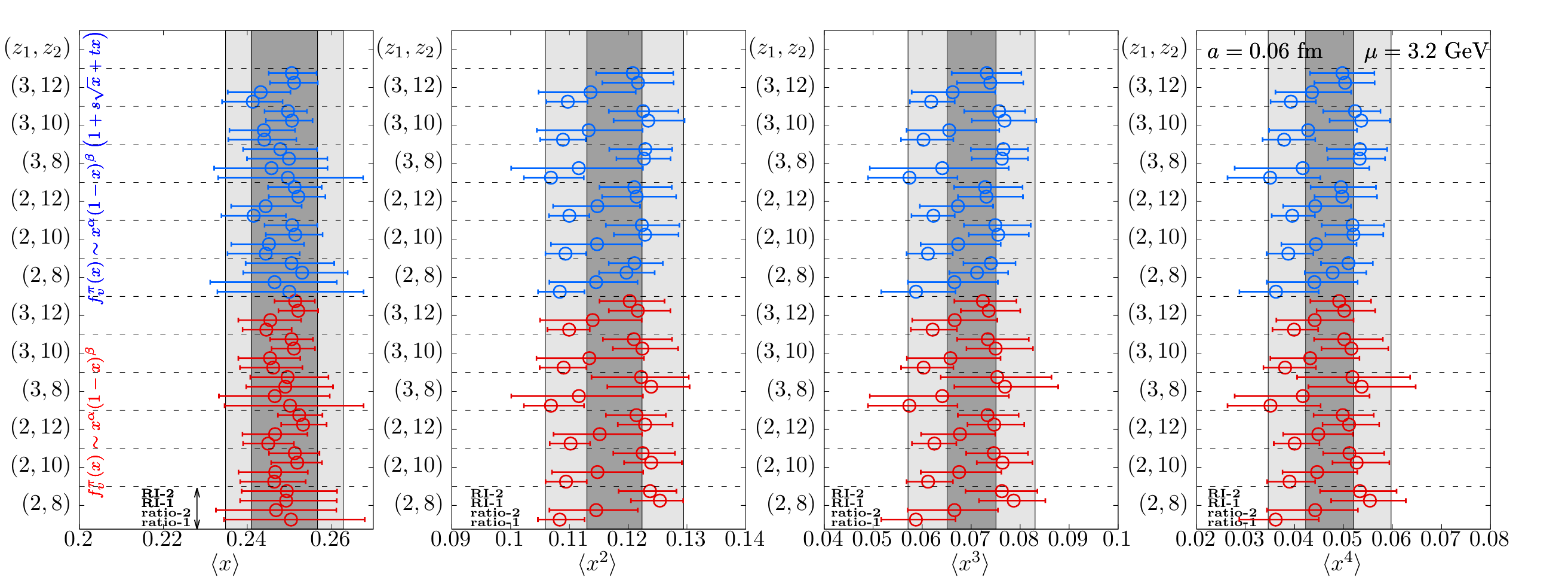}
\includegraphics[scale=0.65]{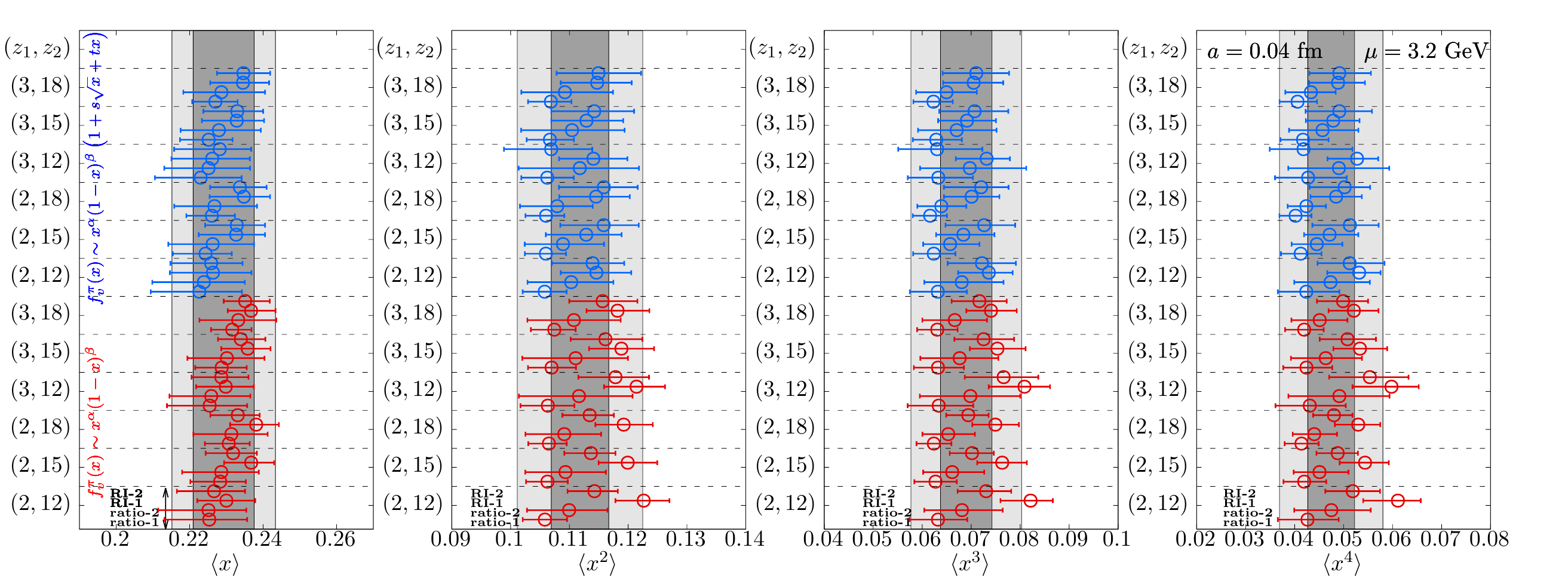}
\caption{The first four valence PDF moments $\langle x^n\rangle$
as inferred from the best estimates of PDFs $f_v^\pi(x)$ that best
describes the equal-time pion matrix elements in ratio and RI-MOM renormalization schemes. The top panels
are for $a=0.06$ fm and bottom ones for $a=0.04$ fm.  The dependence
of $\langle x^n\rangle$ on all the variables in the lattice analysis
is summarized in the above plot. The foremost variable is the range
of quark-antiquark separations used in the fits $z\in[z_1,z_2]$.
Such variations are bunched together as blocks separated by the
dashed lines along the y-axis. The second variable factor is the
renormalization scheme of the matrix elements: it could be RI-MOM scheme
or ratio scheme at reference scale $P_z^0$. At fixed $z\in[z_1,z_2]$,
four different renormalization points are shown:  ratio scheme at
$n_z^0=1$ (ratio-1), $n_z^0=2$ (ratio-2),
RI-MOM scheme at two different scales $P^R$, denoted as RI-1 and RI-2. 
This scheme and scale variations are bunched together
within the dotted lines. The tertiary variable is the fit ansatz:
the results obtained using the ansatz $f_v^\pi(x)={\cal
N}x^\alpha(1-x)^\beta$ are shown in red and those using $f_v^\pi(x)={\cal
N}x^\alpha(1-x)^\beta(1+s\sqrt{x}+tx)$ are shown in blue.  }
\eefs{fitx1}

\section{Discussion on large-$x$ behavior}\label{sec:largex}

\befs
\centering
\includegraphics[scale=0.65]{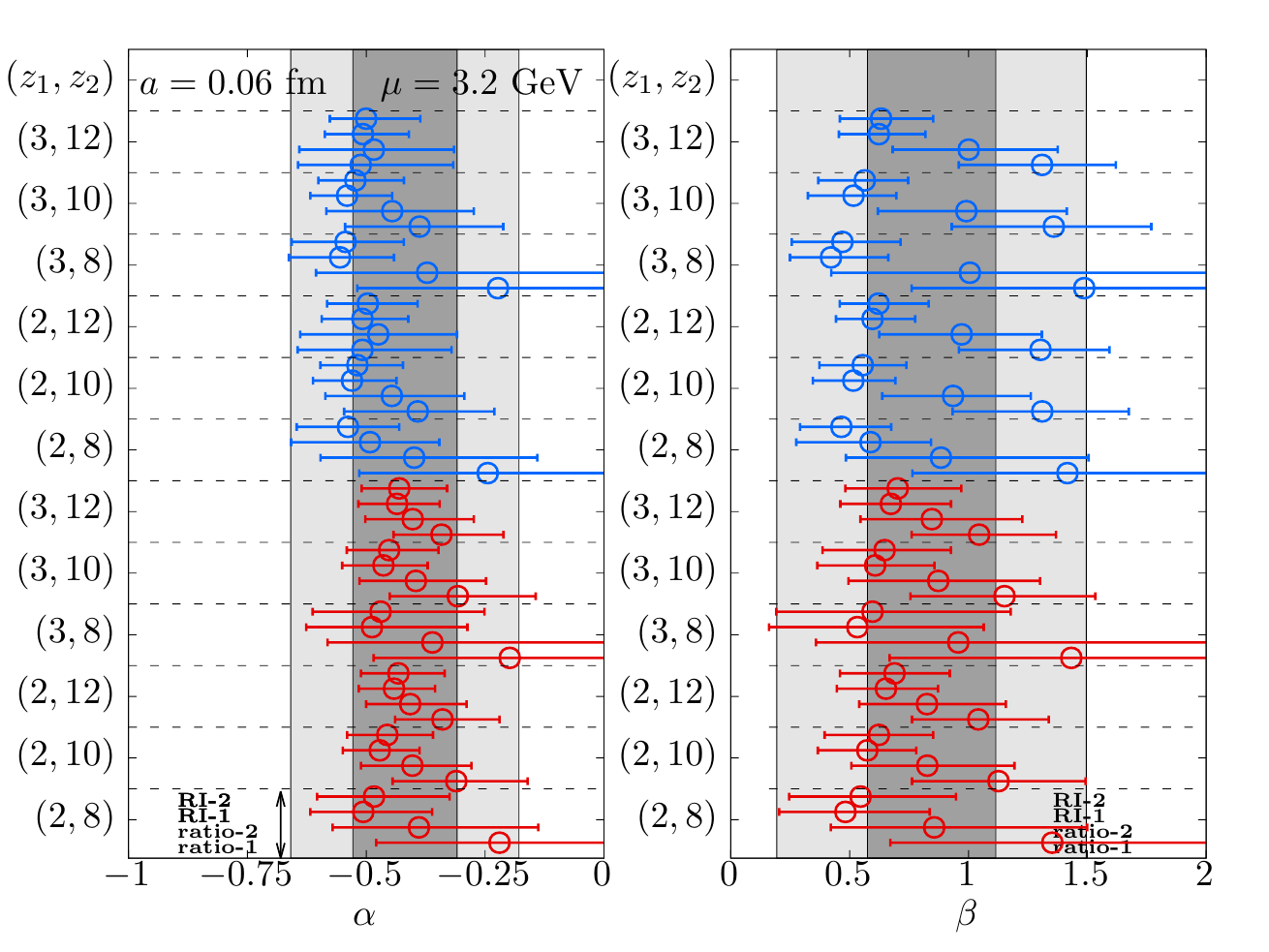}
\includegraphics[scale=0.65]{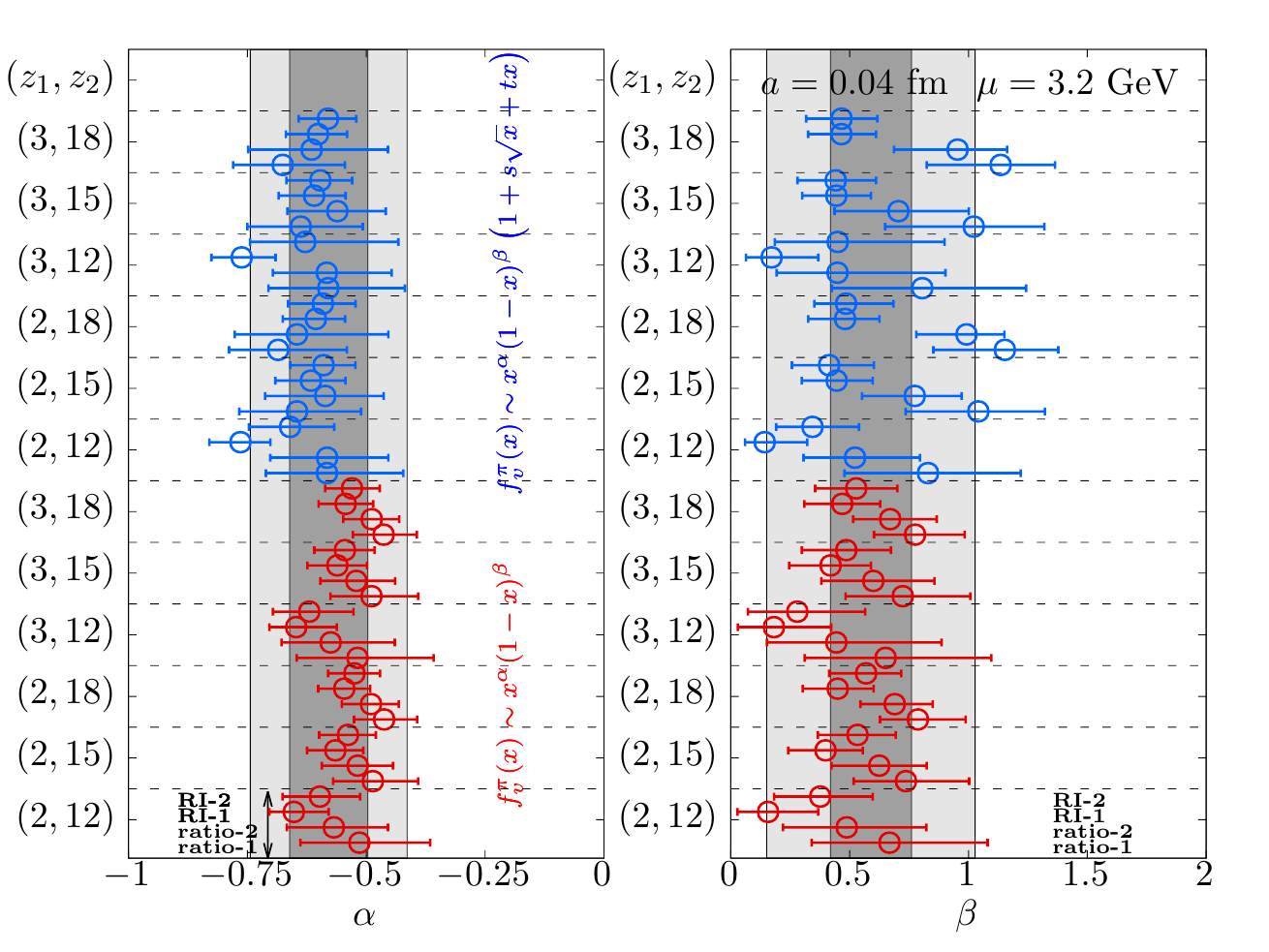}
\caption{
The exponents $\alpha$ and $\beta$ inferred from the estimates
of PDFs $f_v^\pi(x)=x^\alpha(1-x)^\beta(1+\ldots)$
that best describes the ratio and RI-MOM real space data.  The
first two panels are for $a=0.06$ fm and last two for $a=0.04$ fm.
The dependence of the exponents on all the variables in the lattice
analysis is summarized in the above plot.  The notation is similar
to \fgn{fitx1}.  
}
\eefs{fitalpha}

\bef
\centering
\includegraphics[scale=0.95]{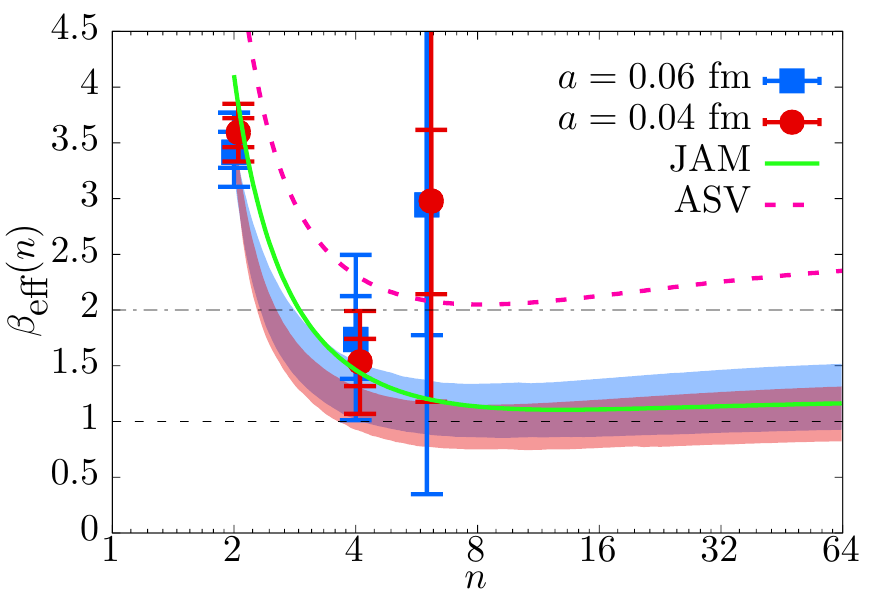}
\caption{
    The plot shows the effective large-$x$ exponent $\beta_{\rm
    eff}(n)$ as a function of $n$.  The data points are obtained
    by using values of moments $\langle x^n\rangle_v$ obtained from
    the model independent combined fits. The smaller error bar is
    only the statistical error and larger error bar is statistical
    plus systematic error.  The red and blue points are the results
    using our $a=0.06$ fm and $0.04$ fm respectively.  The red and
    blue bands are the expectation for the behavior of $\beta_{\rm
    eff}(n)$ as obtained from the model-dependent fits to the pion matrix elements.  The
    green curve is the expectation using the JAM data~\cite{Barry:2018ort} and the purple dashed curve 
    is obtained using the ASV analysis~\cite{Aicher:2010cb}.}

\eef{beff}

The large-$x$ behavior of PDFs are of the form
\beq
f(x)\sim (1-x)^\beta,
\eeq{largex}
characterizing how $f(x)$ vanishes in the $x\to 1$ limit.  The
exponent $\beta$ is hadron dependent, and one uses the Brodsky-Farrar
quark counting rule~\cite{Brodsky:1973kr} to find the typical value
of $\beta$ for a hadron; $\beta=$ 2 for the pion and 3 for the
unpolarized nucleon valence PDFs respectively from this counting
rule.  In fact, for the proton, one does find the value of $\beta$
to be close to 3 for the $u$-quark valence PDF, and motivates the
usage of quark counting rule to predict the value of $\beta$ for
other hadrons.  But, the value of $\beta$ from the
analysis~\cite{Conway:1989fs} of E-0615 Fermilab data was found to
be about 1 as a contradiction to the quark counting rule.  The
importance of soft gluon resummation in the analysis of DIS data
close to $x\to 1$ limit was pointed in~\cite{Aicher:2010cb}, and
consequent reanalysis of Fermilab result suggested a value of
$\beta\approx 2$.  The recent global Monte Carlo analysis of
experimental data from JAM collaboration~\cite{Barry:2018ort}
suggests $\beta=1.2$, and concurred by another analysis using
xFitter~\cite{Novikov:2020snp}.  Nevertheless, quark counting rules
are not direct predictions of nonperturbative QCD and there have
been lot of recent works on computing $\beta$ for the pion that
relies on alternative nonperturbative arguments, such as DSE, BSE
and light-front quantization methods; many such recent
attempts~\cite{Nguyen:2011jy,Chen:2016sno} suggest a value $\beta\approx
2$, but
some~\cite{deTeramond:2018ecg,RuizArriola:2002wr,Broniowski:2017wbr,Lan:2019rba}
of them suggest values close to 1, or a cross-over from $\beta$=1
to 2 behavior very close to $x=1$~\cite{Bednar:2018mtf}. Thus, the
issue of the value of $\beta$ for $f_v^\pi$ is still not settled
and the lattice computations, as the present one, can play an
important role.

\subsection{Model dependent estimate of $\beta$}
The recent lattice
computations~\cite{Sufian:2019bol,Sufian:2020vzb,Izubuchi:2019lyk} of
pion PDF, including our previous work, have attempted to address
the issue of $\beta$ based on the assumption of ansatz of the type
in \eqn{pdfansatz} for $f_v^\pi(x)$. In a similar way, we summarize
our results on the large-$x$ exponent $\beta$ and the small-$x$
exponent $\alpha$ in \fgn{fitalpha} based on the model dependent analysis
that we presented in the last section.  The notation and
the arrangement of data points in \fgn{fitalpha} is the same as outlined
in \fgn{fitx1} for the moments.  From the plots for $\beta$ in both
the lattice spacings, we find that the fits prefer a value of around
1, and sometimes even smaller than 1.  As suggested
in~\cite{Sufian:2020vzb}, the usage of the 4-parameter ansatz does
lead to somewhat larger values of $\beta$ than obtained using the
2-parameter ansatz, but these values are still closer to 1. Even
though the moments that correspond to these fits showed little
renormalization scheme dependence, the exponents themselves show a
larger sensitivity to the lattice renormalization scheme used, with
a tendency for the RI-MOM scheme to consistently give lesser values
of $\beta$ compared to those from ratio schemes. One can also notice
a somewhat increasing tendency of $\beta$ when larger fit range
$[z_1,z_2]$ is used.  It is possible that the favored value of
$\beta$ could be slightly larger than our estimates if we were to
include data at larger values of $z P_z$, but we have restricted $z_2$
to be less than 0.72 fm to remain close to the perturbative regime.
A naive argument would suggest the LaMET formalism does not permit
one to access smaller values of $x<\Lambda_{\rm QCD}/P_z$, we find
that the model dependent analysis clearly gives robust values of
$\alpha\sim -0.5$. This is because the pion matrix elements constrain the few low
moments, which in turn are functions of both $\alpha$ and $\beta$
due to the model used.  To summarize, the overall shape of the PDF
and its first few moments are well determined by the usage of
phenomenology motivated ansatz, but the exponents $\alpha$ and
$\beta$ themselves show sensitivity to the renormalization schemes
as well as the range of $z$ used. Nevertheless, the estimates of
large-$x$ exponents from this analysis have a tendency to lie closer
to 1 rather than 2.  Quantitatively, we estimate
\beqa
\alpha&=&\begin{cases} -0.43^{+(12)(13)}_{-(10)(13)}\ ,\ \ a=0.06\text{\ fm}\cr
                     -0.58^{+(08)(08)}_{-(08)(08)}\ ,\ \ a=0.04\text{\ fm}\quad\text{\ (all schemes)}
       \end{cases}\cr
\beta&=&\begin{cases}\ \ 0.82^{+(30)(38)}_{-(24)(38)}\ ,\ \ a=0.06\text{\ fm}\cr
                     \ \ 0.58^{+(18)(27)}_{-(15)(27)}\ ,\ \ a=0.04\text{\ fm}\quad\text{\ (all schemes)}
       \end{cases}
\eeqa{expdat1}
taking into account the different fit ranges and ansatz dependences
in the ratio schemes as well as RI-MOM scheme.  These are the bands shown in 
\fgn{fitalpha}. Since RI-MOM scheme
has a tendency to obtain smaller $\beta$ systematically, and since
we found the ratio scheme performs better at suppressing the
higher-twist effects, we also give the estimates below using only
the $n_z^0=1$ and $n_z^0=2$ ratio schemes
\beqa
\alpha&=&\begin{cases} -0.37^{+(16)(13)}_{-(11)(13)}\ ,\ \ a=0.06\text{\ fm}\cr
                     -0.55^{+(11)(09)}_{-(08)(09)}\ ,\ \ a=0.04\text{\ fm}\quad\text{\ (only ratio)}
       \end{cases}\cr
\beta&=&\begin{cases} \ \ 1.05^{+(42)(30)}_{-(42)(33)}\ ,\ \ a=0.06\text{\ fm}\cr
                     \ \ 0.76^{+(22)(24)}_{-(20)(24)}\ ,\ \ a=0.04\text{\ fm}\quad\text{\ (only ratio)}.
       \end{cases}
\eeqa{expdat2}
Indeed, leaving out RI-MOM scheme results leads to slightly larger
$\beta$, but still around 1.  For comparison, the JAM global fits
at the same $\mu$ give $\alpha=-0.37$ and $\beta=1.20$.

The downside of the above model dependent analysis is the question
of whether by using a sufficiently general functional form $f_v^\pi(x)$,
it is possible to find $\beta\approx 2$.  For example, if we performed the 
analysis with $\beta=2$ fixed, the $\chi^2/{\rm dof}$ was between 1.5 and 2 
as opposed to the global minimum between 0.5 and 1 when $\beta$ was allowed as 
a free parameter. We discuss such an analysis at fixed $\beta=2$ in~\apx{b2}. 
A recent lattice study~\cite{Sufian:2020vzb}
using the good lattice cross-section approach found $\beta=2$ is
not ruled out when an ansatz, which we refer to as the 4-parameter
ansatz here, is used while $\beta\approx 1$ was preferred when the
2-parameter ansatz was used. 
Another recent study~\cite{Broniowski:2020had} found that with the 
limited sensitivity of the lattice calculations to higher moments,
it is difficult to make definite conclusions about the large-$x$ behavior.
Therefore, we discuss a novel model
independent way to find $\beta$.

\subsection{Model independent estimate of $\beta$}
We note that the higher moments get more contribution from larger
$x$, and hence, are more sensitive to the exponent $\beta$.
Consequently, one finds that the moments $\langle x^n\rangle$
approach zero in the large-$n$ limit in a manner dependent only on
$\beta$ as
\beq
\langle x^n\rangle \propto n^{-\beta-1}\left(1+{\cal O}(1/n)\right).
\eeq{largen}
The exponent is universal, and independent of the small-$x$ (i.e.,
$x^{\alpha}$) or intermediate-$x$ (i.e., ${\cal G}(x)$) behaviors,
but the constant of proportionality in \eqn{largen} does depend on
the details of the PDF.  We outline a proof of this behavior in
\apx{largen}.  The asymptotic behavior of large-$n$ moments was
also considered in the context of evolution of $\beta$ with scale in 
Refs~\cite{RuizArriola:2004ui,Cui:2020dlm}.
Thus, one can determine $\beta$ in a model independent
way by taking the $\log$-derivative of the above behavior,
\beq
\beta+1=-\frac{d\log\left(\langle x^n\rangle\right)}{d\log\left(n\right)}+{\cal O}(1/n).
\eeq{betan}
A discretised form of the above expression that is suitable for a
practical implementation is by defining an effective value of $\beta$
at finite $n$ as
\beq
\beta_{\rm eff}(n)\equiv -1+\frac{\langle x^{n-2}\rangle - \langle x^{n+2}\rangle}{\langle x^n\rangle}\frac{n}{4}.
\eeq{betaeff}
As one uses moments at larger values of $n$ in the above equation,
one will find $\beta_{\rm eff}(n)$ to plateau at the value of
large-$x$ exponent $\beta$.  While this method is straight forward,
it also points to the challenge of addressing the large-$x$ exponent
--- one needs to compute larger moments for a reliable estimate,
and puts a limit on what lattice studies can actually address about
$\beta$ without a modeling bias.

In \scn{mom}, we determined the first few even moments in a model
independent manner. We used these estimates of $\langle x^2\rangle_v,
\langle x^4\rangle_v, \langle x^6\rangle_v$ and $\langle x^8\rangle_v$
from the model independent analysis using \eqn{betaeff}. Since, the
larger moments are required we used the values from the fits where
prior was imposed on $\langle x^2\rangle_v$. Even though the larger
moments are relatively noisier, the ratios of moments that enter
\eqn{betaeff} are better determined owing to their correlations.
We estimated the central values of $\beta_{\rm eff}$ and its
statistical and systematic error by the outlined procedure to take
care of the variations in $[z_1,z_2]$ and $P_z^0$. We show the
result of $\beta_{\rm eff}(n)$ as a function of $n$ in \fgn{beff}
from the two lattice spacings (red and blue data points).  We notice
that it is possible at the most to use data up to $\beta_{\rm
eff}(n=6)$. As one would expect, the higher order $1/n$ corrections
to the $n^{-\beta-1}$ behavior to be the largest for $n=2$,
and hence, we find $\beta_{\rm eff}(n=2)$ to have larger value around
3.5.  For $n=4$ and $n=6$, the value of $\beta_{\rm eff}$ decreases
and stays around 1.5; the errors are large enough to see any $n$
dependence beyond $n=2$. Thus taking the well determined estimate
at $n=4$ as a proxy for the $\beta$, we find
\beq
\beta_{\rm eff}(n=4)=\begin{cases} 1.73^{+(39)(37)}_{-(35)(37)}\ ,\ a=0.06\text{\ fm}\cr
                                   1.53^{+(21)(25)}_{-(21)(25)}\ ,\ a=0.04\text{\ fm}.
                     \end{cases}
\eeq{effvals}
Thus, we find some evidence for $\beta$ to be between 1 and 2,
consistent with both our model dependent findings of $\beta\approx
1$ and with quark-counting rule expectation of 2.  Thus, we are
unable to rule out $\beta=1$ or 2 simply from this model independent
analysis.  Also, apriori, it is not clear what {\sl large}-$n$
means; whether one will observe an approximate plateau for $\beta_{\rm
eff}(n)$ at $n\sim {\cal O}(1)$ or ${\cal O}(100)$. Using our model
dependent fits as well as JAM result, we will show some evidence
below that the plateau is likely to develop for $n\sim {\cal O}(1)$
for pion.

In order to see the expected behavior of $\beta_{\rm eff}$ for
$n>6$, we simply use $\langle x^n\rangle(\alpha,\beta,s,t)$ from
our PDF fits in the last section to compute the corresponding
$\beta_{\rm eff}(n)$.  In this case, we know that $\beta_{\rm
eff}(\alpha,\beta,s,t)\to \beta$ in the large-$n$ limit, and we
already found such model dependent analysis predict $\beta\approx
1$.  We show this resulting  $\beta_{\rm eff}(n;\alpha,\beta,s,t)$
as the red and blue bands in \fgn{beff}. We used the parameters as
obtained from combined fits to ${\cal M}(z,P_z,n_z^0=1)$ using a
fit range $[z_1,z_2]=[a,0.72\text{\ fm}]$ for the case shown.  We
also show the expected result for $\beta_{\rm eff}$ using the JAM
result as the green curve.  The important observation here is that
the models of the type in \eqn{pdfansatz} predict that $\beta_{\rm
eff}(n)$ is almost plateaued by $n=4$, and makes $\beta_{\rm eff}(n\ge
4)$ to be meaningful estimators of $\beta$.  To contrast with the JAM expectation for $\beta_{\rm eff}$, 
we also plot $\beta_{\rm eff}$ as expected using ASV soft-gluon resummed analysis as the purple dashed curve (in order 
to infer the higher moments for the ASV result, we interpolated their result evolved to $\mu=3.2$ GeV 
with $f(x)=1.091x^{-0.443}(1-x)^{2.484}(1-1.842\sqrt{x}+4.959 x)$.) 
The $\beta_{\rm eff}$ for ASV never goes below 2, and approaches its plateau value at $2.48$ from below.

\subsection{A semi-model-independent analysis of pion matrix element and exponent $\beta$}
Based on the asymptotic behavior of large-$n$ moments, we propose 
a new way to fit the moments to $zP_z$ and $z^2$ dependence of pion matrix elements and, at the same time, obtain 
the value of $\beta$ in a manner that is not dependent on PDF ansatz.
We fit low moments up to an order $N_{\rm asym}$ in the usual manner and 
use the asymptotic expression for the moments beyond the order $N_{\rm asym}$
using \eqn{largen} with some $1/n$ corrections, as
\beq
\langle x^n\rangle_v \equiv \begin{cases} a_n \ ,\quad n<N_{\rm asym}\cr
                                        n^{-\beta}\left(\frac{A_0}{n}+\frac{A_1}{n^2}+\frac{A_2}{n^3}\right)\ ,\quad n\ge N_{\rm asym},
\end{cases}
\eeq{modelmom}
The fit parameters are the lower moments $a_2, a_4, \ldots,a_{N_{\rm
asym}-2}$, and the parameters $\beta, A_0, A_1, A_2$ that model the
large-$n$ moments.  We input the constraint that $a_2>a_4>\ldots>a_{N_{\rm
asym}-2}>\langle x^{N_{\rm asym}}\rangle(\beta,A_0,A_1,A_2)$.  Using
this model for the moments in \eqn{ratgenexp}, we fit the parameters
to best describe ${\cal M}(z,P_z,P_z^0)$ in the same way as we
described in \scn{mom}, but we use $z$ up to 0.72 fm in this analysis.
Some analysis bias comes from the choices of $N_{\rm asym}$ and the
order of $1/n$ corrections to use. We used $N_{\rm asym}=2,4$ and
6 in our fits, and any usage of more than $A_2$ in the fits made
the fits unstable and did not converge properly (it is an asymptotic
series after all).  It was quite surprising that we were able to
even use $N_{\rm asym}=2$ in the analysis to get good fits, i.e.,
using the hypothesis that moments starting from $\langle x^2
\rangle_v$ can be described by the asymptotic expression in
\eqn{modelmom}. The resulting values of $\langle x^n\rangle_v$ using
this method compares well within errors with the moments obtained
in \scn{mom} and \scn{pdf}. For example , using the ratio scheme
with $n_z^0=1$ in the $a=0.06$ fm lattice and with $N_{\rm asym}=4$,
we get $\langle x^2\rangle_v=0.110(3)$, $\langle x^4\rangle_v=0.038(5)$,
$\langle x^6\rangle_v=0.016(4)$ which compares well with the other
determinations (and for this case, the other parameters were
$\beta=1.20(21), A_0=0.87(21), A_1=-0.51(47), A_2=-0.28(28)$ and
$\chi^2/{\rm dof}=46.5/48$). The novel outcome of this analysis is
the estimate of $\beta$ in addition to moments, and for the $n_z^0=1$
ratio matrix element using fits from $z=a$ to 0.72 fm, we get
\beqa
&&\beta=\begin{cases}0.93^{+11}_{-10}\ ,\ N_{\rm asym}=2\cr
                   1.20^{+20}_{-26}\ ,\ N_{\rm asym}=4\cr 
                   1.79^{+59}_{-36}\ ,\ N_{\rm asym}=6\end{cases},\ \text{for\ } a=0.06\text{\ fm},
                   \cr
&&\beta=\begin{cases}0.85^{+11}_{-15}\ ,\ N_{\rm asym}=2\cr
                   1.02^{+16}_{-12}\ ,\ N_{\rm asym}=4\cr 
                   1.82^{+51}_{-40}\ ,\ N_{\rm asym}=6\end{cases},\ \text{for\ } a=0.04\text{\ fm}.
\eeqa{betaasym}
As one relaxes the order $N_{\rm asym}$ where large-$n$ asymptotic
behavior sets in from $N_{\rm asym}=4$ to 6, the best fit values
of $\beta$ changes from a smaller value $\approx 1.0(2)$ to $\approx
1.8(5)$. Thus, this analysis suggests that the values of $\beta$
around $1$ seem preferred when one is aggressive on the order $N_{\rm
asym}$, but $\beta$ is consistent within 1-$\sigma$ error (albeit
a noisier estimate) if one uses conservatively larger $N_{\rm
asym}\ge 6$. We reach the same conclusion again; in order to obtain
a conclusive result on $\beta$, we need even more precise data at
larger $P_z z$ to be sensitive to higher moments.

\section{Continuum estimates}\label{sec:cont}

\begingroup
\renewcommand{\arraystretch}{1.2}
\bets
\resizebox{18cm}{!}{%
\centering
\begin{tabular}{|l|c|l|l|l|l|l|l|l|l|}
\hline
\hline
    Method & $a$ (fm) & $\langle x\rangle_v$& $\langle x^2\rangle_v$ & $\langle x^3\rangle_v$ & $\langle x^4\rangle_v$ & $\alpha$ & $\beta$ & $s$ & $t$ \cr
\hline
    (a) Model independent analysis &  0.06 &      & 0.1088(48)(58) &       & 0.0346(57)(73) &   &   &   &    \cr
                &  0.04 &      & 0.1050(43)(39) &       & 0.0382(44)(54)  &   &   &   &    \cr
                &  $a\to0$ &      & 0.0993(71)(54) &       & 0.0356(39)(60)  &   &   &   &    \cr
\hline
    (b) 2-parameter &  0.06 & 0.2470(92)(52) & 0.1122(54)(51) & 0.0649(53)(62) & 0.0423(52)(60) & -0.33(15)(11) & 1.02(37)(32) &    &     \cr
                        &  0.04 & 0.2289(96)(44) & 0.1083(47)(34) & 0.0652(49)(36) & 0.0444(48)(34) & -0.51(10)(05) & 0.66(24)(20)&     &     \cr
                        &  $a\to0$ & 0.216(19)(08) & 0.1008(69)(43) & 0.0604(39)(46) & 0.0408(37)(44) & -0.55(15)(08) & 0.66(34)(22)&     &     \cr
\hline
    (c) 4-parameter  &  0.06 & 0.2457(92)(61) & 0.1121(54)(50) & 0.0649(53)(62) & 0.0420(51)(59) & -0.40(16)(14) & 1.11(41)(34) & -0.14(16)(20)   & 1.0(1.0)(1.2)  \cr
                        &  0.04 & 0.2253(98)(45) & 0.1080(46)(34) & 0.0647(47)(38) & 0.0436(43)(38) & -0.61(13)(06) & 0.86(22)(25)&  -0.20(24)(19)   &  2.5(1.9)(2.5) \cr
                        &  $a\to 0$ & 0.213(19)(08) & 0.1009(68)(42) & 0.0607(40)(47) & 0.0410(40)(47) & -0.61(16)(08) & 0.77(26)(30)&  -0.19(27)(17)   &  1.5(2.0)(1.7) \cr
\hline
    (d) large-$n$  asymptotics &  0.06 &       & 0.1093(48)(53) &   &  0.0365(44)(58)              &                &                1.40(25)(30)  &  & \cr
                           &  0.04 &       & 0.1050(49)(37)  &        &  0.0392(38)(43)              &                &    1.12(24)(20)            &    &    \cr
                           &  $a\to 0$ &       & 0.0996(71)(61)  &        &  0.0386(56)(58)              &                &    1.15(23)(22)            &    &    \cr
\hline
    (e) Effective $\beta$  &   0.06 &                  &                &                &                &               &  1.73(39)(37)  &       &     \cr
                     &   0.04 &                  &                &                &                &               &  1.53(21)(25)  &       &     \cr
                     &   $a\to 0$ &                  &                &                &                &               &  1.55(34)(27)  &       &     \cr
\hline
\hline
\end{tabular}
}
\caption{Summary of results from analyses presented this paper at
$\mu=3.2$ GeV. Each row is a type of analysis, namely ---
(a) the model-independent estimates of the moments
(b,c) fits to 2- and 4-parameter PDF ansatz,
(d) a semi-model independent analysis based on modeling $\langle
x^n\rangle_v$ by the asymptotic formula for $n\ge 4$, and (e) the
estimate of exponent $\beta$ from $\beta_{\rm eff}(n=4)$.
The columns are
the outcomes; namely, the value of the first four valence moments
$\langle x^n\rangle_v$, the parameters $\alpha,\beta,s,t$ in the
PDF ansatz $f_v^\pi(x)\sim x^\alpha(1-x)^\beta\left(1+s\sqrt{x}+t
x\right)$. For each analysis, values from two different lattice
spacings $a$ are given, and also our continuum expectations, denoted
by $a\to 0$, based on $a^2$ extrapolation are also given. For these
estimates, we used the ratio scheme with $n_z^0=1,2$.
}
\eets{tabsummary}
\endgroup

\bef
\centering
\includegraphics[scale=0.8]{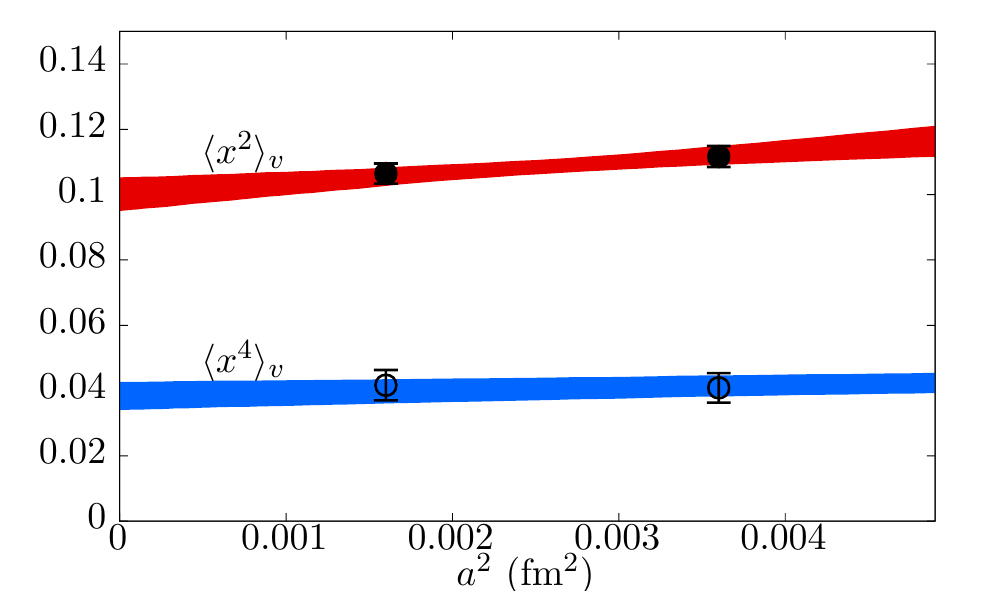}
\caption{Estimate of continuum extrapolation of $\langle x^2\rangle_v$
and $\langle x^4\rangle_v$ from combined fits to $a=0.04$ fm and
$a=0.06$ fm data using the ansatz in \eqn{ratgenexp3}. For the case
shown, $n_z^0=1$, $[z_1,z_2]=[a,0.5\text{\ fm}]$, and analyzed using
4-parameter PDF ansatz. The black circles are the data from the
analysis at the two fixed $a$. The bands are the $a^2$ extrapolations
using the combined fits.}
\eef{context}

In the last part of the paper, we discuss the continuum estimates of
the PDF and its moments. The estimates are speculative because we
only have two lattice spacings, nevertheless both very fine. We
should note that we already demonstrated the presence of lattice
spacing effects of the type $(P_z a)^2$
at distances of 
the order of few lattice spacings and took care of them in our
analysis. Once the $(P_z a)^2$ artifacts were removed, it was
possible to describe the boosted pion matrix elements at any finite lattice spacing using
the twist-2 OPE expressions.  Therefore, we assume that any additional
lattice spacing effects will simply affect the values of the extracted
moments themselves.  That is, we model the moments $\langle
x^n\rangle(a)$ at any fixed lattice spacing $a$ to behave as
\beq
\langle x^n \rangle_v(a)=\langle x^n\rangle_v + d_n a^2,
\eeq{lateffect}
where $\langle x^n\rangle_v$ is the continuum value and $d_n$ are
numerical coefficients that can be fit to the data.  One should note that there could be 
residual ${\cal O}(\alpha_s a)$ corrections as well, which we are implicitly assuming to be small
compared to tree-level ${\cal O}(a^2)$ artifacts in the fine lattices we are using.
Such dependences on fit forms need to checked in a precise analysis of continuum extrapolations
that can be performed with data from multiple lattice spacings, but the extrapolations
presented here are meant to be only rough estimates.
We repeated
all the analysis (model-independent estimates of even moments, fits to model PDFs, the semi-model
dependent analyses) presented in the previous sections using combined
fits to both $a=0.06$ fm and 0.04 fm data (for fixed physical values
of $z_2$, and keeping $n_z^0$ to be the same between the two lattice
spacings) using the above ansatz for $\langle x^n \rangle_v(a)$ in
the twist-2 OPE expressions; concretely, we used fits of the type
\beqa
&&{\cal M}(z,P_z,P^0_z; a)=\cr&&\ \frac{\sum_n c_n(z^2\mu^2) \left(\langle x^n \rangle+a^2 d_n\right) \frac{(-i P_z z)^n}{n!}+r (a P_z)^2}{\sum_n c_n(z^2\mu^2) \left(\langle x^n \rangle+a^2 d_n\right) \frac{(-i P^0_z z)^n}{n!}+r (a P_z^0)^2}),
\eeqa{ratgenexp3}
for the ratio schemes with $P_z^0\ne 0$ for this analysis. We found
using $d_2$ and $d_4$ as additional fit parameters was sufficient
to describe the data at both the lattice spacings with $\chi^2/{\rm
dof}$ between 0.5 and 1 depending on the range of fits and analysis
type. Since we found that the ratio scheme succeeded in reducing
higher-twist effect well, we focus on this scheme in order to
discuss our best estimates and their continuum estimates. As a sample
result from this analysis, in \fgn{context}, we show the $a^2$
extrapolation of $\langle x^2\rangle_v$ and $\langle x^4\rangle_v$
as obtained from the above analysis at fixed $z_2=0.5$ fm and $z_1=a$
using $n_z^0=1$ for both lattice spacings. For the case shown, we
used the 4-parameter PDF ansatz for the combined fit. The two data
points in the plot are the values for the same case at fixed $a=0.04$
fm and 0.06 fm. We remind the reader that this is not a straight-line
fit to the two data point, but rather an outcome of the combined
analysis as described above (with $d_2=3.1(1.8)$ fm$^{-2}$,
$d_4=0.79(62)$ fm$^{-2}$, and $\chi^2/{\rm dof}=60.5/92$.)

In \tbn{tabsummary}, we tabulate all our estimates from different
kinds of analysis from the previous sections using only the ratio
schemes with $n_z^0=1,2$, where we expect the higher-twist corrections
to be even milder. Along with the estimates at two fixed $a$, we also
tabulate our continuum estimates based on the above analysis for
each quantity. 
There is only little effect from including
$a^2$ corrections. We find that removing the lattice spacing effect
have a slight tendency to bring the moments closer to the JAM
values of 0.223, 0.095, 0.052, 0.032 for the first four moments.
The best fit values of the large-$x$  exponent $\beta$ from the
fits, however, continue to remain closer to 1, and thus, the lattice
spacing effects might not be an issue in our results.

As a check, we also used an ansatz similar to \eqn{lateffect} to 
include only ${\cal O}(a)$ correction (that could result from operators 
being used) to moments instead of ${\cal O}(a^2)$
correction as done above. Such results were consistent with the ${\cal O}(a^2)$
results presented above, albeit with larger error bars. For example,
for the case of 4-parameter PDF fit, we obtained $(\alpha,\beta,s,t)=
[-0.70(17)(09), 0.65(30)(32),-0.20(20)(14),0.9(1.4)(1.3)]$ which 
corresponds to first four moments being 0.194(31)(15), 0.093(11)(07), 
0.057(5)(6),0.039(3)(5) respectively. A careful study using both 
${\cal O}(a)$ and ${\cal O}(a^2)$ terms in the extrapolation can be 
made in a future study with more that two lattice spacings.

\section{Conclusions}\label{sec:conclusion}

\bef
\centering
\includegraphics[scale=0.6]{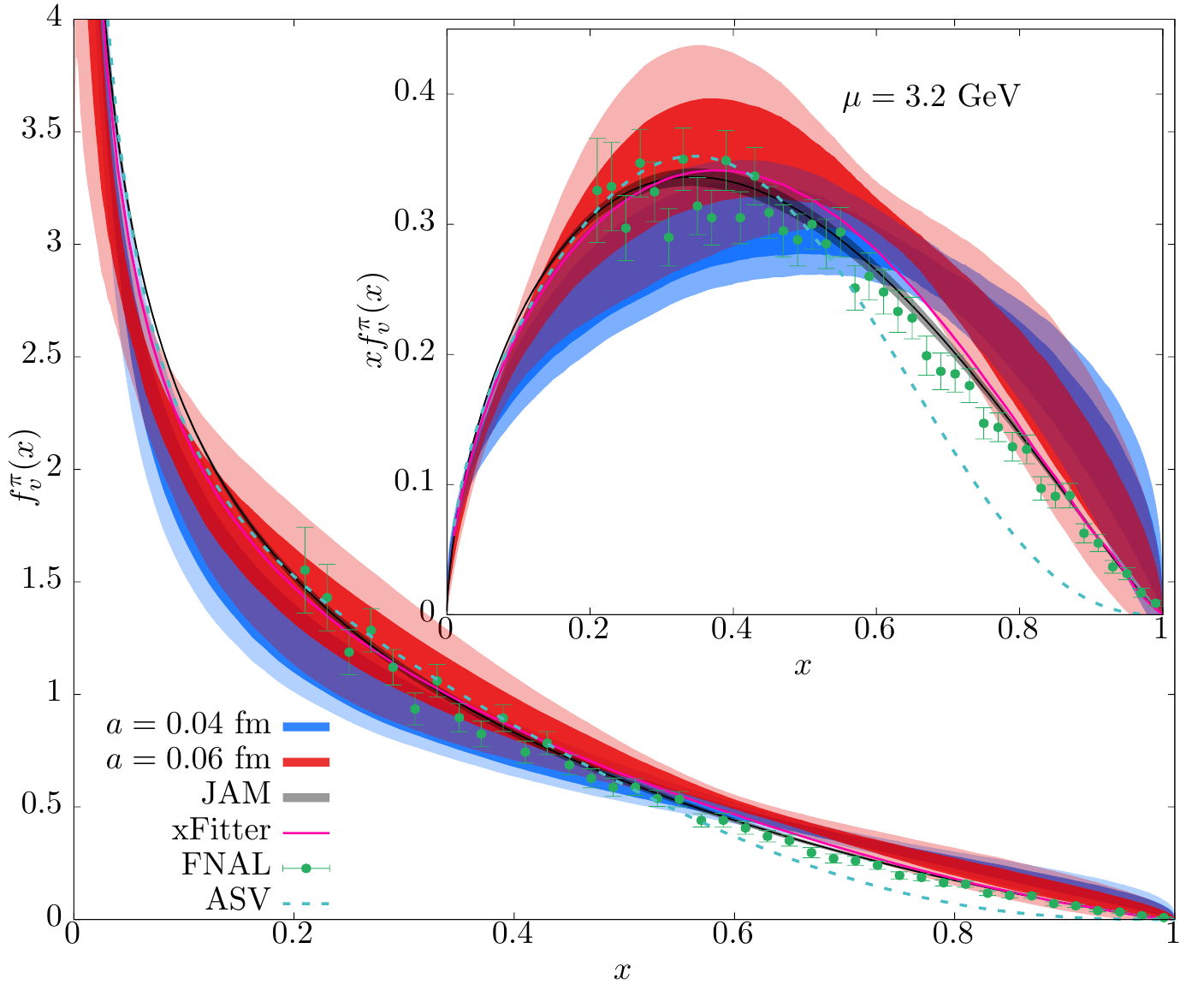}
\includegraphics[scale=0.6]{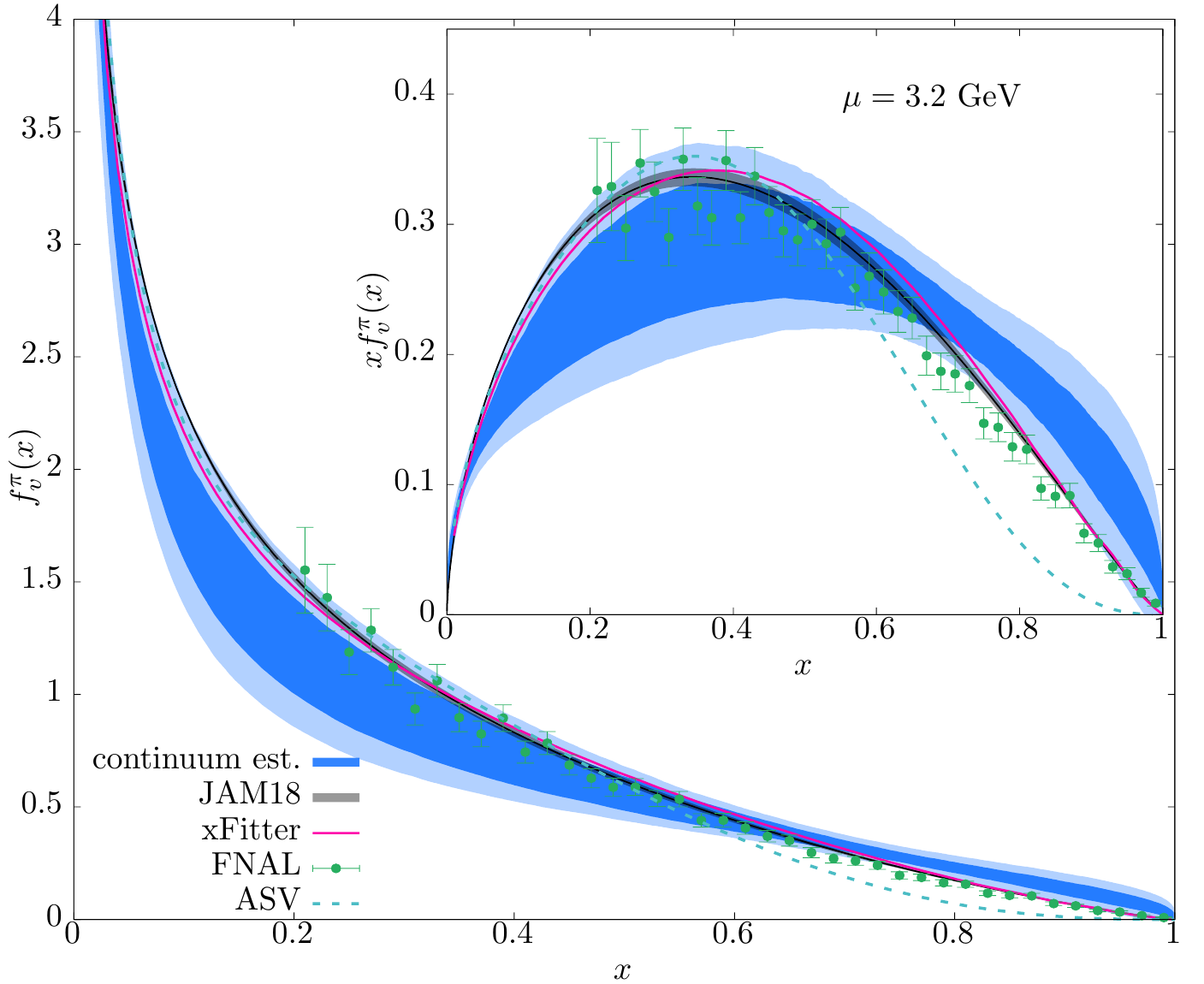}
\caption{ Our determinations of valence PDF of pion, $f_v^\pi(x,\mu)$,
at factorization scale $\mu=3.2\text{\ GeV}$. (\textbf{Top panel})
The PDF determination from $a=0.04$ fm  data (red) and $a=0.04$ fm
data (blue). (\textbf{Bottom panel}) Our estimate of PDF in the
continuum limit (blue).  In both top and bottom panels, the darker
inner band includes only the statistical error.  The lighter outer
band includes both statistical error as well as the systematic
errors.  Our estimates are compared with the FNAL E-0615
estimate~\cite{Conway:1989fs} (green symbols), ASV
estimate~\cite{Aicher:2010cb} (green dashed line), JAM
estimate~\cite{Barry:2018ort} (black band) and xFitter
analysis~\cite{Novikov:2020snp} (purple line).  Insets: the same
data are replotted as the traditional $x f_v^\pi(x)$ versus $x$.  }
\eef{finalpdf}

In this paper, we presented a lattice computation of the $\msbar$
isovector $u-d$ parton distribution function of 300 MeV pion and
its moments using the recently proposed twist-2 perturbative 
matching framework (Large Momentum Effective
Theory (LaMET) framework / short-distance factorization framework).  Using isospin symmetry, we related the
properties of isovector pion PDF to the valence $u-\overline{u}$ PDF
of pion, $\pi^+$.

In order to access the short distance physics required for the perturbative 
twist-2
framework, we used two lattices ensembles with very fine lattice
spacings of $a=0.06$ fm and $0.04$ fm for the first time in such
pion PDF computations. Using high statistics, we were able to compute
the required equal-time bilocal quark bilinear matrix elements evaluated
with pions boosted up to $2.42$ GeV. Thus, a major advancement
resulting from this work is the demonstration that current lattice
calculations can satisfy both the theoretical requirement of sub-fermi
separations (in order to be consistent with the OPE-based framework
reliant on naive power counting for operator hierarchy), and the
requirement of large hadron momentum (in order for a controlled
truncation of the OPE at leading-twist).
As a handle on quantifying perturbative uncertainties and other
higher-twist systematics, we used multiple renormalization schemes
for the equal-time matrix element, namely RI-MOM, ratio scheme and new
variants thereof with the advantage of reducing higher-twist effects.
As a technical elaboration, we proposed and used the pion matrix element at zero pion
momentum as a suitable quantity to study higher-twist effects and
demonstrated practically as to why the ratio renormalization scheme
effectively eliminates higher-twist effects to a good accuracy (with 
respect to typical errors in lattice data at larger $z$) even up
to 1 fm distances.

From the renormalized boosted pion matrix elements, we performed two kinds of analysis.
In the first kind, we obtained the first few even valence moments
$\langle x^n\rangle_v$ by fitting both $z^2$ and $P_z z$ dependence
of the matrix elements making use of perturbative matching coefficients
at 1-loop order. Though the twist-2 perturbative matching 
methodologies help us access higher
moments without the problem of mixing, especially with the usage
of priors on lower moments, it comes at a cost of introducing dependencies on
the range of $z$ and $P_z z$ used in the fits, and we discussed them
in this work. Folding in such dependencies in the systematic error, we
estimated the $\msbar$ moments $\langle x^2\rangle_v, \langle
x^4\rangle_v$ and $\langle x^6\rangle_v$ at $\mu=3.2\text{\ GeV}$.
In the second kind of analysis, we reconstructed the $x$-dependent
valence pion PDF by modeling the PDF via $x^\alpha(1-x)^\beta {\cal
G}(x)$ type ansatz, and fitting the parameters of the model so as
to best describe both $z$ and $zP_z$ dependence of pion matrix elements in various
renormalization schemes with $z$ restricted to sub-fermi values.
We summarize our reconstructed valence PDFs at $\mu=3.2\text{\
GeV}$ from the two lattice spacing in the top panel of \fgn{finalpdf} -- the
estimate using only statistical error is shown as darker band, and
statistical and systematic error (coming from fit ranges, renormalization
scheme used, and the PDF ansatz used) is shown as the
lighter band.  This model dependent method lets us access the odd
moments of valence PDF as well. We also provided estimates of the values of 
moments as well as the PDF in the continuum limit based on the results at 
the two lattice spacings; the numerical results from various 
analysis approaches are summarized in \tbn{tabsummary}. 
In the bottom panel of \fgn{finalpdf}, we have shown our estimate for the PDF in the 
continuum limit for the 300 MeV pion.

We discussed the large-$x$ behavior using our model-dependent PDF
that we reconstructed from fits.  We found that even though the
overall $x$ dependence of the reconstructed PDFs remained the same
with variations coming from the fit range, renormalization scheme
and PDF model used, the specific details such as the value
of the large-$x$ exponent $\beta$ showed a larger dependence on
such analysis choices, but largely showed a tendency to be close
to 1. To avoid such issues, we proposed a new model independent
observable, constructed out of moments, that converges to the
large-$x$ exponent $\beta$ as one uses larger moments. At present,
our computed matrix elements are sensitive only to moments up to order 4 or 6, and
given this limitation, we find the effective value of the exponent
to be between 1 and 2, but ruling out neither.  However, it is at
present hard to conclude if such an estimate would remain unchanged
as one includes larger moments, which would be possible with increased statistics
in the future. But, with this work, the computation that one should
perform to reach a model independent robust conclusion about $\beta$
is clear.

Finally, we compare our PDF determinations with other global fit
analysis in the summary plot in \fgn{finalpdf}.  Along with our
determinations of valence PDFs at the two lattice spacing (top
panel) and our estimate of valence PDF in the continuum limit (bottom
panel), we have also shown the Fermilab E-0615
estimate~\cite{Conway:1989fs} (green symbols), the ASV reanalysis
of the Fermilab result after taking soft-gluon resummation into
account~\cite{Aicher:2010cb} (dashed green line), the recent JAM
Monte-Carlo global analysis~\cite{Barry:2018ort} (black band) and
the result from analysis using xFitter~\cite{Novikov:2020snp} (purple
line), all evolved to the same scale $\mu=3.2\text{\ GeV}$. One
can find an overall agreement of our determinations with the
phenomenological results; with better agreement with JAM, xFitter
and the initial E-0615 estimates, than with the ASV result.  Some
caveats are clear --- firstly, our computation is for 300 MeV pion,
and hence a future computation with physical pion mass is crucial.
Secondly, we used 1-loop matching coefficients to match the lattice
results to $\msbar$ PDF, and it is at present unclear what the
effect of adding higher-loop perturbative terms in the matching
kernel (and also ASV-type resummation of soft-gluon contribution
in the matching kernel, if at all possible) on the extracted PDFs
and moments will be (very recently, works~\cite{Chen:2020arf,Chen:2020iqi,Li:2020xml} related to 2-loop matching
appeared as the present manuscript was being completed).  We leave these questions
for the future.

\section*{Acknowledgments}

This material is based upon work supported by: (i) The U.S. Department
of Energy, Office of Science, Office of Nuclear Physics through the Contract No. DE-SC0012704; (ii) The U.S.
Department of Energy, Office of Science, Office of Nuclear Physics
and Office of Advanced Scientific Computing Research within the
framework of Scientific Discovery through Advance Computing (ScIDAC)
award Computing the Properties of Matter with Leadership Computing
Resources; (iii) The Brookhaven National Laboratory's Laboratory
Directed Research and Development (LDRD) project No. 16-37.  (iv)
X.G. is partially supported by the NSFC Grant Number 11890712. (v)
S.S. is supported by the National Science Foundation under CAREER
Award PHY-1847893. (vi) L.J. acknowledges support from U.S. DOE
Grant No. DE-SC0010339.  
(vii) C.K. was supported in part by U.S. DOE grant No. DE-FG02-04ER41302 
and in part by the Office of Nuclear Physics, Scientific Discovery through Advanced Computing (SciDAC) program.
(viii)  Y.Z. is partially supported by the
U.S. Department of Energy, Office of Science, Office of Nuclear
Physics, within the framework of the TMD Topical Collaboration.
(ix) This research used awards of computer time provided by the
INCITE and ALCC programs at Oak Ridge Leadership Computing Facility,
a DOE Office of Science User Facility operated under Contract No.
DE-AC05- 00OR22725. (x) Computations for this work were carried
out in part on facilities of the USQCD Collaboration, which are
funded by the Office of Science of the U.S. Department of Energy.

\appendix 

\section{Effect of lattice periodicity on $P_z=0$ three-point function}\label{sec:wrap}

The value of $m_\pi L_t$ measures quantitatively the magnitude of
{\sl wrap-around effects} due to lattice periodicity in the temporal
direction. Its value on the $48^3\times64$ lattice is 5.85 and on
$64^4$ lattice is 3.9. Thus, we expect the wrap-around effect in
correlation functions to be more in our finer lattice than in the
coarser lattice.  We take care of this wrap-around effect in two-point
function by replacing $A_0 \exp\left(-m_\pi t\right)$ with
$A_0\left[\exp(-m_\pi t)+\exp(-m_\pi(L_t-t))\right]$. But there are
additional effects in the three-point function itself.
To quantify the effects of finite $L_t$ on the three point function we
recall that finite $L_t$ can
be interpreted as inverse temperature, and therefore, we can write
\beq
C_{\rm 3pt}(t_s,\tau)={\rm Tr}\left(e^{-(L_t-t_s)H} \pi e^{-(t_s-\tau) H} O_{\gamma_t} e^{-\tau H} \pi^{\dagger}\right),
\eeq{transfer}
with $H$ being the QCD Hamiltonian.
Inserting the complete set of energy eigenstates
we can also write the above expression as
\beqa
C_{\rm 3pt}(t_s,\tau)&=&\sum_{m,n,k}\mel**{m}{\pi}{n}\mel**{n}{O_{\gamma_t}}{k}\mel**{k}{\pi^\dagger}{m} e^{-\tau E_n}\cr&& e^{-(t_s-\tau)E_k}e^{-(L_t-t_s)E_m}.\cr
&&\quad
\eeqa{spec1}
We can split the above sum into two parts; namely, a part without
any wrap-around effect in which the state $|m\rangle$ is the vacuum
$|0\rangle$, and the states $|n\rangle$ and $|k\rangle$ run
over excited states with quantum number of pion. The second part
captures the wrap-around effect, and in which case, the state
$|m\rangle$ is the pion state, while the states $|n\rangle$ and
$|k\rangle$ run over states with vacuum quantum numbers. In the
discussion below, we restrict these states to include $|0\rangle$
and the first lightest iso-singlet, G-parity positive state,
$|S\rangle$, with energy $E_S$.  That is, we write the spectral
decomposition of the three-point function as
\beqa
&&C_{\rm 3pt}(t_s,\tau)=\bigg{(}\sum_{n,k\in \text{iso-triplet}}\mel**{0}{\pi}{n}\mel**{n}{O_{\gamma_t}}{k}\mel**{k}{\pi^\dagger}{0}\cr
&&\qquad e^{-\tau E_n} e^{-(t_s-\tau)E_k}\bigg{)}\cr
\quad&+& \mel**{\pi}{\pi}{0} \mel**{0}{O_{\gamma_t}}{S}\mel**{S}{\pi^\dagger}{\pi}e^{-(t_s-\tau)E_S} e^{-(L_t-t_s)E_\pi}\cr
\quad&+& \mel**{\pi}{\pi}{S} \mel**{S}{O_{\gamma_t}}{S}\mel**{S}{\pi^\dagger}{\pi} e^{-t_s E_{S}} e^{-(L_t-t_s)E_\pi}.\cr &&\quad
\eeqa{spec2}
The terms in the sum within the brackets is the part without
wrap-around effect, and we used this part in the main text to extract
the matrix element. The two terms below that are due to finite
$L_t$.

We focus on $O_{\gamma_t}(z=0)$ now, where we can make well-motivated
estimates of the wrap-around effect.  In this case, the term in the
second line in \eqn{spec2} involving the vacuum vanishes. Further,
we make the following assumptions in order to estimate the wrap-around
effect:
\begin{enumerate}
\item The state $|S\rangle$ is a two-pion state $|\pi,\pi\rangle$
with both the pions with zero relative momentum, and projected to
be iso-singlet, G-parity positive.
\item The energy of the state $E_{\pi,\pi}\approx 2E_\pi$.
\item The amplitude $\mel**{\pi}{\pi}{\pi,\pi}\approx \mel**{0}{\pi}{\pi}$.
\item The matrix element \\ $\mel**{\pi,\pi}{O_{\gamma_t}(z=0)}{\pi,\pi} \approx 2 \mel**{\pi}{O(z=0)}{\pi}$.
\end{enumerate}
With these assumptions and using \eqn{spec2}, 
the ratio $R=C_{\rm 3pt}/C_{\rm 2pt}$ becomes
\beqa
&&R(z=0,t_s,\tau)\approx\langle \pi| O_{\gamma_t}(z=0)|\pi\rangle \left(\frac{1+2e^{-E_\pi L_t}}{1+e^{-E_\pi L_t}}\right)\cr&&+(t_s,\tau)\text{dependent excited state terms}.
\eeqa{est1}
Thus, we have to correct our estimated value for $\langle \pi|
O_{\gamma_t}(z=0)|\pi\rangle$ from excited state fits by the factor
above. For non-zero $P_z$, the values of $E_\pi$ are large and the
correction factor is almost 1. For zero $P_z$, this wrap around
effect is the highest.  For $a=0.06$ fm lattice, this factor is 1.0028,
which almost unity. However, for the finer $a=0.04$ fm lattice, the
correction factor at zero $P_z$ is 1.020, which is comparable to
the estimated value of matrix element at $z=0$ itself, and hence,
it cannot be neglected.  In \scn{3pt}, we estimated $\langle \pi|
O_{\gamma_t}(z=0)|\pi\rangle=1.045(1)$ without taking wrap-around
effect in the three-point function into account. We estimate the
corrected value to be
\beq
\langle \pi(P_z=0)| O_{\gamma_t}(z=0)|\pi(P_z=0)\rangle=1.024(1),
\eeq{correctedb}
for $a=0.04$ fm lattice. This is comparable to other values of $z=0$
bare matrix elements at non-zero $P_z$ for $a=0.04$ fm lattice (refer \fgn{z0dep}).
Thus, we understand quantitatively the underlying issue in $P_z=0$
matrix element for the finer lattice, and hence we avoided the usage
of it in the analyses discussed in the main text.

\section{Discussion on $Z_V$}\label{sec:zv}

\bef
\centering
\includegraphics[width=7.5cm]{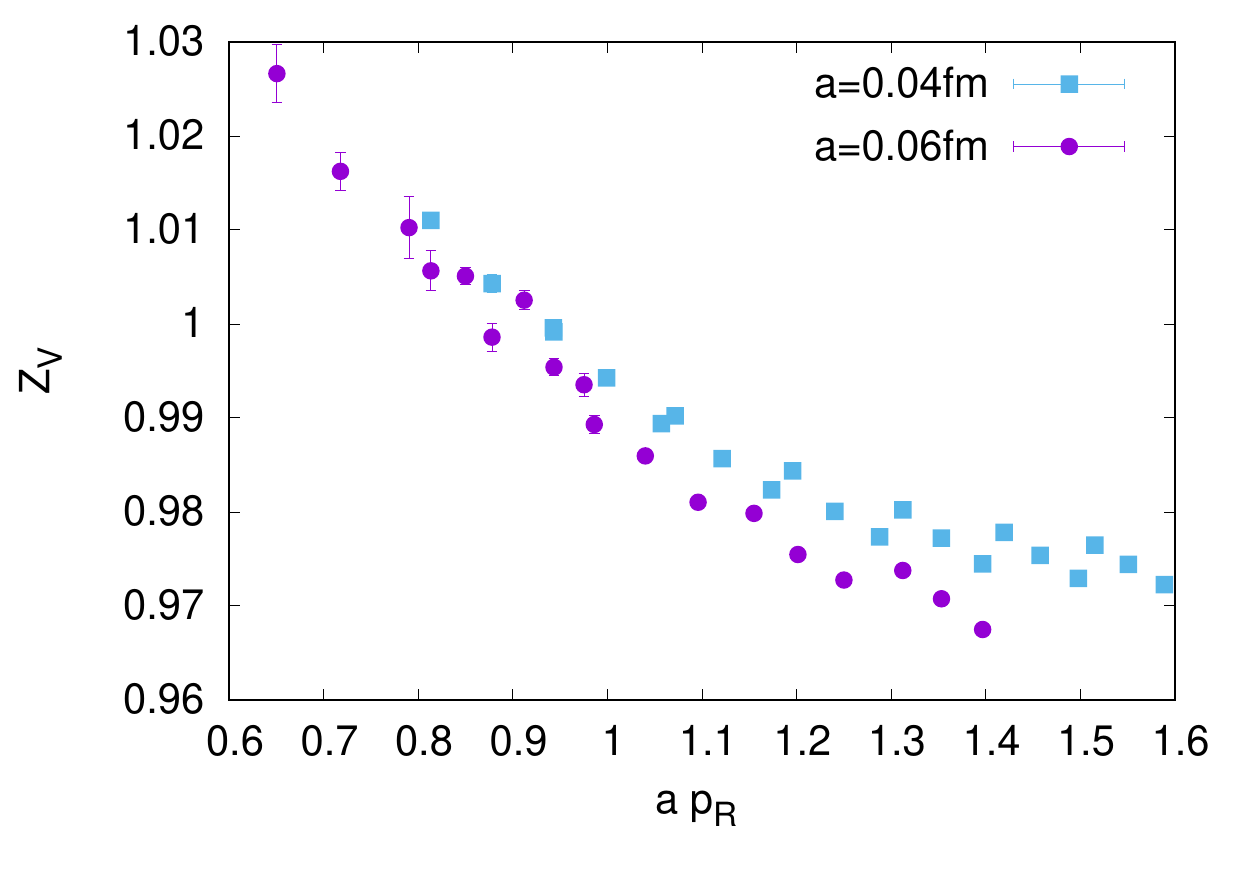}
\caption{
The vector current renormalization factor $Z_V$ as function of $p_R$ in 
lattice units.
}
\eef{ZV}

In the main text, we normalized the $z=0$ renormalized pion matrix elements to 1, thereby avoiding the 
issue of vector current renormalization factor. Here,
we provide details of the renormalization constant $Z_V$ of the vector current
$O_{\mu}=\bar \psi \gamma_{\mu} \psi$ in the RI-MOM scheme for our two lattices. We can write $Z_V=Z_{\gamma_t \gamma_t} Z_q$,
where $Z_{\gamma_t \gamma_t}$ is the renormalization of the vertex function for $\gamma_{\mu}=\gamma_{t}$
and $Z_q$ is the renormalization of the quark field. We use the same notation as in Ref. \cite{Izubuchi:2019lyk},
where the details of RI-MOM renormalization can be found.
In Fig. \ref{fig:ZV}, we show $Z_V$ for the two lattice spacings used in our study, $a=0.04$ fm and $a=0.06$ fm
as function of the renormalization point $p_R$. To minimize the discretization effects,  the lattice momenta
\begin{equation}
a p_{\mu}=\frac{2 \pi}{L_{\mu}} (n_{\mu} + \frac{1}{2} \delta_{\mu,0})
\end{equation}
are substituted by $p_{\mu}'=\sin (a p_{\mu})$, so $p_R^2=\sum_{\mu=1,4} (p'_{\mu})^2$.
The vector current renormalization constant should not depend on $p_R$, because in the $a \rightarrow 0$ limit the
local current is conserved. Nevertheless, we see a significant dependence on $p_R$. This dependence can be caused by
lattice artifacts as well as by non-perturbative effects that for large values of $p_R$ can be parametrized by
local condensates. As we use off-shell quark states in Landau gauge in the RI-MOM renormalization procedure 
the lowest dimension local condensate is the dimension two gluon condensate $\langle A^2 \rangle$ 
\cite{Chetyrkin:2009kh,Blossier:2010ky}. Lattice artifacts show up as breaking of the rotational symmetry on the lattice. We see
from Fig. \ref{fig:ZV} the fish-bone structure in the lattice data at the level much larger than the statistical errors
on $Z_V$. All these effects need to be taken into account if one wants to extract $Z_V$. These effects are easier to understand
by analyzing $Z_q$ and $Z_{\gamma_t \gamma_t}$, separately as discussed below.
\bef
\centering
\includegraphics[width=7.5cm]{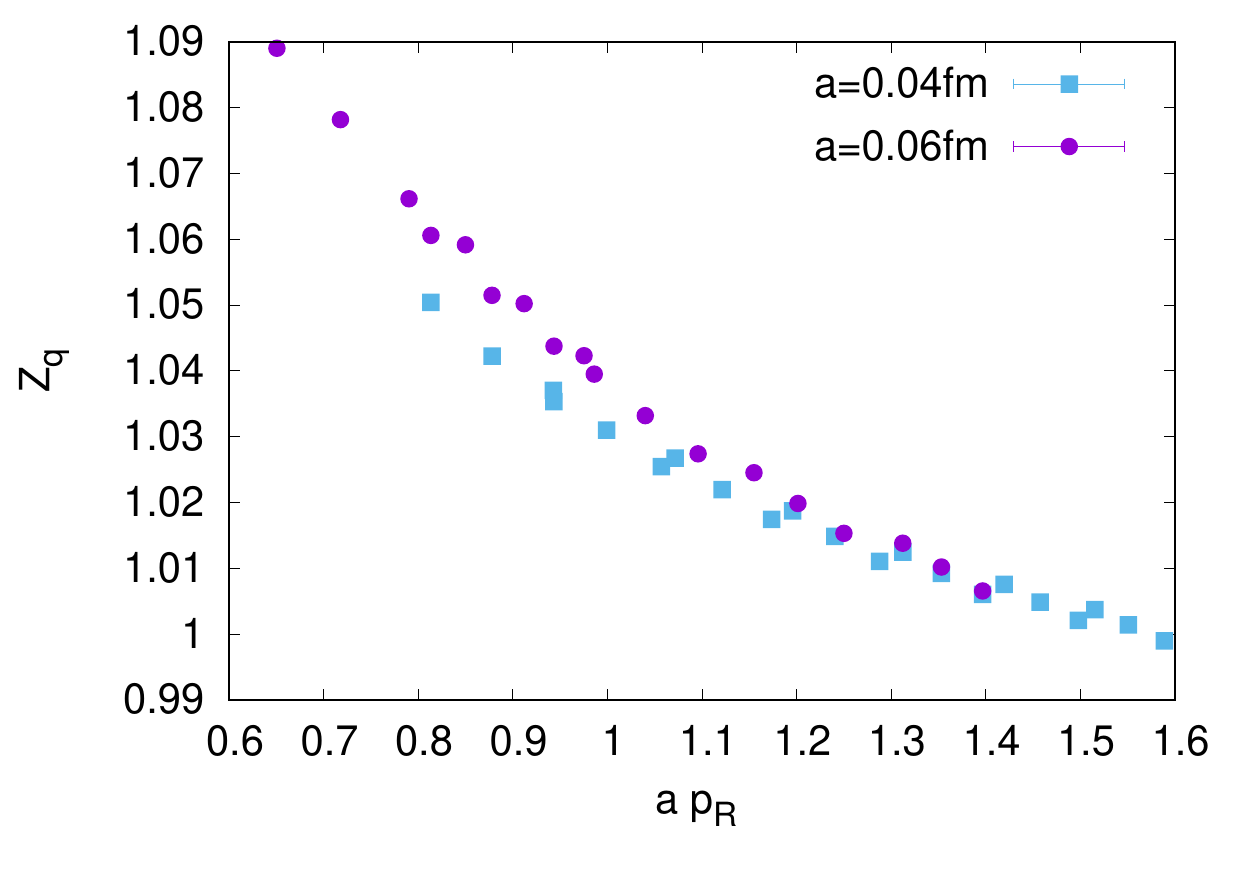}
\caption{
The quark field renormalization factor $Z_q$ as function of $p_R$ in
lattice units.
}
\eef{Zq}
\bef
\centering
\includegraphics[width=7.5cm]{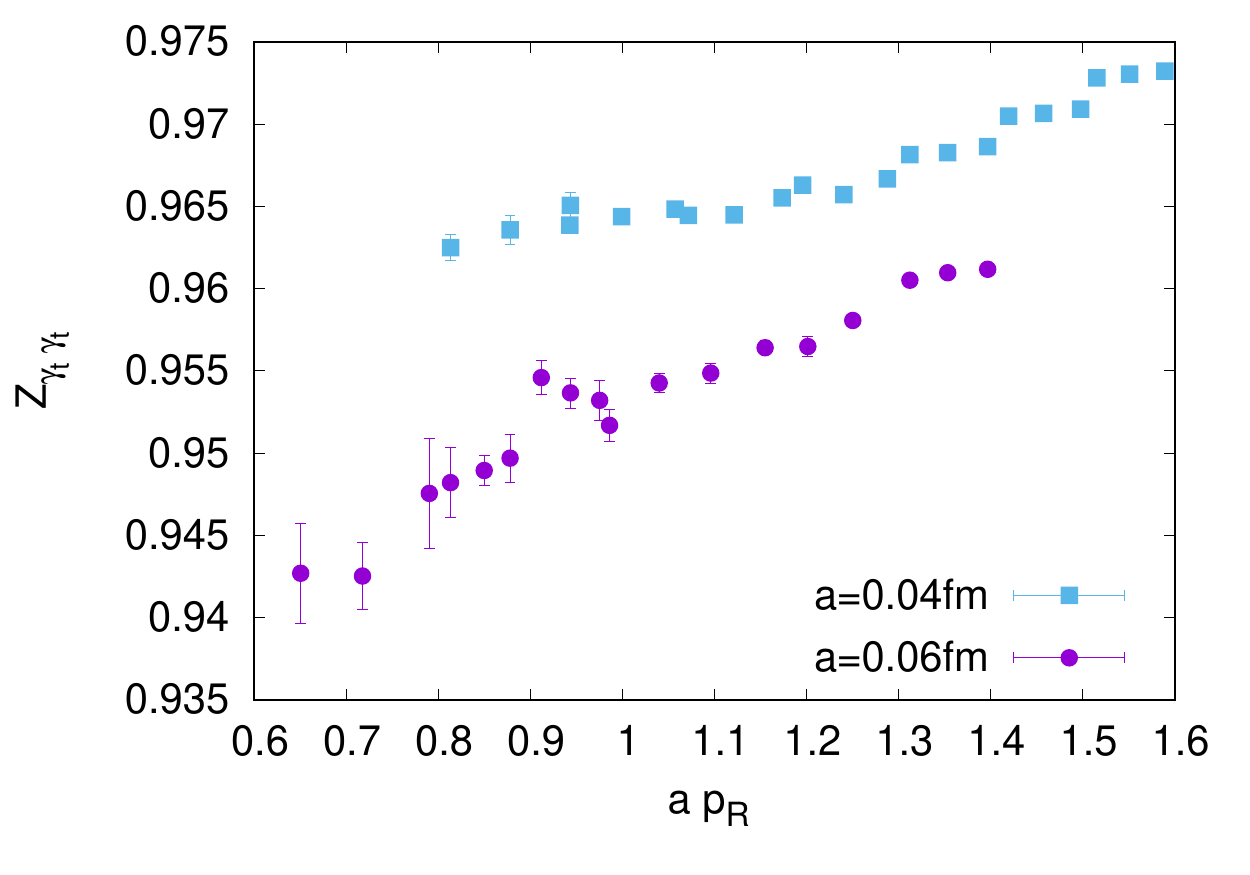}
\caption{
The renormalization factor $Z_{\gamma_t \gamma_t}$ as function of $p_R$ in
lattice units.
}
\eef{Zt}

In Fig. \ref{fig:Zq} and \ref{fig:Zt}, we show the numerical results $Z_q$ and $Z_{\gamma_t \gamma_t}$
as function of $p_R$. The numerical results on $Z_q$ have much smaller statistical errors compared
to $Z_{\gamma_t \gamma_t}$. The relative statistical errors on $Z_q$ are always smaller than $4.5 \cdot 10^{-4}$ 
and for large $p_R$ are smaller than $5 \cdot 10^{-5}$. We parametrize the $p_R$ dependence of $Z_q$ and $Z_{\gamma_t \gamma_t}$
by the following form
\begin{eqnarray}
    Z_i&=&Z_i^{0}+B/(ap_R)^2+ C\cdot (ap_R)^k \left(1+C_4 \Delta^{(4)} + C_6 \Delta^{(6)} \right),\cr && i=q, \gamma_t,
\end{eqnarray}
where 
\begin{equation}
\Delta^{(4)}=\frac{\sum_{\mu}( p_{\mu}' )^4}{p_R^4},~\Delta^{(6)}=\frac{\sum_{\mu}( p_{\mu}' )^6}{p_R^6}.
\end{equation}
This form is motivated by the 1-loop lattice perturbation theory \cite{Constantinou:2009tr,Constantinou:2010gr}
and the perturbative analysis with dimension two gluon condensate \cite{Lytle:2018evc}.
For the non-perturbative clover action $k=2$, while for Wilson action $k=1$. For HISQ smeared clover action with
tapdole improved value of $c_{sw}$ we expect ${\cal O}(a)$ discretization errors to be proportional to $\alpha_s^2$
with a very small coefficient, so it is reasonable to assume that the dominant cutoff effects scale like $a^2$.
Nevertheless, we also perform fits using $k=1$. 
In Ref. \cite{Constantinou:2010gr} the condensate contribution was ignored but it was included 
in the analysis of PNDME collaboration \cite{Bhattacharya:2015wna}
when fitting the $p_R$ dependence of the renormalization constant.
For $Z_{\gamma_t \gamma_t}$ and $a=0.06$ fm the fits with $k=1$ and $k=2$ work well, while
for $a=0.04$ fm the fits with $k=2$ work better. Fits of $Z_{\gamma_t \gamma_t}$ with $k=1$ give $\chi^2/df$ that is
around 2 for $a=0.04$fm . It is obvious from Figs. \ref{fig:Zq}
and \ref{fig:Zt} that the condensate contribution to $Z_{\gamma_t\gamma_t}$ is quite small, while it is large for $Z_q$.

The fits of $Z_q$ have very large $\chi^2/{\rm dof}$, most likely because of the very small statistical errors.
The value of the condensate obtained from the fit is compatible with the value $g^2 \langle A^2 \rangle \sim 4 ~\rm{GeV}^2$
found in Ref. \cite{Blossier:2010ky} for $a=0.06$ fm. For the smaller lattice spacing it is, however, is twice
larger, which could be due to instabilities in the fits.
Therefore we fix the condensate to the above value in order to stabilize the fits.
From the fits, we obtain the values of $Z_i^0$ which can serve as estimates for $Z_q$ and $Z_{\gamma_t \gamma_t}$ 
for the two lattice spacings. Multiplying these two renormalization constants we obtain $Z_V$. The results
of our analysis are summarized in Table \ref{tab:tabz}. The large uncertainties in $Z_q$ come from the differences
in the fits with the condensate contribution being fixed and treated as the fit parameter.
We can compare our result for $Z_V$ at $a=0.06$ fm with the value $Z_V=0.945(15)$ from  the PNDME collaboration.
Our result is slightly larger.
\begin{table}
\begin{tabular}{|c|ccc|ccc|c|}
\hline
a [fm]  &        &   $k=2$                  &           &          &  $k=1$                   &               \\
\hline
        & $Z_q$   &  $Z_{\gamma_t \gamma_t}$ & $Z_V$     &  $Z_q$   & $Z_{\gamma_t \gamma_t}$  & $Z_V$     \\
\hline
0.06    & 1.02(2) &  0.944(1)                & 0.963(20) &  1.04(1) &  0.930(1)                & 0.967(10) \\
0.04    & 1.03(3) &  0.950(3)                & 0.980(30) &  1.05(3) &  0.920(3)                & 0.966(30) \\
\hline
\end{tabular}
    \caption{The values of the renormalization constants obtained from the different fits.}
\label{tab:tabz}
\end{table}

\section{Leading twist target mass correction}\label{sec:tmc}

Unlike the light-cone ITD, the terms in the twist-2 OPE of the
equal-time bilocal bilinear have trace terms, which are proportional to powers of
hadron mass. At the level of twist-2 trace terms, such target mass
effects have been calculated
explicitly~\cite{Chen:2016utp,Radyushkin:2017ffo}. For the case of
pion matrix element, such target mass corrected expressions are
obtained from \eqn{t2ope},\eqn{t2operi} and \eqn{ratgenexp}  given
by the replacement
\beq
(P_z z)^n \to (P_z z)^n \sum_{k=0}^{n/2} \frac{(n-k)!}{k!(n-2k)!}\left(\frac{m_\pi^2}{4P_z^2}\right)^k,
\eeq{tmcreduced}
where $n$ are even integer valued for the $u-d$ pion PDF case.
Including such correction terms in our analysis did not change the
results (i.e., effectively the inferred values of valence pion moments
from twist-2 OPE) well within their errors. However, there could
be unaccounted target mass effects that originate from higher-twist
terms and it is an expectation that their coefficients are ${\cal
O}(1)$ or smaller, and hence suppressed as simple powers of
$m_\pi^2/P_z^2$. We ignore any such effect in this computation.

\section{Goodness of fits}\label{sec:chisq}

\bef
\centering
\includegraphics[scale=0.5]{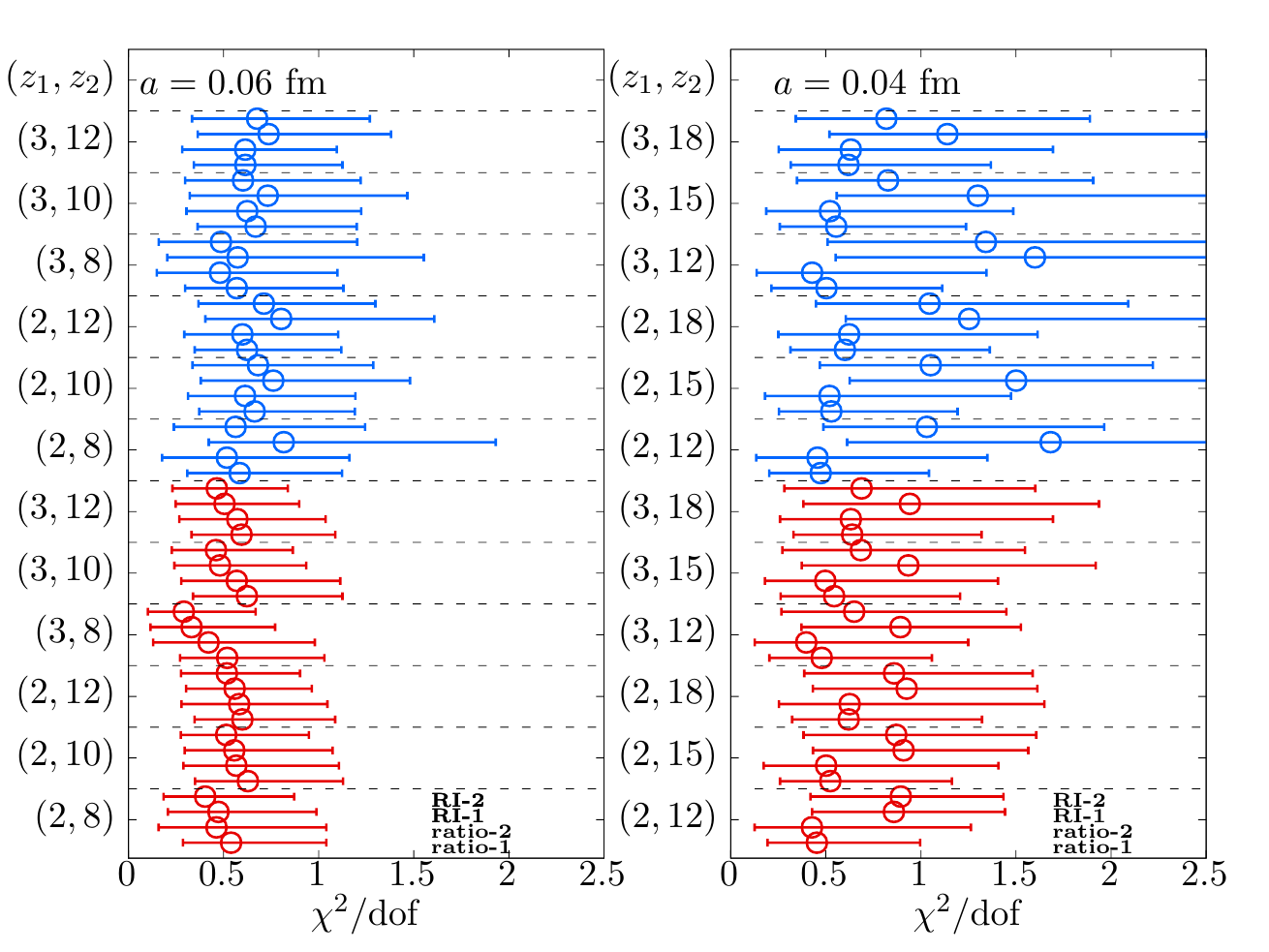}
\caption{$\chi^2/{\rm dof}$ for the 2-parameter (red) and 4-parameter
(blue) fits is shown from various fit ranges and renormalization
schemes. This plot accompanies \fgn{fitx1} and \fgn{fitalpha}.  The
description of the plot is similar to \fgn{fitx1}. The left panel
is for $a=0.06$ fm and the right one for $a=0.04$ fm.}
\eef{chiplot}

In \fgn{chiplot}, we plot the $\chi^2/{\rm dof}$ for our fits to the
2-parameter and 4-parameter PDF ansatz, \eqn{pdfansatz} in the main
text. The description of the plot is similar to \fgn{fitx1}. For
each case ($[z_1,z_2]$, renormalization scheme and ansatz), the
$\chi^2/{\rm dof}$ as sampled during the bootstrap is shown. The
definition of $\chi^2$ includes both statistical error as well as
perturbative uncertainties via \eqn{chisqsys}.
\section{Dependence of large-$x$ exponent on factorization scale}\label{sec:betaevol}

The parameters used in 2- and 4-parameter PDF ansatz are dependent
on the factorization scale $\mu$ used in the Wilson coefficients
$c_n(\mu^2 z^2)$ that enter the twist-2 OPE.  In \fgn{betaevol},
we address the dependence of the large-$x$ exponent $\beta(\mu^2)$
on the factorization scale $\mu$ (c.f.,~\cite{Ball:2016spl} for a similar 
analysis on phenomenological PDFs).  We repeated the analysis of matrix elements in ratio
scheme with $n_z^0=1$ at fixed $[z_1,z_2]=[a,0.5 \text{\ fm}]$ using
different values of $\mu$, varying by few factors around the typical
momentum scales of $\sim 3$ GeV as set by the momenta $P_z$ we used,
and the lattice spacing used. This is so as to keep logarithms
of $\mu/P_z$ and $\mu z$ small, and be consistent with the fixed order
calculation.  From each such $\mu$, we obtained the best fit values
of $\beta(\mu^2)$. In \fgn{betaevol}, we show $\beta(\mu^2)$ as a
function of $\mu^2$ from the two lattice spacings. In the main text,
we presented results at $\mu^2=10.2$ GeV$^2$. The variation with
$\mu$ is mild, perhaps logarithmic in the range of $\mu$ used. Thus,
we do not expect the results to change drastically due to the choice
of $\mu$ used.

\bef
\centering
\includegraphics[scale=0.6]{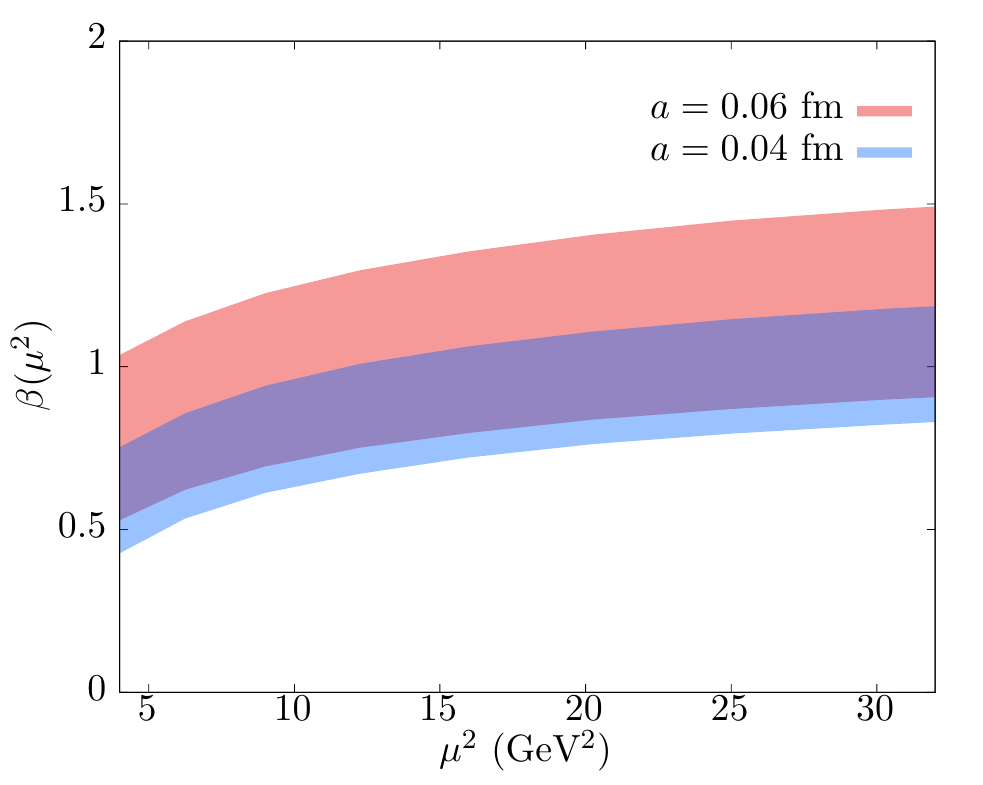}
\caption{The value of large-$x$ exponent, $\beta(\mu^2)$, for the valence PDF at 
the $\msbar$ scale $\mu^2$.}
\eef{betaevol}

\section{Asymptotic expansion of $\langle x^n\rangle$ at large-$n$}\label{sec:largen}

We consider PDFs of the form,
\beq
f(x)={\cal N} x^{\alpha}(1-x)^\beta {\cal G}(x),
\eeq{pdf1}
with ${\cal G}(x)$ being a smooth well behaved function that does not vanish 
between 0 and 1. Then, the $n$-th moment is
\beq
\langle x^n\rangle ={\cal N} \int_0^1 x^{\alpha+n}(1-x)^\beta{\cal G}(x)dx\equiv {\cal N} \int_0^1 e^{F(x)} dx,
\eeq{nmom}
for $F(x)=(n+\alpha)\log(x)+\beta\log(1-x)+\log{\cal G}(x)$. Now,
we proceed towards doing a saddle point approximation in order to
evaluate the leading term in the above integral in the limit of
infinite $n$. The maximum of $F(x)$ occurs at $x=x_0=1-\beta/n+{\cal
O}(1/n^2)$, which is less than 1 and hence within the domain of
integration, and on the real axis. Thus, $F(x)$ in the proximity
of $x=x_0$ is
\beq
F(x)\approx\log\left(\frac{\beta^\beta}{n^\beta}\right)+n \log\left(1-\frac{\beta}{n}\right)+\log{\cal G}(1)-\frac{n^2}{2\beta} (x-x_0)^2.
\eeq{saddle}
Thus, the saddle point approximation gives the asymptotic dependence on $n$,
\beq
\langle x^n\rangle \propto \frac{1}{n^{\beta+1}},
\eeq{asymn}
from the first term in \eqn{saddle} and an extra $n$ from the change
of variables in the last term of \eqn{saddle} to perform the remaining
Gaussian integral. The asymptotic series for $\langle x^n\rangle$
in the limit of large-$n$ is given by the standard multiplication
correction factor which is a series in $1/n$.

\section{Attractor at $x=x^*(\beta)$ for family of PDFs at fixed $\langle x\rangle$}\label{sec:xfp}

\bef
\centering
\includegraphics[scale=0.72]{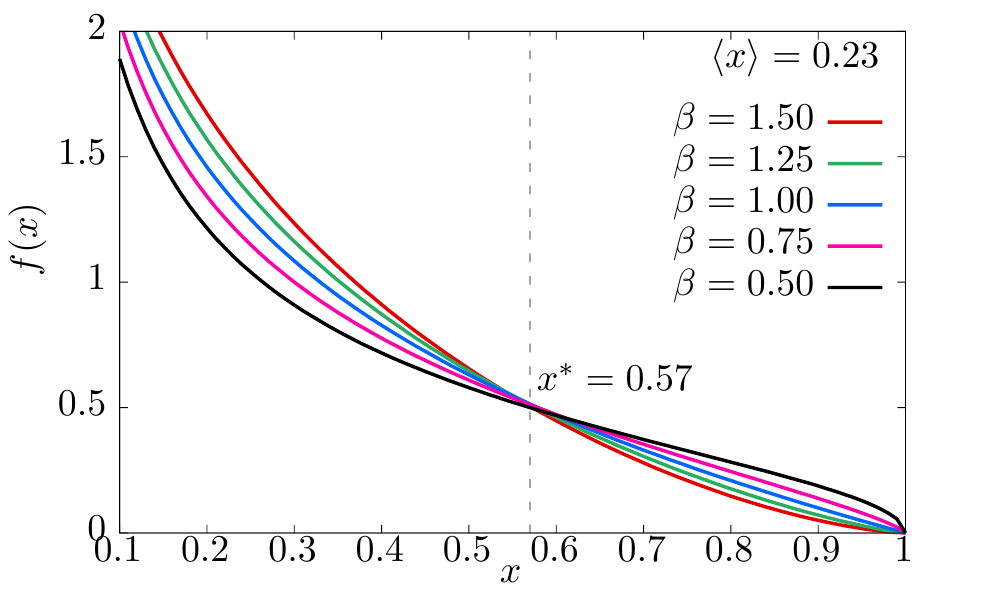}
\caption{The plot shows different set of 2-parameter PDFs, $f(x,\alpha,\beta)$,
that have fixed $\langle x\rangle=0.23$. The different colored curves correspond
to different values of $\beta$ centered around $\beta=1$. The dashed line 
is the value $x^*=0.57$ which is the fixed point for the PDFs.}
\eef{pdfscatter}

In this appendix, we explore a curious feature of the
extracted PDFs in \fgn{pdffit}, \fgn{pdfri}, \fgn{pdfansatz} and
\fgn{finalpdf}; in all these figures, one can observe that the
error-bands for the best fit PDFs pinch at around
$x\approx 0.6$.  We understand this to arise due to a weak attractor
in $x$-$f(x)$ plane, once we specify a value of the first moment
$\langle x\rangle$ and the value of $\beta$. In \fgn{pdfscatter},
we plot PDFs of the form $f(x,\alpha,\beta)={\cal N}x^\alpha (1-x)^\beta$
that have fixed $\langle x\rangle=0.23$ (the typical value for pion)
and with $\beta$ around 1. It is interesting to observe that all
these PDFs have a tendency to converge around $x=0.6$ as in our
PDFs in the main text. Thus, the pinch observed around $x=0.6$ is actually
a robust feature arising of $\langle x\rangle\approx 0.23$ and $\beta\approx
1$ in our calculations.

One way to understand this behavior is the following. Once we specify
$\langle x\rangle=a_0$, for some $a_0$, then it induces a relation
$\alpha=\alpha(\beta,a_0)$; for the 2-parameter ansatz, it is
$\alpha(\beta,a_0)=(1-2 a_0-a_0\beta)/(a_0-1)$. Therefore, the PDF
is also of the form $f(x,\alpha(a_0,\beta),\beta)$. For there to
be a basin of attraction at $x=x^*$, it satisfies $\frac{\partial
f(x,\alpha(a_0,\beta),\beta)}{\partial \beta}\bigg{|}_{x=x^*}=0$.
One can demonstrate numerically that there exists such a solution
at $x=x^*\approx 0.57$ when $\beta=1$ and $\langle x\rangle= 0.23$.
Therefore, this feature of PDF ties directly to our various estimates,
and forms yet another consistency check of our observations.
However, it is not clear if such an effect would also persist when
deviation from 2-parameter PDF ansatz is significant. It is worth
pointing out that there could be other robust features of valence
PDF, for example, a study~\cite{Leon:2020nfb} found a near constant
relationship of peak position and height on the factorization scale
$\mu$. It would be interesting to make use of such features in the
future analysis.

\section{Results using the pion matrix element in ratio scheme with $P_z^0=0$ (reduced ITD) for $a=0.06$ fm lattice}\label{sec:kx0}

\bef
\centering
\includegraphics[scale=0.6]{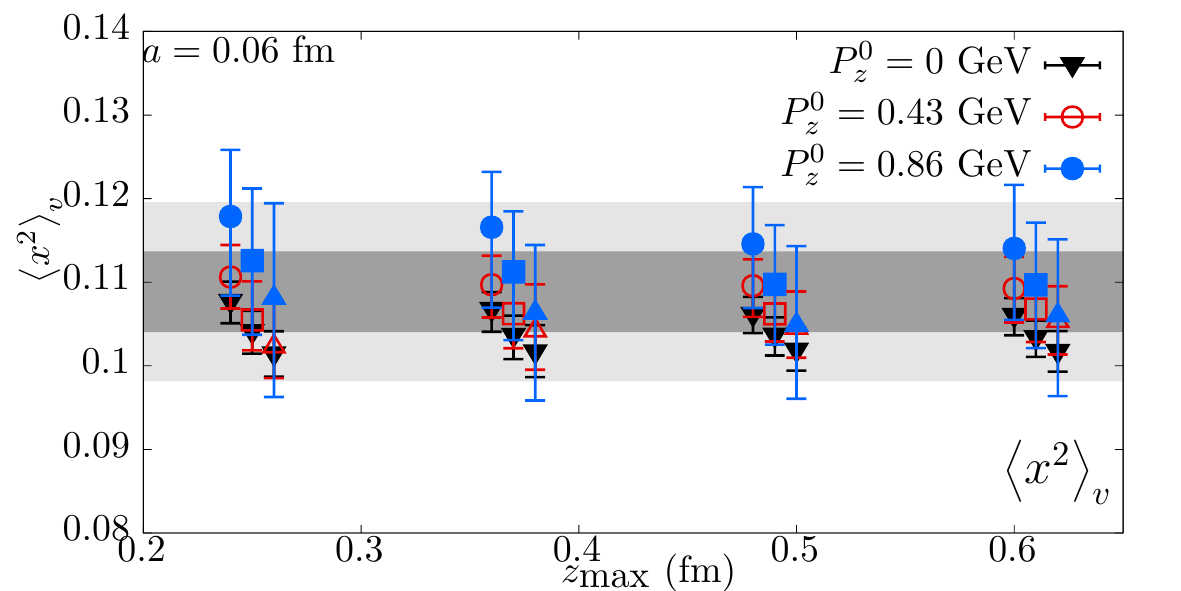}
\includegraphics[scale=0.6]{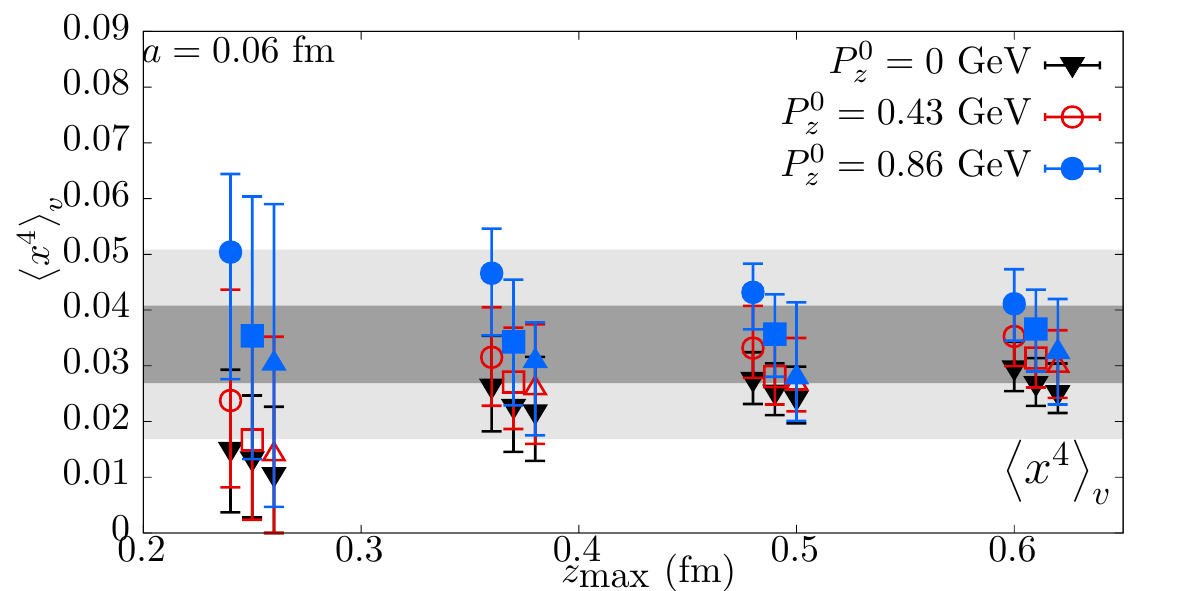}
\includegraphics[scale=0.6]{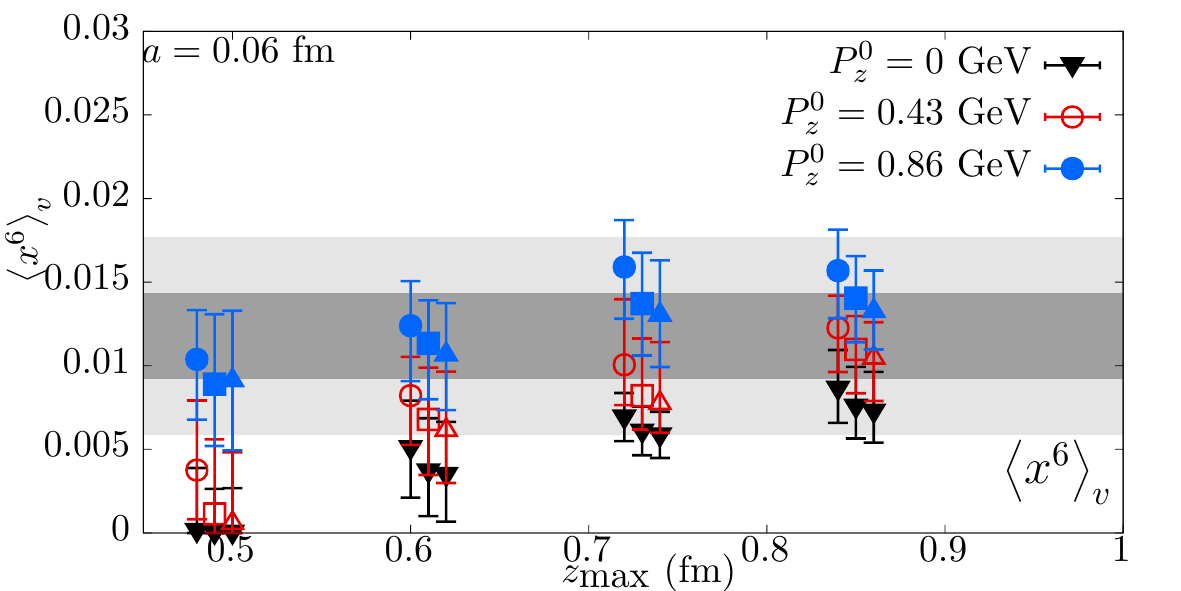}
\caption{Plot shows the first three even moments obtained on $a=0.06$
fm lattice in a model independent manner as described in \scn{mom}.
The description of the three plots is similar to that in \fgn{combinedx2},
\fgn{combinedx4} and \fgn{combinedx6}.  In addition to the two
values of $P_z^0\ne 0$, this plot includes $P_z^0=0$ reference scale
for the ratio.  }
\eef{kx0ope}

\bef
\centering
\includegraphics[scale=0.8]{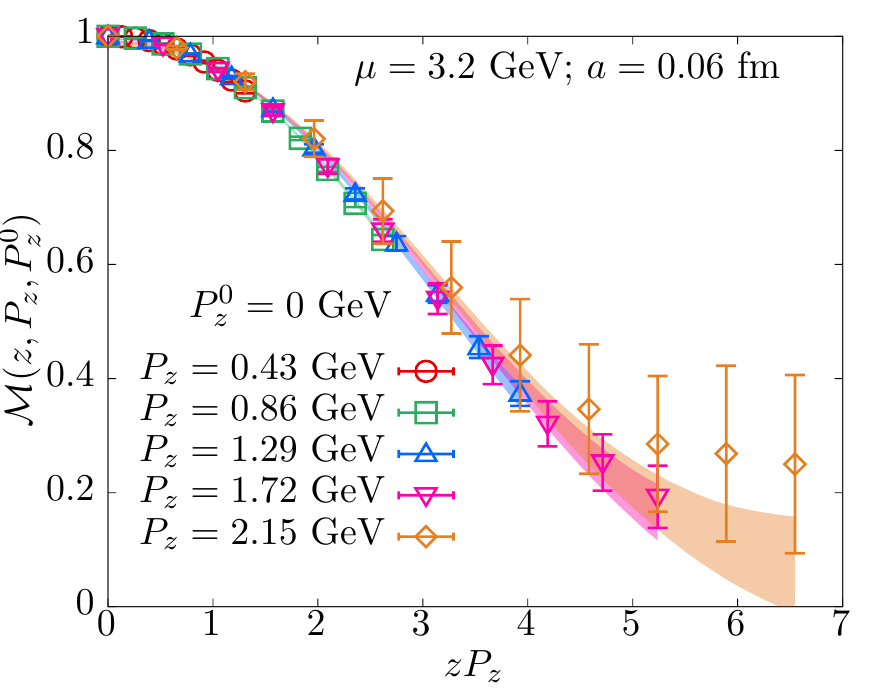}
\includegraphics[scale=0.8]{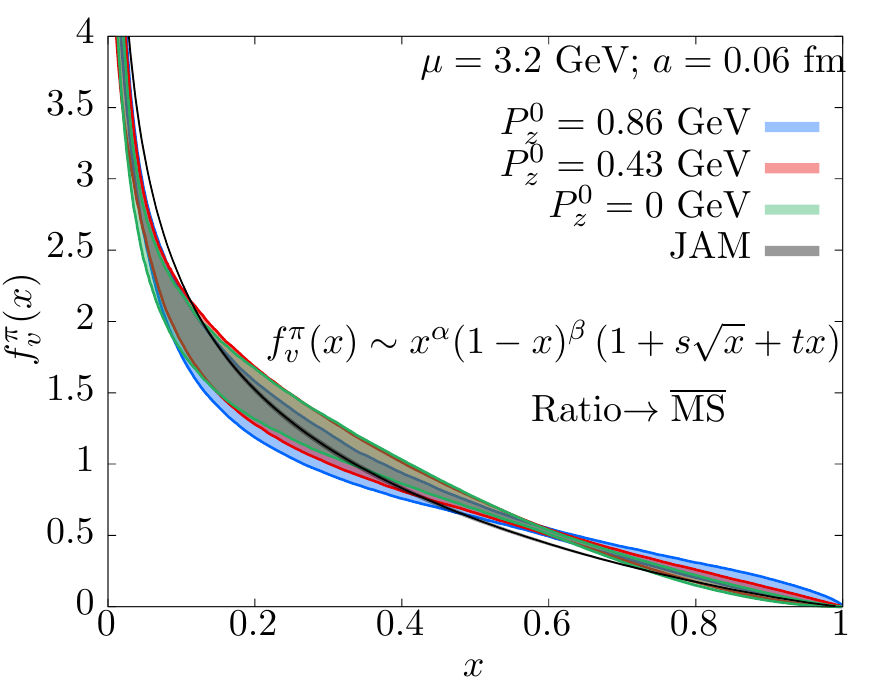}
\caption{
The plot is on the 4-parameter PDF ansatz fits to the $P_z^0=0$
ratio data on $a=0.06$ fm lattice. The top panel shows the 
ratio data a function of $P_z z$ along with the fits. The bottom 
panel shows the PDF extracted from $P_z^0=0$ ratio with 
other ratio data at non-zero $P_z^0$ discussed in the main text.
The description of the plots are similar to the ones in 
\fgn{ritdfit} and \fgn{pdffit}.
}
\eef{kx0pdf}

In the main text, we utilized ratio scheme with non-zero reference
scale $P_z^0$. We had two reasons to do this; first, by using all
the momenta $P_z$ as well as the reference $P_z^0$ greater than the
$\Lambda_{\rm QCD}$ scale, we adhered to the twist-2 perturbative  matching
framework closely. This is to be contrasted with the usage of $P_z^0=0$ in the 
ratio scheme (reduced ITD) which relies on the cancellation of higher twist effects, 
which our data also supports, but nevertheless lacks a firm theoretical basis. 
Second, we observed a large lattice periodicity
effect in the $P_z=0$ three-point function in the fine lattice (see
\apx{wrap}). Therefore, we avoided the traditional $P_z^0=0$ reference scale. 
In this appendix, for completeness sake, we present results
including the $P_z^0=0$ ratio for the $a=0.06$ fm lattice, where at
least there is no issue with the lattice wrap-around effect. In
addition, in \scn{matching}, we provided empirical evidence and
rationale behind the validity of using $P_z^0=0$ in defining
renormalized ratios that is consistent with twist-2 matching framework, given
the statistical errors in the data.  Therefore, we include the
results from  $P_z^0=0$ ratio with the other two non-zero $P_z^0$ presented
in the main text.

In \fgn{kx0ope}, we include $P_z^0=0$ ratio results (black
symbols) for lowest three even moments along with other two non-zero
$P_z^0$ results that we presented in \scn{mom}. We refer the reader
to \scn{mom} and the captions of the figures therein for detailed
explanations. There is a slight tendency for the extracted moments
using $P_z^0=0$ to be smaller than at higher non-zero $P_z^0$. If
we include the $P_z^0=0$ results along with the other results, we
estimate the moments in a model-independent way as $\langle
x^2\rangle_v= 0.1071^{+(33)(54)}_{-(37)(54)}$, $\langle x^4\rangle_v=
0.0317^{+(50)(75)}_{-(50)(75)}$ and $\langle x^6\rangle_v=
0.0102^{+(23)(39)}_{-(20)(39)}$ with the same methodology as in
\scn{mom}.

In \fgn{kx0pdf}, we present results using fits to PDF ansatz. We
refer the reader to \scn{pdf} for explanations and methodology. In
the top panel, the ratio matrix element with $P_z^0=0$ is shown along with the bands
resulting from fits to the 4-parameter ansatz. In the bottom panel,
the best fit PDF using 4-parameter ansatz is shown (green) and
compared to results using other non-zero $P_z^0$. The PDFs using
different $P_z^0$ remain more or less the same.  The values of the
exponent including results from $P_z^0$ are
$\alpha=-0.40^{+(14)(17)}_{-(14)(17)}$ and
$\beta=1.30^{+(35)(46)}_{(35)(46)}$. These results are to be compared
with the entries in \tbn{tabsummary}. There is a tendency for
$P_z^0=0$ to pull the result for $\beta$ higher; an opposite 
behavior wherein $\beta$ increased when $P_z^0$ is increased would have 
been more desirable.
The inferred
moments from model-dependent ansatz including $P_z^0=0$ ratio with
other ratio schemes give $\langle x\rangle_v=
0.2470^{+(93)(66)}_{-(94)(66)}$, $\langle x^2\rangle_v=
0.1100^{+(38)(57)}_{-(36)(57)}$, $\langle
x^3\rangle_v=0.0617^{+(44)(70)}_{-(42)(70)}$ and $\langle x^4\rangle_v=
0.0393^{+(44)(67)}_{-(43)(67)}$. Using the analysis using asymptotic
expansion for moments with $N_{\rm asym}=4$ as presented in
\scn{largex}, we get $\beta=1.47^{+(27)(30)}_{-(23)(30)}$. The
analysis of effective $\beta$ that we discussed in \scn{largex}
gives $\beta_{\rm eff}(n=4)=2.04^{+(33)(50)}_{(35)(50)}$ that is
again consistent with both $\beta=1$ and $\beta=2$. 

\section{Analysis imposing $\beta=2$ in PDF ansatz}\label{sec:b2}

\bef
\centering
\includegraphics[scale=0.8]{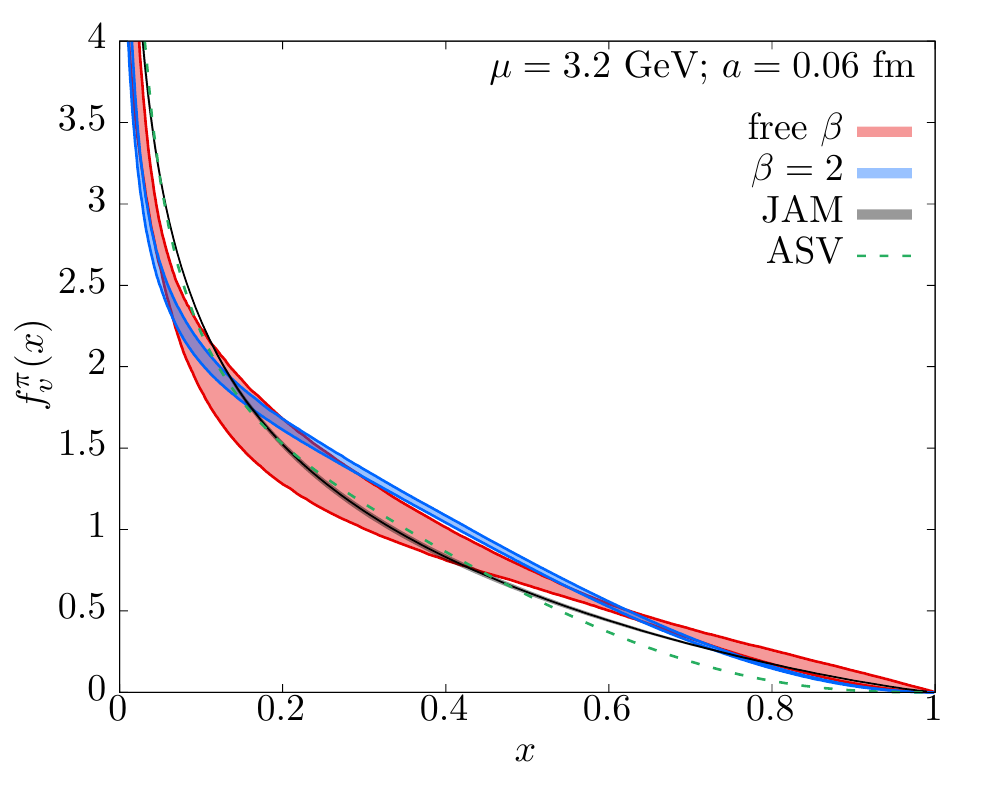}
\caption{The valence PDF with $\beta$ being a free parameter (red) and
$\beta$ being fixed at a value 2 (blue) in the ansatz $f_v^\pi(x)={\cal
N} x^\alpha(1-x)^\beta(1+s \sqrt{x}+t x)$.  For comparison, the
JAM18~\cite{Barry:2018ort} curve (black) and ASV~\cite{Aicher:2010cb}
result (green) are also shown.}
\eef{pdfbeta2}

In \scn{pdf}, we used 2- and 4-parameter PDF ansatz in \eqn{pdfansatz}
to reconstruct the PDF that best describes the real-space lattice data.
In that analysis, we kept $\beta$ as a free parameter.  Through
that analysis, we found PDFs with $\beta\approx 1$ or less to best
describe the data.  Here, we do the following; we take $\beta=2$
as if it is a well established fact, and impose the constraint
$\beta=2$ in the 4-parameter ansatz and fit only $\alpha,s$ and $t$
to minimize $\chi^2$.  That is, even though there is a set of PDF that better
describe the lattice data in the space of $(\alpha,\beta,s,t)$, we
restrict now to a subspace $(\alpha,\beta=2,s,t)$ and ask what PDFs
within this subspace best describes the data.

Let us take a specific case of $P_z^0=0.43$ GeV ratio matrix element on
$a=0.06$ fm lattice. We fit the lattice data in the range $z\in[a,0.5\text{\
fm}]$. The resultant $\chi^2/{\rm dof}$ for this 3-parameter ansatz
was of course larger compared to the 4-parameter ansatz, but not
very large either; for the 4-parameter ansatz for the same case,
$\chi^2/{\rm dof}\approx 25/36$ while it was $56/37$ for the
3-parameter one tried here. The resulting PDF is shown as the blue
band in \fgn{pdfbeta2}. For comparison, the unconstrained 4-parameter
PDF result is also shown. The $\beta=2$ result closely hugs the
best-fit result and tries to lie within the 1-$\sigma$ vicinity of
it. The error bar on the constrained PDF is small because there only
exists a small set of PDF with $\beta=2$ that have a decent $\chi^2$.
One might wonder if imposing $\beta=2$ took our result closer to
the ASV result~\cite{Aicher:2010cb}. For this, the ASV result is shown as the green
dashed line \fgn{pdfbeta2}. The $\beta=2$ result actually misses
the ASV result badly at intermediate $x$ at the expense of agreeing
well very close to $x=1$. It is fascinating that this is actually
due to robust tendency of the PDFs to pass through an approximate
fixed-point at $x=x^*$ ($\approx 0.6$ specific to our data) that
we discussed in \apx{xfp} and determined by the first moment. 
Not surprisingly, the $\beta=2$ fits
resulted in values of first four moments as $0.254(5), 0.108(3),
0.057(2)$ and $0.034(2)$ for the case discussed, which compares well
with the first four moments for the same case obtained using a full
4-parameter fit; namely, $0.245(8), 0.111(3), 0.064(4)$ and $0.040(5)$.
This analysis for fixed $\beta=2$ assumes a very specific functional form for 
${\cal G}(x)=1+s\sqrt{x}+t x$. Thus, it is very much a possibility that 
by choosing some other flexible functional form for ${\cal G}(x)$, one might 
still be able to get $\beta\approx 2$ and get better $\chi^2$. We do not 
explore this any further in this paper since effective $\beta$ analysis 
addresses this in a better manner.

\input{pap.bbl}
%\bibliography{pap.bib}

\end{document}

%% file: pap.bbl

%% file: pap.bbl
\begin{thebibliography}{95}%
\makeatletter
\providecommand \@ifxundefined [1]{%
 \@ifx{#1\undefined}
}%
\providecommand \@ifnum [1]{%
 \ifnum #1\expandafter \@firstoftwo
 \else \expandafter \@secondoftwo
 \fi
}%
\providecommand \@ifx [1]{%
 \ifx #1\expandafter \@firstoftwo
 \else \expandafter \@secondoftwo
 \fi
}%
\providecommand \natexlab [1]{#1}%
\providecommand \enquote  [1]{``#1''}%
\providecommand \bibnamefont  [1]{#1}%
\providecommand \bibfnamefont [1]{#1}%
\providecommand \citenamefont [1]{#1}%
\providecommand \href@noop [0]{\@secondoftwo}%
\providecommand \href [0]{\begingroup \@sanitize@url \@href}%
\providecommand \@href[1]{\@@startlink{#1}\@@href}%
\providecommand \@@href[1]{\endgroup#1\@@endlink}%
\providecommand \@sanitize@url [0]{\catcode `\\12\catcode `\$12\catcode
  `\&12\catcode `\#12\catcode `\^12\catcode `\_12\catcode `\%12\relax}%
\providecommand \@@startlink[1]{}%
\providecommand \@@endlink[0]{}%
\providecommand \url  [0]{\begingroup\@sanitize@url \@url }%
\providecommand \@url [1]{\endgroup\@href {#1}{\urlprefix }}%
\providecommand \urlprefix  [0]{URL }%
\providecommand \Eprint [0]{\href }%
\providecommand \doibase [0]{http://dx.doi.org/}%
\providecommand \selectlanguage [0]{\@gobble}%
\providecommand \bibinfo  [0]{\@secondoftwo}%
\providecommand \bibfield  [0]{\@secondoftwo}%
\providecommand \translation [1]{[#1]}%
\providecommand \BibitemOpen [0]{}%
\providecommand \bibitemStop [0]{}%
\providecommand \bibitemNoStop [0]{.\EOS\space}%
\providecommand \EOS [0]{\spacefactor3000\relax}%
\providecommand \BibitemShut  [1]{\csname bibitem#1\endcsname}%
\let\auto@bib@innerbib\@empty
%</preamble>
\bibitem [{\citenamefont {Collins}\ \emph {et~al.}(1989)\citenamefont
  {Collins}, \citenamefont {Soper},\ and\ \citenamefont
  {Sterman}}]{Collins:1989gx}%
  \BibitemOpen
  \bibfield  {author} {\bibinfo {author} {\bibfnamefont {J.~C.}\ \bibnamefont
  {Collins}}, \bibinfo {author} {\bibfnamefont {D.~E.}\ \bibnamefont {Soper}},
  \ and\ \bibinfo {author} {\bibfnamefont {G.~F.}\ \bibnamefont {Sterman}},\
  }\enquote {\bibinfo {title} {{Factorization of Hard Processes in QCD}},}\ \
  (\bibinfo {year} {1989})\ pp.\ \bibinfo {pages} {1--91},\ \Eprint
  {http://arxiv.org/abs/hep-ph/0409313} {arXiv:hep-ph/0409313} \BibitemShut
  {NoStop}%
\bibitem [{\citenamefont {Soper}(1977)}]{Soper:1976jc}%
  \BibitemOpen
  \bibfield  {author} {\bibinfo {author} {\bibfnamefont {D.~E.}\ \bibnamefont
  {Soper}},\ }\href {\doibase 10.1103/PhysRevD.15.1141} {\bibfield  {journal}
  {\bibinfo  {journal} {Phys. Rev. D}\ }\textbf {\bibinfo {volume} {15}},\
  \bibinfo {pages} {1141} (\bibinfo {year} {1977})}\BibitemShut {NoStop}%
\bibitem [{\citenamefont {Martinelli}\ and\ \citenamefont
  {Sachrajda}(1987)}]{Martinelli:1987zd}%
  \BibitemOpen
  \bibfield  {author} {\bibinfo {author} {\bibfnamefont {G.}~\bibnamefont
  {Martinelli}}\ and\ \bibinfo {author} {\bibfnamefont {C.~T.}\ \bibnamefont
  {Sachrajda}},\ }\href {\doibase 10.1016/0370-2693(87)90601-0} {\bibfield
  {journal} {\bibinfo  {journal} {Phys. Lett. B}\ }\textbf {\bibinfo {volume}
  {196}},\ \bibinfo {pages} {184} (\bibinfo {year} {1987})}\BibitemShut
  {NoStop}%
\bibitem [{\citenamefont {Ji}(2013)}]{Ji:2013dva}%
  \BibitemOpen
  \bibfield  {author} {\bibinfo {author} {\bibfnamefont {X.}~\bibnamefont
  {Ji}},\ }\href {\doibase 10.1103/PhysRevLett.110.262002} {\bibfield
  {journal} {\bibinfo  {journal} {Phys. Rev. Lett.}\ }\textbf {\bibinfo
  {volume} {110}},\ \bibinfo {pages} {262002} (\bibinfo {year} {2013})},\
  \Eprint {http://arxiv.org/abs/1305.1539} {arXiv:1305.1539 [hep-ph]}
  \BibitemShut {NoStop}%
\bibitem [{\citenamefont {Ji}(2014)}]{Ji:2014gla}%
  \BibitemOpen
  \bibfield  {author} {\bibinfo {author} {\bibfnamefont {X.}~\bibnamefont
  {Ji}},\ }\href {\doibase 10.1007/s11433-014-5492-3} {\bibfield  {journal}
  {\bibinfo  {journal} {Sci. China Phys. Mech. Astron.}\ }\textbf {\bibinfo
  {volume} {57}},\ \bibinfo {pages} {1407} (\bibinfo {year} {2014})},\ \Eprint
  {http://arxiv.org/abs/1404.6680} {arXiv:1404.6680 [hep-ph]} \BibitemShut
  {NoStop}%
\bibitem [{\citenamefont
  {Radyushkin}(2017{\natexlab{a}})}]{Radyushkin:2017cyf}%
  \BibitemOpen
  \bibfield  {author} {\bibinfo {author} {\bibfnamefont {A.}~\bibnamefont
  {Radyushkin}},\ }\href {\doibase 10.1103/PhysRevD.96.034025} {\bibfield
  {journal} {\bibinfo  {journal} {Phys. Rev. D}\ }\textbf {\bibinfo {volume}
  {96}},\ \bibinfo {pages} {034025} (\bibinfo {year} {2017}{\natexlab{a}})},\
  \Eprint {http://arxiv.org/abs/1705.01488} {arXiv:1705.01488 [hep-ph]}
  \BibitemShut {NoStop}%
\bibitem [{\citenamefont {Orginos}\ \emph {et~al.}(2017)\citenamefont
  {Orginos}, \citenamefont {Radyushkin}, \citenamefont {Karpie},\ and\
  \citenamefont {Zafeiropoulos}}]{Orginos:2017kos}%
  \BibitemOpen
  \bibfield  {author} {\bibinfo {author} {\bibfnamefont {K.}~\bibnamefont
  {Orginos}}, \bibinfo {author} {\bibfnamefont {A.}~\bibnamefont {Radyushkin}},
  \bibinfo {author} {\bibfnamefont {J.}~\bibnamefont {Karpie}}, \ and\ \bibinfo
  {author} {\bibfnamefont {S.}~\bibnamefont {Zafeiropoulos}},\ }\href {\doibase
  10.1103/PhysRevD.96.094503} {\bibfield  {journal} {\bibinfo  {journal} {Phys.
  Rev. D}\ }\textbf {\bibinfo {volume} {96}},\ \bibinfo {pages} {094503}
  (\bibinfo {year} {2017})},\ \Eprint {http://arxiv.org/abs/1706.05373}
  {arXiv:1706.05373 [hep-ph]} \BibitemShut {NoStop}%
\bibitem [{\citenamefont {Braun}\ and\ \citenamefont
  {M\"uller}(2008)}]{Braun:2007wv}%
  \BibitemOpen
  \bibfield  {author} {\bibinfo {author} {\bibfnamefont {V.}~\bibnamefont
  {Braun}}\ and\ \bibinfo {author} {\bibfnamefont {D.}~\bibnamefont
  {M\"uller}},\ }\href {\doibase 10.1140/epjc/s10052-008-0608-4} {\bibfield
  {journal} {\bibinfo  {journal} {Eur. Phys. J. C}\ }\textbf {\bibinfo {volume}
  {55}},\ \bibinfo {pages} {349} (\bibinfo {year} {2008})},\ \Eprint
  {http://arxiv.org/abs/0709.1348} {arXiv:0709.1348 [hep-ph]} \BibitemShut
  {NoStop}%
\bibitem [{\citenamefont {Braun}\ \emph {et~al.}(1995)\citenamefont {Braun},
  \citenamefont {Gornicki},\ and\ \citenamefont {Mankiewicz}}]{Braun:1994jq}%
  \BibitemOpen
  \bibfield  {author} {\bibinfo {author} {\bibfnamefont {V.}~\bibnamefont
  {Braun}}, \bibinfo {author} {\bibfnamefont {P.}~\bibnamefont {Gornicki}}, \
  and\ \bibinfo {author} {\bibfnamefont {L.}~\bibnamefont {Mankiewicz}},\
  }\href {\doibase 10.1103/PhysRevD.51.6036} {\bibfield  {journal} {\bibinfo
  {journal} {Phys. Rev. D}\ }\textbf {\bibinfo {volume} {51}},\ \bibinfo
  {pages} {6036} (\bibinfo {year} {1995})},\ \Eprint
  {http://arxiv.org/abs/hep-ph/9410318} {arXiv:hep-ph/9410318} \BibitemShut
  {NoStop}%
\bibitem [{\citenamefont {Musch}\ \emph {et~al.}(2011)\citenamefont {Musch},
  \citenamefont {Hagler}, \citenamefont {Negele},\ and\ \citenamefont
  {Schafer}}]{Musch:2010ka}%
  \BibitemOpen
  \bibfield  {author} {\bibinfo {author} {\bibfnamefont {B.~U.}\ \bibnamefont
  {Musch}}, \bibinfo {author} {\bibfnamefont {P.}~\bibnamefont {Hagler}},
  \bibinfo {author} {\bibfnamefont {J.~W.}\ \bibnamefont {Negele}}, \ and\
  \bibinfo {author} {\bibfnamefont {A.}~\bibnamefont {Schafer}},\ }\href
  {\doibase 10.1103/PhysRevD.83.094507} {\bibfield  {journal} {\bibinfo
  {journal} {Phys. Rev. D}\ }\textbf {\bibinfo {volume} {83}},\ \bibinfo
  {pages} {094507} (\bibinfo {year} {2011})},\ \Eprint
  {http://arxiv.org/abs/1011.1213} {arXiv:1011.1213 [hep-lat]} \BibitemShut
  {NoStop}%
\bibitem [{\citenamefont {Ji}\ \emph {et~al.}(2018)\citenamefont {Ji},
  \citenamefont {Zhang},\ and\ \citenamefont {Zhao}}]{Ji:2017oey}%
  \BibitemOpen
  \bibfield  {author} {\bibinfo {author} {\bibfnamefont {X.}~\bibnamefont
  {Ji}}, \bibinfo {author} {\bibfnamefont {J.-H.}\ \bibnamefont {Zhang}}, \
  and\ \bibinfo {author} {\bibfnamefont {Y.}~\bibnamefont {Zhao}},\ }\href
  {\doibase 10.1103/PhysRevLett.120.112001} {\bibfield  {journal} {\bibinfo
  {journal} {Phys. Rev. Lett.}\ }\textbf {\bibinfo {volume} {120}},\ \bibinfo
  {pages} {112001} (\bibinfo {year} {2018})},\ \Eprint
  {http://arxiv.org/abs/1706.08962} {arXiv:1706.08962 [hep-ph]} \BibitemShut
  {NoStop}%
\bibitem [{\citenamefont {Ishikawa}\ \emph {et~al.}(2017)\citenamefont
  {Ishikawa}, \citenamefont {Ma}, \citenamefont {Qiu},\ and\ \citenamefont
  {Yoshida}}]{Ishikawa:2017faj}%
  \BibitemOpen
  \bibfield  {author} {\bibinfo {author} {\bibfnamefont {T.}~\bibnamefont
  {Ishikawa}}, \bibinfo {author} {\bibfnamefont {Y.-Q.}\ \bibnamefont {Ma}},
  \bibinfo {author} {\bibfnamefont {J.-W.}\ \bibnamefont {Qiu}}, \ and\
  \bibinfo {author} {\bibfnamefont {S.}~\bibnamefont {Yoshida}},\ }\href
  {\doibase 10.1103/PhysRevD.96.094019} {\bibfield  {journal} {\bibinfo
  {journal} {Phys. Rev. D}\ }\textbf {\bibinfo {volume} {96}},\ \bibinfo
  {pages} {094019} (\bibinfo {year} {2017})},\ \Eprint
  {http://arxiv.org/abs/1707.03107} {arXiv:1707.03107 [hep-ph]} \BibitemShut
  {NoStop}%
\bibitem [{\citenamefont {Green}\ \emph {et~al.}(2018)\citenamefont {Green},
  \citenamefont {Jansen},\ and\ \citenamefont {Steffens}}]{Green:2017xeu}%
  \BibitemOpen
  \bibfield  {author} {\bibinfo {author} {\bibfnamefont {J.}~\bibnamefont
  {Green}}, \bibinfo {author} {\bibfnamefont {K.}~\bibnamefont {Jansen}}, \
  and\ \bibinfo {author} {\bibfnamefont {F.}~\bibnamefont {Steffens}},\ }\href
  {\doibase 10.1103/PhysRevLett.121.022004} {\bibfield  {journal} {\bibinfo
  {journal} {Phys. Rev. Lett.}\ }\textbf {\bibinfo {volume} {121}},\ \bibinfo
  {pages} {022004} (\bibinfo {year} {2018})},\ \Eprint
  {http://arxiv.org/abs/1707.07152} {arXiv:1707.07152 [hep-lat]} \BibitemShut
  {NoStop}%
\bibitem [{\citenamefont {Ji}\ \emph {et~al.}(2020)\citenamefont {Ji},
  \citenamefont {Liu}, \citenamefont {Liu}, \citenamefont {Zhang},\ and\
  \citenamefont {Zhao}}]{Ji:2020ect}%
  \BibitemOpen
  \bibfield  {author} {\bibinfo {author} {\bibfnamefont {X.}~\bibnamefont
  {Ji}}, \bibinfo {author} {\bibfnamefont {Y.-S.}\ \bibnamefont {Liu}},
  \bibinfo {author} {\bibfnamefont {Y.}~\bibnamefont {Liu}}, \bibinfo {author}
  {\bibfnamefont {J.-H.}\ \bibnamefont {Zhang}}, \ and\ \bibinfo {author}
  {\bibfnamefont {Y.}~\bibnamefont {Zhao}},\ }\href@noop {} {\  (\bibinfo
  {year} {2020})},\ \Eprint {http://arxiv.org/abs/2004.03543} {arXiv:2004.03543
  [hep-ph]} \BibitemShut {NoStop}%
\bibitem [{\citenamefont {Constantinou}\ and\ \citenamefont
  {Panagopoulos}(2017)}]{Constantinou:2017sej}%
  \BibitemOpen
  \bibfield  {author} {\bibinfo {author} {\bibfnamefont {M.}~\bibnamefont
  {Constantinou}}\ and\ \bibinfo {author} {\bibfnamefont {H.}~\bibnamefont
  {Panagopoulos}},\ }\href {\doibase 10.1103/PhysRevD.96.054506} {\bibfield
  {journal} {\bibinfo  {journal} {Phys. Rev. D}\ }\textbf {\bibinfo {volume}
  {96}},\ \bibinfo {pages} {054506} (\bibinfo {year} {2017})},\ \Eprint
  {http://arxiv.org/abs/1705.11193} {arXiv:1705.11193 [hep-lat]} \BibitemShut
  {NoStop}%
\bibitem [{\citenamefont {Alexandrou}\ \emph {et~al.}(2017)\citenamefont
  {Alexandrou}, \citenamefont {Cichy}, \citenamefont {Constantinou},
  \citenamefont {Hadjiyiannakou}, \citenamefont {Jansen}, \citenamefont
  {Panagopoulos},\ and\ \citenamefont {Steffens}}]{Alexandrou:2017huk}%
  \BibitemOpen
  \bibfield  {author} {\bibinfo {author} {\bibfnamefont {C.}~\bibnamefont
  {Alexandrou}}, \bibinfo {author} {\bibfnamefont {K.}~\bibnamefont {Cichy}},
  \bibinfo {author} {\bibfnamefont {M.}~\bibnamefont {Constantinou}}, \bibinfo
  {author} {\bibfnamefont {K.}~\bibnamefont {Hadjiyiannakou}}, \bibinfo
  {author} {\bibfnamefont {K.}~\bibnamefont {Jansen}}, \bibinfo {author}
  {\bibfnamefont {H.}~\bibnamefont {Panagopoulos}}, \ and\ \bibinfo {author}
  {\bibfnamefont {F.}~\bibnamefont {Steffens}},\ }\href {\doibase
  10.1016/j.nuclphysb.2017.08.012} {\bibfield  {journal} {\bibinfo  {journal}
  {Nucl. Phys. B}\ }\textbf {\bibinfo {volume} {923}},\ \bibinfo {pages} {394}
  (\bibinfo {year} {2017})},\ \Eprint {http://arxiv.org/abs/1706.00265}
  {arXiv:1706.00265 [hep-lat]} \BibitemShut {NoStop}%
\bibitem [{\citenamefont {Stewart}\ and\ \citenamefont
  {Zhao}(2018)}]{Stewart:2017tvs}%
  \BibitemOpen
  \bibfield  {author} {\bibinfo {author} {\bibfnamefont {I.~W.}\ \bibnamefont
  {Stewart}}\ and\ \bibinfo {author} {\bibfnamefont {Y.}~\bibnamefont {Zhao}},\
  }\href {\doibase 10.1103/PhysRevD.97.054512} {\bibfield  {journal} {\bibinfo
  {journal} {Phys. Rev. D}\ }\textbf {\bibinfo {volume} {97}},\ \bibinfo
  {pages} {054512} (\bibinfo {year} {2018})},\ \Eprint
  {http://arxiv.org/abs/1709.04933} {arXiv:1709.04933 [hep-ph]} \BibitemShut
  {NoStop}%
\bibitem [{\citenamefont {Izubuchi}\ \emph {et~al.}(2018)\citenamefont
  {Izubuchi}, \citenamefont {Ji}, \citenamefont {Jin}, \citenamefont
  {Stewart},\ and\ \citenamefont {Zhao}}]{Izubuchi:2018srq}%
  \BibitemOpen
  \bibfield  {author} {\bibinfo {author} {\bibfnamefont {T.}~\bibnamefont
  {Izubuchi}}, \bibinfo {author} {\bibfnamefont {X.}~\bibnamefont {Ji}},
  \bibinfo {author} {\bibfnamefont {L.}~\bibnamefont {Jin}}, \bibinfo {author}
  {\bibfnamefont {I.~W.}\ \bibnamefont {Stewart}}, \ and\ \bibinfo {author}
  {\bibfnamefont {Y.}~\bibnamefont {Zhao}},\ }\href {\doibase
  10.1103/PhysRevD.98.056004} {\bibfield  {journal} {\bibinfo  {journal} {Phys.
  Rev. D}\ }\textbf {\bibinfo {volume} {98}},\ \bibinfo {pages} {056004}
  (\bibinfo {year} {2018})},\ \Eprint {http://arxiv.org/abs/1801.03917}
  {arXiv:1801.03917 [hep-ph]} \BibitemShut {NoStop}%
\bibitem [{\citenamefont {Liu}\ \emph {et~al.}(2020)\citenamefont {Liu} \emph
  {et~al.}}]{Liu:2018uuj}%
  \BibitemOpen
  \bibfield  {author} {\bibinfo {author} {\bibfnamefont {Y.-S.}\ \bibnamefont
  {Liu}} \emph {et~al.} (\bibinfo {collaboration} {Lattice Parton}),\ }\href
  {\doibase 10.1103/PhysRevD.101.034020} {\bibfield  {journal} {\bibinfo
  {journal} {Phys. Rev. D}\ }\textbf {\bibinfo {volume} {101}},\ \bibinfo
  {pages} {034020} (\bibinfo {year} {2020})},\ \Eprint
  {http://arxiv.org/abs/1807.06566} {arXiv:1807.06566 [hep-lat]} \BibitemShut
  {NoStop}%
\bibitem [{\citenamefont {Radyushkin}(2018)}]{Radyushkin:2017lvu}%
  \BibitemOpen
  \bibfield  {author} {\bibinfo {author} {\bibfnamefont {A.}~\bibnamefont
  {Radyushkin}},\ }\href {\doibase 10.1016/j.physletb.2018.04.023} {\bibfield
  {journal} {\bibinfo  {journal} {Phys. Lett. B}\ }\textbf {\bibinfo {volume}
  {781}},\ \bibinfo {pages} {433} (\bibinfo {year} {2018})},\ \Eprint
  {http://arxiv.org/abs/1710.08813} {arXiv:1710.08813 [hep-ph]} \BibitemShut
  {NoStop}%
\bibitem [{\citenamefont {Zhang}\ \emph {et~al.}(2018)\citenamefont {Zhang},
  \citenamefont {Chen},\ and\ \citenamefont {Monahan}}]{Zhang:2018ggy}%
  \BibitemOpen
  \bibfield  {author} {\bibinfo {author} {\bibfnamefont {J.-H.}\ \bibnamefont
  {Zhang}}, \bibinfo {author} {\bibfnamefont {J.-W.}\ \bibnamefont {Chen}}, \
  and\ \bibinfo {author} {\bibfnamefont {C.}~\bibnamefont {Monahan}},\ }\href
  {\doibase 10.1103/PhysRevD.97.074508} {\bibfield  {journal} {\bibinfo
  {journal} {Phys. Rev. D}\ }\textbf {\bibinfo {volume} {97}},\ \bibinfo
  {pages} {074508} (\bibinfo {year} {2018})},\ \Eprint
  {http://arxiv.org/abs/1801.03023} {arXiv:1801.03023 [hep-ph]} \BibitemShut
  {NoStop}%
\bibitem [{\citenamefont {Chen}\ \emph
  {et~al.}(2020{\natexlab{a}})\citenamefont {Chen}, \citenamefont {Wang},\ and\
  \citenamefont {Zhu}}]{Chen:2020arf}%
  \BibitemOpen
  \bibfield  {author} {\bibinfo {author} {\bibfnamefont {L.-B.}\ \bibnamefont
  {Chen}}, \bibinfo {author} {\bibfnamefont {W.}~\bibnamefont {Wang}}, \ and\
  \bibinfo {author} {\bibfnamefont {R.}~\bibnamefont {Zhu}},\ }\href {\doibase
  10.1103/PhysRevD.102.011503} {\bibfield  {journal} {\bibinfo  {journal}
  {Phys. Rev. D}\ }\textbf {\bibinfo {volume} {102}},\ \bibinfo {pages}
  {011503} (\bibinfo {year} {2020}{\natexlab{a}})},\ \Eprint
  {http://arxiv.org/abs/2005.13757} {arXiv:2005.13757 [hep-ph]} \BibitemShut
  {NoStop}%
\bibitem [{\citenamefont {Chen}\ \emph
  {et~al.}(2020{\natexlab{b}})\citenamefont {Chen}, \citenamefont {Wang},\ and\
  \citenamefont {Zhu}}]{Chen:2020iqi}%
  \BibitemOpen
  \bibfield  {author} {\bibinfo {author} {\bibfnamefont {L.-B.}\ \bibnamefont
  {Chen}}, \bibinfo {author} {\bibfnamefont {W.}~\bibnamefont {Wang}}, \ and\
  \bibinfo {author} {\bibfnamefont {R.}~\bibnamefont {Zhu}},\ }\href@noop {} {\
   (\bibinfo {year} {2020}{\natexlab{b}})},\ \Eprint
  {http://arxiv.org/abs/2006.10917} {arXiv:2006.10917 [hep-ph]} \BibitemShut
  {NoStop}%
\bibitem [{\citenamefont {Li}\ \emph {et~al.}(2020)\citenamefont {Li},
  \citenamefont {Ma},\ and\ \citenamefont {Qiu}}]{Li:2020xml}%
  \BibitemOpen
  \bibfield  {author} {\bibinfo {author} {\bibfnamefont {Z.-Y.}\ \bibnamefont
  {Li}}, \bibinfo {author} {\bibfnamefont {Y.-Q.}\ \bibnamefont {Ma}}, \ and\
  \bibinfo {author} {\bibfnamefont {J.-W.}\ \bibnamefont {Qiu}},\ }\href@noop
  {} {\  (\bibinfo {year} {2020})},\ \Eprint {http://arxiv.org/abs/2006.12370}
  {arXiv:2006.12370 [hep-ph]} \BibitemShut {NoStop}%
\bibitem [{\citenamefont {Ma}\ and\ \citenamefont
  {Qiu}(2018{\natexlab{a}})}]{Ma:2014jla}%
  \BibitemOpen
  \bibfield  {author} {\bibinfo {author} {\bibfnamefont {Y.-Q.}\ \bibnamefont
  {Ma}}\ and\ \bibinfo {author} {\bibfnamefont {J.-W.}\ \bibnamefont {Qiu}},\
  }\href {\doibase 10.1103/PhysRevD.98.074021} {\bibfield  {journal} {\bibinfo
  {journal} {Phys. Rev. D}\ }\textbf {\bibinfo {volume} {98}},\ \bibinfo
  {pages} {074021} (\bibinfo {year} {2018}{\natexlab{a}})},\ \Eprint
  {http://arxiv.org/abs/1404.6860} {arXiv:1404.6860 [hep-ph]} \BibitemShut
  {NoStop}%
\bibitem [{\citenamefont {Ma}\ and\ \citenamefont
  {Qiu}(2018{\natexlab{b}})}]{Ma:2017pxb}%
  \BibitemOpen
  \bibfield  {author} {\bibinfo {author} {\bibfnamefont {Y.-Q.}\ \bibnamefont
  {Ma}}\ and\ \bibinfo {author} {\bibfnamefont {J.-W.}\ \bibnamefont {Qiu}},\
  }\href {\doibase 10.1103/PhysRevLett.120.022003} {\bibfield  {journal}
  {\bibinfo  {journal} {Phys. Rev. Lett.}\ }\textbf {\bibinfo {volume} {120}},\
  \bibinfo {pages} {022003} (\bibinfo {year} {2018}{\natexlab{b}})},\ \Eprint
  {http://arxiv.org/abs/1709.03018} {arXiv:1709.03018 [hep-ph]} \BibitemShut
  {NoStop}%
\bibitem [{\citenamefont {Zhao}(2019)}]{Zhao:2018fyu}%
  \BibitemOpen
  \bibfield  {author} {\bibinfo {author} {\bibfnamefont {Y.}~\bibnamefont
  {Zhao}},\ }\href {\doibase 10.1142/S0217751X18300338} {\bibfield  {journal}
  {\bibinfo  {journal} {Int. J. Mod. Phys. A}\ }\textbf {\bibinfo {volume}
  {33}},\ \bibinfo {pages} {1830033} (\bibinfo {year} {2019})},\ \Eprint
  {http://arxiv.org/abs/1812.07192} {arXiv:1812.07192 [hep-ph]} \BibitemShut
  {NoStop}%
\bibitem [{\citenamefont {Cichy}\ and\ \citenamefont
  {Constantinou}(2019)}]{Cichy:2018mum}%
  \BibitemOpen
  \bibfield  {author} {\bibinfo {author} {\bibfnamefont {K.}~\bibnamefont
  {Cichy}}\ and\ \bibinfo {author} {\bibfnamefont {M.}~\bibnamefont
  {Constantinou}},\ }\href {\doibase 10.1155/2019/3036904} {\bibfield
  {journal} {\bibinfo  {journal} {Adv. High Energy Phys.}\ }\textbf {\bibinfo
  {volume} {2019}},\ \bibinfo {pages} {3036904} (\bibinfo {year} {2019})},\
  \Eprint {http://arxiv.org/abs/1811.07248} {arXiv:1811.07248 [hep-lat]}
  \BibitemShut {NoStop}%
\bibitem [{\citenamefont {Monahan}(2018)}]{Monahan:2018euv}%
  \BibitemOpen
  \bibfield  {author} {\bibinfo {author} {\bibfnamefont {C.}~\bibnamefont
  {Monahan}},\ }\href {\doibase 10.22323/1.334.0018} {\bibfield  {journal}
  {\bibinfo  {journal} {PoS}\ }\textbf {\bibinfo {volume} {LATTICE2018}},\
  \bibinfo {pages} {018} (\bibinfo {year} {2018})},\ \Eprint
  {http://arxiv.org/abs/1811.00678} {arXiv:1811.00678 [hep-lat]} \BibitemShut
  {NoStop}%
\bibitem [{\citenamefont {Liu}\ and\ \citenamefont {Dong}(1994)}]{Liu:1993cv}%
  \BibitemOpen
  \bibfield  {author} {\bibinfo {author} {\bibfnamefont {K.-F.}\ \bibnamefont
  {Liu}}\ and\ \bibinfo {author} {\bibfnamefont {S.-J.}\ \bibnamefont {Dong}},\
  }\href {\doibase 10.1103/PhysRevLett.72.1790} {\bibfield  {journal} {\bibinfo
   {journal} {Phys. Rev. Lett.}\ }\textbf {\bibinfo {volume} {72}},\ \bibinfo
  {pages} {1790} (\bibinfo {year} {1994})},\ \Eprint
  {http://arxiv.org/abs/hep-ph/9306299} {arXiv:hep-ph/9306299} \BibitemShut
  {NoStop}%
\bibitem [{\citenamefont {Detmold}\ and\ \citenamefont
  {Lin}(2006)}]{Detmold:2005gg}%
  \BibitemOpen
  \bibfield  {author} {\bibinfo {author} {\bibfnamefont {W.}~\bibnamefont
  {Detmold}}\ and\ \bibinfo {author} {\bibfnamefont {C.}~\bibnamefont {Lin}},\
  }\href {\doibase 10.1103/PhysRevD.73.014501} {\bibfield  {journal} {\bibinfo
  {journal} {Phys. Rev. D}\ }\textbf {\bibinfo {volume} {73}},\ \bibinfo
  {pages} {014501} (\bibinfo {year} {2006})},\ \Eprint
  {http://arxiv.org/abs/hep-lat/0507007} {arXiv:hep-lat/0507007} \BibitemShut
  {NoStop}%
\bibitem [{\citenamefont {Nguyen}\ \emph {et~al.}(2011)\citenamefont {Nguyen},
  \citenamefont {Bashir}, \citenamefont {Roberts},\ and\ \citenamefont
  {Tandy}}]{Nguyen:2011jy}%
  \BibitemOpen
  \bibfield  {author} {\bibinfo {author} {\bibfnamefont {T.}~\bibnamefont
  {Nguyen}}, \bibinfo {author} {\bibfnamefont {A.}~\bibnamefont {Bashir}},
  \bibinfo {author} {\bibfnamefont {C.~D.}\ \bibnamefont {Roberts}}, \ and\
  \bibinfo {author} {\bibfnamefont {P.~C.}\ \bibnamefont {Tandy}},\ }\href
  {\doibase 10.1103/PhysRevC.83.062201} {\bibfield  {journal} {\bibinfo
  {journal} {Phys. Rev. C}\ }\textbf {\bibinfo {volume} {83}},\ \bibinfo
  {pages} {062201} (\bibinfo {year} {2011})},\ \Eprint
  {http://arxiv.org/abs/1102.2448} {arXiv:1102.2448 [nucl-th]} \BibitemShut
  {NoStop}%
\bibitem [{\citenamefont {Chen}\ \emph
  {et~al.}(2016{\natexlab{a}})\citenamefont {Chen}, \citenamefont {Chang},
  \citenamefont {Roberts}, \citenamefont {Wan},\ and\ \citenamefont
  {Zong}}]{Chen:2016sno}%
  \BibitemOpen
  \bibfield  {author} {\bibinfo {author} {\bibfnamefont {C.}~\bibnamefont
  {Chen}}, \bibinfo {author} {\bibfnamefont {L.}~\bibnamefont {Chang}},
  \bibinfo {author} {\bibfnamefont {C.~D.}\ \bibnamefont {Roberts}}, \bibinfo
  {author} {\bibfnamefont {S.}~\bibnamefont {Wan}}, \ and\ \bibinfo {author}
  {\bibfnamefont {H.-S.}\ \bibnamefont {Zong}},\ }\href {\doibase
  10.1103/PhysRevD.93.074021} {\bibfield  {journal} {\bibinfo  {journal} {Phys.
  Rev. D}\ }\textbf {\bibinfo {volume} {93}},\ \bibinfo {pages} {074021}
  (\bibinfo {year} {2016}{\natexlab{a}})},\ \Eprint
  {http://arxiv.org/abs/1602.01502} {arXiv:1602.01502 [nucl-th]} \BibitemShut
  {NoStop}%
\bibitem [{\citenamefont {Bednar}\ \emph {et~al.}(2020)\citenamefont {Bednar},
  \citenamefont {Clo\"et},\ and\ \citenamefont {Tandy}}]{Bednar:2018mtf}%
  \BibitemOpen
  \bibfield  {author} {\bibinfo {author} {\bibfnamefont {K.~D.}\ \bibnamefont
  {Bednar}}, \bibinfo {author} {\bibfnamefont {I.~C.}\ \bibnamefont {Clo\"et}},
  \ and\ \bibinfo {author} {\bibfnamefont {P.~C.}\ \bibnamefont {Tandy}},\
  }\href {\doibase 10.1103/PhysRevLett.124.042002} {\bibfield  {journal}
  {\bibinfo  {journal} {Phys. Rev. Lett.}\ }\textbf {\bibinfo {volume} {124}},\
  \bibinfo {pages} {042002} (\bibinfo {year} {2020})},\ \Eprint
  {http://arxiv.org/abs/1811.12310} {arXiv:1811.12310 [nucl-th]} \BibitemShut
  {NoStop}%
\bibitem [{\citenamefont {Aicher}\ \emph {et~al.}(2010)\citenamefont {Aicher},
  \citenamefont {Schafer},\ and\ \citenamefont {Vogelsang}}]{Aicher:2010cb}%
  \BibitemOpen
  \bibfield  {author} {\bibinfo {author} {\bibfnamefont {M.}~\bibnamefont
  {Aicher}}, \bibinfo {author} {\bibfnamefont {A.}~\bibnamefont {Schafer}}, \
  and\ \bibinfo {author} {\bibfnamefont {W.}~\bibnamefont {Vogelsang}},\ }\href
  {\doibase 10.1103/PhysRevLett.105.252003} {\bibfield  {journal} {\bibinfo
  {journal} {Phys. Rev. Lett.}\ }\textbf {\bibinfo {volume} {105}},\ \bibinfo
  {pages} {252003} (\bibinfo {year} {2010})},\ \Eprint
  {http://arxiv.org/abs/1009.2481} {arXiv:1009.2481 [hep-ph]} \BibitemShut
  {NoStop}%
\bibitem [{\citenamefont {Ruiz~Arriola}(2002)}]{RuizArriola:2002wr}%
  \BibitemOpen
  \bibfield  {author} {\bibinfo {author} {\bibfnamefont {E.}~\bibnamefont
  {Ruiz~Arriola}},\ }\href@noop {} {\bibfield  {journal} {\bibinfo  {journal}
  {Acta Phys. Polon. B}\ }\textbf {\bibinfo {volume} {33}},\ \bibinfo {pages}
  {4443} (\bibinfo {year} {2002})},\ \Eprint
  {http://arxiv.org/abs/hep-ph/0210007} {arXiv:hep-ph/0210007} \BibitemShut
  {NoStop}%
\bibitem [{\citenamefont {Broniowski}\ and\ \citenamefont
  {Ruiz~Arriola}(2017)}]{Broniowski:2017wbr}%
  \BibitemOpen
  \bibfield  {author} {\bibinfo {author} {\bibfnamefont {W.}~\bibnamefont
  {Broniowski}}\ and\ \bibinfo {author} {\bibfnamefont {E.}~\bibnamefont
  {Ruiz~Arriola}},\ }\href {\doibase 10.1016/j.physletb.2017.08.055} {\bibfield
   {journal} {\bibinfo  {journal} {Phys. Lett. B}\ }\textbf {\bibinfo {volume}
  {773}},\ \bibinfo {pages} {385} (\bibinfo {year} {2017})},\ \Eprint
  {http://arxiv.org/abs/1707.09588} {arXiv:1707.09588 [hep-ph]} \BibitemShut
  {NoStop}%
\bibitem [{\citenamefont {de~Teramond}\ \emph {et~al.}(2018)\citenamefont
  {de~Teramond}, \citenamefont {Liu}, \citenamefont {Sufian}, \citenamefont
  {Dosch}, \citenamefont {Brodsky},\ and\ \citenamefont
  {Deur}}]{deTeramond:2018ecg}%
  \BibitemOpen
  \bibfield  {author} {\bibinfo {author} {\bibfnamefont {G.~F.}\ \bibnamefont
  {de~Teramond}}, \bibinfo {author} {\bibfnamefont {T.}~\bibnamefont {Liu}},
  \bibinfo {author} {\bibfnamefont {R.~S.}\ \bibnamefont {Sufian}}, \bibinfo
  {author} {\bibfnamefont {H.~G.}\ \bibnamefont {Dosch}}, \bibinfo {author}
  {\bibfnamefont {S.~J.}\ \bibnamefont {Brodsky}}, \ and\ \bibinfo {author}
  {\bibfnamefont {A.}~\bibnamefont {Deur}} (\bibinfo {collaboration} {HLFHS}),\
  }\href {\doibase 10.1103/PhysRevLett.120.182001} {\bibfield  {journal}
  {\bibinfo  {journal} {Phys. Rev. Lett.}\ }\textbf {\bibinfo {volume} {120}},\
  \bibinfo {pages} {182001} (\bibinfo {year} {2018})},\ \Eprint
  {http://arxiv.org/abs/1801.09154} {arXiv:1801.09154 [hep-ph]} \BibitemShut
  {NoStop}%
\bibitem [{\citenamefont {Ding}\ \emph {et~al.}(2020)\citenamefont {Ding},
  \citenamefont {Raya}, \citenamefont {Binosi}, \citenamefont {Chang},
  \citenamefont {Roberts},\ and\ \citenamefont {Schmidt}}]{Ding:2019qlr}%
  \BibitemOpen
  \bibfield  {author} {\bibinfo {author} {\bibfnamefont {M.}~\bibnamefont
  {Ding}}, \bibinfo {author} {\bibfnamefont {K.}~\bibnamefont {Raya}}, \bibinfo
  {author} {\bibfnamefont {D.}~\bibnamefont {Binosi}}, \bibinfo {author}
  {\bibfnamefont {L.}~\bibnamefont {Chang}}, \bibinfo {author} {\bibfnamefont
  {C.~D.}\ \bibnamefont {Roberts}}, \ and\ \bibinfo {author} {\bibfnamefont
  {S.~M.}\ \bibnamefont {Schmidt}},\ }\href {\doibase
  10.1088/1674-1137/44/3/031002} {\bibfield  {journal} {\bibinfo  {journal}
  {Chin. Phys. C}\ }\textbf {\bibinfo {volume} {44}},\ \bibinfo {pages}
  {031002} (\bibinfo {year} {2020})},\ \Eprint
  {http://arxiv.org/abs/1912.07529} {arXiv:1912.07529 [hep-ph]} \BibitemShut
  {NoStop}%
\bibitem [{\citenamefont {Sufian}\ \emph {et~al.}(2019)\citenamefont {Sufian},
  \citenamefont {Karpie}, \citenamefont {Egerer}, \citenamefont {Orginos},
  \citenamefont {Qiu},\ and\ \citenamefont {Richards}}]{Sufian:2019bol}%
  \BibitemOpen
  \bibfield  {author} {\bibinfo {author} {\bibfnamefont {R.~S.}\ \bibnamefont
  {Sufian}}, \bibinfo {author} {\bibfnamefont {J.}~\bibnamefont {Karpie}},
  \bibinfo {author} {\bibfnamefont {C.}~\bibnamefont {Egerer}}, \bibinfo
  {author} {\bibfnamefont {K.}~\bibnamefont {Orginos}}, \bibinfo {author}
  {\bibfnamefont {J.-W.}\ \bibnamefont {Qiu}}, \ and\ \bibinfo {author}
  {\bibfnamefont {D.~G.}\ \bibnamefont {Richards}},\ }\href {\doibase
  10.1103/PhysRevD.99.074507} {\bibfield  {journal} {\bibinfo  {journal} {Phys.
  Rev. D}\ }\textbf {\bibinfo {volume} {99}},\ \bibinfo {pages} {074507}
  (\bibinfo {year} {2019})},\ \Eprint {http://arxiv.org/abs/1901.03921}
  {arXiv:1901.03921 [hep-lat]} \BibitemShut {NoStop}%
\bibitem [{\citenamefont {Sufian}\ \emph {et~al.}(2020)\citenamefont {Sufian},
  \citenamefont {Egerer}, \citenamefont {Karpie}, \citenamefont {Edwards},
  \citenamefont {Jo\'o}, \citenamefont {Ma}, \citenamefont {Orginos},
  \citenamefont {Qiu},\ and\ \citenamefont {Richards}}]{Sufian:2020vzb}%
  \BibitemOpen
  \bibfield  {author} {\bibinfo {author} {\bibfnamefont {R.~S.}\ \bibnamefont
  {Sufian}}, \bibinfo {author} {\bibfnamefont {C.}~\bibnamefont {Egerer}},
  \bibinfo {author} {\bibfnamefont {J.}~\bibnamefont {Karpie}}, \bibinfo
  {author} {\bibfnamefont {R.~G.}\ \bibnamefont {Edwards}}, \bibinfo {author}
  {\bibfnamefont {B.}~\bibnamefont {Jo\'o}}, \bibinfo {author} {\bibfnamefont
  {Y.-Q.}\ \bibnamefont {Ma}}, \bibinfo {author} {\bibfnamefont
  {K.}~\bibnamefont {Orginos}}, \bibinfo {author} {\bibfnamefont {J.-W.}\
  \bibnamefont {Qiu}}, \ and\ \bibinfo {author} {\bibfnamefont {D.~G.}\
  \bibnamefont {Richards}},\ }\href {\doibase 10.1103/PhysRevD.102.054508}
  {\bibfield  {journal} {\bibinfo  {journal} {Phys. Rev. D}\ }\textbf {\bibinfo
  {volume} {102}},\ \bibinfo {pages} {054508} (\bibinfo {year} {2020})},\
  \Eprint {http://arxiv.org/abs/2001.04960} {arXiv:2001.04960 [hep-lat]}
  \BibitemShut {NoStop}%
\bibitem [{\citenamefont {Izubuchi}\ \emph {et~al.}(2019)\citenamefont
  {Izubuchi}, \citenamefont {Jin}, \citenamefont {Kallidonis}, \citenamefont
  {Karthik}, \citenamefont {Mukherjee}, \citenamefont {Petreczky},
  \citenamefont {Shugert},\ and\ \citenamefont {Syritsyn}}]{Izubuchi:2019lyk}%
  \BibitemOpen
  \bibfield  {author} {\bibinfo {author} {\bibfnamefont {T.}~\bibnamefont
  {Izubuchi}}, \bibinfo {author} {\bibfnamefont {L.}~\bibnamefont {Jin}},
  \bibinfo {author} {\bibfnamefont {C.}~\bibnamefont {Kallidonis}}, \bibinfo
  {author} {\bibfnamefont {N.}~\bibnamefont {Karthik}}, \bibinfo {author}
  {\bibfnamefont {S.}~\bibnamefont {Mukherjee}}, \bibinfo {author}
  {\bibfnamefont {P.}~\bibnamefont {Petreczky}}, \bibinfo {author}
  {\bibfnamefont {C.}~\bibnamefont {Shugert}}, \ and\ \bibinfo {author}
  {\bibfnamefont {S.}~\bibnamefont {Syritsyn}},\ }\href {\doibase
  10.1103/PhysRevD.100.034516} {\bibfield  {journal} {\bibinfo  {journal}
  {Phys. Rev. D}\ }\textbf {\bibinfo {volume} {100}},\ \bibinfo {pages}
  {034516} (\bibinfo {year} {2019})},\ \Eprint
  {http://arxiv.org/abs/1905.06349} {arXiv:1905.06349 [hep-lat]} \BibitemShut
  {NoStop}%
\bibitem [{\citenamefont {Badier}\ \emph {et~al.}(1983)\citenamefont {Badier}
  \emph {et~al.}}]{Badier:1983mj}%
  \BibitemOpen
  \bibfield  {author} {\bibinfo {author} {\bibfnamefont {J.}~\bibnamefont
  {Badier}} \emph {et~al.} (\bibinfo {collaboration} {NA3}),\ }\href {\doibase
  10.1007/BF01573728} {\bibfield  {journal} {\bibinfo  {journal} {Z. Phys. C}\
  }\textbf {\bibinfo {volume} {18}},\ \bibinfo {pages} {281} (\bibinfo {year}
  {1983})}\BibitemShut {NoStop}%
\bibitem [{\citenamefont {Betev}\ \emph {et~al.}(1985)\citenamefont {Betev}
  \emph {et~al.}}]{Betev:1985pf}%
  \BibitemOpen
  \bibfield  {author} {\bibinfo {author} {\bibfnamefont {B.}~\bibnamefont
  {Betev}} \emph {et~al.} (\bibinfo {collaboration} {NA10}),\ }\href {\doibase
  10.1007/BF01550243} {\bibfield  {journal} {\bibinfo  {journal} {Z. Phys. C}\
  }\textbf {\bibinfo {volume} {28}},\ \bibinfo {pages} {9} (\bibinfo {year}
  {1985})}\BibitemShut {NoStop}%
\bibitem [{\citenamefont {Conway}\ \emph {et~al.}(1989)\citenamefont {Conway}
  \emph {et~al.}}]{Conway:1989fs}%
  \BibitemOpen
  \bibfield  {author} {\bibinfo {author} {\bibfnamefont {J.}~\bibnamefont
  {Conway}} \emph {et~al.},\ }\href {\doibase 10.1103/PhysRevD.39.92}
  {\bibfield  {journal} {\bibinfo  {journal} {Phys. Rev. D}\ }\textbf {\bibinfo
  {volume} {39}},\ \bibinfo {pages} {92} (\bibinfo {year} {1989})}\BibitemShut
  {NoStop}%
\bibitem [{\citenamefont {Owens}(1984)}]{Owens:1984zj}%
  \BibitemOpen
  \bibfield  {author} {\bibinfo {author} {\bibfnamefont {J.}~\bibnamefont
  {Owens}},\ }\href {\doibase 10.1103/PhysRevD.30.943} {\bibfield  {journal}
  {\bibinfo  {journal} {Phys. Rev. D}\ }\textbf {\bibinfo {volume} {30}},\
  \bibinfo {pages} {943} (\bibinfo {year} {1984})}\BibitemShut {NoStop}%
\bibitem [{\citenamefont {Sutton}\ \emph {et~al.}(1992)\citenamefont {Sutton},
  \citenamefont {Martin}, \citenamefont {Roberts},\ and\ \citenamefont
  {Stirling}}]{Sutton:1991ay}%
  \BibitemOpen
  \bibfield  {author} {\bibinfo {author} {\bibfnamefont {P.}~\bibnamefont
  {Sutton}}, \bibinfo {author} {\bibfnamefont {A.~D.}\ \bibnamefont {Martin}},
  \bibinfo {author} {\bibfnamefont {R.}~\bibnamefont {Roberts}}, \ and\
  \bibinfo {author} {\bibfnamefont {W.}~\bibnamefont {Stirling}},\ }\href
  {\doibase 10.1103/PhysRevD.45.2349} {\bibfield  {journal} {\bibinfo
  {journal} {Phys. Rev. D}\ }\textbf {\bibinfo {volume} {45}},\ \bibinfo
  {pages} {2349} (\bibinfo {year} {1992})}\BibitemShut {NoStop}%
\bibitem [{\citenamefont {Gluck}\ \emph {et~al.}(1992)\citenamefont {Gluck},
  \citenamefont {Reya},\ and\ \citenamefont {Vogt}}]{Gluck:1991ey}%
  \BibitemOpen
  \bibfield  {author} {\bibinfo {author} {\bibfnamefont {M.}~\bibnamefont
  {Gluck}}, \bibinfo {author} {\bibfnamefont {E.}~\bibnamefont {Reya}}, \ and\
  \bibinfo {author} {\bibfnamefont {A.}~\bibnamefont {Vogt}},\ }\href {\doibase
  10.1007/BF01559743} {\bibfield  {journal} {\bibinfo  {journal} {Z. Phys. C}\
  }\textbf {\bibinfo {volume} {53}},\ \bibinfo {pages} {651} (\bibinfo {year}
  {1992})}\BibitemShut {NoStop}%
\bibitem [{\citenamefont {Gluck}\ \emph {et~al.}(1999)\citenamefont {Gluck},
  \citenamefont {Reya},\ and\ \citenamefont {Schienbein}}]{Gluck:1999xe}%
  \BibitemOpen
  \bibfield  {author} {\bibinfo {author} {\bibfnamefont {M.}~\bibnamefont
  {Gluck}}, \bibinfo {author} {\bibfnamefont {E.}~\bibnamefont {Reya}}, \ and\
  \bibinfo {author} {\bibfnamefont {I.}~\bibnamefont {Schienbein}},\ }\href
  {\doibase 10.1007/s100529900124} {\bibfield  {journal} {\bibinfo  {journal}
  {Eur. Phys. J. C}\ }\textbf {\bibinfo {volume} {10}},\ \bibinfo {pages} {313}
  (\bibinfo {year} {1999})},\ \Eprint {http://arxiv.org/abs/hep-ph/9903288}
  {arXiv:hep-ph/9903288} \BibitemShut {NoStop}%
\bibitem [{\citenamefont {Wijesooriya}\ \emph {et~al.}(2005)\citenamefont
  {Wijesooriya}, \citenamefont {Reimer},\ and\ \citenamefont
  {Holt}}]{Wijesooriya:2005ir}%
  \BibitemOpen
  \bibfield  {author} {\bibinfo {author} {\bibfnamefont {K.}~\bibnamefont
  {Wijesooriya}}, \bibinfo {author} {\bibfnamefont {P.}~\bibnamefont {Reimer}},
  \ and\ \bibinfo {author} {\bibfnamefont {R.}~\bibnamefont {Holt}},\ }\href
  {\doibase 10.1103/PhysRevC.72.065203} {\bibfield  {journal} {\bibinfo
  {journal} {Phys. Rev. C}\ }\textbf {\bibinfo {volume} {72}},\ \bibinfo
  {pages} {065203} (\bibinfo {year} {2005})},\ \Eprint
  {http://arxiv.org/abs/nucl-ex/0509012} {arXiv:nucl-ex/0509012} \BibitemShut
  {NoStop}%
\bibitem [{\citenamefont {Zhang}\ \emph {et~al.}(2019)\citenamefont {Zhang},
  \citenamefont {Chen}, \citenamefont {Jin}, \citenamefont {Lin}, \citenamefont
  {Sch\"afer},\ and\ \citenamefont {Zhao}}]{Chen:2018fwa}%
  \BibitemOpen
  \bibfield  {author} {\bibinfo {author} {\bibfnamefont {J.-H.}\ \bibnamefont
  {Zhang}}, \bibinfo {author} {\bibfnamefont {J.-W.}\ \bibnamefont {Chen}},
  \bibinfo {author} {\bibfnamefont {L.}~\bibnamefont {Jin}}, \bibinfo {author}
  {\bibfnamefont {H.-W.}\ \bibnamefont {Lin}}, \bibinfo {author} {\bibfnamefont
  {A.}~\bibnamefont {Sch\"afer}}, \ and\ \bibinfo {author} {\bibfnamefont
  {Y.}~\bibnamefont {Zhao}},\ }\href {\doibase 10.1103/PhysRevD.100.034505}
  {\bibfield  {journal} {\bibinfo  {journal} {Phys. Rev. D}\ }\textbf {\bibinfo
  {volume} {100}},\ \bibinfo {pages} {034505} (\bibinfo {year} {2019})},\
  \Eprint {http://arxiv.org/abs/1804.01483} {arXiv:1804.01483 [hep-lat]}
  \BibitemShut {NoStop}%
\bibitem [{\citenamefont {Jo\'o}\ \emph
  {et~al.}(2019{\natexlab{a}})\citenamefont {Jo\'o}, \citenamefont {Karpie},
  \citenamefont {Orginos}, \citenamefont {Radyushkin}, \citenamefont
  {Richards}, \citenamefont {Sufian},\ and\ \citenamefont
  {Zafeiropoulos}}]{Joo:2019bzr}%
  \BibitemOpen
  \bibfield  {author} {\bibinfo {author} {\bibfnamefont {B.}~\bibnamefont
  {Jo\'o}}, \bibinfo {author} {\bibfnamefont {J.}~\bibnamefont {Karpie}},
  \bibinfo {author} {\bibfnamefont {K.}~\bibnamefont {Orginos}}, \bibinfo
  {author} {\bibfnamefont {A.~V.}\ \bibnamefont {Radyushkin}}, \bibinfo
  {author} {\bibfnamefont {D.~G.}\ \bibnamefont {Richards}}, \bibinfo {author}
  {\bibfnamefont {R.~S.}\ \bibnamefont {Sufian}}, \ and\ \bibinfo {author}
  {\bibfnamefont {S.}~\bibnamefont {Zafeiropoulos}},\ }\href {\doibase
  10.1103/PhysRevD.100.114512} {\bibfield  {journal} {\bibinfo  {journal}
  {Phys. Rev. D}\ }\textbf {\bibinfo {volume} {100}},\ \bibinfo {pages}
  {114512} (\bibinfo {year} {2019}{\natexlab{a}})},\ \Eprint
  {http://arxiv.org/abs/1909.08517} {arXiv:1909.08517 [hep-lat]} \BibitemShut
  {NoStop}%
\bibitem [{\citenamefont {Lin}\ \emph {et~al.}(2020)\citenamefont {Lin},
  \citenamefont {Chen}, \citenamefont {Fan}, \citenamefont {Zhang},\ and\
  \citenamefont {Zhang}}]{Lin:2020ssv}%
  \BibitemOpen
  \bibfield  {author} {\bibinfo {author} {\bibfnamefont {H.-W.}\ \bibnamefont
  {Lin}}, \bibinfo {author} {\bibfnamefont {J.-W.}\ \bibnamefont {Chen}},
  \bibinfo {author} {\bibfnamefont {Z.}~\bibnamefont {Fan}}, \bibinfo {author}
  {\bibfnamefont {J.-H.}\ \bibnamefont {Zhang}}, \ and\ \bibinfo {author}
  {\bibfnamefont {R.}~\bibnamefont {Zhang}},\ }\href@noop {} {\  (\bibinfo
  {year} {2020})},\ \Eprint {http://arxiv.org/abs/2003.14128} {arXiv:2003.14128
  [hep-lat]} \BibitemShut {NoStop}%
\bibitem [{\citenamefont {Chen}\ \emph {et~al.}(2018)\citenamefont {Chen},
  \citenamefont {Ishikawa}, \citenamefont {Jin}, \citenamefont {Lin},
  \citenamefont {Yang}, \citenamefont {Zhang},\ and\ \citenamefont
  {Zhao}}]{Chen:2017mzz}%
  \BibitemOpen
  \bibfield  {author} {\bibinfo {author} {\bibfnamefont {J.-W.}\ \bibnamefont
  {Chen}}, \bibinfo {author} {\bibfnamefont {T.}~\bibnamefont {Ishikawa}},
  \bibinfo {author} {\bibfnamefont {L.}~\bibnamefont {Jin}}, \bibinfo {author}
  {\bibfnamefont {H.-W.}\ \bibnamefont {Lin}}, \bibinfo {author} {\bibfnamefont
  {Y.-B.}\ \bibnamefont {Yang}}, \bibinfo {author} {\bibfnamefont {J.-H.}\
  \bibnamefont {Zhang}}, \ and\ \bibinfo {author} {\bibfnamefont
  {Y.}~\bibnamefont {Zhao}},\ }\href {\doibase 10.1103/PhysRevD.97.014505}
  {\bibfield  {journal} {\bibinfo  {journal} {Phys. Rev. D}\ }\textbf {\bibinfo
  {volume} {97}},\ \bibinfo {pages} {014505} (\bibinfo {year} {2018})},\
  \Eprint {http://arxiv.org/abs/1706.01295} {arXiv:1706.01295 [hep-lat]}
  \BibitemShut {NoStop}%
\bibitem [{\citenamefont {Bazavov}\ \emph {et~al.}(2014)\citenamefont {Bazavov}
  \emph {et~al.}}]{Bazavov:2014pvz}%
  \BibitemOpen
  \bibfield  {author} {\bibinfo {author} {\bibfnamefont {A.}~\bibnamefont
  {Bazavov}} \emph {et~al.} (\bibinfo {collaboration} {HotQCD}),\ }\href
  {\doibase 10.1103/PhysRevD.90.094503} {\bibfield  {journal} {\bibinfo
  {journal} {Phys. Rev. D}\ }\textbf {\bibinfo {volume} {90}},\ \bibinfo
  {pages} {094503} (\bibinfo {year} {2014})},\ \Eprint
  {http://arxiv.org/abs/1407.6387} {arXiv:1407.6387 [hep-lat]} \BibitemShut
  {NoStop}%
\bibitem [{\citenamefont {Follana}\ \emph {et~al.}(2007)\citenamefont
  {Follana}, \citenamefont {Mason}, \citenamefont {Davies}, \citenamefont
  {Hornbostel}, \citenamefont {Lepage}, \citenamefont {Shigemitsu},
  \citenamefont {Trottier},\ and\ \citenamefont {Wong}}]{Follana:2006rc}%
  \BibitemOpen
  \bibfield  {author} {\bibinfo {author} {\bibfnamefont {E.}~\bibnamefont
  {Follana}}, \bibinfo {author} {\bibfnamefont {Q.}~\bibnamefont {Mason}},
  \bibinfo {author} {\bibfnamefont {C.}~\bibnamefont {Davies}}, \bibinfo
  {author} {\bibfnamefont {K.}~\bibnamefont {Hornbostel}}, \bibinfo {author}
  {\bibfnamefont {G.}~\bibnamefont {Lepage}}, \bibinfo {author} {\bibfnamefont
  {J.}~\bibnamefont {Shigemitsu}}, \bibinfo {author} {\bibfnamefont
  {H.}~\bibnamefont {Trottier}}, \ and\ \bibinfo {author} {\bibfnamefont
  {K.}~\bibnamefont {Wong}} (\bibinfo {collaboration} {HPQCD, UKQCD}),\ }\href
  {\doibase 10.1103/PhysRevD.75.054502} {\bibfield  {journal} {\bibinfo
  {journal} {Phys. Rev. D}\ }\textbf {\bibinfo {volume} {75}},\ \bibinfo
  {pages} {054502} (\bibinfo {year} {2007})},\ \Eprint
  {http://arxiv.org/abs/hep-lat/0610092} {arXiv:hep-lat/0610092} \BibitemShut
  {NoStop}%
\bibitem [{\citenamefont {Hasenfratz}\ and\ \citenamefont
  {Knechtli}(2001)}]{Hasenfratz:2001hp}%
  \BibitemOpen
  \bibfield  {author} {\bibinfo {author} {\bibfnamefont {A.}~\bibnamefont
  {Hasenfratz}}\ and\ \bibinfo {author} {\bibfnamefont {F.}~\bibnamefont
  {Knechtli}},\ }\href {\doibase 10.1103/PhysRevD.64.034504} {\bibfield
  {journal} {\bibinfo  {journal} {Phys. Rev. D}\ }\textbf {\bibinfo {volume}
  {64}},\ \bibinfo {pages} {034504} (\bibinfo {year} {2001})},\ \Eprint
  {http://arxiv.org/abs/hep-lat/0103029} {arXiv:hep-lat/0103029} \BibitemShut
  {NoStop}%
\bibitem [{\citenamefont {Bali}\ \emph {et~al.}(2016)\citenamefont {Bali},
  \citenamefont {Lang}, \citenamefont {Musch},\ and\ \citenamefont
  {Sch\"afer}}]{Bali:2016lva}%
  \BibitemOpen
  \bibfield  {author} {\bibinfo {author} {\bibfnamefont {G.~S.}\ \bibnamefont
  {Bali}}, \bibinfo {author} {\bibfnamefont {B.}~\bibnamefont {Lang}}, \bibinfo
  {author} {\bibfnamefont {B.~U.}\ \bibnamefont {Musch}}, \ and\ \bibinfo
  {author} {\bibfnamefont {A.}~\bibnamefont {Sch\"afer}},\ }\href {\doibase
  10.1103/PhysRevD.93.094515} {\bibfield  {journal} {\bibinfo  {journal} {Phys.
  Rev. D}\ }\textbf {\bibinfo {volume} {93}},\ \bibinfo {pages} {094515}
  (\bibinfo {year} {2016})},\ \Eprint {http://arxiv.org/abs/1602.05525}
  {arXiv:1602.05525 [hep-lat]} \BibitemShut {NoStop}%
\bibitem [{\citenamefont {Brannick}\ \emph {et~al.}(2008)\citenamefont
  {Brannick}, \citenamefont {Brower}, \citenamefont {Clark}, \citenamefont
  {Osborn},\ and\ \citenamefont {Rebbi}}]{Brannick:2007ue}%
  \BibitemOpen
  \bibfield  {author} {\bibinfo {author} {\bibfnamefont {J.}~\bibnamefont
  {Brannick}}, \bibinfo {author} {\bibfnamefont {R.}~\bibnamefont {Brower}},
  \bibinfo {author} {\bibfnamefont {M.}~\bibnamefont {Clark}}, \bibinfo
  {author} {\bibfnamefont {J.}~\bibnamefont {Osborn}}, \ and\ \bibinfo {author}
  {\bibfnamefont {C.}~\bibnamefont {Rebbi}},\ }\href {\doibase
  10.1103/PhysRevLett.100.041601} {\bibfield  {journal} {\bibinfo  {journal}
  {Phys. Rev. Lett.}\ }\textbf {\bibinfo {volume} {100}},\ \bibinfo {pages}
  {041601} (\bibinfo {year} {2008})},\ \Eprint {http://arxiv.org/abs/0707.4018}
  {arXiv:0707.4018 [hep-lat]} \BibitemShut {NoStop}%
\bibitem [{\citenamefont {Clark}\ \emph {et~al.}(2010)\citenamefont {Clark},
  \citenamefont {Babich}, \citenamefont {Barros}, \citenamefont {Brower},\ and\
  \citenamefont {Rebbi}}]{Clark:2009wm}%
  \BibitemOpen
  \bibfield  {author} {\bibinfo {author} {\bibfnamefont {M.}~\bibnamefont
  {Clark}}, \bibinfo {author} {\bibfnamefont {R.}~\bibnamefont {Babich}},
  \bibinfo {author} {\bibfnamefont {K.}~\bibnamefont {Barros}}, \bibinfo
  {author} {\bibfnamefont {R.}~\bibnamefont {Brower}}, \ and\ \bibinfo {author}
  {\bibfnamefont {C.}~\bibnamefont {Rebbi}},\ }\href {\doibase
  10.1016/j.cpc.2010.05.002} {\bibfield  {journal} {\bibinfo  {journal}
  {Comput. Phys. Commun.}\ }\textbf {\bibinfo {volume} {181}},\ \bibinfo
  {pages} {1517} (\bibinfo {year} {2010})},\ \Eprint
  {http://arxiv.org/abs/0911.3191} {arXiv:0911.3191 [hep-lat]} \BibitemShut
  {NoStop}%
\bibitem [{\citenamefont {Babich}\ \emph {et~al.}(2011)\citenamefont {Babich},
  \citenamefont {Clark}, \citenamefont {Joo}, \citenamefont {Shi},
  \citenamefont {Brower},\ and\ \citenamefont {Gottlieb}}]{Babich:2011np}%
  \BibitemOpen
  \bibfield  {author} {\bibinfo {author} {\bibfnamefont {R.}~\bibnamefont
  {Babich}}, \bibinfo {author} {\bibfnamefont {M.}~\bibnamefont {Clark}},
  \bibinfo {author} {\bibfnamefont {B.}~\bibnamefont {Joo}}, \bibinfo {author}
  {\bibfnamefont {G.}~\bibnamefont {Shi}}, \bibinfo {author} {\bibfnamefont
  {R.}~\bibnamefont {Brower}}, \ and\ \bibinfo {author} {\bibfnamefont
  {S.}~\bibnamefont {Gottlieb}},\ }in\ \href {\doibase 10.1145/2063384.2063478}
  {\emph {\bibinfo {booktitle} {{SC11 International Conference for High
  Performance Computing, Networking, Storage and Analysis}}}}\ (\bibinfo {year}
  {2011})\ \Eprint {http://arxiv.org/abs/1109.2935} {arXiv:1109.2935 [hep-lat]}
  \BibitemShut {NoStop}%
\bibitem [{\citenamefont {Clark}\ \emph {et~al.}(2016)\citenamefont {Clark},
  \citenamefont {Jo\'o}, \citenamefont {Strelchenko}, \citenamefont {Cheng},
  \citenamefont {Gambhir},\ and\ \citenamefont {Brower}}]{Clark:2016rdz}%
  \BibitemOpen
  \bibfield  {author} {\bibinfo {author} {\bibfnamefont {M.}~\bibnamefont
  {Clark}}, \bibinfo {author} {\bibfnamefont {B.}~\bibnamefont {Jo\'o}},
  \bibinfo {author} {\bibfnamefont {A.}~\bibnamefont {Strelchenko}}, \bibinfo
  {author} {\bibfnamefont {M.}~\bibnamefont {Cheng}}, \bibinfo {author}
  {\bibfnamefont {A.}~\bibnamefont {Gambhir}}, \ and\ \bibinfo {author}
  {\bibfnamefont {R.}~\bibnamefont {Brower}},\ }\href@noop {} {\  (\bibinfo
  {year} {2016})},\ \Eprint {http://arxiv.org/abs/1612.07873} {arXiv:1612.07873
  [hep-lat]} \BibitemShut {NoStop}%
\bibitem [{\citenamefont {Gusken}\ \emph {et~al.}(1989)\citenamefont {Gusken},
  \citenamefont {Low}, \citenamefont {Mutter}, \citenamefont {Sommer},
  \citenamefont {Patel},\ and\ \citenamefont {Schilling}}]{Gusken:1989ad}%
  \BibitemOpen
  \bibfield  {author} {\bibinfo {author} {\bibfnamefont {S.}~\bibnamefont
  {Gusken}}, \bibinfo {author} {\bibfnamefont {U.}~\bibnamefont {Low}},
  \bibinfo {author} {\bibfnamefont {K.}~\bibnamefont {Mutter}}, \bibinfo
  {author} {\bibfnamefont {R.}~\bibnamefont {Sommer}}, \bibinfo {author}
  {\bibfnamefont {A.}~\bibnamefont {Patel}}, \ and\ \bibinfo {author}
  {\bibfnamefont {K.}~\bibnamefont {Schilling}},\ }\href {\doibase
  10.1016/S0370-2693(89)80034-6} {\bibfield  {journal} {\bibinfo  {journal}
  {Phys. Lett. B}\ }\textbf {\bibinfo {volume} {227}},\ \bibinfo {pages} {266}
  (\bibinfo {year} {1989})}\BibitemShut {NoStop}%
\bibitem [{\citenamefont {Shintani}\ \emph {et~al.}(2015)\citenamefont
  {Shintani}, \citenamefont {Arthur}, \citenamefont {Blum}, \citenamefont
  {Izubuchi}, \citenamefont {Jung},\ and\ \citenamefont
  {Lehner}}]{Shintani:2014vja}%
  \BibitemOpen
  \bibfield  {author} {\bibinfo {author} {\bibfnamefont {E.}~\bibnamefont
  {Shintani}}, \bibinfo {author} {\bibfnamefont {R.}~\bibnamefont {Arthur}},
  \bibinfo {author} {\bibfnamefont {T.}~\bibnamefont {Blum}}, \bibinfo {author}
  {\bibfnamefont {T.}~\bibnamefont {Izubuchi}}, \bibinfo {author}
  {\bibfnamefont {C.}~\bibnamefont {Jung}}, \ and\ \bibinfo {author}
  {\bibfnamefont {C.}~\bibnamefont {Lehner}},\ }\href {\doibase
  10.1103/PhysRevD.91.114511} {\bibfield  {journal} {\bibinfo  {journal} {Phys.
  Rev. D}\ }\textbf {\bibinfo {volume} {91}},\ \bibinfo {pages} {114511}
  (\bibinfo {year} {2015})},\ \Eprint {http://arxiv.org/abs/1402.0244}
  {arXiv:1402.0244 [hep-lat]} \BibitemShut {NoStop}%
\bibitem [{\citenamefont {Fan}\ \emph {et~al.}(2020)\citenamefont {Fan},
  \citenamefont {Gao}, \citenamefont {Li}, \citenamefont {Lin}, \citenamefont
  {Karthik}, \citenamefont {Mukherjee}, \citenamefont {Petreczky},
  \citenamefont {Syritsyn}, \citenamefont {Yang},\ and\ \citenamefont
  {Zhang}}]{Fan:2020nzz}%
  \BibitemOpen
  \bibfield  {author} {\bibinfo {author} {\bibfnamefont {Z.}~\bibnamefont
  {Fan}}, \bibinfo {author} {\bibfnamefont {X.}~\bibnamefont {Gao}}, \bibinfo
  {author} {\bibfnamefont {R.}~\bibnamefont {Li}}, \bibinfo {author}
  {\bibfnamefont {H.-W.}\ \bibnamefont {Lin}}, \bibinfo {author} {\bibfnamefont
  {N.}~\bibnamefont {Karthik}}, \bibinfo {author} {\bibfnamefont
  {S.}~\bibnamefont {Mukherjee}}, \bibinfo {author} {\bibfnamefont
  {P.}~\bibnamefont {Petreczky}}, \bibinfo {author} {\bibfnamefont
  {S.}~\bibnamefont {Syritsyn}}, \bibinfo {author} {\bibfnamefont {Y.-B.}\
  \bibnamefont {Yang}}, \ and\ \bibinfo {author} {\bibfnamefont
  {R.}~\bibnamefont {Zhang}},\ }\href@noop {} {\  (\bibinfo {year} {2020})},\
  \Eprint {http://arxiv.org/abs/2005.12015} {arXiv:2005.12015 [hep-lat]}
  \BibitemShut {NoStop}%
\bibitem [{\citenamefont {Lepage}\ \emph {et~al.}(2002)\citenamefont {Lepage},
  \citenamefont {Clark}, \citenamefont {Davies}, \citenamefont {Hornbostel},
  \citenamefont {Mackenzie}, \citenamefont {Morningstar},\ and\ \citenamefont
  {Trottier}}]{Lepage:2001ym}%
  \BibitemOpen
  \bibfield  {author} {\bibinfo {author} {\bibfnamefont {G.}~\bibnamefont
  {Lepage}}, \bibinfo {author} {\bibfnamefont {B.}~\bibnamefont {Clark}},
  \bibinfo {author} {\bibfnamefont {C.}~\bibnamefont {Davies}}, \bibinfo
  {author} {\bibfnamefont {K.}~\bibnamefont {Hornbostel}}, \bibinfo {author}
  {\bibfnamefont {P.}~\bibnamefont {Mackenzie}}, \bibinfo {author}
  {\bibfnamefont {C.}~\bibnamefont {Morningstar}}, \ and\ \bibinfo {author}
  {\bibfnamefont {H.}~\bibnamefont {Trottier}},\ }\href {\doibase
  10.1016/S0920-5632(01)01638-3} {\bibfield  {journal} {\bibinfo  {journal}
  {Nucl. Phys. B Proc. Suppl.}\ }\textbf {\bibinfo {volume} {106}},\ \bibinfo
  {pages} {12} (\bibinfo {year} {2002})},\ \Eprint
  {http://arxiv.org/abs/hep-lat/0110175} {arXiv:hep-lat/0110175} \BibitemShut
  {NoStop}%
\bibitem [{\citenamefont {Tanabashi}\ \emph {et~al.}(2018)\citenamefont
  {Tanabashi} \emph {et~al.}}]{Tanabashi:2018oca}%
  \BibitemOpen
  \bibfield  {author} {\bibinfo {author} {\bibfnamefont {M.}~\bibnamefont
  {Tanabashi}} \emph {et~al.} (\bibinfo {collaboration} {Particle Data
  Group}),\ }\href {\doibase 10.1103/PhysRevD.98.030001} {\bibfield  {journal}
  {\bibinfo  {journal} {Phys. Rev. D}\ }\textbf {\bibinfo {volume} {98}},\
  \bibinfo {pages} {030001} (\bibinfo {year} {2018})}\BibitemShut {NoStop}%
\bibitem [{\citenamefont {Maiani}\ \emph {et~al.}(1987)\citenamefont {Maiani},
  \citenamefont {Martinelli}, \citenamefont {Paciello},\ and\ \citenamefont
  {Taglienti}}]{Maiani:1987by}%
  \BibitemOpen
  \bibfield  {author} {\bibinfo {author} {\bibfnamefont {L.}~\bibnamefont
  {Maiani}}, \bibinfo {author} {\bibfnamefont {G.}~\bibnamefont {Martinelli}},
  \bibinfo {author} {\bibfnamefont {M.}~\bibnamefont {Paciello}}, \ and\
  \bibinfo {author} {\bibfnamefont {B.}~\bibnamefont {Taglienti}},\ }\href
  {\doibase 10.1016/0550-3213(87)90078-2} {\bibfield  {journal} {\bibinfo
  {journal} {Nucl. Phys. B}\ }\textbf {\bibinfo {volume} {293}},\ \bibinfo
  {pages} {420} (\bibinfo {year} {1987})}\BibitemShut {NoStop}%
\bibitem [{\citenamefont {Bhattacharya}\ \emph {et~al.}(2015)\citenamefont
  {Bhattacharya}, \citenamefont {Cirigliano}, \citenamefont {Cohen},
  \citenamefont {Gupta}, \citenamefont {Joseph}, \citenamefont {Lin},\ and\
  \citenamefont {Yoon}}]{Bhattacharya:2015wna}%
  \BibitemOpen
  \bibfield  {author} {\bibinfo {author} {\bibfnamefont {T.}~\bibnamefont
  {Bhattacharya}}, \bibinfo {author} {\bibfnamefont {V.}~\bibnamefont
  {Cirigliano}}, \bibinfo {author} {\bibfnamefont {S.}~\bibnamefont {Cohen}},
  \bibinfo {author} {\bibfnamefont {R.}~\bibnamefont {Gupta}}, \bibinfo
  {author} {\bibfnamefont {A.}~\bibnamefont {Joseph}}, \bibinfo {author}
  {\bibfnamefont {H.-W.}\ \bibnamefont {Lin}}, \ and\ \bibinfo {author}
  {\bibfnamefont {B.}~\bibnamefont {Yoon}} (\bibinfo {collaboration} {PNDME}),\
  }\href {\doibase 10.1103/PhysRevD.92.094511} {\bibfield  {journal} {\bibinfo
  {journal} {Phys. Rev. D}\ }\textbf {\bibinfo {volume} {92}},\ \bibinfo
  {pages} {094511} (\bibinfo {year} {2015})},\ \Eprint
  {http://arxiv.org/abs/1506.06411} {arXiv:1506.06411 [hep-lat]} \BibitemShut
  {NoStop}%
\bibitem [{\citenamefont {Bazavov}\ \emph {et~al.}(2018)\citenamefont
  {Bazavov}, \citenamefont {Brambilla}, \citenamefont {Petreczky},
  \citenamefont {Vairo},\ and\ \citenamefont {Weber}}]{Bazavov:2018wmo}%
  \BibitemOpen
  \bibfield  {author} {\bibinfo {author} {\bibfnamefont {A.}~\bibnamefont
  {Bazavov}}, \bibinfo {author} {\bibfnamefont {N.}~\bibnamefont {Brambilla}},
  \bibinfo {author} {\bibfnamefont {P.}~\bibnamefont {Petreczky}}, \bibinfo
  {author} {\bibfnamefont {A.}~\bibnamefont {Vairo}}, \ and\ \bibinfo {author}
  {\bibfnamefont {J.~H.}\ \bibnamefont {Weber}} (\bibinfo {collaboration}
  {TUMQCD}),\ }\href {\doibase 10.1103/PhysRevD.98.054511} {\bibfield
  {journal} {\bibinfo  {journal} {Phys. Rev. D}\ }\textbf {\bibinfo {volume}
  {98}},\ \bibinfo {pages} {054511} (\bibinfo {year} {2018})},\ \Eprint
  {http://arxiv.org/abs/1804.10600} {arXiv:1804.10600 [hep-lat]} \BibitemShut
  {NoStop}%
\bibitem [{\citenamefont {Alexandrou}\ \emph {et~al.}(2011)\citenamefont
  {Alexandrou}, \citenamefont {Constantinou}, \citenamefont {Korzec},
  \citenamefont {Panagopoulos},\ and\ \citenamefont
  {Stylianou}}]{Alexandrou:2010me}%
  \BibitemOpen
  \bibfield  {author} {\bibinfo {author} {\bibfnamefont {C.}~\bibnamefont
  {Alexandrou}}, \bibinfo {author} {\bibfnamefont {M.}~\bibnamefont
  {Constantinou}}, \bibinfo {author} {\bibfnamefont {T.}~\bibnamefont
  {Korzec}}, \bibinfo {author} {\bibfnamefont {H.}~\bibnamefont
  {Panagopoulos}}, \ and\ \bibinfo {author} {\bibfnamefont {F.}~\bibnamefont
  {Stylianou}},\ }\href {\doibase 10.1103/PhysRevD.83.014503} {\bibfield
  {journal} {\bibinfo  {journal} {Phys. Rev. D}\ }\textbf {\bibinfo {volume}
  {83}},\ \bibinfo {pages} {014503} (\bibinfo {year} {2011})},\ \Eprint
  {http://arxiv.org/abs/1006.1920} {arXiv:1006.1920 [hep-lat]} \BibitemShut
  {NoStop}%
\bibitem [{\citenamefont {Chen}\ \emph {et~al.}(2019)\citenamefont {Chen},
  \citenamefont {Ishikawa}, \citenamefont {Jin}, \citenamefont {Lin},
  \citenamefont {Zhang},\ and\ \citenamefont {Zhao}}]{Chen:2017mie}%
  \BibitemOpen
  \bibfield  {author} {\bibinfo {author} {\bibfnamefont {J.-W.}\ \bibnamefont
  {Chen}}, \bibinfo {author} {\bibfnamefont {T.}~\bibnamefont {Ishikawa}},
  \bibinfo {author} {\bibfnamefont {L.}~\bibnamefont {Jin}}, \bibinfo {author}
  {\bibfnamefont {H.-W.}\ \bibnamefont {Lin}}, \bibinfo {author} {\bibfnamefont
  {J.-H.}\ \bibnamefont {Zhang}}, \ and\ \bibinfo {author} {\bibfnamefont
  {Y.}~\bibnamefont {Zhao}} (\bibinfo {collaboration} {LP3}),\ }\href {\doibase
  10.1088/1674-1137/43/10/103101} {\bibfield  {journal} {\bibinfo  {journal}
  {Chin. Phys. C}\ }\textbf {\bibinfo {volume} {43}},\ \bibinfo {pages}
  {103101} (\bibinfo {year} {2019})},\ \Eprint
  {http://arxiv.org/abs/1710.01089} {arXiv:1710.01089 [hep-lat]} \BibitemShut
  {NoStop}%
\bibitem [{\citenamefont {Chen}\ \emph
  {et~al.}(2016{\natexlab{b}})\citenamefont {Chen}, \citenamefont {Cohen},
  \citenamefont {Ji}, \citenamefont {Lin},\ and\ \citenamefont
  {Zhang}}]{Chen:2016utp}%
  \BibitemOpen
  \bibfield  {author} {\bibinfo {author} {\bibfnamefont {J.-W.}\ \bibnamefont
  {Chen}}, \bibinfo {author} {\bibfnamefont {S.~D.}\ \bibnamefont {Cohen}},
  \bibinfo {author} {\bibfnamefont {X.}~\bibnamefont {Ji}}, \bibinfo {author}
  {\bibfnamefont {H.-W.}\ \bibnamefont {Lin}}, \ and\ \bibinfo {author}
  {\bibfnamefont {J.-H.}\ \bibnamefont {Zhang}},\ }\href {\doibase
  10.1016/j.nuclphysb.2016.07.033} {\bibfield  {journal} {\bibinfo  {journal}
  {Nucl. Phys. B}\ }\textbf {\bibinfo {volume} {911}},\ \bibinfo {pages} {246}
  (\bibinfo {year} {2016}{\natexlab{b}})},\ \Eprint
  {http://arxiv.org/abs/1603.06664} {arXiv:1603.06664 [hep-ph]} \BibitemShut
  {NoStop}%
\bibitem [{\citenamefont
  {Radyushkin}(2017{\natexlab{b}})}]{Radyushkin:2017ffo}%
  \BibitemOpen
  \bibfield  {author} {\bibinfo {author} {\bibfnamefont {A.}~\bibnamefont
  {Radyushkin}},\ }\href {\doibase 10.1016/j.physletb.2017.05.024} {\bibfield
  {journal} {\bibinfo  {journal} {Phys. Lett. B}\ }\textbf {\bibinfo {volume}
  {770}},\ \bibinfo {pages} {514} (\bibinfo {year} {2017}{\natexlab{b}})},\
  \Eprint {http://arxiv.org/abs/1702.01726} {arXiv:1702.01726 [hep-ph]}
  \BibitemShut {NoStop}%
\bibitem [{\citenamefont {Martinelli}(1999)}]{Martinelli:1998hz}%
  \BibitemOpen
  \bibfield  {author} {\bibinfo {author} {\bibfnamefont {G.}~\bibnamefont
  {Martinelli}},\ }\href {\doibase 10.1016/S0920-5632(99)85007-5} {\bibfield
  {journal} {\bibinfo  {journal} {Nucl. Phys. B Proc. Suppl.}\ }\textbf
  {\bibinfo {volume} {73}},\ \bibinfo {pages} {58} (\bibinfo {year} {1999})},\
  \Eprint {http://arxiv.org/abs/hep-lat/9810013} {arXiv:hep-lat/9810013}
  \BibitemShut {NoStop}%
\bibitem [{\citenamefont {Karpie}\ \emph {et~al.}(2018)\citenamefont {Karpie},
  \citenamefont {Orginos},\ and\ \citenamefont
  {Zafeiropoulos}}]{Karpie:2018zaz}%
  \BibitemOpen
  \bibfield  {author} {\bibinfo {author} {\bibfnamefont {J.}~\bibnamefont
  {Karpie}}, \bibinfo {author} {\bibfnamefont {K.}~\bibnamefont {Orginos}}, \
  and\ \bibinfo {author} {\bibfnamefont {S.}~\bibnamefont {Zafeiropoulos}},\
  }\href {\doibase 10.1007/JHEP11(2018)178} {\bibfield  {journal} {\bibinfo
  {journal} {JHEP}\ }\textbf {\bibinfo {volume} {11}},\ \bibinfo {pages} {178}
  (\bibinfo {year} {2018})},\ \Eprint {http://arxiv.org/abs/1807.10933}
  {arXiv:1807.10933 [hep-lat]} \BibitemShut {NoStop}%
\bibitem [{\citenamefont {Jo\'o}\ \emph
  {et~al.}(2019{\natexlab{b}})\citenamefont {Jo\'o}, \citenamefont {Karpie},
  \citenamefont {Orginos}, \citenamefont {Radyushkin}, \citenamefont
  {Richards},\ and\ \citenamefont {Zafeiropoulos}}]{Joo:2019jct}%
  \BibitemOpen
  \bibfield  {author} {\bibinfo {author} {\bibfnamefont {B.}~\bibnamefont
  {Jo\'o}}, \bibinfo {author} {\bibfnamefont {J.}~\bibnamefont {Karpie}},
  \bibinfo {author} {\bibfnamefont {K.}~\bibnamefont {Orginos}}, \bibinfo
  {author} {\bibfnamefont {A.}~\bibnamefont {Radyushkin}}, \bibinfo {author}
  {\bibfnamefont {D.}~\bibnamefont {Richards}}, \ and\ \bibinfo {author}
  {\bibfnamefont {S.}~\bibnamefont {Zafeiropoulos}},\ }\href {\doibase
  10.1007/JHEP12(2019)081} {\bibfield  {journal} {\bibinfo  {journal} {JHEP}\
  }\textbf {\bibinfo {volume} {12}},\ \bibinfo {pages} {081} (\bibinfo {year}
  {2019}{\natexlab{b}})},\ \Eprint {http://arxiv.org/abs/1908.09771}
  {arXiv:1908.09771 [hep-lat]} \BibitemShut {NoStop}%
\bibitem [{\citenamefont {Jo\'o}\ \emph {et~al.}(2020)\citenamefont {Jo\'o},
  \citenamefont {Karpie}, \citenamefont {Orginos}, \citenamefont {Radyushkin},
  \citenamefont {Richards},\ and\ \citenamefont {Zafeiropoulos}}]{Joo:2020spy}%
  \BibitemOpen
  \bibfield  {author} {\bibinfo {author} {\bibfnamefont {B.}~\bibnamefont
  {Jo\'o}}, \bibinfo {author} {\bibfnamefont {J.}~\bibnamefont {Karpie}},
  \bibinfo {author} {\bibfnamefont {K.}~\bibnamefont {Orginos}}, \bibinfo
  {author} {\bibfnamefont {A.~V.}\ \bibnamefont {Radyushkin}}, \bibinfo
  {author} {\bibfnamefont {D.~G.}\ \bibnamefont {Richards}}, \ and\ \bibinfo
  {author} {\bibfnamefont {S.}~\bibnamefont {Zafeiropoulos}},\ }\href@noop {}
  {\  (\bibinfo {year} {2020})},\ \Eprint {http://arxiv.org/abs/2004.01687}
  {arXiv:2004.01687 [hep-lat]} \BibitemShut {NoStop}%
\bibitem [{\citenamefont {Barry}\ \emph {et~al.}(2018)\citenamefont {Barry},
  \citenamefont {Sato}, \citenamefont {Melnitchouk},\ and\ \citenamefont
  {Ji}}]{Barry:2018ort}%
  \BibitemOpen
  \bibfield  {author} {\bibinfo {author} {\bibfnamefont {P.}~\bibnamefont
  {Barry}}, \bibinfo {author} {\bibfnamefont {N.}~\bibnamefont {Sato}},
  \bibinfo {author} {\bibfnamefont {W.}~\bibnamefont {Melnitchouk}}, \ and\
  \bibinfo {author} {\bibfnamefont {C.-R.}\ \bibnamefont {Ji}},\ }\href
  {\doibase 10.1103/PhysRevLett.121.152001} {\bibfield  {journal} {\bibinfo
  {journal} {Phys. Rev. Lett.}\ }\textbf {\bibinfo {volume} {121}},\ \bibinfo
  {pages} {152001} (\bibinfo {year} {2018})},\ \Eprint
  {http://arxiv.org/abs/1804.01965} {arXiv:1804.01965 [hep-ph]} \BibitemShut
  {NoStop}%
\bibitem [{\citenamefont {Alexandrou}\ \emph {et~al.}(2019)\citenamefont
  {Alexandrou}, \citenamefont {Cichy}, \citenamefont {Constantinou},
  \citenamefont {Hadjiyiannakou}, \citenamefont {Jansen}, \citenamefont
  {Scapellato},\ and\ \citenamefont {Steffens}}]{Alexandrou:2019lfo}%
  \BibitemOpen
  \bibfield  {author} {\bibinfo {author} {\bibfnamefont {C.}~\bibnamefont
  {Alexandrou}}, \bibinfo {author} {\bibfnamefont {K.}~\bibnamefont {Cichy}},
  \bibinfo {author} {\bibfnamefont {M.}~\bibnamefont {Constantinou}}, \bibinfo
  {author} {\bibfnamefont {K.}~\bibnamefont {Hadjiyiannakou}}, \bibinfo
  {author} {\bibfnamefont {K.}~\bibnamefont {Jansen}}, \bibinfo {author}
  {\bibfnamefont {A.}~\bibnamefont {Scapellato}}, \ and\ \bibinfo {author}
  {\bibfnamefont {F.}~\bibnamefont {Steffens}},\ }\href {\doibase
  10.1103/PhysRevD.99.114504} {\bibfield  {journal} {\bibinfo  {journal} {Phys.
  Rev. D}\ }\textbf {\bibinfo {volume} {99}},\ \bibinfo {pages} {114504}
  (\bibinfo {year} {2019})},\ \Eprint {http://arxiv.org/abs/1902.00587}
  {arXiv:1902.00587 [hep-lat]} \BibitemShut {NoStop}%
\bibitem [{\citenamefont {Karpie}\ \emph {et~al.}(2019)\citenamefont {Karpie},
  \citenamefont {Orginos}, \citenamefont {Rothkopf},\ and\ \citenamefont
  {Zafeiropoulos}}]{Karpie:2019eiq}%
  \BibitemOpen
  \bibfield  {author} {\bibinfo {author} {\bibfnamefont {J.}~\bibnamefont
  {Karpie}}, \bibinfo {author} {\bibfnamefont {K.}~\bibnamefont {Orginos}},
  \bibinfo {author} {\bibfnamefont {A.}~\bibnamefont {Rothkopf}}, \ and\
  \bibinfo {author} {\bibfnamefont {S.}~\bibnamefont {Zafeiropoulos}},\ }\href
  {\doibase 10.1007/JHEP04(2019)057} {\bibfield  {journal} {\bibinfo  {journal}
  {JHEP}\ }\textbf {\bibinfo {volume} {04}},\ \bibinfo {pages} {057} (\bibinfo
  {year} {2019})},\ \Eprint {http://arxiv.org/abs/1901.05408} {arXiv:1901.05408
  [hep-lat]} \BibitemShut {NoStop}%
\bibitem [{\citenamefont {Bhat}\ \emph {et~al.}(2020)\citenamefont {Bhat},
  \citenamefont {Cichy}, \citenamefont {Constantinou},\ and\ \citenamefont
  {Scapellato}}]{Bhat:2020ktg}%
  \BibitemOpen
  \bibfield  {author} {\bibinfo {author} {\bibfnamefont {M.}~\bibnamefont
  {Bhat}}, \bibinfo {author} {\bibfnamefont {K.}~\bibnamefont {Cichy}},
  \bibinfo {author} {\bibfnamefont {M.}~\bibnamefont {Constantinou}}, \ and\
  \bibinfo {author} {\bibfnamefont {A.}~\bibnamefont {Scapellato}},\
  }\href@noop {} {\  (\bibinfo {year} {2020})},\ \Eprint
  {http://arxiv.org/abs/2005.02102} {arXiv:2005.02102 [hep-lat]} \BibitemShut
  {NoStop}%
\bibitem [{\citenamefont {Brodsky}\ and\ \citenamefont
  {Farrar}(1973)}]{Brodsky:1973kr}%
  \BibitemOpen
  \bibfield  {author} {\bibinfo {author} {\bibfnamefont {S.~J.}\ \bibnamefont
  {Brodsky}}\ and\ \bibinfo {author} {\bibfnamefont {G.~R.}\ \bibnamefont
  {Farrar}},\ }\href {\doibase 10.1103/PhysRevLett.31.1153} {\bibfield
  {journal} {\bibinfo  {journal} {Phys. Rev. Lett.}\ }\textbf {\bibinfo
  {volume} {31}},\ \bibinfo {pages} {1153} (\bibinfo {year}
  {1973})}\BibitemShut {NoStop}%
\bibitem [{\citenamefont {Novikov}\ \emph {et~al.}(2020)\citenamefont {Novikov}
  \emph {et~al.}}]{Novikov:2020snp}%
  \BibitemOpen
  \bibfield  {author} {\bibinfo {author} {\bibfnamefont {I.}~\bibnamefont
  {Novikov}} \emph {et~al.},\ }\href {\doibase 10.1103/PhysRevD.102.014040}
  {\bibfield  {journal} {\bibinfo  {journal} {Phys. Rev. D}\ }\textbf {\bibinfo
  {volume} {102}},\ \bibinfo {pages} {014040} (\bibinfo {year} {2020})},\
  \Eprint {http://arxiv.org/abs/2002.02902} {arXiv:2002.02902 [hep-ph]}
  \BibitemShut {NoStop}%
\bibitem [{\citenamefont {Lan}\ \emph {et~al.}(2020)\citenamefont {Lan},
  \citenamefont {Mondal}, \citenamefont {Jia}, \citenamefont {Zhao},\ and\
  \citenamefont {Vary}}]{Lan:2019rba}%
  \BibitemOpen
  \bibfield  {author} {\bibinfo {author} {\bibfnamefont {J.}~\bibnamefont
  {Lan}}, \bibinfo {author} {\bibfnamefont {C.}~\bibnamefont {Mondal}},
  \bibinfo {author} {\bibfnamefont {S.}~\bibnamefont {Jia}}, \bibinfo {author}
  {\bibfnamefont {X.}~\bibnamefont {Zhao}}, \ and\ \bibinfo {author}
  {\bibfnamefont {J.~P.}\ \bibnamefont {Vary}},\ }\href {\doibase
  10.1103/PhysRevD.101.034024} {\bibfield  {journal} {\bibinfo  {journal}
  {Phys. Rev. D}\ }\textbf {\bibinfo {volume} {101}},\ \bibinfo {pages}
  {034024} (\bibinfo {year} {2020})},\ \Eprint
  {http://arxiv.org/abs/1907.01509} {arXiv:1907.01509 [nucl-th]} \BibitemShut
  {NoStop}%
\bibitem [{\citenamefont {Broniowski}\ and\ \citenamefont
  {Ruiz~Arriola}(2020)}]{Broniowski:2020had}%
  \BibitemOpen
  \bibfield  {author} {\bibinfo {author} {\bibfnamefont {W.}~\bibnamefont
  {Broniowski}}\ and\ \bibinfo {author} {\bibfnamefont {E.}~\bibnamefont
  {Ruiz~Arriola}},\ }\href@noop {} {\  (\bibinfo {year} {2020})},\ \Eprint
  {http://arxiv.org/abs/2006.03832} {arXiv:2006.03832 [hep-ph]} \BibitemShut
  {NoStop}%
\bibitem [{\citenamefont {Ruiz~Arriola}\ and\ \citenamefont
  {Broniowski}(2004)}]{RuizArriola:2004ui}%
  \BibitemOpen
  \bibfield  {author} {\bibinfo {author} {\bibfnamefont {E.}~\bibnamefont
  {Ruiz~Arriola}}\ and\ \bibinfo {author} {\bibfnamefont {W.}~\bibnamefont
  {Broniowski}},\ }\href {\doibase 10.1103/PhysRevD.70.034012} {\bibfield
  {journal} {\bibinfo  {journal} {Phys. Rev. D}\ }\textbf {\bibinfo {volume}
  {70}},\ \bibinfo {pages} {034012} (\bibinfo {year} {2004})},\ \Eprint
  {http://arxiv.org/abs/hep-ph/0404008} {arXiv:hep-ph/0404008} \BibitemShut
  {NoStop}%
\bibitem [{\citenamefont {Cui}\ \emph {et~al.}(2020)\citenamefont {Cui},
  \citenamefont {Ding}, \citenamefont {Gao}, \citenamefont {Raya},
  \citenamefont {Binosi}, \citenamefont {Chang}, \citenamefont {Roberts},
  \citenamefont {Rodr\'\i{}guez-Quintero},\ and\ \citenamefont
  {Schmidt}}]{Cui:2020dlm}%
  \BibitemOpen
  \bibfield  {author} {\bibinfo {author} {\bibfnamefont {Z.-F.}\ \bibnamefont
  {Cui}}, \bibinfo {author} {\bibfnamefont {M.}~\bibnamefont {Ding}}, \bibinfo
  {author} {\bibfnamefont {F.}~\bibnamefont {Gao}}, \bibinfo {author}
  {\bibfnamefont {K.}~\bibnamefont {Raya}}, \bibinfo {author} {\bibfnamefont
  {D.}~\bibnamefont {Binosi}}, \bibinfo {author} {\bibfnamefont
  {L.}~\bibnamefont {Chang}}, \bibinfo {author} {\bibfnamefont {C.~D.}\
  \bibnamefont {Roberts}}, \bibinfo {author} {\bibfnamefont {J.}~\bibnamefont
  {Rodr\'\i{}guez-Quintero}}, \ and\ \bibinfo {author} {\bibfnamefont {S.~M.}\
  \bibnamefont {Schmidt}},\ }\href@noop {} {\  (\bibinfo {year} {2020})},\
  \Eprint {http://arxiv.org/abs/2006.14075} {arXiv:2006.14075 [hep-ph]}
  \BibitemShut {NoStop}%
\bibitem [{\citenamefont {Chetyrkin}\ and\ \citenamefont
  {Maier}(2010)}]{Chetyrkin:2009kh}%
  \BibitemOpen
  \bibfield  {author} {\bibinfo {author} {\bibfnamefont {K.}~\bibnamefont
  {Chetyrkin}}\ and\ \bibinfo {author} {\bibfnamefont {A.}~\bibnamefont
  {Maier}},\ }\href {\doibase 10.1007/JHEP01(2010)092} {\bibfield  {journal}
  {\bibinfo  {journal} {JHEP}\ }\textbf {\bibinfo {volume} {01}},\ \bibinfo
  {pages} {092} (\bibinfo {year} {2010})},\ \Eprint
  {http://arxiv.org/abs/0911.0594} {arXiv:0911.0594 [hep-ph]} \BibitemShut
  {NoStop}%
\bibitem [{\citenamefont {Blossier}\ \emph {et~al.}(2010)\citenamefont
  {Blossier}, \citenamefont {Boucaud}, \citenamefont {De~soto}, \citenamefont
  {Morenas}, \citenamefont {Gravina}, \citenamefont {Pene},\ and\ \citenamefont
  {Rodriguez-Quintero}}]{Blossier:2010ky}%
  \BibitemOpen
  \bibfield  {author} {\bibinfo {author} {\bibfnamefont {B.}~\bibnamefont
  {Blossier}}, \bibinfo {author} {\bibfnamefont {P.}~\bibnamefont {Boucaud}},
  \bibinfo {author} {\bibfnamefont {F.}~\bibnamefont {De~soto}}, \bibinfo
  {author} {\bibfnamefont {V.}~\bibnamefont {Morenas}}, \bibinfo {author}
  {\bibfnamefont {M.}~\bibnamefont {Gravina}}, \bibinfo {author} {\bibfnamefont
  {O.}~\bibnamefont {Pene}}, \ and\ \bibinfo {author} {\bibfnamefont
  {J.}~\bibnamefont {Rodriguez-Quintero}} (\bibinfo {collaboration} {ETM}),\
  }\href {\doibase 10.1103/PhysRevD.82.034510} {\bibfield  {journal} {\bibinfo
  {journal} {Phys. Rev. D}\ }\textbf {\bibinfo {volume} {82}},\ \bibinfo
  {pages} {034510} (\bibinfo {year} {2010})},\ \Eprint
  {http://arxiv.org/abs/1005.5290} {arXiv:1005.5290 [hep-lat]} \BibitemShut
  {NoStop}%
\bibitem [{\citenamefont {Constantinou}\ \emph {et~al.}(2009)\citenamefont
  {Constantinou}, \citenamefont {Lubicz}, \citenamefont {Panagopoulos},\ and\
  \citenamefont {Stylianou}}]{Constantinou:2009tr}%
  \BibitemOpen
  \bibfield  {author} {\bibinfo {author} {\bibfnamefont {M.}~\bibnamefont
  {Constantinou}}, \bibinfo {author} {\bibfnamefont {V.}~\bibnamefont
  {Lubicz}}, \bibinfo {author} {\bibfnamefont {H.}~\bibnamefont
  {Panagopoulos}}, \ and\ \bibinfo {author} {\bibfnamefont {F.}~\bibnamefont
  {Stylianou}},\ }\href {\doibase 10.1088/1126-6708/2009/10/064} {\bibfield
  {journal} {\bibinfo  {journal} {JHEP}\ }\textbf {\bibinfo {volume} {10}},\
  \bibinfo {pages} {064} (\bibinfo {year} {2009})},\ \Eprint
  {http://arxiv.org/abs/0907.0381} {arXiv:0907.0381 [hep-lat]} \BibitemShut
  {NoStop}%
\bibitem [{\citenamefont {Constantinou}\ \emph {et~al.}(2010)\citenamefont
  {Constantinou} \emph {et~al.}}]{Constantinou:2010gr}%
  \BibitemOpen
  \bibfield  {author} {\bibinfo {author} {\bibfnamefont {M.}~\bibnamefont
  {Constantinou}} \emph {et~al.} (\bibinfo {collaboration} {ETM}),\ }\href
  {\doibase 10.1007/JHEP08(2010)068} {\bibfield  {journal} {\bibinfo  {journal}
  {JHEP}\ }\textbf {\bibinfo {volume} {08}},\ \bibinfo {pages} {068} (\bibinfo
  {year} {2010})},\ \Eprint {http://arxiv.org/abs/1004.1115} {arXiv:1004.1115
  [hep-lat]} \BibitemShut {NoStop}%
\bibitem [{\citenamefont {Lytle}\ \emph {et~al.}(2018)\citenamefont {Lytle},
  \citenamefont {Davies}, \citenamefont {Hatton}, \citenamefont {Lepage},\ and\
  \citenamefont {Sturm}}]{Lytle:2018evc}%
  \BibitemOpen
  \bibfield  {author} {\bibinfo {author} {\bibfnamefont {A.}~\bibnamefont
  {Lytle}}, \bibinfo {author} {\bibfnamefont {C.}~\bibnamefont {Davies}},
  \bibinfo {author} {\bibfnamefont {D.}~\bibnamefont {Hatton}}, \bibinfo
  {author} {\bibfnamefont {G.}~\bibnamefont {Lepage}}, \ and\ \bibinfo {author}
  {\bibfnamefont {C.}~\bibnamefont {Sturm}} (\bibinfo {collaboration}
  {HPQCD}),\ }\href {\doibase 10.1103/PhysRevD.98.014513} {\bibfield  {journal}
  {\bibinfo  {journal} {Phys. Rev. D}\ }\textbf {\bibinfo {volume} {98}},\
  \bibinfo {pages} {014513} (\bibinfo {year} {2018})},\ \Eprint
  {http://arxiv.org/abs/1805.06225} {arXiv:1805.06225 [hep-lat]} \BibitemShut
  {NoStop}%
\bibitem [{\citenamefont {Ball}\ \emph {et~al.}(2016)\citenamefont {Ball},
  \citenamefont {Nocera},\ and\ \citenamefont {Rojo}}]{Ball:2016spl}%
  \BibitemOpen
  \bibfield  {author} {\bibinfo {author} {\bibfnamefont {R.~D.}\ \bibnamefont
  {Ball}}, \bibinfo {author} {\bibfnamefont {E.~R.}\ \bibnamefont {Nocera}}, \
  and\ \bibinfo {author} {\bibfnamefont {J.}~\bibnamefont {Rojo}},\ }\href
  {\doibase 10.1140/epjc/s10052-016-4240-4} {\bibfield  {journal} {\bibinfo
  {journal} {Eur. Phys. J. C}\ }\textbf {\bibinfo {volume} {76}},\ \bibinfo
  {pages} {383} (\bibinfo {year} {2016})},\ \Eprint
  {http://arxiv.org/abs/1604.00024} {arXiv:1604.00024 [hep-ph]} \BibitemShut
  {NoStop}%
\bibitem [{\citenamefont {Leon}\ \emph {et~al.}(2020)\citenamefont {Leon},
  \citenamefont {Sargsian},\ and\ \citenamefont {Vera}}]{Leon:2020nfb}%
  \BibitemOpen
  \bibfield  {author} {\bibinfo {author} {\bibfnamefont {C.}~\bibnamefont
  {Leon}}, \bibinfo {author} {\bibfnamefont {M.~M.}\ \bibnamefont {Sargsian}},
  \ and\ \bibinfo {author} {\bibfnamefont {F.}~\bibnamefont {Vera}},\
  }\href@noop {} {\  (\bibinfo {year} {2020})},\ \Eprint
  {http://arxiv.org/abs/2003.12902} {arXiv:2003.12902 [hep-ph]} \BibitemShut
  {NoStop}%
\end{thebibliography}
